\documentclass[a4paper,11pt]{article}
\pdfoutput=1 

\usepackage{jinstpub} 
\usepackage{siunitx}
\usepackage{comment}
\usepackage{xspace}
\usepackage{multicol}
\usepackage{multirow}
\usepackage{graphicx}
\usepackage{caption}
\usepackage{subcaption}
\usepackage{url}
\usepackage{mhchem}

\newcommand{\numu}{\ensuremath{\nu_{\mu}}\xspace}
\newcommand{\nutau}{\ensuremath{\nu_{\tau}}\xspace}
\newcommand{\nue}{\ensuremath{\nu_{e}}\xspace}
\newcommand{\anue}{\ensuremath{\bar{\nu}_{e}}\xspace}
\newcommand{\anumu}{\ensuremath{\bar{\nu}_{\mu}}\xspace}
\newcommand{\anutau}{\ensuremath{\bar{\nu}_{\tau}}\xspace}
\newcommand{\dvlem}{\ensuremath{\Delta V_{\mbox{LEM}}}\xspace}

\newcommand{\dcp}{\ensuremath{\delta_{CP}}\xspace}
\newcommand{\um}{\ensuremath{\mu~m} \xspace}

\newcommand{\pddp}{ProtoDUNE-DP\xspace}
\newcommand{\Pddp}{ProtoDUNE-DP\xspace}
\newcommand{\pdsp}{ProtoDUNE-SP\xspace}

\title{Operation and performance of ProtoDUNE Dual Phase liquid argon time projection chamber}

\usepackage{orcidlink}
\collaboration{The DUNE Collaboration}
\affiliation[0]{ Indian Institute of Science, Bengaluru, India, CV Raman Road, Bengaluru, Karnataka, 560012, India}
\affiliation[1]{University of Albany, SUNY, Albany, NY 12222, USA}
\affiliation[2]{Institute of Nuclear Physics at Almaty, Almaty 050032, Kazakhstan
}
\affiliation[3]{University of Amsterdam, NL-1098 XG Amsterdam, The Netherlands}
\affiliation[4]{Antalya Bilim University, 07190 D\"o{\c s}emealtı/Antalya, Turkey}
\affiliation[5]{University of Antananarivo, Antananarivo 101, Madagascar}
\affiliation[6]{University of Antioquia, Medell\'in, Colombia}
\affiliation[7]{Universidad Antonio Nari\~no, Bogot\'a, Colombia}
\affiliation[8]{Argonne National Laboratory, Argonne, IL 60439, USA}
\affiliation[9]{University of Arizona, Tucson, AZ 85721, USA}
\affiliation[10]{Universidad Nacional de Asunci\'on, San Lorenzo, Paraguay}
\affiliation[11]{University of Athens, Zografou GR 157 84, Greece}
\affiliation[12]{Universidad del Atl\'antico, Barranquilla, Atl\'antico, Colombia}
\affiliation[13]{Augustana University, Sioux Falls, SD 57197, USA}
\affiliation[14]{University of Bern, CH-3012 Bern, Switzerland}
\affiliation[15]{Beykent University, Istanbul, Turkey}
\affiliation[16]{University of Birmingham, Birmingham B15 2TT, United Kingdom}
\affiliation[17]{Universit\`a di Bologna, 40127 Bologna, Italy}
\affiliation[18]{Boston University, Boston, MA 02215, USA}
\affiliation[19]{University of Bristol, Bristol BS8 1TL, United Kingdom}
\affiliation[20]{Brookhaven National Laboratory, Upton, NY 11973, USA}
\affiliation[21]{University of Bucharest, Bucharest, Romania}
\affiliation[22]{University of California Berkeley, Berkeley, CA 94720, USA}
\affiliation[23]{University of California Davis, Davis, CA 95616, USA}
\affiliation[24]{University of California Irvine, Irvine, CA 92697, USA}
\affiliation[25]{University of California Riverside, Riverside CA 92521, USA}
\affiliation[26]{University of California Santa Barbara, Santa Barbara, CA 93106, USA}
\affiliation[27]{California Institute of Technology, Pasadena, CA 91125, USA}
\affiliation[28]{University of Cambridge, Cambridge CB3 0HE, United Kingdom}
\affiliation[29]{Universidade Estadual de Campinas, Campinas - SP, 13083-970, Brazil}
\affiliation[30]{Universit\`a di Catania, 2 - 95131 Catania, Italy}
\affiliation[31]{Universidad Cat\'olica del Norte, Antofagasta, Chile}
\affiliation[32]{Centro Brasileiro de Pesquisas F\'isicas, Rio de Janeiro, RJ 22290-180, Brazil}
\affiliation[33]{IRFU, CEA, Universit\'e Paris-Saclay, F-91191 Gif-sur-Yvette, France}
\affiliation[34]{CERN, The European Organization for Nuclear Research, 1211 Meyrin, Switzerland}
\affiliation[35]{Institute of Particle and Nuclear Physics of the Faculty of Mathematics and Physics of the Charles University, 180 00 Prague 8, Czech Republic }
\affiliation[36]{University of Chicago, Chicago, IL 60637, USA}
\affiliation[37]{Chung-Ang University, Seoul 06974, South Korea}
\affiliation[38]{CIEMAT, Centro de Investigaciones Energ\'eticas, Medioambientales y Tecnol\'ogicas, E-28040 Madrid, Spain}
\affiliation[39]{University of Cincinnati, Cincinnati, OH 45221, USA}
\affiliation[40]{Centro de Investigaci\'on y de Estudios Avanzados del Instituto Polit\'ecnico Nacional (Cinvestav), Mexico City, Mexico}
\affiliation[41]{Universidad de Colima, Colima, Mexico}
\affiliation[42]{University of Colorado Boulder, Boulder, CO 80309, USA}
\affiliation[43]{Colorado State University, Fort Collins, CO 80523, USA}
\affiliation[44]{Columbia University, New York, NY 10027, USA}
\affiliation[45]{Comisi\'on Nacional de Investigaci\'on y Desarrollo Aeroespacial, Lima, Peru}
\affiliation[46]{Centro de Tecnologia da Informacao Renato Archer, Amarais - Campinas, SP - CEP 13069-901}
\affiliation[47]{Central University of South Bihar, Gaya, 824236, India
}
\affiliation[48]{Institute of Physics, Czech Academy of Sciences, 182 00 Prague 8, Czech Republic}
\affiliation[49]{Czech Technical University, 115 19 Prague 1, Czech Republic}
\affiliation[50]{Laboratoire d'Annecy de Physique des Particules, Universit\'e Savoie Mont Blanc, CNRS, LAPP-IN2P3, 74000 Annecy, France}
\affiliation[51]{Daresbury Laboratory, Cheshire WA4 4AD, United Kingdom}
\affiliation[52]{Dordt University, Sioux Center, IA 51250, USA}
\affiliation[53]{Drew University, Madison, NJ 07940, USA}
\affiliation[54]{Drexel University, Philadelphia, PA 19104, USA}
\affiliation[55]{Duke University, Durham, NC 27708, USA}
\affiliation[56]{Durham University, Durham DH1 3LE, United Kingdom}
\affiliation[57]{University of Edinburgh, Edinburgh EH8 9YL, United Kingdom}
\affiliation[58]{Universidad EIA, Envigado, Antioquia, Colombia}
\affiliation[59]{E\"otv\"os Lor\'and University, 1053 Budapest, Hungary}
\affiliation[60]{Erciyes University, Kayseri, Turkey}
\affiliation[61]{Faculdade de Ci\^encias da Universidade de Lisboa, 1749-016 Lisboa, Portugal}
\affiliation[62]{Universidade Federal de Alfenas, Po{\c c}os de Caldas - MG, 37715-400, Brazil}
\affiliation[63]{Universidade Federal de Goias, Goiania, GO 74690-900, Brazil}
\affiliation[64]{Universidade Federal do ABC, Santo Andr\'e - SP, 09210-580, Brazil}
\affiliation[65]{Universidade Federal do Rio de Janeiro, Rio de Janeiro - RJ, 21941-901, Brazil}
\affiliation[66]{Fermi National Accelerator Laboratory, Batavia, IL 60510, USA}
\affiliation[67]{University of Ferrara, Ferrara, Italy}
\affiliation[68]{University of Florida, Gainesville, FL 32611-8440, USA}
\affiliation[69]{Florida State University, Tallahassee, FL, 32306 USA}
\affiliation[70]{Fluminense Federal University, 9 Icara\'i Niter\'oi - RJ, 24220-900, Brazil }
\affiliation[71]{Georgia Institute of Technology, North Avenue
Atlanta, GA 30332}
\affiliation[72]{Universit\`a degli Studi di Genova, Genova, Italy}
\affiliation[73]{Georgian Technical University, Tbilisi, Georgia}
\affiliation[74]{University of Granada \& CAFPE, 18002 Granada, Spain}
\affiliation[75]{Gran Sasso Science Institute, L'Aquila, Italy}
\affiliation[76]{Laboratori Nazionali del Gran Sasso, L'Aquila AQ, Italy}
\affiliation[77]{University Grenoble Alpes, CNRS, Grenoble INP, LPSC-IN2P3, 38000 Grenoble, France}
\affiliation[78]{Universidad de Guanajuato, Guanajuato, C.P. 37000, Mexico}
\affiliation[79]{Harish-Chandra Research Institute, Jhunsi, Allahabad 211 019, India}
\affiliation[80]{University of Hawaii, Honolulu, HI 96822, USA}
\affiliation[81]{Hong Kong University of Science and Technology, Kowloon, Hong Kong, China}
\affiliation[82]{University of Houston, Houston, TX 77204, USA}
\affiliation[83]{University of  Hyderabad, Gachibowli, Hyderabad - 500 046, India}
\affiliation[84]{Idaho State University, Pocatello, ID 83209, USA}
\affiliation[85]{Instituto de F\'isica Corpuscular, CSIC and Universitat de Val\`encia, 46980 Paterna, Valencia, Spain}
\affiliation[86]{Instituto Galego de F\'isica de Altas Enerx\'ias, University of Santiago de Compostela, Santiago de Compostela, 15782, Spain}
\affiliation[87]{Institute of High Energy Physics, Chinese Academy of Sciences, Beijing, China}
\affiliation[88]{Indian Institute of Technology Kanpur, Uttar Pradesh 208016, India}
\affiliation[89]{Jozef Stefan Institute, Jamova cesta 39, 1000 Ljubljana, Slovenia}
\affiliation[90]{Illinois Institute of Technology, Chicago, IL 60616, USA}
\affiliation[91]{Imperial College of Science, Technology and Medicine, London SW7 2BZ, United Kingdom}
\affiliation[92]{Indian Institute of Technology Guwahati, Guwahati, 781 039, India}
\affiliation[93]{Indian Institute of Technology Hyderabad, Hyderabad, 502285, India}
\affiliation[94]{Indiana University, Bloomington, IN 47405, USA}
\affiliation[95]{Istituto Nazionale di Fisica Nucleare Sezione di Bologna, 40127 Bologna BO, Italy}
\affiliation[96]{Istituto Nazionale di Fisica Nucleare Sezione di Catania, I-95123 Catania, Italy}
\affiliation[97]{Istituto Nazionale di Fisica Nucleare Sezione di Ferrara, I-44122 Ferrara, Italy}
\affiliation[98]{Istituto Nazionale di Fisica Nucleare Laboratori Nazionali di Frascati, Frascati, Roma, Italy}
\affiliation[99]{Istituto Nazionale di Fisica Nucleare Sezione di Genova, 16146 Genova GE, Italy}
\affiliation[100]{Istituto Nazionale di Fisica Nucleare Sezione di Lecce, 73100 - Lecce, Italy}
\affiliation[101]{Istituto Nazionale di Fisica Nucleare Sezione di Milano Bicocca, 3 - I-20126 Milano, Italy}
\affiliation[102]{Istituto Nazionale di Fisica Nucleare Sezione di Milano, 20133 Milano, Italy}
\affiliation[103]{Istituto Nazionale di Fisica Nucleare Sezione di Napoli, I-80126 Napoli, Italy}
\affiliation[104]{Istituto Nazionale di Fisica Nucleare Sezione di Padova, 35131 Padova, Italy}
\affiliation[105]{Istituto Nazionale di Fisica Nucleare Sezione di Pavia,  I-27100 Pavia, Italy}
\affiliation[106]{Istituto Nazionale di Fisica Nucleare Laboratori Nazionali di Pisa, Pisa PI, Italy}
\affiliation[107]{Istituto Nazionale di Fisica Nucleare Sezione di Roma, 00185 Roma RM, Italy}
\affiliation[108]{Istituto Nazionale di Fisica Nucleare Roma Tor Vergata , 00133 Roma RM, Italy}
\affiliation[109]{Istituto Nazionale di Fisica Nucleare Laboratori Nazionali del Sud, 95123 Catania, Italy}
\affiliation[110]{Istituto Nazionale di Fisica Nucleare, Sezione di Torino, Turin, Italy}
\affiliation[111]{Universidad Nacional de Ingenier\'ia, Lima 25, Per\'u}
\affiliation[112]{University of Insubria, Via Ravasi, 2, 21100 Varese VA, Italy}
\affiliation[113]{University of Iowa, Iowa City, IA 52242, USA}
\affiliation[114]{Iowa State University, Ames, Iowa 50011, USA}
\affiliation[115]{Institut de Physique des 2 Infinis de Lyon, 69622 Villeurbanne, France}
\affiliation[116]{Institute for Research in Fundamental Sciences, Tehran, Iran}
\affiliation[117]{Particle Physics and Cosmology International Research Laboratory	, Chicago IL,  60637 USA}
\affiliation[118]{Instituto Superior T\'ecnico da Universidade de Lisboa, Universidade de Lisboa, 1049-001 Lisboa, Portugal}
\affiliation[119]{Instituto Tecnol\'ogico de Aeron\'autica, Sao Jose dos Campos, Brazil}
\affiliation[120]{Institute for Theoretical Physics and Modeling, Yerevan 0036, Armenia}
\affiliation[121]{Iwate University, Morioka, Iwate 020-8551, Japan}
\affiliation[122]{Jackson State University, Jackson, MS 39217, USA}
\affiliation[123]{Jawaharlal Nehru University, New Delhi 110067, India}
\affiliation[124]{Jeonbuk National University, Jeonrabuk-do 54896, South Korea}
\affiliation[125]{Jyv\"askyl\"a University, FI-40014 Jyv\"askyl\"a, Finland}
\affiliation[126]{Kansas State University, Manhattan, KS 66506, USA}
\affiliation[127]{Kavli Institute for the Physics and Mathematics of the Universe, Kashiwa, Chiba 277-8583, Japan}
\affiliation[128]{High Energy Accelerator Research Organization (KEK), Ibaraki, 305-0801, Japan}
\affiliation[129]{Korea Institute of Science and Technology Information, Daejeon, 34141, South Korea}
\affiliation[130]{Taras Shevchenko National University of Kyiv, 01601 Kyiv, Ukraine}
\affiliation[131]{Lancaster University, Lancaster LA1 4YB, United Kingdom}
\affiliation[132]{Lawrence Berkeley National Laboratory, Berkeley, CA 94720, USA}
\affiliation[133]{Laborat\'orio de Instrumenta{\c c}\~ao e F\'isica Experimental de Part\'iculas, 1649-003 Lisboa, 3004-516 Coimbra, and 4710-057 Braga Portugal}
\affiliation[134]{University of Liverpool, L69 7ZE, Liverpool, United Kingdom}
\affiliation[135]{National Laboratory for Astrophysics, Rua dos Estados Unidos, 154
Bairro das Na{\c c}ões
Itajub\'a / MG - 37.504-364
Brasil}
\affiliation[136]{Los Alamos National Laboratory, Los Alamos, NM 87545, USA}
\affiliation[137]{Louisiana State University, Baton Rouge, LA 70803, USA}
\affiliation[138]{Laboratoire de Physique des Deux Infinis Bordeaux - IN2P3, F-33175 Gradignan, Bordeaux, France, }
\affiliation[139]{University of Lucknow, Uttar Pradesh 226007, India}
\affiliation[140]{Johannes Gutenberg-Universit\"at Mainz, 55122 Mainz, Germany}
\affiliation[141]{University of Manchester, Manchester M13 9PL, United Kingdom}
\affiliation[142]{Marmara University, Marmara Üniversitesi G\"oztepe Yerle{\c s}kesi 34722 Kadık\"oy - İstanbul}
\affiliation[143]{Massachusetts Institute of Technology, Cambridge, MA 02139, USA}
\affiliation[144]{University of Medell\'in, Medell\'in, 050026 Colombia }
\affiliation[145]{University of Michigan, Ann Arbor, MI 48109, USA}
\affiliation[146]{Michigan State University, East Lansing, MI 48824, USA}
\affiliation[147]{Universit\`a di Milano Bicocca , 20126 Milano, Italy}
\affiliation[148]{Universit\`a degli Studi di Milano, I-20133 Milano, Italy}
\affiliation[149]{Universidade do Minho, R. da Universidade, 4710-057 Braga, Portugal}
\affiliation[150]{University of Minnesota Duluth, Duluth, MN 55812, USA}
\affiliation[151]{University of Minnesota Twin Cities, Minneapolis, MN 55455, USA}
\affiliation[152]{University of Mississippi, University, MS 38677 USA}
\affiliation[153]{Universit\`a degli Studi di Napoli Federico II , 80138 Napoli NA, Italy}
\affiliation[154]{Nikhef National Institute of Subatomic Physics, 1098 XG Amsterdam, Netherlands}
\affiliation[155]{National Institute of Science Education and Research, An OCC of Homi Bhabha National Institute, Bhubaneswar, Odisha, India, National Institute of Science Education and Research, An OCC of Homi Bhabha National Institute, Odisha 752050, India}
\affiliation[156]{University of North Dakota, Grand Forks, ND 58202-8357, USA}
\affiliation[157]{Northern Illinois University, DeKalb, IL 60115, USA}
\affiliation[158]{Northwestern University, Evanston, Il 60208, USA}
\affiliation[159]{University of Notre Dame, Notre Dame, IN 46556, USA}
\affiliation[160]{University of Novi Sad, 21102 Novi Sad, Serbia}
\affiliation[161]{Ohio State University, Columbus, OH 43210, USA}
\affiliation[162]{Oregon State University, Corvallis, OR 97331, USA}
\affiliation[163]{University of Oxford, Oxford, OX1 3RH, United Kingdom}
\affiliation[164]{Pacific Northwest National Laboratory, Richland, WA 99352, USA}
\affiliation[165]{Universt\`a degli Studi di Padova, I-35131 Padova, Italy}
\affiliation[166]{Panjab University, Chandigarh, 160014, India}
\affiliation[167]{Universit\'e Paris-Saclay, CNRS/IN2P3, IJCLab, 91405 Orsay, France}
\affiliation[168]{Universit\'e Paris Cit\'e, CNRS, Astroparticule et Cosmologie, Paris, France}
\affiliation[169]{University of Parma,  43121 Parma PR, Italy}
\affiliation[170]{Universit\`a degli Studi di Pavia, 27100 Pavia PV, Italy}
\affiliation[171]{University of Pennsylvania, Philadelphia, PA 19104, USA}
\affiliation[172]{Pennsylvania State University, University Park, PA 16802, USA}
\affiliation[173]{Physical Research Laboratory, Ahmedabad 380 009, India}
\affiliation[174]{Universit\`a di Pisa, I-56127 Pisa, Italy}
\affiliation[175]{University of Pittsburgh, Pittsburgh, PA 15260, USA}
\affiliation[176]{Pontificia Universidad Cat\'olica del Per\'u, Lima, Per\'u}
\affiliation[177]{University of Puerto Rico, Mayaguez 00681, Puerto Rico, USA}
\affiliation[178]{Punjab Agricultural University, Ludhiana 141004, India}
\affiliation[179]{Queen Mary University of London, London E1 4NS, United Kingdom
}
\affiliation[180]{Radboud University, NL-6525 AJ Nijmegen, Netherlands}
\affiliation[181]{Rice University, Houston, TX 77005}
\affiliation[182]{University of Rochester, Rochester, NY 14627, USA}
\affiliation[183]{Royal Holloway College London, London, TW20 0EX, United Kingdom}
\affiliation[184]{Rutgers University, Piscataway, NJ, 08854, USA}
\affiliation[185]{STFC Rutherford Appleton Laboratory, Didcot OX11 0QX, United Kingdom}
\affiliation[186]{Universit\`a del Salento, 73100 Lecce, Italy}
\affiliation[187]{Universidade do Estado de Santa Catarina , Santa Catarina, 89219-710, Brazil}
\affiliation[188]{Universidad del Magdalena, Santa Marta - Colombia}
\affiliation[189]{Sapienza University of Rome, 00185 Roma RM, Italy}
\affiliation[190]{Universidad Sergio Arboleda, 11022 Bogot\'a, Colombia}
\affiliation[191]{University of Sheffield, Sheffield S3 7RH, United Kingdom}
\affiliation[192]{SLAC National Accelerator Laboratory, Menlo Park, CA 94025, USA}
\affiliation[193]{University of South Carolina, Columbia, SC 29208, USA}
\affiliation[194]{South Dakota School of Mines and Technology, Rapid City, SD 57701, USA}
\affiliation[195]{South Dakota State University, Brookings, SD 57007, USA}
\affiliation[196]{Stony Brook University, SUNY, Stony Brook, NY 11794, USA}
\affiliation[197]{Sanford Underground Research Facility, Lead, SD, 57754, USA}
\affiliation[198]{University of Sussex, Brighton, BN1 9RH, United Kingdom}
\affiliation[199]{Syracuse University, Syracuse, NY 13244, USA}
\affiliation[200]{Universidade Tecnol\'ogica Federal do Paran\'a, Curitiba, Brazil}
\affiliation[201]{Tel Aviv University, Tel Aviv-Yafo, Israel}
\affiliation[202]{Texas A\&M University, College Station, Texas 77840}
\affiliation[203]{Texas A\&M University - Corpus Christi, Corpus Christi, TX 78412, USA}
\affiliation[204]{University of Texas at Arlington, Arlington, TX 76019, USA}
\affiliation[205]{University of Texas at Austin, Austin, TX 78712, USA}
\affiliation[206]{University of Toronto, Toronto, Ontario M5S 1A1, Canada}
\affiliation[207]{Tufts University, Medford, MA 02155, USA}
\affiliation[208]{Universidade Federal de S\~ao Paulo, 09913-030, S\~ao Paulo, Brazil}
\affiliation[209]{University College London, London, WC1E 6BT, United Kingdom}
\affiliation[210]{University of Kansas, Lawrence, KS 66045}
\affiliation[211]{Universidad Nacional Mayor de San Marcos, Lima, Peru}
\affiliation[212]{Valley City State University, Valley City, ND 58072, USA}
\affiliation[213]{University of Vigo, E- 36310 Vigo Spain}
\affiliation[214]{Virginia Tech, Blacksburg, VA 24060, USA}
\affiliation[215]{University of Warsaw, 02-093 Warsaw, Poland}
\affiliation[216]{University of Warwick, Coventry CV4 7AL, United Kingdom}
\affiliation[217]{Wellesley College, Wellesley, MA 02481, USA}
\affiliation[218]{Wichita State University, Wichita, KS 67260, USA}
\affiliation[219]{William and Mary, Williamsburg, VA 23187, USA}
\affiliation[220]{University of Wisconsin Madison, Madison, WI 53706, USA}
\affiliation[221]{Yale University, New Haven, CT 06520, USA}
\affiliation[222]{York University, Toronto M3J 1P3, Canada}
\author[116]{S.~Abbaslu,}
\author[82]{F.~Abd Alrahman,}
\author[34]{A.~Abed Abud,}
\author[34]{R.~Acciarri,}
\author[200]{L.~P.~Accorsi,}
\author[12]{M.~A.~Acero,}
\author[200]{M.~R.~Adames,}
\author[73]{G.~Adamov,}
\author[66]{M.~Adamowski,}
\author[131]{K.~Adhikari,}
% ...

\emailAdd{C. Cuesta (clara.cuesta@ciemat.es)}

\abstract{
\Pddp, with a total mass of the active volume of \SI{300}{\tonne}, is the largest Liquid Argon Time Projection Chamber (LArTPC) built to operate in a Dual-Phase (DP) mode where the ionisation charge produced in the liquid bulk by charged particles traversing the detector is transferred for the readout in the gaseous phase above. Intended as a prototype of a far detector module for the DUNE neutrino oscillation experiment, \pddp was operated at the CERN Neutrino Platform from 2019 to 2022. This paper describes various \pddp detector components, discusses the commissioning and operation experiences, and presents the results on the detector performance from an analysis of collected cosmic ray data.  
Some key deliverables of the detector were achieved, including stable delivery of \SI{300}{\kilo\volt} to the TPC cathode, operation of the charge readout with an effective gain exceeding 5 before charging-up, stable functioning of the photon detection system with a relative time accuracy among PMTs below \SI{16}{\nano\second}. The long term operation of the charge readout system in dual-phase, however, was limited by the liquid surface and electrical instabilities. 
The ProtoDUNE-DP experience contributed to the final design of the DUNE Vertical Drift LArTPC.

}
\begin{document}
\maketitle
\flushbottom



\section{Introduction}
\label{sec:intro}
This paper describes the Dual-Phase ProtoDUNE detector, \pddp, and its performance using cosmic-ray data collected at the CERN Neutrino Platform. The manuscript is organised as follows. After a general introduction on DUNE and the ProtoDUNE programme, the detector components are described in section~\ref{sec:detoverview} with particular emphasis on those needed for the operation of the TPC in dual-phase. In section~\ref{sec:operation} the observations and general performance is described. The detector collected cosmic-ray data and the results of the analysis showing the TPC response are presented in section~\ref{sec:tpcresp}, while performance specific to the photon detection system is discussed in section~\ref{sec:photon}. The summary of the findings are given in section~\ref{sec:conclusion} highlighting the principal outcomes.

\subsection {The DUNE Experiment}
The Deep Underground Neutrino Experiment (DUNE) is a next generation neutrino experiment~\cite{DUNE:2020lwj} hosted by the U.S. Department of Energy's Fermilab, Illinois, U.S.  At Fermilab, a high power wide-energy band beam operating in neutrino (anti-neutrino) mode will be produced and its flux and flavour characterised by the near detector complex. At a baseline of 1,300\,km and 1.5\,km underground at the Sanford Underground Research Facility (SURF, South Dakota, U.S.), four large far detector modules will measure \numu (\anumu) disappearance, \nue (\anue) and \nutau (\anutau) appearance.  The experiment is conceived to be highly sensitive to the matter effect with its long baseline. At this distance, DUNE will be able to unambiguously determine both the neutrino mass ordering and the CP violating phase, $\dcp$, as well as measure precisely the oscillation parameters and test the 3-flavour paradigm~\cite{DUNE_LBL}. With large detectors deep underground, DUNE will also search for Beyond the Standard Model Physics and will be prepared to detect low energy neutrinos from a Galactic core-collapse supernova and the sun~\cite{DUNE_BSM, DUNE_SNB}.

The DUNE Far Detectors will be liquid argon (LAr) time projection chambers (TPCs) since this technology permits the construction of large detectors of dimensions $\sim$60~m long, 12\,m high and 12\,m wide, to achieve a total active mass of 40~kt with four detectors each with excellent calorimetric and spatial resolving power. The DUNE TPCs are conceived to make fine-grained (with resolutions of a few mm) reconstruction of the ionisation tracks from the products of neutrino interactions in the LAr. 

Two different technologies were originally proposed for the read-out of the drifted ionisation charge, which can be measured either directly in the liquid (Single Phase detector with horizontal drift~\cite{SP_TDR}) or after extraction and multiplication in a layer of gaseous argon (Dual Phase detector with vertical drift). Two large-scale prototypes, known as ProtoDUNEs, were constructed at the CERN neutrino platform in order to test these two technologies at the kiloton scale. The Single Phase technology was successfully demonstrated with \pdsp~\cite{Abi_2020,abud:hal-03519598}, validating the design which will be used for the horizontal drift Far Detector module. Unfortunately, the Dual Phase technology demonstrator, \pddp, suffered major technical problems such that the DUNE collaboration decided to no longer pursue this design for a Far Detector module. 

Based on the experience gained from the ProtoDUNEs, a revised TPC design for one DUNE Far Detector module was successfully developed. The new design, including successfully demonstrated design features from both Single and Dual Phase prototypes, is known as Vertical Drift~\cite{VD_TDR}. This new detector design relies on the successful validation of a High Voltage (HV) system capable of operation at \SI{-300}{\kilo\volt} which was first made in \pddp and is described in this paper. One of the principal aims of this paper is to report the experience from running the largest dual-phase TPC ever constructed.

\subsection{The dual-phase liquid argon TPC} 
\label{sec:DualPhase}
\begin{figure}
\begin{centering} 
\includegraphics[ width=7cm]{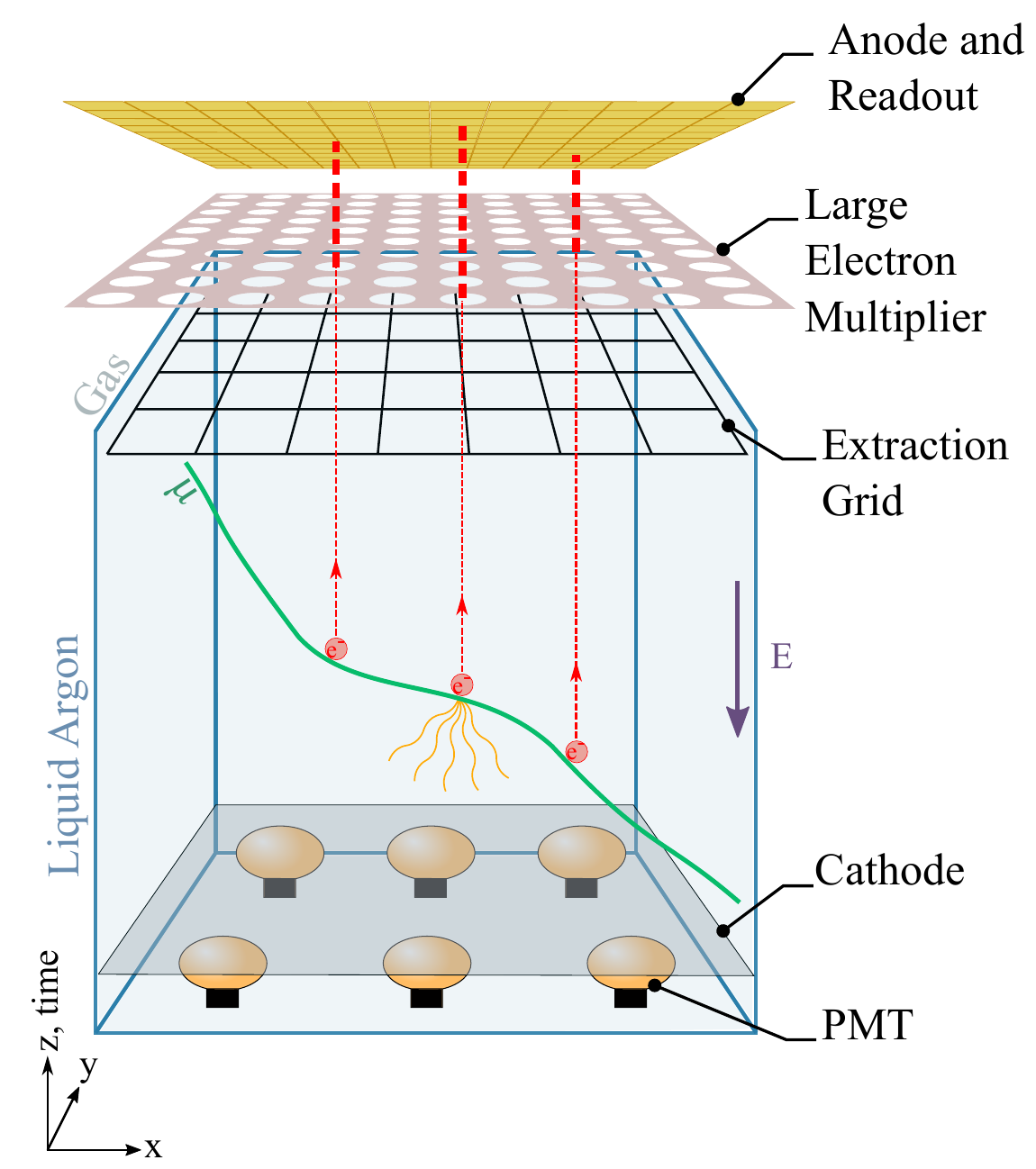}
\caption{\label{fig:twophasediag} Schematic of the interactions produced by charged particles crossing the Dual-Phase LArTPC.  Scintillation light is detected by photomultiplier tubes at the bottom, and the charge tracks are imaged at the top by the LEM/Anode structure.}
\end{centering}
\end{figure}

\begin{figure}
\begin{centering}
\includegraphics[width=0.9\textwidth]{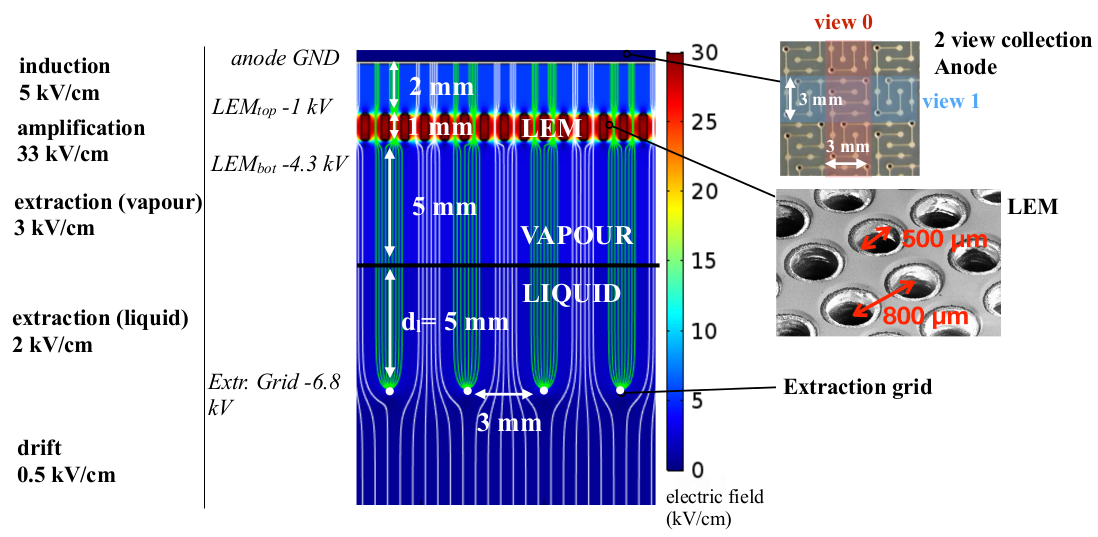}
\caption{\label{fig:LEM} Schematic of the applied potentials and resulting electric field strengths~\cite{311_technical}. Field lines (shown in white) indicate the electron drift paths. Also shown are zoomed images of the LEM (bottom right) and Anode (top right).}
\end{centering}
\end{figure}

The Dual Phase technology uses LAr as a detection medium covered with a thin layer of gaseous argon under an applied electric field, as shown in figure~\ref{fig:twophasediag}. 
As charged particles interact with LAr, pairs of positively charged argon ions (Ar$^{+}$) and free electrons are created. Also excited argon atoms (Ar$^{\ast}$) are produced, that rapidly couple to other argon atoms, producing excimers. The excimers then decay emitting scintillation VUV (Vacuum Ultra Violet) photons. Argon ions also couple to neutral argon atoms, producing the molecular state Ar$^{+}_{2}$. In the absence of a drift field, Ar$^{+}_{2}$ eventually recombines with an electron producing Ar$^{\ast}_{2}$, which decays producing more VUV photons. This recombination is suppressed by the drift field. The VUV photons emitted as scintillation can be detected by the Photo-Multiplier Tubes (PMT) which line the bottom of the detector. The arrival time of the scintillation light signal provides the particle interaction time.

The drift field causes ionisation electrons to move upwards towards the liquid surface. Measurements of electron drift velocity in liquid argon have been parameterised as a function of liquid temperature (87$<$T$<$94 K) and electric field strength (0.5$\le$E$\le$\SI{12.6}{\kilo\volt\per\cm}) \cite{WALKOWIAK2000288}. The parameterisation has been extended to lower electric field strengths (0.1$\le$E$\le$\SI{0.5}{\kilo\volt\per\cm}) at a fixed temperature of \SI{89}{\kelvin} \cite{AMORUSO200468} with electron drift velocity in \SI{0.5}{\kilo\volt\per\cm} and at \SI{87}{\kelvin} predicted to be around \SI{1.6}{\mm\per\micro\second}.

The readout of the charge signal is made in the gas phase with the Charge Readout Plane (CRP) which consists of an extraction grid and Charge Readout Detectors (CRD). Each CRD consists of a Large Electron Multiplier (LEM) and anode. Figure~\ref{fig:LEM} provides the schematic illustration of CRP indicating the reference distances between different elements and the design target values of electric field strengths in various regions. 

The ionisation electrons arriving at the surface are transferred into the gas phase with the help of an extraction field formed by an applied potential difference between the bottom electrode of the LEM and the extraction grid (figure~\ref{fig:LEM}). The extraction field strength in the liquid phase is given by  
\begin{equation}
    E_{extr} = \frac{ \Delta V_{\rm{GL}} }{d_l \left(1- \frac{ \epsilon_l}{\epsilon_g}\right) + L \frac{ \epsilon_l}{\epsilon_g} },
\label{eqn:Eextr}    
\end{equation}
where $\Delta V_{\rm{GL}}$ is the voltage difference between the grid and LEM bottom, $L$ (\SI{=1}{\cm}) is the grid to LEM distance, $d_l$ (\SI{=0.5}{\cm}) is the height of the liquid argon above the grid, and $\epsilon_l$ ($\epsilon_g$) is the liquid (gas) argon dielectric constant, respectively. The proportion of electrons extracted into the gas phase, or extraction efficiency, has first been reported in \cite{Gushchin:1982}.  At the nominal extraction field of 2\,kV/cm in liquid, 90\% of drifting electrons cross the liquid-gas boundary.

The extracted charge is then amplified in LEMs. The LEM, also known as a Thick GEM (THGEM) \cite{Breskin_2009}, is a \SI{1}{\mm} thick PCB plate, covered by copper layers and drilled with a hole density of the order of 150 holes/cm$^2$.  A high potential difference applied across ($\dvlem\SI{>3.0}{\kilo\volt}$) provides an electric field strength within the holes sufficient to allow the passing electrons to start ionising the medium thereby igniting a chain reaction, Townsend avalanche, where the added free electrons can librate further charge. This gain of electrons through the avalanche process is dependent on the argon gas density and the electric field strength (e.g., \cite{Sauli:1977zz} and reference within). The first influences the mean free path of the drifting electrons between atomic collisions, while the second sets how much energy they can gain before the next collision occurs. In addition to the ionisation, the argon atom excitation can also occur resulting in emissions of VUV photons (electroluminescence) \cite{MONTEIRO2008167,BOLOZDYNYA1999314} detectable with the PMTs. 

During avalanches within the LEM, charge (both electrons and positive ions) can accumulate on the dialectic surfaces inside the holes thereby reducing the field strength within. This leads to a gradual decrease in the gain over time until it eventually stabilises at the point where no charge can be further deposited on the dialectic. The speed of this process, referred to as charging-up, is dependent on the geometry of the holes as well as the experimental conditions, in particularly the electric field strength within the LEM (gain) and the rate of ionisation charge traversing it. 

The amplified charge exiting the holes is guided to a segmented anode placed \SI{2}{\mm} above by an induction field defined by the potential different between the top LEM electrode (\SI{-1}{\kilo\volt} nominal) and the anode plane (at ground). The moving charge in the induction region induces uni-polar signals on the anode electrodes (often referred to as a collection signals). The electrode strips are interleaved in two orthogonal directions (View 0 and 1) such that the charge is equally shared between each view. 

The amount of charge arriving at the anode from the initial interaction in the liquid argon is dependent on several factors. To simplify, it can be divided into two parts. The first contains the energy loss of the initial interacting particle, the quantity of electrons liberated by the applied electric field and the proportion surviving the drift to the liquid surface. The second part, is then the effective gain  ($G_\mathrm{eff}$)  of the 
CRD, which is the ratio of the number of electrons collected on both views to the number arriving at the liquid surface.  It comprises the efficiency of extraction from liquid to gas, the gain of the LEM which includes the transparency of the LEM and the multiplication of the electrons within the LEM holes, and the efficiency to collect the electrons on the anode strips. 

The Dual-Phase technique means to achieve high Signal/Noise levels on the strip readout due to the signal amplification of the LEM (expected to be around $\sim$20) which compensates for signal losses due to the long drift, transparencies and inefficiencies. The LEM gain can be changed by modifying the potentials applied across it, allowing to increase or decrease the gain on individual CRDs. 
As the charge is read out at the top of the detector, custom signal feedthrough chimneys were conceived that allow removal and replacement of the cryogenic analogue electronics. 

It is worth noting the performance of large volume LArTPCs at the surface is hindered by the accumulation of positive ions, produced by the very large rate of cosmic ray interactions in the detector, which drift slowly towards the cathode. This accumulated charges modifies the electric field both in strength and direction. This can result in  shifts in the reconstructed positions, bending tracks for example, and changes to the measured ionisation yield.  The space charge effect, as it is known, is dependent on the cosmic ray rate, the TPC drift distance, and the flow or movement of LAr within the detector. For deep underground LArTPCs, this effect is expected to be negligible.

The Dual-Phase design has an additional source of positive ions, known as positive ion backflow, in which positive ions generated in the LEM avalanches flow slowly through the gas layer. Such ions could sweep up electrons on their path, their paths could end at the bottom of the LEM, or they could build up (or even traverse) the liquid surface. A first simulation study of Dual-Phase has been made considering both ion backflow and electroluminescence, and finding that the ion backflow can be expected to be large \cite{Lux_2019}.

\subsection{Small-scale demonstrators}
 
The Dual-Phase design has a number of assets for a large volume LAr TPC. The ability to access the electronics in case of problems is an advantage for an experiment that intends to operate for more than 20 years.  The PCB-based charge read-out is easier to manufacture and manipulate, particularly for large dimensions, than the traditional wire readout.  The two-view readout minimises the number of channels, lowering costs.  A monolithic, homogenous detector is another advantage, with no dead volumes which could render event reconstruction challenging.  However, extrapolating to a DUNE Far Detector module, the drift length increases to 12\,m which would be a major technical challenge. 

Dual-Phase LAr TPCs operating with LEMs have been previously demonstrated at small scales ~\cite{Resnati_2011, Badertscher_2011, Cantini_2014, Badertscher_2013}.  The world's first large-scale Dual-Phase detector so far was the 4-tonne demonstrator~\cite{311_technical} at CERN, whose performance and results are described in~\cite{311_performance}.

Two important technical issues prevented the 4-tonne demonstrator from validating the Dual-Phase concept for future large-scale experiments: the failure to achieve sufficient LEM gains and the operation instability of the extraction grid. Both of these issues were addressed with modifications to the \pddp LEM and extraction grid designs. In addition, a dedicated test facility, known as the coldbox, was constructed at the CERN Neutrino Platform in order to test the components of the CRPs  before their installation in \pddp.


\subsection{The ProtoDUNE detectors}
\begin{figure}
\begin{centering}
\includegraphics[width=0.7\textwidth]{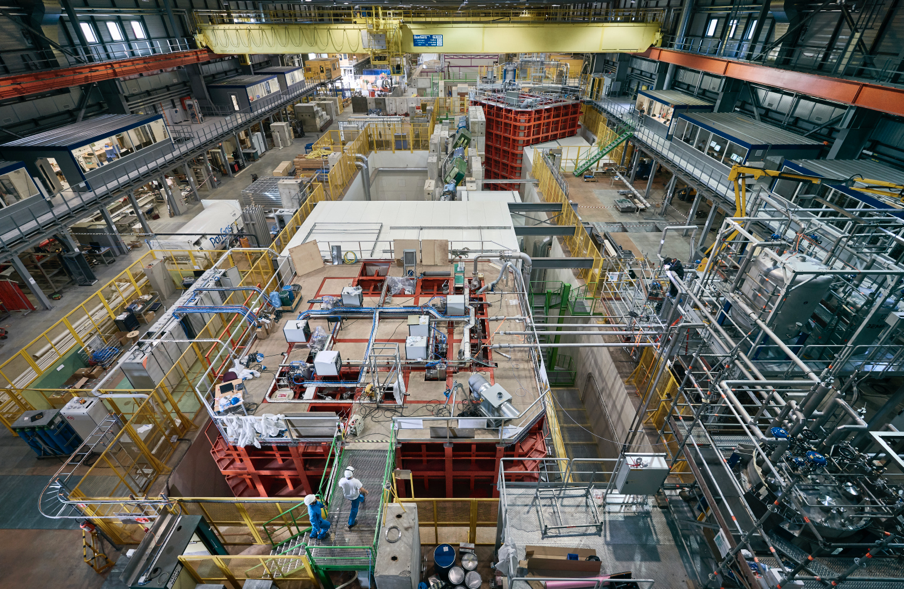}
\caption{\label{fig:neutrinoplatform} The ProtoDUNE detectors installed at the CERN neutrino platform. \pddp is towards the back of the hall.}
\end{centering}
\end{figure}

The ProtoDUNEs are housed at the CERN Neutrino Platform Facility, a dedicated extension of building 887 (EHN1) located at the CERN Prévessin site.  Using identical membrane cryostats, denominated NP02 and NP04, and with TPC active masses of 430\,ton and 300\,ton for Single Phase and Dual Phase respectively, these are to-date the largest LArTPCs ever constructed. The ProtoDUNEs were constructed using full-size components intended for the DUNE Far Detector modules. As well as testing individual components, these detectors would also provide the necessary step towards construction of the Far Detector, testing realistic engineering solutions and installation procedures. Figure~\ref{fig:neutrinoplatform} gives a view of the hall showing both detectors. The design, operation, and performance of \pddp are the topics of this publication.



\section{Detector description}
\label{sec:detoverview}

\begin{figure} 
\begin{centering}
\includegraphics[width=10cm]{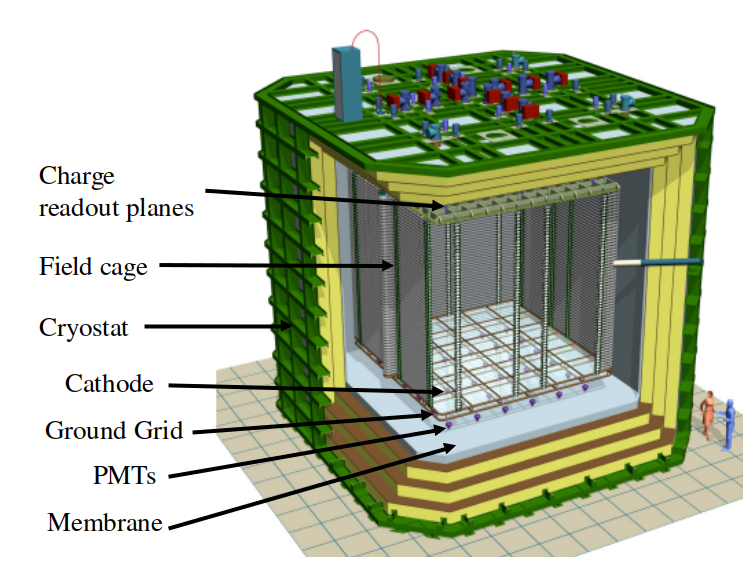}
\caption{\label{fig:detector} \pddp detector schematic showing major components.}
\end{centering}
\end{figure}

A model representation of the \pddp detector \cite{WA105} is shown in figure~\ref{fig:detector}. It consists of a membrane cryostat, described in section~\ref{ssec:det:cryo}, which can contain up to \SI{800}{\tonne} of LAr. The majority of the connections, entries and exits, to the detector are made through the top of the cryostat.  The detector includes several monitoring systems pertaining to the cryostat structure, the internal temperatures, LAr purity (see section~\ref{ssec:det:purity}), and real-time visual aid with the inclusion of cryogenic video cameras (see section~\ref{ssec:det:camera}).

The TPC structure, described in section~\ref{ssec:det:tpc}, comprises the cathode grid and field cage assembly, which is suspended from the roof of the cryostat. Achieving the necessary field strength to drift the ionisation charge in the TPC volume requires powering the cathode at \SI{300}{\kilo\volt}. The description of the HV delivery system enabling to do this is given in section~\ref{ssec:det:hvft}. Four CRPs (of $3\times3$\,m$^2$ each), described in section~\ref{ssec:det:crp}, are positioned at the top of the detector to collect the ionisation charge. The anode strips are read with a dedicated custom-designed electronics (section~\ref{ssec:det:elec}) and data acquisition system (section~\ref{ssec:det:daq}).  The floor of the cryostat is lined with an array of PMTs, which form the Photon Detection System described in section~\ref{ssec:det:pds}, to detect LAr scintillation and electroluminescence light. A grounded grid is placed directly above the PMT array to protect them from potential discharges of the HV applied to the TPC cathode.  Two Cosmic Ray Taggers (CRT), section~\ref{ssec:det:crt},  are installed on the external cryostat walls to provide a potential trigger for muons crossing the active volume. 

\subsection{Cryostat and cryogenic system}
\label{ssec:det:cryo}

The two identical cryostats housing both ProtoDUNE detectors were built based on the technology developed for the storage and transport of liquefied natural gas. Its application for building non-evacuated massive cryostats to house the neutrino detectors has been a subject of extensive studies and prototyping efforts finally culminating with the construction of the two large cryostats at CERN. 

The cryostat inner volume, with dimensions $8.5\times8.5\times7.9$\,m$^3$, is defined by a \SI{1.2}{\mm} thick stainless steel membrane shell. This membrane\footnote{A proprietary development by GTT (Gaztransport \& Technigaz)} has a two dimensional corrugated structure allowing it elastically expand or contract with changes in temperature and loads. A \SI{800}{\mm} thick thermal insulation shell surrounding the cryostat inner membrane vessel consists of two \SI{390}{\mm} thick layers of an expanded polyurethane foam sandwiched between plywood panels. For safety, a containment shell is inserted between two insulation panels. This secondary membrane is made from a proprietary composite\footnote{Triplex material produced by GTT} of thin-sheet aluminium laminated on both sides with glass cloth and resin. A steel outer structure, with dimensions $11.4\times11.4\times10.8$\,m$^3$, provides the mechanical support for the entire assembly. 

The cryostat design integrates a resealable large opening, called Temporary Construction Opening (TCO), in one of the walls. It is needed to bring in the material for the construction of the cryostat itself as well as the large detector elements for the TPC assembly later on. The closure of the TCO is the last step in the detector assembly after which all the subsequent access to the interior can only be done  though the cryostat roof, which has a sealable access hatch for personnel. All the service feedthroughs (with the exception of the LAr outlet for re-circulation located in one of the walls close to floor of the inner vessel) for the different subsystems (cryogenics, instrumentation, CRP, signal readout, etc.) are integrated in the penetrations in the cryostat roof. 

The cryogenic system design, identical for both ProtoDUNE detectors, is described in~\cite{Bremer_2022} with some operational aspects covered further in~\cite{abud:hal-03519598}. Under normal operating conditions, the cryostat is operated in a closed-circuit loop with LAr continuously re-circulated though a liquid purification filter to remove \ce{O2} and \ce{H2O} contaminations with the help of a pump, external to the cryostat. The liquid flow rate though this re-circulation and purification circuit is approximately \SI{7}{\tonne\per\hour}. A complete LAr volume turnover in the detector thus requires about 5 days. In order to facilitate the mixing of the purified LAr within the detector volume, the liquid after the purification filter passes though a phase-separator. Its pressure is regulated such that the exiting liquid, returned to the bottom of the cryostat, is slightly warmer than the one inside. 

The liquid purification filter consists of a series of three volumes: the upstream (in the sense of the liquid flow) one filled with molecular sieve\footnote{Type 4A} for \ce{H2O} capture, and the two downstream ones containing a bed of copper pellets\footnote{BASF Cu-0226S} for \ce{O2} capture. The filter input and output are protected with a fine mesh (mechanical filters) to block the circulation of micro-particulates. The total filter weight of about \SI{650}{\kg} has been dimensioned to handle the commercial liquid argon---\ce{O2}/\ce{N2}/\ce{H2O} contaminations were required to be below \SI{2}{ppm}/\SI{2}{ppm}/\SI{1}{ppm} for the deliveries---during the initial filling, where the delivered LAr would pass through the purification filter before entering the cryostat volume. The copper media capacity for \ce{O2} capture is \SI{0.5}{\gram\per\kg} of the filter bed mass, while that of the molecular sieve is less certain at cryogenic temperature and a conservative value of \SI{22}{\gram\per\kg} (about ten times lower than the value obtained at room temperatures) was assumed in the filter mass calculations. The filter was sized to undergo no more than one regeneration during cryostat filling. 

In the normal operation mode, the boil-off Ar gas (the cryostat as well as the phase-separator) is re-circulated. The gas is filtered though a gas filter---a lighter, \SI{30}{\kg}, version of the liquid filter---before being re-condensed in a liquid nitrogen condenser. This liquid is then returned to the cryostat via the liquid purification circuit (i.e., by passing though the purification filter and the phase-separator).  The absolute gas pressure inside the cryostat ullage can be precisely set and maintained (within \SI{1}{\milli\bar}) by regulating the flow rate of \ce{LN2} in the condenser and/or of the boil-off gas argon (e.g., venting it if necessary).

To assure controlled and safe operation throughout various phases (cooling, filling, emptying, etc.), the cryostat is instrumented with a large variety of different types of sensors. Their complete inventory is outside of the scope of this paper. However, some, more relevant to the subsequent discussions, will be briefly described.

\subsubsection{Temperature and liquid level monitoring}
\label{ssec:det:levelmon}

The temperature profile of the liquid and gas is monitored by 58 temperature resistive temperature probes that were installed vertically along two corners of the cryostat interior. The absolute positions of each sensors is known to a few millimetres allowing also for their use for additional cross-check of the LAr level during filling.  

The argon gas temperature in the ullage is monitored by 24 temperature sensors installed near the top of the cryostat. These measurements are needed to estimate the argon gas density near the LEMs in order to compute the effect on their gain. Another 28 temperature sensors, installed on the CRP frame structure, provide a measure of the local thermal gradients there with the goal of helping in the estimation of the CRP deformations.

During the filling (or emptying), the changing liquid level inside the cryostat is monitored by two coaxial \SI{4}{\meter} long capacitive level meters to about a centimetre precision. These level meters are vertically staggered to provide a complete coverage over the entire height of the cryostat. Two additional planar \SI{60}{\mm}-long level meters installed on the cryostat walls at the nominal height of the liquid provide more precise (about \SI{0.6}{\mm} resolution) monitoring of the level once the filling is complete. 

\subsubsection{Purity monitoring}
\label{ssec:det:purity}

\begin{figure}
\begin{centering}
\includegraphics[height=12pc, width=12pc]{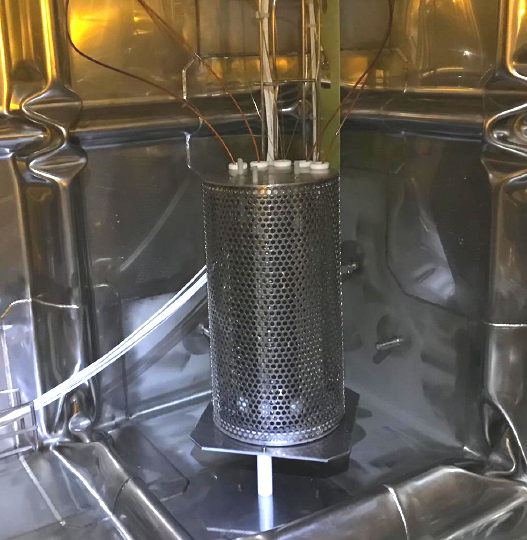}
\includegraphics[height=12pc, width=12pc]{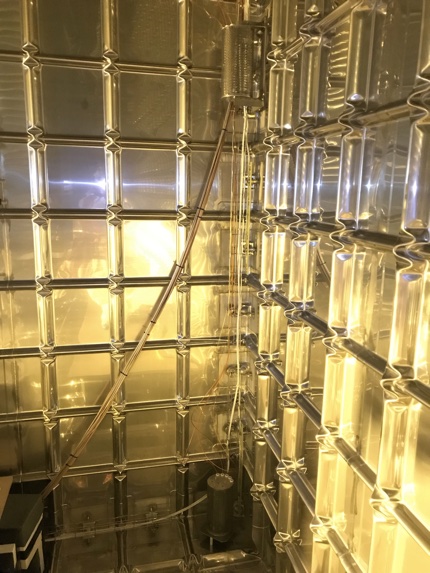}
\includegraphics[height=12pc]{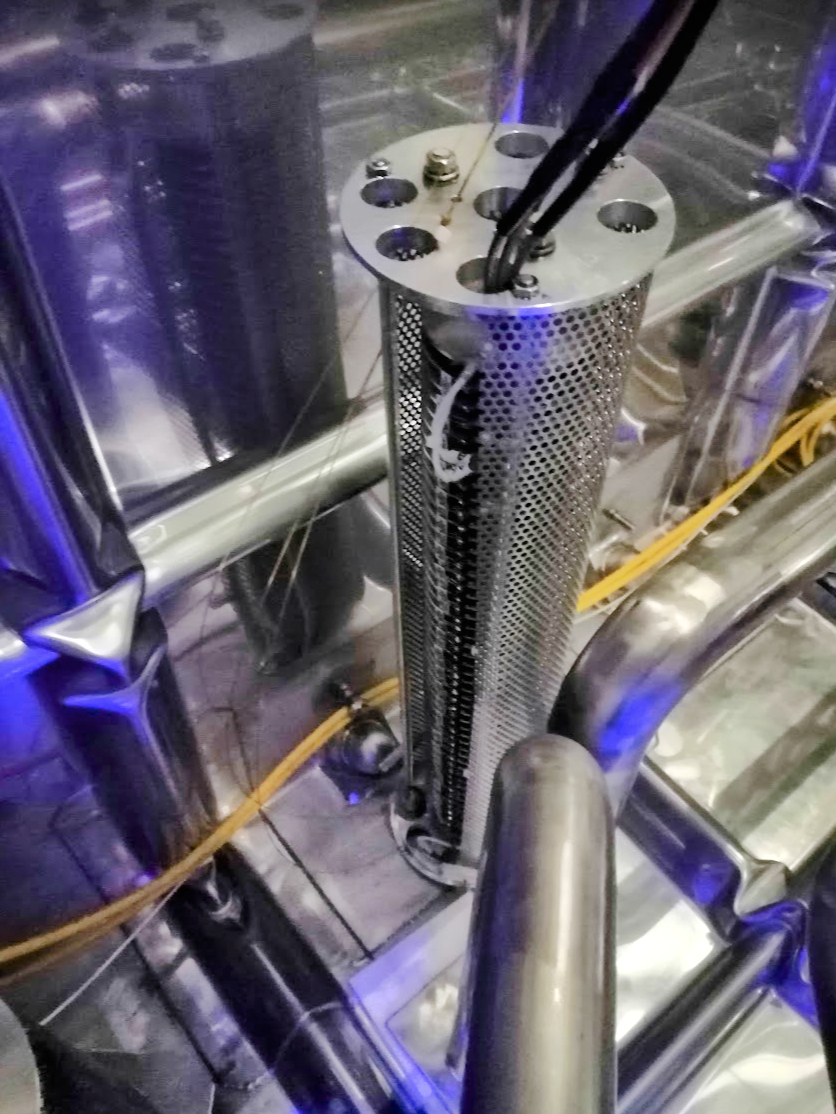}
\caption{\label{fig:dp_shortpuritymonitor} (Left) The bottom short purity monitor in its Faraday cage enclosure with the fibres for flashing the cathode entering from the top. (Centre) Both short purity monitors. (Right) The long purity monitor.}
\end{centering}
\end{figure}

The LAr purity inside the cryostat is monitored with three purity monitors, shown in figure~\ref{fig:dp_shortpuritymonitor}. These are independent self-contained detectors whose purpose is to measure the electron lifetime in LAr. The monitors work along the same principle: they measure the amount of charge generated at the cathode and that arriving at the anode. The electron lifetime can then be estimated by comparing the two signals. Two of the monitors (short) are \SI{17}{\cm} long~\cite{Manenti:2020gzi}: one located at the bottom of the cryostat in the corner and the other placed \SI{2.5}{\meter} above. The third, a \SI{48}{\cm} long purity monitor ~\cite{Adamowski_2014}, is located next to the bottom short purity monitor. 

A single xenon flash lamp, on the cryostat roof, provides a source of UV light, delivered to the purity monitors via quartz fibres, for extracting electrons from the cathode via the photoelectric effect. The emitted electrons are drifted in adjustable electric field (typically in the fields \SIrange[range-phrase = \text{-}, range-units  = single]{10}{100}{\volt\per\cm}) towards the anode. The longer the drift time between the cathode and the anode (i.e. with the longest anode-cathode distance and/or with the lowest drift field) the more sensitive the device is to the amount of the electronegative impurities. The short purity monitors are sensitive to electron lifetime up to a few ms. The long purity monitor is suitable for measuring from \SI{1.5}{\milli\second} to about \SI{10}{\milli\second} electron lifetimes. 

\begin{figure}
\begin{centering} 
\includegraphics[width=0.8\textwidth]{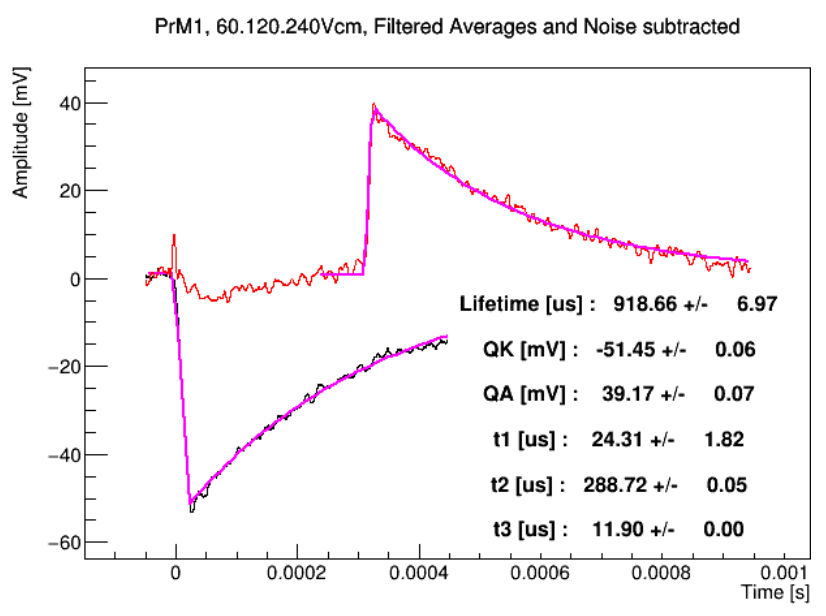}
\caption{Example cathode (anode) signals from one of the purity monitors shown in black (red). The fitted curves modelling the response of the readout electronics and providing the estimates of the charge leaving the cathode ($Q_{K}$) and arriving at the anode $Q_{A}$ as well as the resultant electron lifetime value are overlaid. The fit parameter $t_1$ ($t_3$) is the rise-time of the cathode (anode) signal, while $t_2$ is the time interval from the cathode peak to the start of the anode signal ($t_2 + t_3$ is the peak-to-peak time).}
\label{fig:prm_waveform}
\end{centering}
\end{figure}

Figure~\ref{fig:prm_waveform} illustrates an example of the signals from one of the short purity monitor. The temporal displacement between the cathode (black) and anode signals (red) is given by the drift time. The relative strength of the two signals---extracted from the fit (purple) that takes into account the response of the readout electronics---gives the measure of the charge attenuation. 

\subsubsection{Cryogenic cameras}
\label{ssec:det:camera}

The inner cryostat volume is instrumented with cameras capable of operating at cryogenic temperatures. These are Raspberry Pi $1280\times 960$ pixel Ethernet cameras housed in transparent plexiglass cylinders, vacuum pumped and sealed. In order to operate the cameras in cryogenic conditions, the temperature of the electronic circuit is regulated with a heater and controlled by sensors.

A total of twelve cameras were installed and equipped, depending on their location, with either zoom, normal or wide-angle lenses. The cameras were positioned in strategic locations within the cryostat, providing the views of the LAr level and surface, the position of the CRPs, the points of entry of the LAr during filling, the cathode, and along the path of HV delivery from the feedthrough to the cathode.

\subsection{Time projection chamber}
\label{ssec:det:tpc}

\begin{figure}
	\begin{center}
\begin{minipage}{0.4\textwidth}
	 \includegraphics[width = \textwidth]{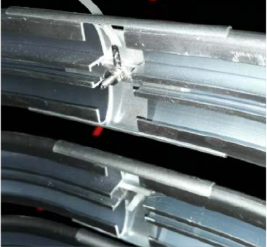}
 \includegraphics[width = \textwidth]{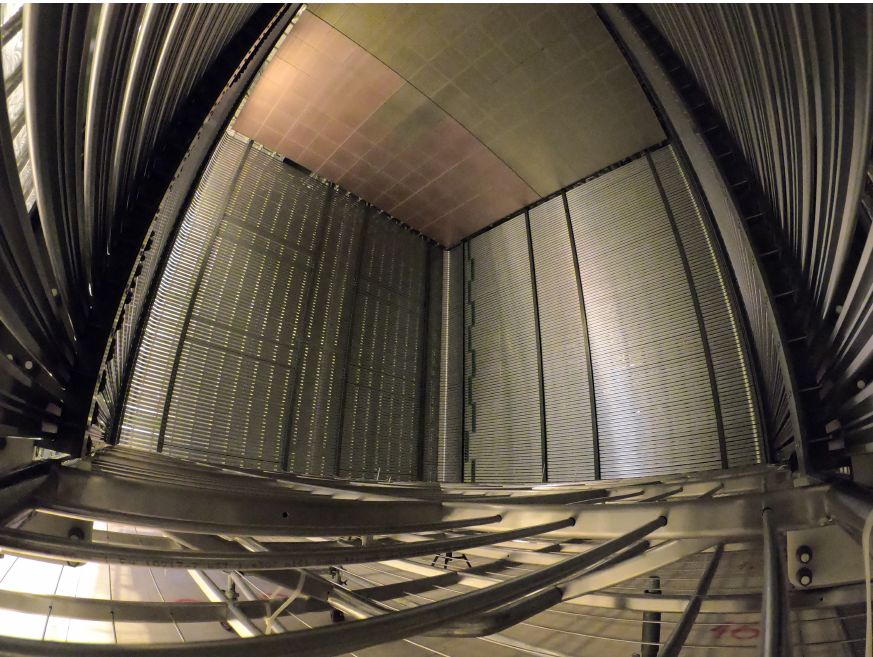}
 \end{minipage}
\hfill
\begin{minipage}{0.55\textwidth}
      \includegraphics[height = 0.34\textheight, angle=-90,origin=]{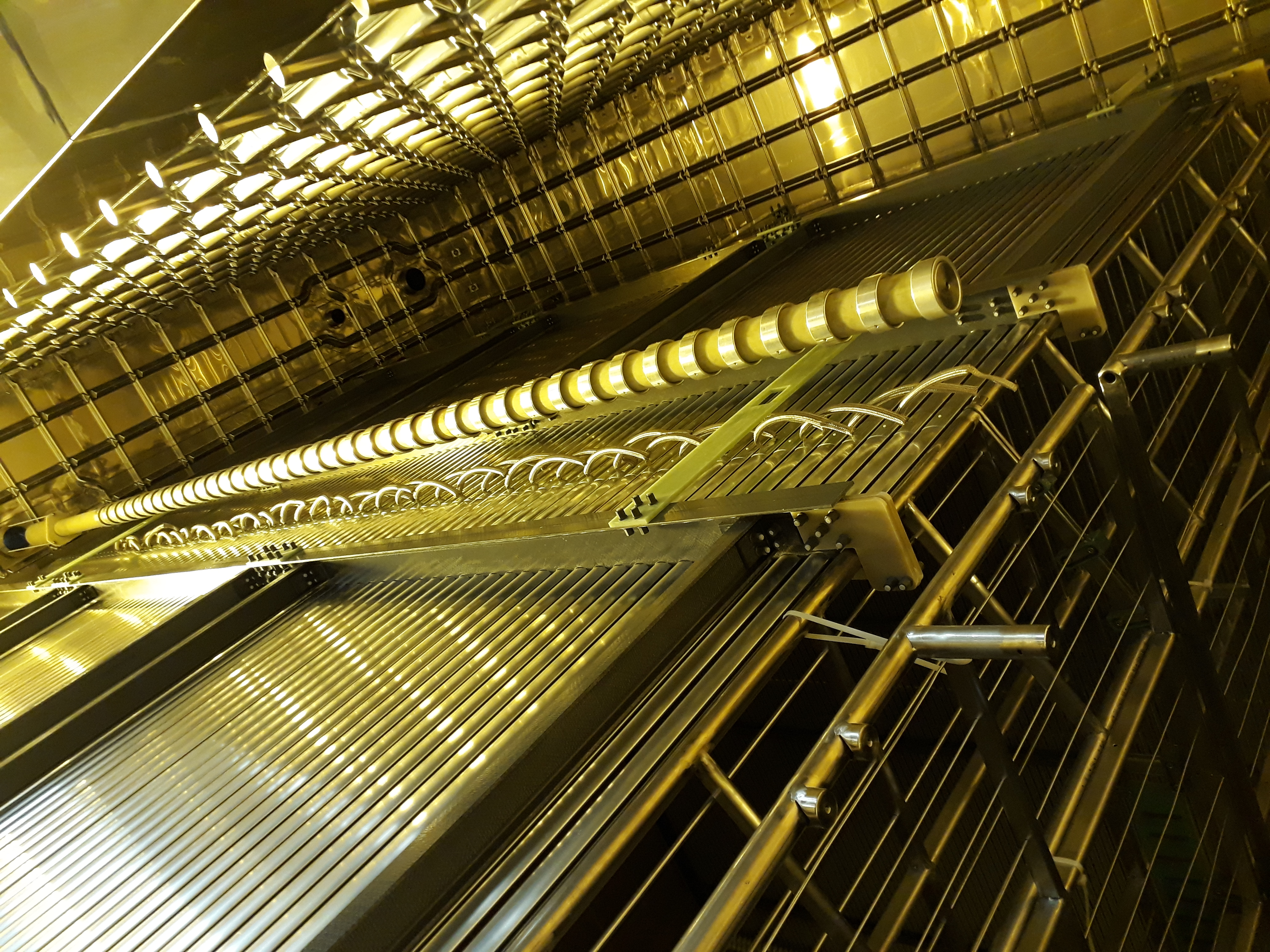}
\end{minipage}
		\caption{\label{fig:fieldcage}Left Top: Close-up of the interior view of the field cage bars showing the bar profile and assembly clip. Left Bottom: Wide angle view of the interior of the TPC showing the field cage and cathode during construction. Right: field cage side with a view of the HV extender, which is part of the cathode HV distribution system. The stainless steel ground grid placed between the cathode and the PMTs is also visible.  At the time of construction it was attached to the cathode, and later it was moved to the floor.}
	\end{center}
\end{figure}

The TPC structure consists of a field cage and cathode which are suspended from the cryostat roof. Figure~\ref{fig:fieldcage} (left bottom) shows a wide-angled view of the TPC interior. The Field Cage is made of 98 Field Shapers, spaced at 6 cm intervals along the vertical direction. Each Field Shaper is formed by aluminum profiles attached by the assembly clips shown in figure~\ref{fig:fieldcage}
(left top), in a square-shape base of 6-m length with rounded corners (radius of curvature 9 cm).

\begin{figure}
\begin{center}
    \includegraphics[width=0.5\textwidth]{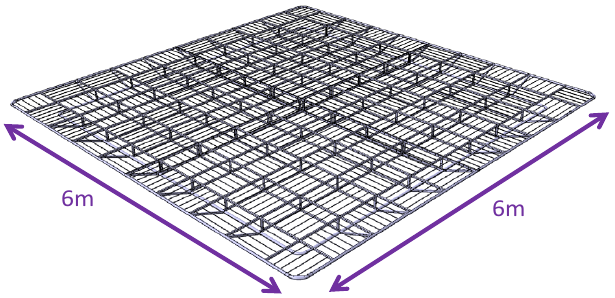}
\caption{The cathode is assembled from 4 identical sections and connected electrically by damping resistors. \label{fig:cathodeandgrid}}
\end{center}
\end{figure}

The cathode is a modular 3x3 m$^2$ metallic structure which was assembled inside the cryostat and hung from the field cage. The design is shown in figure \ref{fig:cathodeandgrid}.  The modules, weighing about \SI{90}{\kg} each,  have a two-floor structure made of stainless steel oval tubes for mechanical rigidity whilst maintaining weight restrictions. The diameter of the tubes was also constrained to keep local electric field strengths at acceptable levels (\SI[per-mode=symbol]{<30}{\kilo\volt\per\cm}).  The modules are connected together mechanically and electrically via resistors in order to slow down the release of stored energy in the case of a discharge. The cathode is designed to achieve 60\% optical transparency necessary for the PMTs located on the cryostat floor.

The HV applied to the cathode (nominally \SI[per-mode=symbol]{-300}{\kilo\volt} for \SI[per-mode=symbol]{0.5}{\kilo\volt\per\cm} drift field strength) is divided by High Voltage Divider Boards (HVDB) to give a series of monotonically increasing voltages which are connected to the field cage rings.  Two rows of HVDBs are implemented for redundancy. Each divider stage of HVDB is formed by two \SI{2}{\giga\ohm} resistors (reference: SM102032007FE) connected in parallel and thus giving a total \SI{0.5}{\giga\ohm} (four \SI{2}{\giga\ohm} in parallel from two HVDBs) resistance between two consecutive field cage rings. The resistors on HVDBs are protected from over-voltage by a parallel branch formed by a series of four metal oxide varistors (reference: ERZV14D182) accommodating up to \SI{6}{\kilo\volt} voltage drop between the field shaping rings.


The upper most ring, First Field Shaper (FFS) of the field cage is terminated to ground via a resistor.  An HV cable connected to FFS is brought out of the cryostat via a custom feedthrough in order to monitor the current flowing through the field cage resistor chain and adjust the voltage drop on the last resistor stage to set the electric field strength in the region below the CRP extraction grids.

In order to protect the PMTs installed on the floor of the cryostat from the very high electric field, a stainless steel ground grid is placed approximately \SI{1}{\meter} below the cathode above the PMTs.  The grid is a free-standing structure consisting of hollow tubes and stands on the corrugated floor of the cryostat.



\subsection{High voltage delivery system}
\label{ssec:det:hvft}

The TPC is designed to be operated with a target drift electric field of \SI[per-mode=symbol]{0.5}{\kilo\volt\per\cm}, which requires a potential of \SI[per-mode=symbol]{-300}{\kilo\volt} applied to the cathode.  Powering the cathode at such HV is accomplished with a dedicated HV delivery system. The HV supply is a Heinzinger PNChp 300000-05-neg High Precision High Voltage Power Supply unit. The HV from the power supply is brought to the cathode via a dedicated high voltage feedthrough (HVFT) coupled to a long \SI{\sim6}{\meter} conductor (HV extender) inside the cryostat responsible for delivering the power at the required depth.

\begin{figure}
    \centering
    \includegraphics[width=0.98\textwidth]{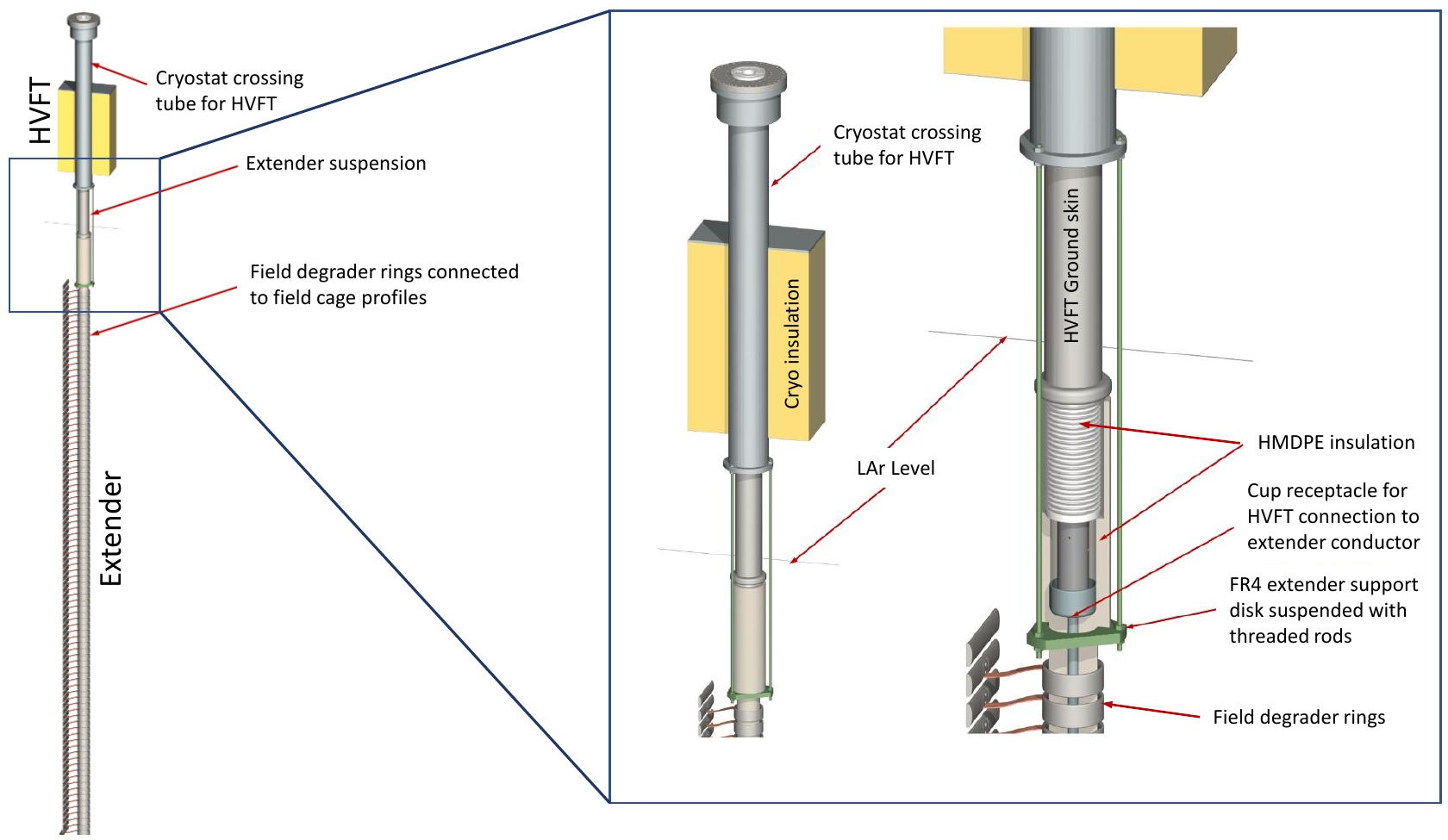}
    \caption{Schematic drawing of the HV feedthrough and the first extender design used in \pddp.}
    \label{fig:hvscheme1}
\end{figure}

Two designs of HVFT and HV extender were successively developed and tested in the detector with the key difference between them being the approach taken for the extender design. The schematic view of the HVFT and the first HV extender is shown in figure~\ref{fig:hvscheme1}. The \SI{2}{\meter} long HVFT is installed in the tube crossing the entire cryostat insulation space. One of its key features is the ease of removing it for replacement or repair even when the cryostat is full. The design \cite{Cantini:2016tfx, AMERIO2004329}, common to both ProtoDUNEs, features a stainless steel conductor (\SI{32}{\mm} in diameter) surrounded by a cylindrical \SI{36}{\mm} thick layer of ultra-high molecular weight polyethylene (UHMWPE) insulator. The insulator is then enclosed by a stainless steel skin which is grounded to the cryostat structure. The assembly must be made at cryogenic temperatures (cryo-fitting) to ensure the feedthrough vacuum tightness. The insulation cylinder extends by about \SI{20}{\cm} beyond this ground skin and its surface is corrugated in this region to increase the surface area. The spring-loaded conductor tip jutting out at the bottom couples to a stainless steel cup receptacle on the HV extender below. As can be seen in the insert in figure~\ref{fig:hvscheme1}, a UHMWPE sleeve---part of the HV extender structure---covers the coupling point and the bottom part of HVFT rising up to the ground skin.   

The \SI{6}{\meter} long HV extender (visible in the image on the right in figure~\ref{fig:fieldcage}) is made by a \SI{20}{\mm} diameter stainless steel tube inner conductor surrounded by a tightly fitted cylindrical \SI{20}{\mm} thick UHMWPE layer. The extender is supported by a FR4 disk (glass-reinforced epoxy laminate, shown in green in figure~\ref{fig:hvscheme1}) fixed via threaded rods to the HVFT crossing tube above. The external part of the extender has 41 aluminium rings (degrader rings) mounted equidistantly. Each ring is connected electrically to a field cage shaper profile closest to it in height with an intended function to minimise the electric field in the portion of the liquid argon volume between the field cage and the extender.

During commissioning of the HV delivery system, after the filling of the cryostat, a fault in the extender insulation developed, resulting in a short between the inner conductor and the first degrader ring and its corresponding field cage profile (see section~\ref{sec:hvcommissioning}). This failure, possibly resulting from a structural weakness caused by a mechanical stress exerted during the installation, prompted a complete re-evaluation of the extender design. A new approach was pursued removing the UHMWPE insulation from the conductor, relying instead on the dielectric rigidity of the liquid argon itself for the insulation. However, to limit the electric field strength around the conductor, its diameter had to also be significantly increased. In addition, this design significantly simplified the extender construction by eliminating the need for precise machining of the long insulation block that had to be tightly fitted onto the conductor.

\begin{figure}
    \centering
    \includegraphics[width=0.98\textwidth]{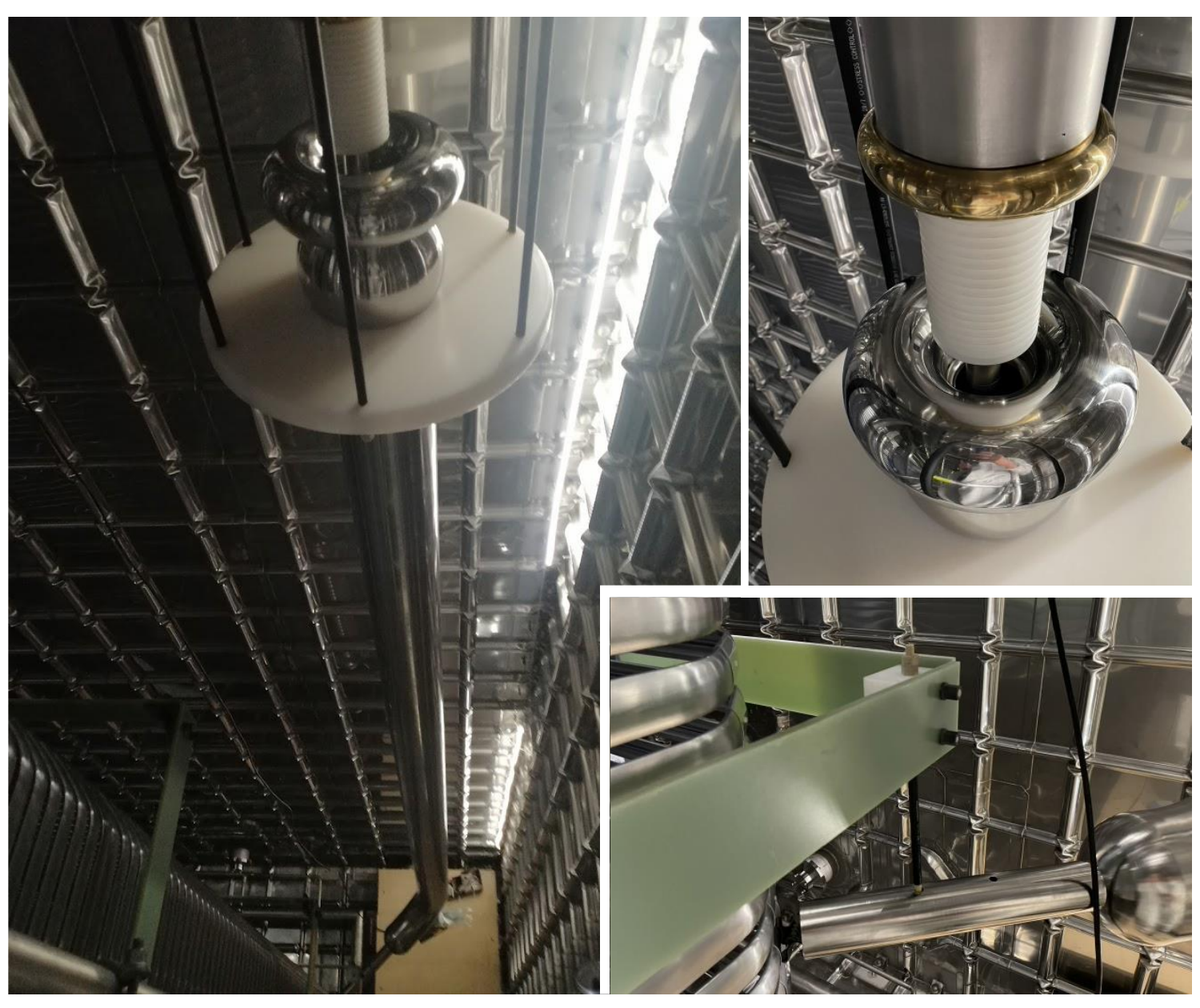}
    \caption{Views of the redesigned HV extender in the \pddp detector. (Left) view of the HV extender from the top. (Bottom-right) zoom on the part of the elbow connection to the cathode. (Top-right) view of the sphere receptacle and the toroidal electrode with the HVFT tip connected. The gold colored ring in the photo is a part of the sleeve to confine the gas bubbles produced by the heat input from the HVFT and evacuate them to the gas phase close to the HVFT ground skin.}
    \label{fig:hvnewpics}
\end{figure}

The new HV extender \cite{VD_TDR}, illustrated in figure~\ref{fig:hvnewpics}, consists of a \SI{\sim5}{\meter} long hollow stainless steel tube \SI{204}{\mm} in diameter. It is terminated at the bottom by a \ang{90} elbow-shaped connection to the cathode. The HVFT receptacle at the top of the extender has a spherical shape with a toroid-shaped electrode mounted on top to mitigate charging of the dielectric near the HVFT conductor tip. The assembly is supported by a \SI{75}{\mm} thick UHMWPE disk \SI{600}{\mm} in diameter, which is suspended from a mounting plate around the HVFT penetration with threaded FR4 rods. 

Apart from the extender, some modifications of the HVFT design were also performed. Particularly the feedthrough length was increased to about \SI{3}{\meter}, essentially extending the part which is at warm above the cryostat insulation. In this region, the cable from the power supply makes a connection to the inner conductor of the feedthrough, and keeping it at warm avoids any possible ice formation in the cable receptacle. A sleeve surrounding the ground skin of the HVFT was also added with the aim to confine any produced gas bubbles due to the feedthrough heat input and funnel them to the gas phase. The new HV delivery system (redesigned HVFT along with the new HV extender) was tested in \pddp detector from Autumn 2021 (section~\ref{sec:fulldrift}).  

\subsection{Charge readout planes}
\label{ssec:det:crp}

The \pddp detector can be read by up to four 3x3 Charge Readout Plane (CRP) modules. Each CRP is a mechanically independent structure that supports an extraction grid and 36 Charge Readout Detectors (CRD). The extraction grid is formed by two interlaced wire planes, made of \SI{3}{\m}-long \SI{0.1}{\mm}-diameter stainless steel wires spaced at a \SI{3.125}{\mm} pitch. 

\begin{figure}[htbp]
	\begin{center}
		\includegraphics[width = 0.365\textwidth]{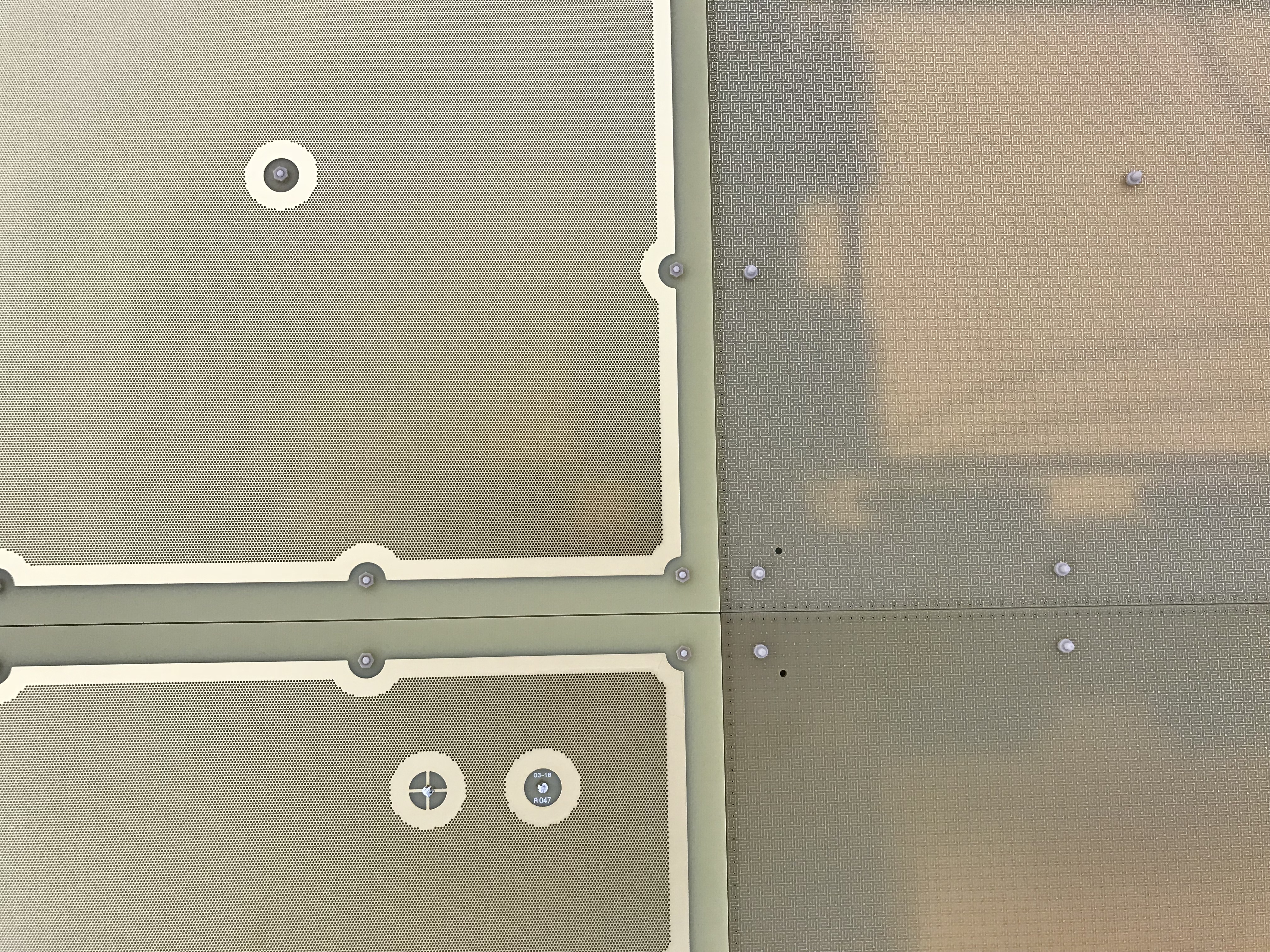}
        \includegraphics[width = 0.625\textwidth]{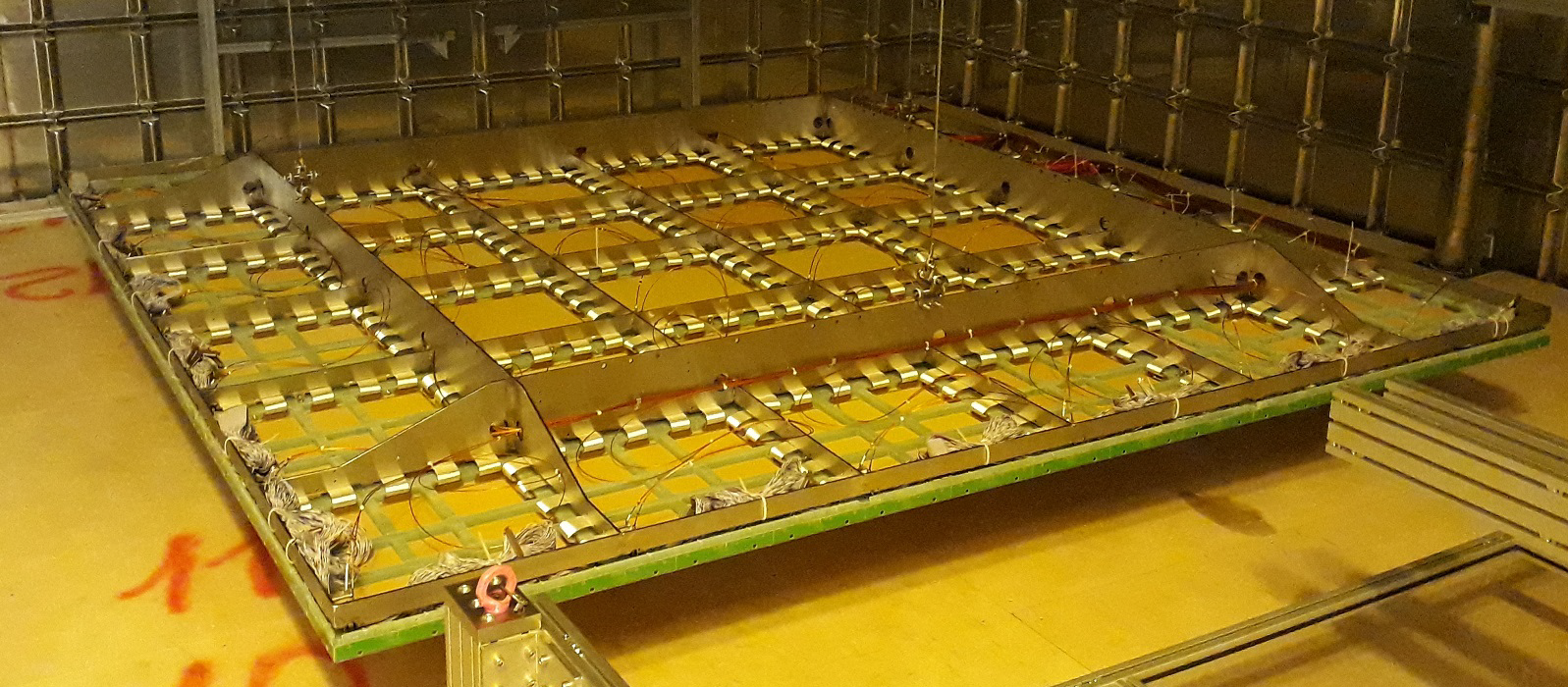}
		\caption{(Left) a close-up on two fully assembled CRDs (left) with the LEM HV connections areas visible for LEM on the bottom left LEM and two anodes with \SI{2}{\mm} spacers (right) mounted on CRP. (Right) an assembled CRP suspended from the Invar frame just above the cryostat floor; the visible silver jumper cables interconnect groups of 32 channels from the anodes of each CRD to form \SI{3}{\meter} long readout strips.}
    \label{fig:crpdetail}
	\end{center}
\end{figure}

Each CRD (figure~\ref{fig:crpdetail} left) is made of a 50x50 m$^2$ LEM attached \SI{2}{\mm} below a Printed Circuit Board (the anode) of the same dimensions with 29 precisely machined spacers which guarantee a uniform induction gap. The HV connection to polarise the bottom and top LEM faces, placed in a corner of each CRD, are made with two through pins soldered on the LEM dedicated copper areas, each one protected inside a Macor cylinder glued on the top LEM top face. Within the CRP, the CRDs are arranged in $6\times 6$ grid. Each row (column) of 6 anodes of given CRDs are interconnected with short jumper cables (figure~\ref{fig:crpdetail} right), bridging groups of 32 channels, and thereby forming the \SI{3}{\meter} long electrode strips which are read by the charge readout electronics. 

Mechanically, the CRP consists of an Invar (FeNi36) frame, which holds the structure made of G10 Fiberglass within which CRDs are integrated. The Invar was chosen due to its extremely low coefficient of thermal expansion, which is about ten times smaller than that of stainless steel. The frame must support the weight of the entire CRP with minimal planar deformation both due to gravity and thermal stresses. The latter arise from the fact that the frame must be positioned in the gas layer and exposed to a temperature gradient of $\sim\SI[per-mode=symbol]{2}{\kelvin\per\cm}$.  Finite element analysis calculations and material deformation measurements at cryogenic temperatures showed that only Invar would be a suitable choice for the frame material. To compensate for the difference in the thermal contraction of the Invar and G10 structures, the connections between the two are made with a decoupling system that allows only lateral (but not vertical) movement. The G10 structure itself is assembled from nine 1x1 m$^2$ sub-frames, each holding four CRDs.

\begin{figure}
\begin{centering}
\includegraphics[ width=0.8\textwidth]{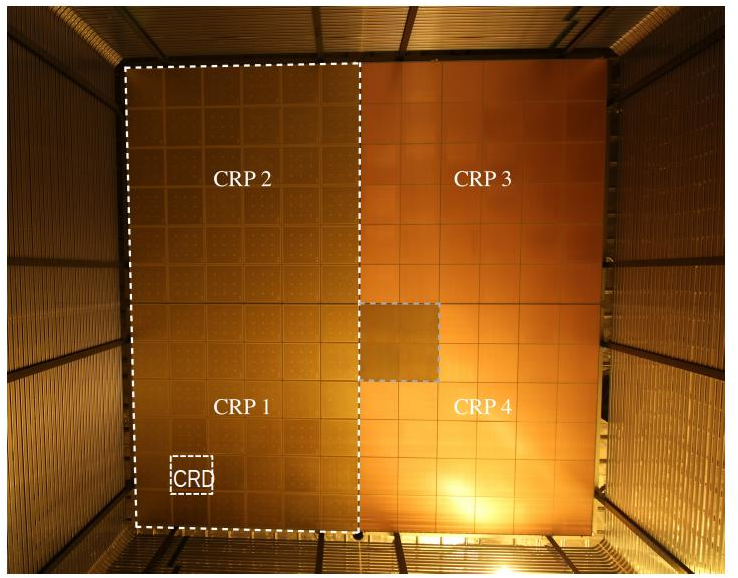}
\caption{\label{fig:photoinside}View from inside the TPC looking upwards at CRPs. A single CRD is highlighted. The two fully instrumented CRPs are labelled CRP1 and CRP2 in the image. A darker area in the corner of CRP4 corresponds to the part instrumented with four CRDs containing anodes only.}
\end{centering}
\end{figure}

The extraction grid wire ends are fixed on support boards, attached along the perimeter of the G10 structure. These boards are shimmed with spacers allowing to tension the wires to a target value during assembly (room temperature). The wire tension is expected to increase when CRP is at cryogenic temperatures due to the difference in the thermal expansion coefficients of the stainless steel and G10. While the wire breaking point is at \SI{15}{\newton}, the tension must, however, be limited to $\SI{<5}{\newton}$ due to potential CRP deformation resulting from forces exerted by the wires on the support structure as computed with a finite element analysis simulation. Comb-like fittings, placed at every metre (boundaries of each 1x1 m$^2$ sub-frame) within G10 structure, further ensure the \SI{10}{\mm} spacing is maintained between the extraction grid wires and the bottom face of the LEMs.
 
Figure \ref{fig:photoinside} offers a view from inside the \pddp TPC, looking upwards to the installed CRPs. Only two fully instrumented CRPs were built (labelled as CRP1 and CRP2 in the figure). The other two, dummy CRPs, although having the same mechanical structure and being equipped with the extraction grid, had all (CRP3) or most (CRP4) of the CRDs replaced with blank grounded copper plates. In the case of CRP4, four CRDs with anodes only covering  1x1 m$^2$ area (also visible in figure~\ref{fig:photoinside}) were installed allowing to measure the extracted charge without the LEM amplification.

The CRP assembly was performed in an ISO-7 clean room at CERN. Each fully assembled CRP was tested in a dedicated cryostat (coldbox) in the dual-phase operation mode prior to their installation in \pddp in February 2019.

\subsubsection{Charge readout detectors}
\label{ssec:det:crd}
The charge detection in \pddp is accomplished by a stack of LEM and anode PCBs integrated in the CRP CRDs. The LEM design is an outcome of several years of optimisation studies performed on small, 10x10 m$^2$, LEM prototypes~\cite{Cantini_2015}. This work led to the following global design specifications: the thickness of \SI{1}{\mm} for the dielectric of the LEM PCB; a hexagonal mechanically drilled hole pattern of \SI{800}{\micro\meter} pitch and \SI{500}{\micro\meter} diameter with a \SI{40}{\micro\meter} annular ring clearance, or rim, around each hole where the copper is removed (see figure~\ref{fig:LEM}) by micro-etching. The function of the rim is to reduce the probability of discharges around the hole edges \cite{Breskin_2009}. To efficiently instrument the required CRP area of \SI{9}{\meter\squared}, the design concept had to be scaled to construct modules of 50x50 cm$^2$ with limited inactive area around the borders of each unit to minimize the dead space. The first such large LEMs were produced for the 4-tonne Dual-Phase detector prototype~\cite{311_technical}. In that design (CFR-34), the LEMs had a 95\% active area coverage. They underwent dedicated stability tests in a vessel filled with room-temperature gaseous argon at \SI{3.3}{\bar}, a pressure resulting in a similar gas density as that of the cold vapour in \pddp. It was subsequently observed in these tests that the LEMs suffered from an increased sparking rate mostly concentrated along edges. Electric field simulations confirmed that this was due to a higher electric field present there. In order to improve the LEM operational stability, a new design, CFR-35, was developed~\cite{Eurin:2025jqt} for \pddp. The \pddp LEMs have an 86\% active area due to significantly larger dead zones around the edge consisting of a \SI{10}{\mm} wide bare FR4 border followed by a \SI{5}{\mm} wide copper strip as well the clearance around the screw holes and the HV connection (\SI{10}{\mm} in diameter opening of bare FR4 surrounded by \SI{10}{\mm} wide copper guard ring). These features are visible in figure~\ref{fig:crpdetail} (left). 

\begin{figure}[h]
     \centering
     \includegraphics[width=0.9\textwidth]{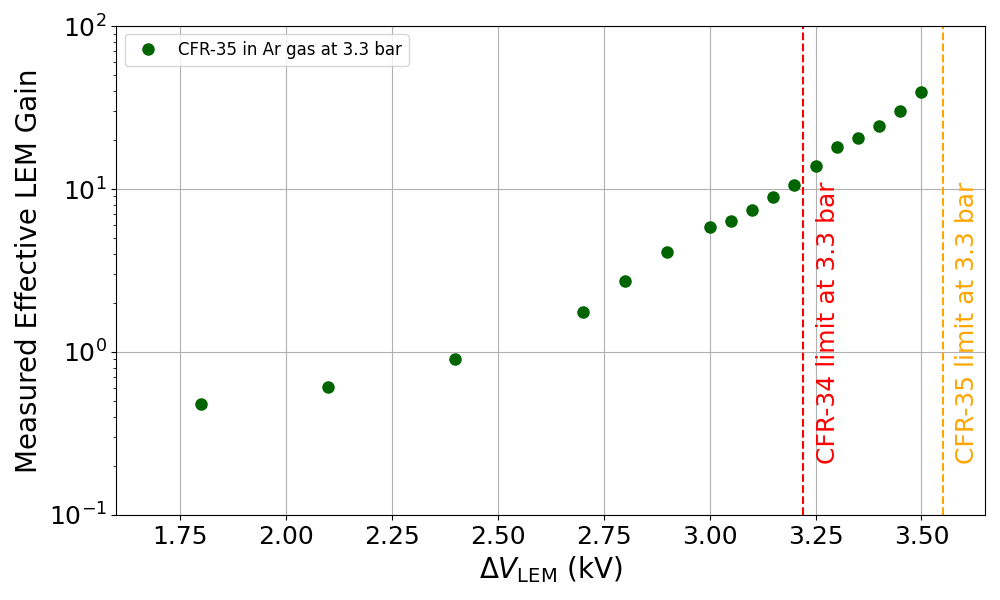}
     \caption{Effective LEM gain measured in gaseous argon at \SI{3.3}{\bar} for one of the CFR-35 LEMs as a function of the voltage difference across LEM in a high pressure chamber. The maximum voltage for each design version is indicated by dashed vertical lines. The gain is measured using \SI{5.5}{\mega\electronvolt} $\alpha$ particles from a collimated americium-241 radioactive source mounted on the cathode. The measurements are performed after the LEM has fully charged up and its effective gain is computed by normalizing the collected charge at the anode by that obtained without LEM. No extraction grid is present in the setup: the drift field is \SI[per-mode=symbol]{500}{\volt\per\cm}, defined by the potential difference between the cathode, and the LEM bottom face and the induction field is \SI[per-mode=symbol]{5}{\kilo\volt\per\cm}. }
    \label{fig:gaincurve}
\end{figure}

The tests in the high pressure gas chamber demonstrated the CFR-35 design could be operated with up to \SI{3.5}{\kilo\volt} applied across the electrodes corresponding to a gain of around 40. For CFR-34 design, this potential difference was limited to $\SI{\sim3.2}{\kilo\volt}$ (a gain of about 10).  An example of the LEM effective gain measured for one of the CFR-35 units as a function of the voltage difference across LEM, \dvlem, is shown in figure~\ref{fig:gaincurve}. 

The design of the \pddp anodes is essentially the same as the one used in ~\cite{311_technical} with some minor modifications related to the ground connections, the diameter of the holes through which the glass-ceramic (Macor) cylinders passing the HV to the LEM are inserted, and the displacement of the copper strips that were deemed too close to the border. The anodes feature a two-dimensional pattern of two sets (views) of strips orthogonal to each other. Within each view the strips have a \SI{3.125}{\mm} pitch. The studies to optimize the strip pattern have been reported in~\cite{Cantini_2014}.

Both LEM and anodes were built within industry by a PCB manufacturer (ELTOS). As LEM gain uniformity is highly sensitive to the electric field strength and shape inside the LEM holes, the production quality control was particularly focused on the LEM thickness, hole diameter, rim quality and rim dimensions. To ensure maximum thickness uniformity of the PCBs used for LEMs the complete production was done with the same batch of raw copper clad double-sided FR4 laminate (PANASONIC R-1566W). From this batch, panels were selected if they met the required specifications of the total LEM mean thickness of $1.15\substack{+0.00 \\ -0.04}$\,mm and the FR4-only thickness of $1.00\substack{+0.00 \\ -0.04}$\,mm. Each LEM was qualified by the manufacturer at critical steps of the production with quality controls performed in thirteen different areas over the LEM surface for the hole diameter ($0.50\substack{+0.01 \\ -0.00}$\,mm) and rim annular width ($40\substack{+0.00 \\ -0.04}$\,\si{\um}). The thicknesses were also measured on two metallographic cuts taken outside the LEM active area. Seventy four LEMs were manufactured over a period of 10 months and the resultant production yield was $100\%$.

Upon reception, anodes and LEMs were subject to a series of dedicated QA/QC tests at CEA/Irfu~\cite{cotte:tel-02382815} prior to their shipment for the integration into the CRP structures at CERN. The anode boards were inspected visually and tested to verify the electrical continuity of each strip. Out of 4480 strips checked for the 76 anodes produced for CRP1, CRP2 and CRP4, only eight (0.2\%) had defective continuity.

 \begin{figure}[h]
     \centering
     \includegraphics[width=0.8\textwidth]{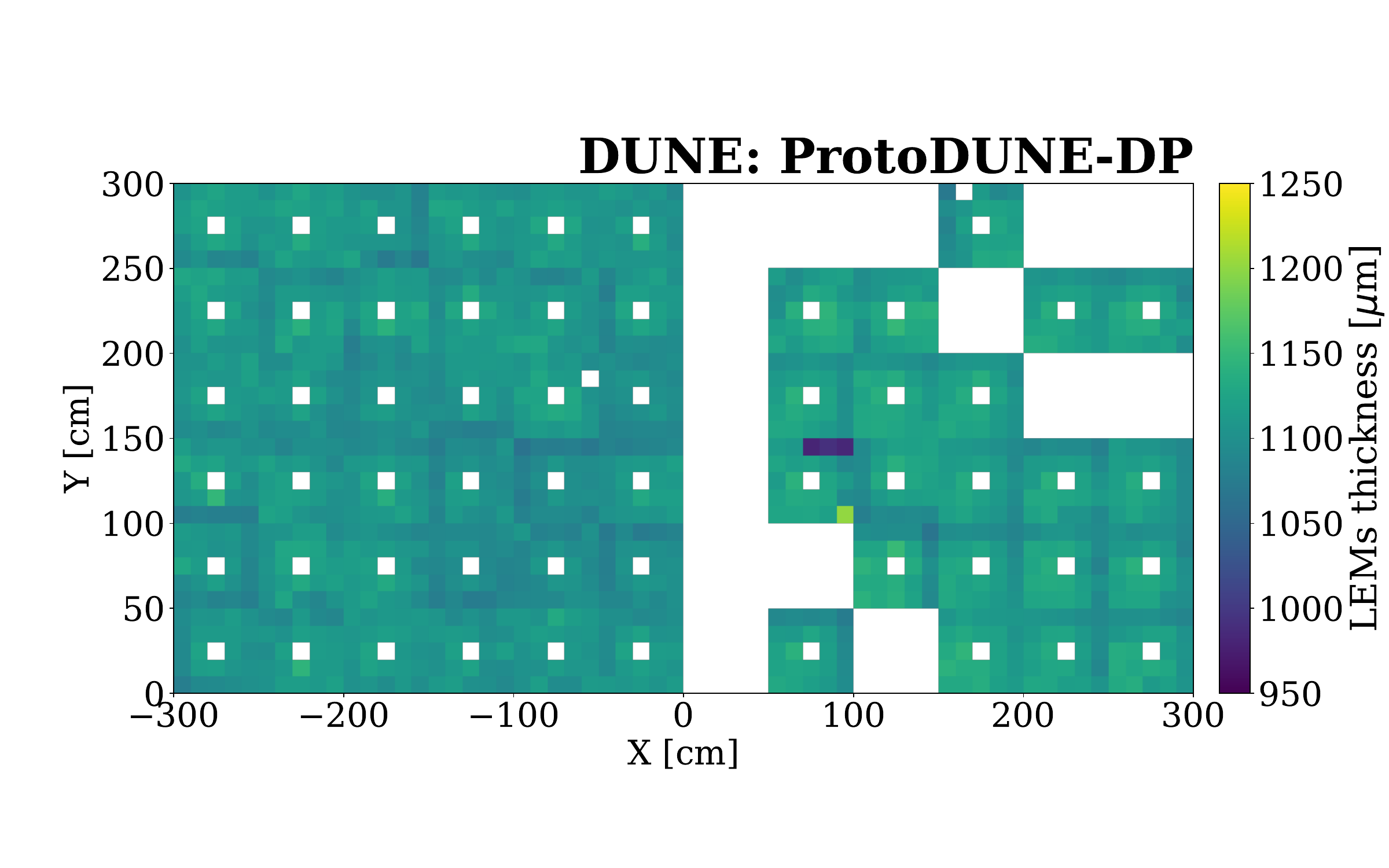}
     \caption{Total thickness (FR4 and copper layers) of the surveyed LEMs and their location in the \pddp detector. The measurements were performed across 24 regions over each LEM; the structure supporting the LEMs during the measurement prevented measurements in the central region. }
     \label{fig:lem_thickness}
 \end{figure}

The QA/QC for the LEMs was more involved. First, a sub-sample of LEMs was surveyed at Irfu on a dedicated optical bench to measure the dielectric and copper thicknesses as well as the hole geometry in 24 regions spread over the LEM surface; a summary of these measurements is shown in figure~\ref{fig:lem_thickness}. All LEMs were then cleaned in an ultrasonic bath of \textit{NGL 17.40 ALU} soap at \SI{65}{\celsius}, followed by a deionised water rinsing and a drying at \SI{80}{\celsius} for 3 hours.  They were then installed (eight LEMs at a time), in a high-pressure vessel for the last step of their qualification with HV operation. The vessel was first filled with dry air in order to check that a bias of at least \SI{4500}{\volt} could be applied across the LEM and maintained for a few hours at this HV for a burn-in of the copper electrodes with sparks to eliminate remaining debris. The dry air was then replaced by a room temperature argon gas (grade 5.7) pressurized at \SI{3.3}{\bar} in order to qualify LEMs for their operation at approximately the same argon gas density as in \pddp, with a potential difference across the LEMs of \SI{3500}{\volt} and a discharge rate of less than 3 sparks per hour (an empirical limit set for acceptance). This working point was chosen to qualify LEMs at the HV corresponding to the maximum achievable gain (figure~\ref{fig:gaincurve}) for which it could be stably operated. 

\subsubsection{Instrumentation and slow control}
\label{ssec:det:crp_sc}

Each CRP is suspended by three points with cables connected to stepper motors on the cryostat roof. The motors have an encoder step size of approximately \SI{50}{\micro\meter} allowing for precise positioning and tilt orientation of each CRP with respect to the liquid surface. Each motor supports a \SI{98}{\mm} range for the vertical adjustments. 

Four (three) parallel plate capacitive level meters (CRP level meters) with a length of \SI{25}{\mm} are installed along the two outer edges of CRP1 and CRP2 (CRP3 and CRP4). These levels meters serve to position the CRPs relative to the liquid surface, as required for the dual-phase operation, once the detector filling is complete. Their measurements can also be used by the CRP motor control system to allow an automatic tracking of the liquid level.

The LEMs are powered using CAEN \SI{8}{\kilo\volt} A1580H 16-channel HV power supplies. Each LEM bottom electrode is connected to a single power supply channel, whereas the top electrodes are connected in groups of six. All LEM power supply connections integrate a \SI{500}{\mega\ohm} current limiting resistor connected in series with the power supply. The extraction grids of each CRP are powered from a 6-channel CAEN \SI{12}{\kilo\volt} module A1524N (single power supply channel per grid).

Dedicated slow control software to control and monitor LEM HV has been developed. The program detects over-current conditions characteristic of LEM discharges and automatically lowers the voltage on the affected LEMs. For a single spark, recovery is on the order of 2 minutes given by the setting of the voltage ramp-up rate on the power supply. If, however, multiple sparks occur, the HV is reduced by up to \SI{2.5}{\kilo\volt} effectively powering down a given LEM with the aim of preventing continuous discharges that can lead to irreversible damages. Recovery time in this case can take up to two hours.

\subsubsection{Qualification in coldbox}
\label{ssec:coldbox}

Prior to their installation in the \pddp cryostat, each assembled CRP underwent testing in a realistic Dual-Phase cryogenic configuration with pure argon vapor to characterise the HV operation of the extraction grid, verify HV stability of each LEM, and test all HV connections internal to the cryostat from the feedthroughs to the LEM and grid HV connectors. For this purpose, a non-evacuable cryostat (coldbox)  with inner dimensions of approximately 3.9x3.9x1 m$^3$ had been constructed at CERN.  The cryogenic system of the cryostat did not include any recirculation and purification circuits (neither for the liquid nor for the boil-off gas argon). The decrease in the liquid level due to evaporation was about \SI{0.7}{\mm/\hour} and it was automatically adjusted by refilling the cryostat with new liquid. For CRP tests the coldbox was filled with liquid argon of commercial grade purity; the contamination of nitrogen and oxygen monitored in the boil-off argon with a gas analyser were below the 100\,ppm level (the limit of the instrument's sensitivity). The CRP qualification campaign in the coldbox lasted from June 2018 to January 2019. All four CRPs were tested. In total seven coldbox test cycles were performed: four with CRP1 and one with each of the remaining three CRPs. 

In view of the past experience with the 4-tonne DP LArTPC demonstrator~\cite{311_technical} where the extraction grid HV could not be raised to its nominal value, one of the key tests for the CRP qualification in the coldbox was therefore to verify the correct operation of the extraction grid. In the design specification, the grid must hold at least \SI{-7.5}{\kilo\volt} and its wires must have a sufficient tension to maintain their position at cold with $<\SI{0.2}{\mm}$ deviation. 

During a very first coldbox test of CRP1, shorts developed between the grid and the LEM bottom face and the maximum reachable voltage on the grid was only about \SI{-2}{\kilo\volt}. It was found to be due to insufficient tensioning of the grid wires with the latter coming in contact with some of the LEMs under the pull of the electrostatic force. The grids were originally tensioned at warm to \SI{0.6}{\newton} per wire during the CRP assembly with the tension at cold extrapolated to \SI{2}{\newton}. As the wires and the CRP G10 frame contract at different rates, this calculation had to rely on the thermal contraction measurements on a samples of a G10 material made at cryogenic temperatures and was, therefore, particularly prone to the uncertainties in the latter in particular since G10 expands at different rates along two perpendicular axes.  

The CRP1 grid tension was increased to \SI{1.5}{\newton} (\SI{3.5}{\newton} estimated at cold) by inserting a thicker spacer between the grid support boards and the G10 support structure. Additionally, the connection of the HV cable to the grid was modified to have no exposed HV line in the vicinity of argon gas. These changes were then validated in the second coldbox test cycle of CRP1, where it was successfully operated for several days in Dual-Phase mode with all LEMs powered and the grid at \SI{-7.5}{\kilo\volt}. These modifications were subsequently applied to the other three CRPs. 

In their respective tests in the coldbox, the grid of each CRP was powered at whichever voltage necessary for the maximum electron extraction efficiency, which corresponds to the voltage setting such that the potential difference between grid and the bottom LEM electrode is $\geq\SI{3}{\kilo\volt}$.

Tests of LEMs in CRP1 and CRP2 were focused on the following: 
\begin{itemize}
\item{measurements of spark rates in different HV CRP configurations;}
\item{measurements of HV recovery times;}
\item{studies of possible crosstalk effects between LEMs or groups of LEMs in an event of discharge.}
\end{itemize}
A given CRP HV configuration was considered stable when on average the LEM spark rate over the entire CRP was about one spark per hour. This translates to an average of one discharge per LEM for 36 hours of continuous operation. With a LEM HV recovery time after discharge of about \SI{10}{\second} and with a very conservative assumption that the entire CRP is inactive for this time, the corresponding dead time is under 0.3\% per CRP. 

The optimal LEM HV configuration was obtained for the top LEM electrode at \SI{-0.5}{\kilo\volt},  corresponding to an induction field of $\SI{2.5}{\kilo\volt/\cm}$ with maximum voltage across LEM reaching around \SI{3.1}{\kilo\volt}, which is a factor of 2 lower than the design target of $\SI{5}{\kilo\volt/\cm}$. The resulting decrease in the charge collection efficiency due to the weaker induction field is about 40\% with respect to the efficiency at the nominal field strength~\cite{cotte:tel-02382815}. On the other hand, the observed limit of $\sim\SI{3.1}{\kilo\volt}$ is well-below the $\sim\SI{3.5}{\kilo\volt}$ achieved in the stand-alone LEM QA/QC tests in the high pressure chamber. This inconsistency has not been resolved.

\begin{figure}[h!]
\begin{centering}
\includegraphics[width=0.8\textwidth]{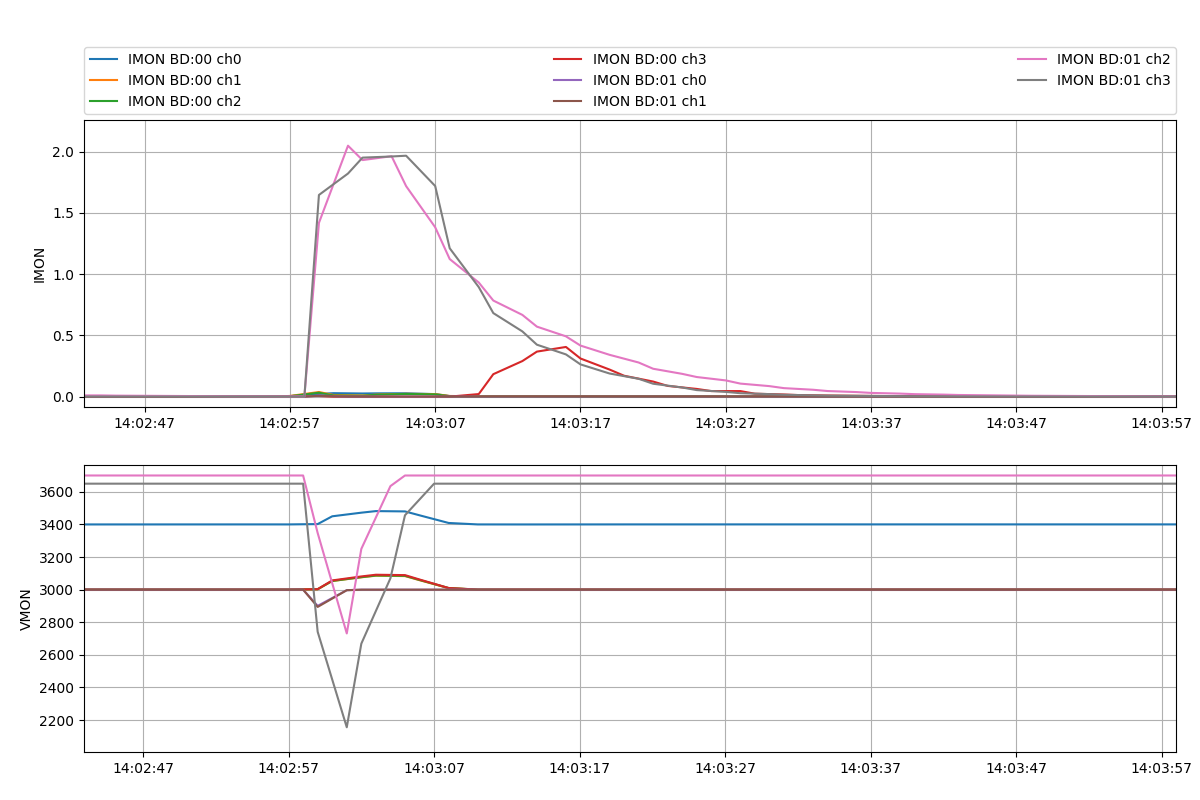}
\includegraphics[width=0.8\textwidth]{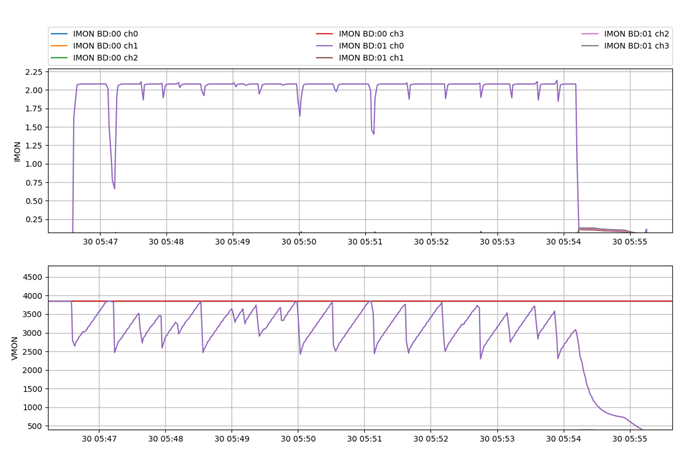}
\caption{LEM HV slow control monitoring plots showing a single discharge event of two nearby LEMs (two panels on the top) and the continuous discharges on a single LEM (two panels on the bottom). The power supply current is shown in microamperes and the voltage in volts.}
\label{fig:lemsc_cb}
\end{centering}
\end{figure}

\begin{figure}
\begin{centering}
\includegraphics[width=0.45\textwidth]{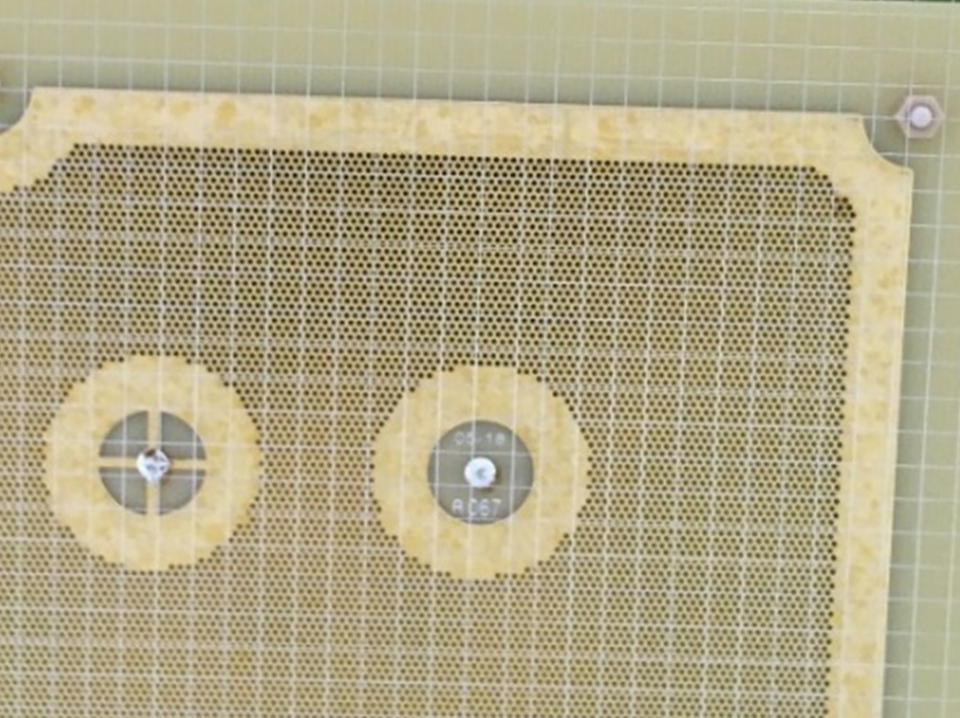}
\includegraphics[width=0.45\textwidth]{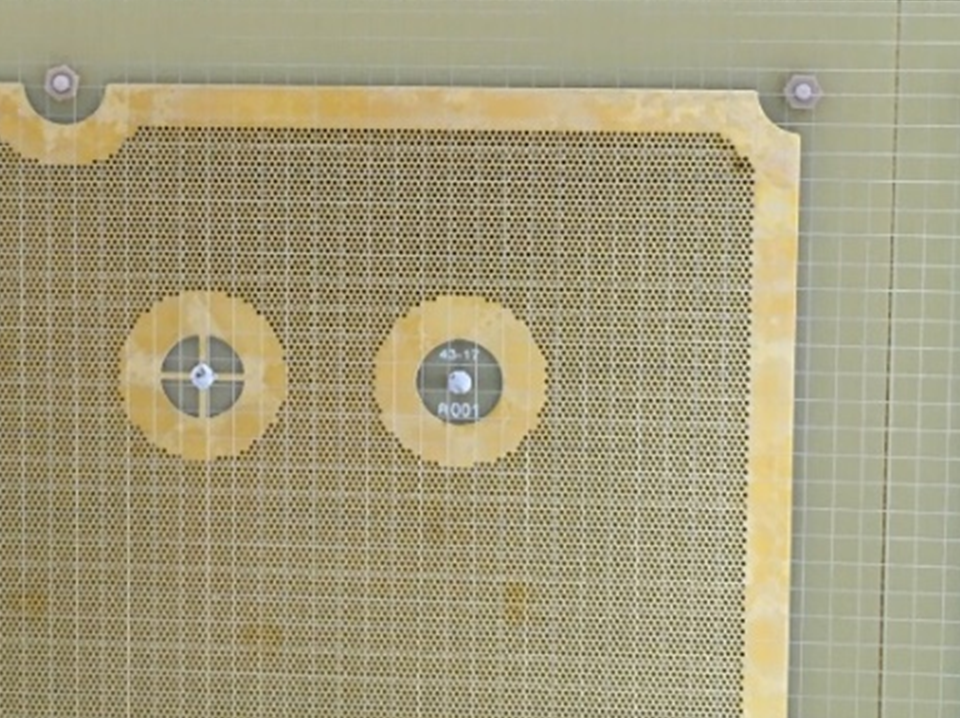}
\caption{Pictures of two LEMs that suffered long lasting continuous discharges during HV tests in the coldbox: dark spots due to carbonisation can be seen for both LEMs in their top-right corner. }
\label{fig:lem_carbon_cb}
\end{centering}
\end{figure}

In the course of the tests, continuous discharges had been observed in a few LEMs. As a consequence, the affected units had to be powered at lower voltages. Figure~\ref{fig:lemsc_cb} shows the slow-control monitoring plots of the power supply current and voltage applied to the bottom LEM face for a selection of LEMs during a single discharge event (two panels on the left) and their pattern in a continuous discharge event (the panels on the right). In total, five faulty LEMs (four in CRP1 and 1 in CRP2) developed this problem. They had been inspected after coldbox opening, and dark spots located in the corner regions of the units have been found (see figure~\ref{fig:lem_carbon_cb}). These spots are due to carbonised dielectric material inside the LEM holes. The fact that these damages are localised in the same regions of the PCB points to a problem in the manufacturing process. A detailed inspection of the damaged LEMs with a microscope found that in the boundary regions (the edges or the corners) where the hole density is no longer uniform in all directions, some holes were surrounded by de-centred rims and had a presence of a copper residue in the etched regions. These defects were understood to be linked to the wet etching process used by the manufacturer to make the $\SI{40}{\micro\meter}$ rims around the holes. 

The damaged LEMs were dismounted from the CRPs. Four of them (in CRP1) were reconditioned by cleaning them in a potassium permanganate bath and one (in CRP2) was replaced with a spare. The cleaned LEMs were re-tested in the pure argon gas at \SI{3.3}{\bar} in the high pressure chamber. With no issues detected, they were re-qualified for CRP installation and CRP1 was subsequently re-tested in the coldbox. As discussed previously, to mitigate the damage from continuous discharging, a dedicated program for the HV slow control system was developed for \pddp. 

\subsection{Charge readout electronics system}
\label{ssec:det:elec}

\begin{figure}
     \centering
     \includegraphics[width=0.9\textwidth]{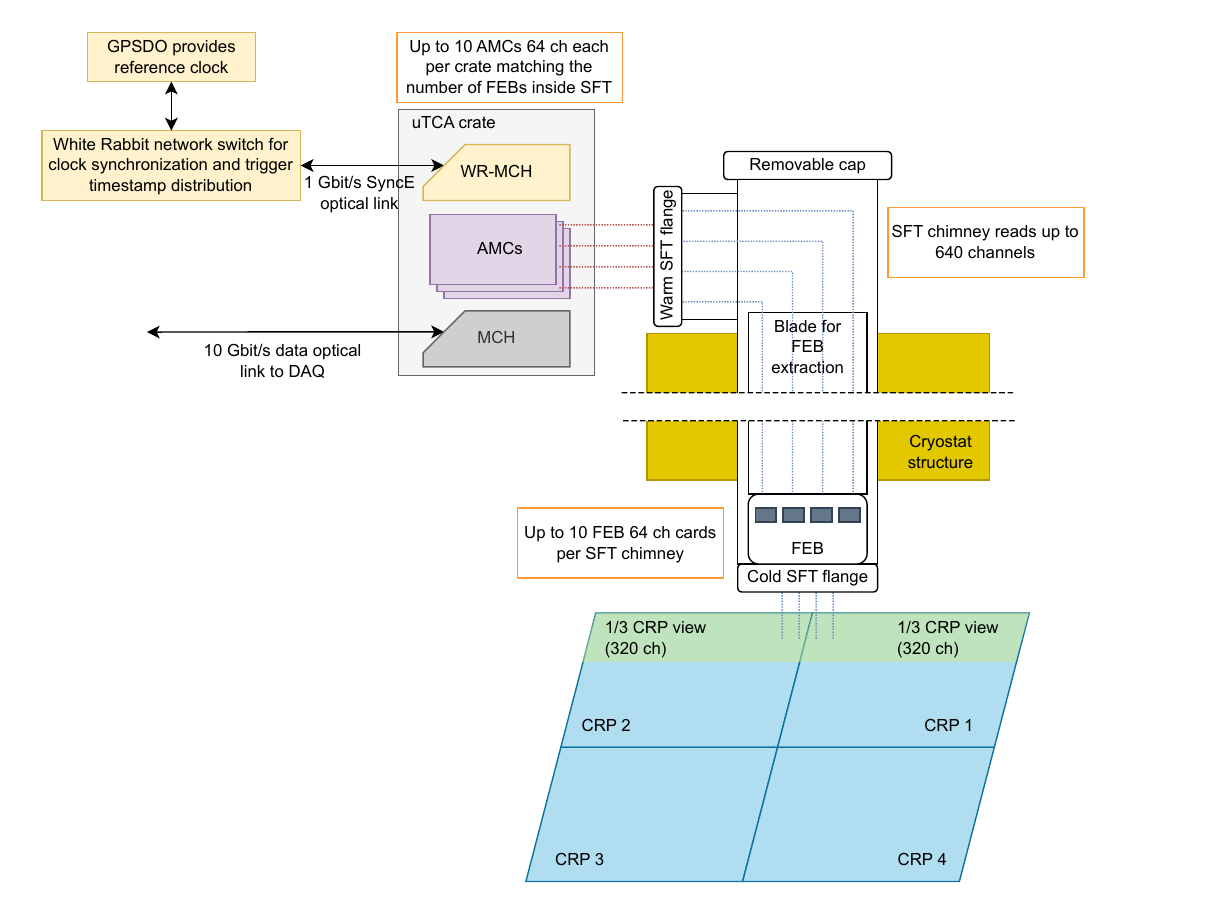}
     \caption{Schematic layout of the readout of the CRP strips by the electronics system. Each SFT chimney serves to read 1/3 of a view from two neighbouring CRPs treating a total of \num{640} channels. Each chimney is connected to a uTCA crate containing the digital electronics.}
     \label{fig:pddp_cro_scheme}
 \end{figure}

\begin{figure}
     \centering
     \includegraphics[width=0.95\textwidth]{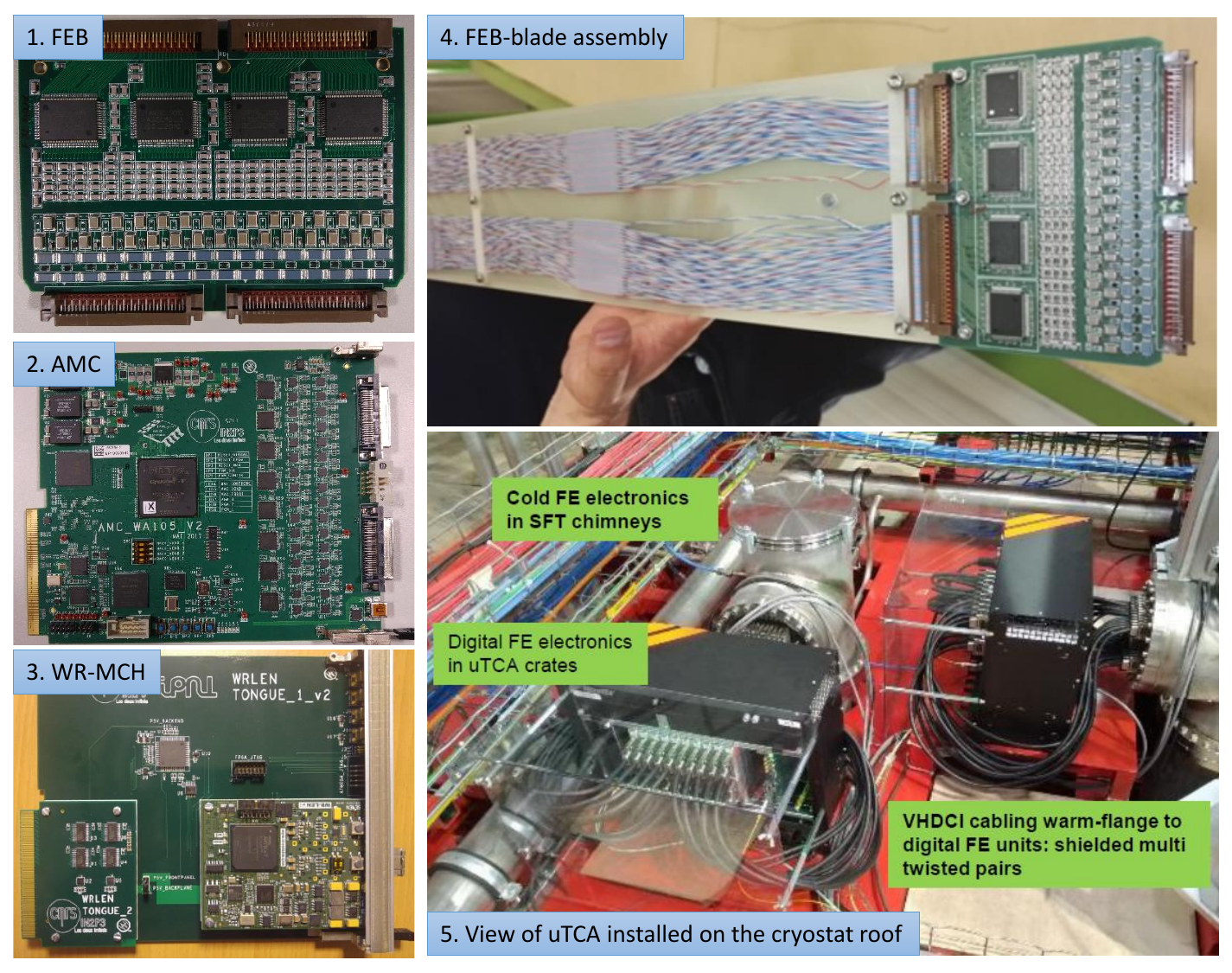}
     \caption{Images of the electronics system for the charge readout: Front End Board (FEB) (1), Advanced Mezzanine Card (AMC) (2), White Rabbit end-node (WR-MCH) (3), FEB-blade assembly (4), and a part of the cryostat roof with a couple of uTCA crates installed (5).}
\label{fig:tde_hw}
\end{figure}

 The electronics for the charge readout is a result of the long-standing development aimed at finding an integrated and cost-effective solution for large LArTPCs. For this detector, the implemented readout scheme was as foreseen for a possible DUNE Far Detector Dual Phase module~\cite{Abi:2018rgm}. The electronics components have also been successfully used to read the 4-tonne DP LArTPC demonstrator in 2017 \cite{311_technical}. Figure~\ref{fig:pddp_cro_scheme} illustrates the principle of the CRP signal readout. The analog signals from the anode strips are amplified by the Front End Board (FEB) analog electronics. The FEBs are connected to the digital electronics implemented in  Advanced Mezzanine Card (AMC) specification and hosted on the roof of the cryostat in uTCA crates. Each FEB reads 64 anode strips and each AMC digitises 64 channels from a given FEB. FEBs operate at cryogenic temperature at the bottom of the Signal Feedthrough (SFT) chimneys, which are \SI{0.25}{\meter} diameter pipes approximately \SI{2}{\meter} long that traverse the entire thickness of the cryostat roof insulation. An SFT chimney can host up to 10 FEBs. A single uTCA crate is associated to each SFT chimney with up to 10 AMCs digitising the data from their respective FEBs. Each crate sends digital data to a DAQ backend system via a dedicated 10\,Gbit/s optical fibre. As shown in figure~\ref{fig:pddp_cro_scheme}, a chimney reads two groups of 320 anode strips (each representing 1/3 of the channels in a given view and CRP) from two nearby CRPs. For a full detector with four active CRPs and \num{7680} anode strips, this would correspond to having 12 chimneys and 12 uTCA crates. In the actual implementation with only two fully active CRPs (CRP1 and CRP2) equipped with LEMs and one partially equipped with only anodes 1x1 m$^2$ area in CRP4), 10 SFT chimneys (uTCA crates) are used with some containing only 5 FEBs (AMCs). Figure~\ref{fig:tde_hw} shows images of several components of the charge readout electronics system. 
 
\subsubsection{Signal feedthrough chimneys} 

\begin{figure}
    \centering
    \includegraphics[width=0.95\textwidth]{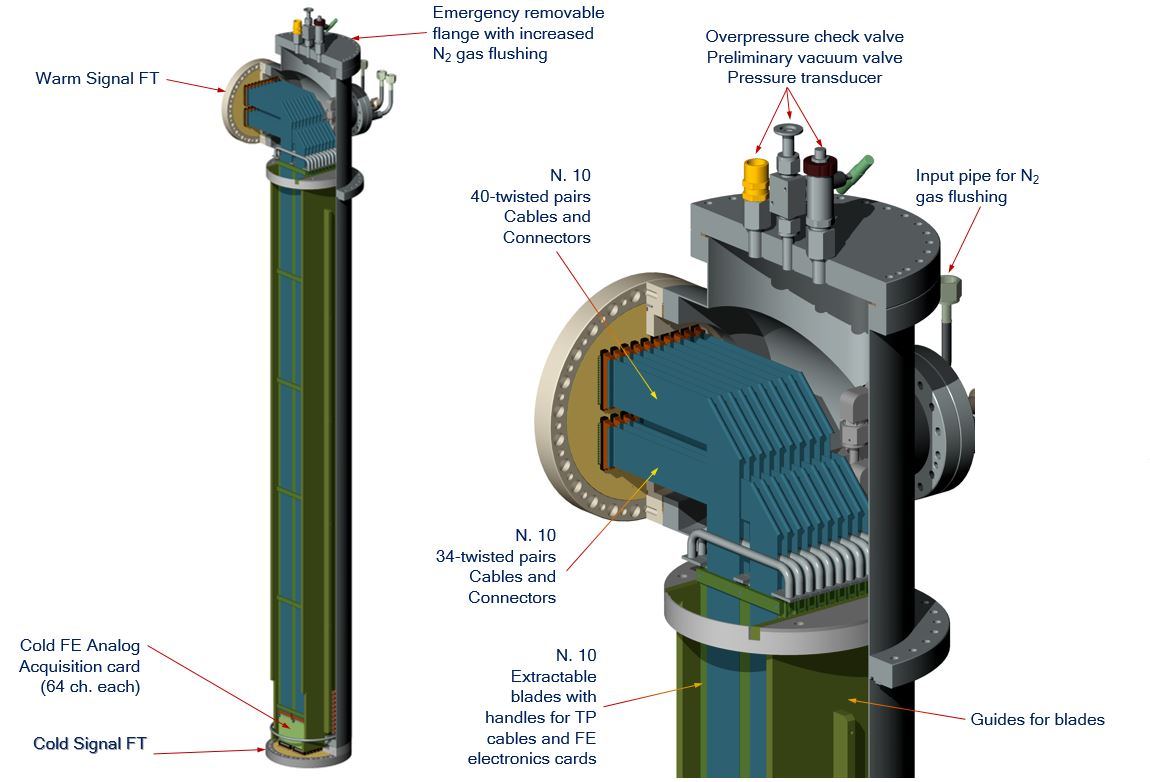}
    \caption{Details of Signal Feedthrough (SFT) chimney design. The left image shows the model of the entire body with the cutout illustrating the elements inside. The zoomed image on the right gives the view of the structure around the warm signal flange.}
    \label{fig:sft_design}
\end{figure}

The SFT chimneys allow placement of the FEBs close to the CRPs limiting the parasitic cable capacitance at the preamplifier input, whilst allowing access for the FEBs replacement/repair at any time throughout the lifetime of the experiment. Various details of the chimney design are illustrated in figure~\ref{fig:sft_design}. The principal part of the body is formed by a machined stainless steel tube \SI{254}{\mm} in (external) diameter and about \SI{1.8}{\meter} long. Once installed, the chimney crosses the entire insulation volume of the roof of the cryostat. It is isolated from the cryostat and exterior by vacuum tight flanges referred to as \textit{cold} (cryostat) and \textit{warm} (exterior) flanges, respectively. These dispatch the signal and slow control lines to/from the FEBs. During operation, the chimney inner volume is filled with nitrogen gas to avoid condensing the humidity on the electronics inside. 

A rail system inside the chimney acts as a guide for each FEB to its connectors on the inner face of the cold flange. The FEBs are mounted on \SI{1.7}{\meter} long FR4 blades (image 4 in figure~\ref{fig:tde_hw}) and the entire assembly glides along the rails. The blades also support flat cables that transmit amplified analog signals from the FEBs to the digital electronics outside of the SFT chimney via the warm flange as well as bring low voltages and slow control lines to FEBs. The FEB extraction and replacement has been performed on many occasions during the detector operation without encountering any mechanical problems or introducing any detectable contamination in the inner cryostat volume.

\subsubsection{Analog chain}

\begin{figure}
     \centering
     \includegraphics[width=0.95\textwidth]{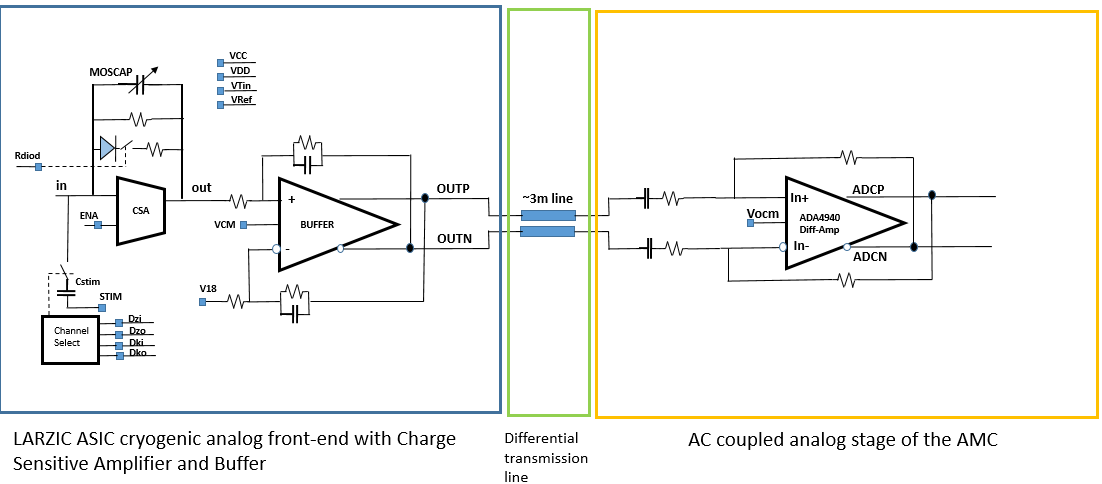}
     \caption{Synopsis of the analog part of the charge readout including the preamplifier ASIC and the analog stage (ADC buffer) on AMC.}
     \label{fig:analog_synopsis}
 \end{figure}

Each FEB contains four 16-channel custom cryogenic preamplifier ASICs implemented in CMOS \SI{0.35}{\micro\meter} technology. Figure~\ref{fig:analog_synopsis} shows a single channel synopsis of the analog part of the readout chain that includes an FEB ASIC channel consisting of Charge Sensitive Amplifier (CSA) and a differential output buffer stage acting as a low-pass filter. The differential analog signals are received in the input analog buffer stage on the AMC side that feeds into Analog-to-Digital Converters (ADCs). 

In order to cope with potentially high gain of the CRP LEMs, the CSA has been developed to have a linear response for input charge of up to \SI{\sim400}{\femto\coulomb} (about $100\times$ the charge deposited by a minimally ionising particle  on a strip) and an approximately logarithmic response in the \SIrange{400}{1200}{\femto\coulomb} range. This change in response is achieved by using a MOSCAP capacitor in the amplifier feedback loop that changes its capacitance above a certain signal threshold. The MOSCAP has a lower gain for input charge above \SI{400}{\femto\coulomb}. The feedback loop also contains a selectable branch (activated with "Rdiode" slow control line) with a resistor in series with a diode. The activation of this branch of the circuit allows to attain similar discharge times for signals below and above \SI{400}{\femto\coulomb} by maintaining the RC of the feedback circuit at about \SI{500}{\nano\second} in both regimes. 

As shown in Figure~\ref{fig:analog_synopsis}, the ASIC integrates a charge injection system consisting of an embedded capacitor ($C_{\text{stim}}$\,=\,\SI{1}{\pico\farad}) and an associated slow control system. The latter allows activating a single channel or a desired configuration of channels for signal injection by connecting them to a stimulus (STIM) 
line delivering the calibration pulse from an external generator. The configuration of the channels for the injection is set with a Serial Peripheral Interface (SPI) bus. Each ASIC has input and output connections for the data (Dzi, Dzo) and clock (Dki, Dko) lines of the SPI bus. These connections allow daisy-chaining on the SPI bus the ASICs mounted on the same front-end card. The possibility of activating the injection single channels allows verifying that the crosstalk between ASIC channels is negligible ($<1\%$).

\subsubsection{Digital stage}
\label{ssec:det:dstage}

The AMCs digitise analog data received via shielded VHDCI (Very-High-Density Cable Interconnect) cables which connect them to the warm flange. Each uTCA crate has a central switch hub, MCH (MicroTCA Carrier Hub), that supports 10\,Gbit/s bandwidth. It is connected via an optical link to the back-end DAQ system (section~\ref{ssec:det:daq}). The AMC transmits the digital data in Ethernet UDP packets to the MCH on its own dedicated 10 Gbit/s XAUI (X Attachment Unit Interface) lane in the crate back-plane. The uTCA crate also hosts a White Rabbit end-node (WR-MCH) that allows for the synchronisation of the AMCs to a global reference clock. The WR-MCH integrates as a mezzanine card on a uTCA adapter board a commercial White Rabbit end-node card (Seven Solutions WR-LEN\footnote{https://sevensols.com/home/timing-products/wr-len/}).
The latter allows the use of the uTCA back-plane lanes, reserved for a possible second uTCA MCH switch, to distribute the reference clock and timing data to the AMCs. The WR-MCH is connected to the White Rabbit network (described in~\ref{ssec:det:daq}) via a dedicated 1\,Gbit/s SyncE optical link. 

\begin{figure}
     \centering
     \includegraphics[width=0.9\textwidth]{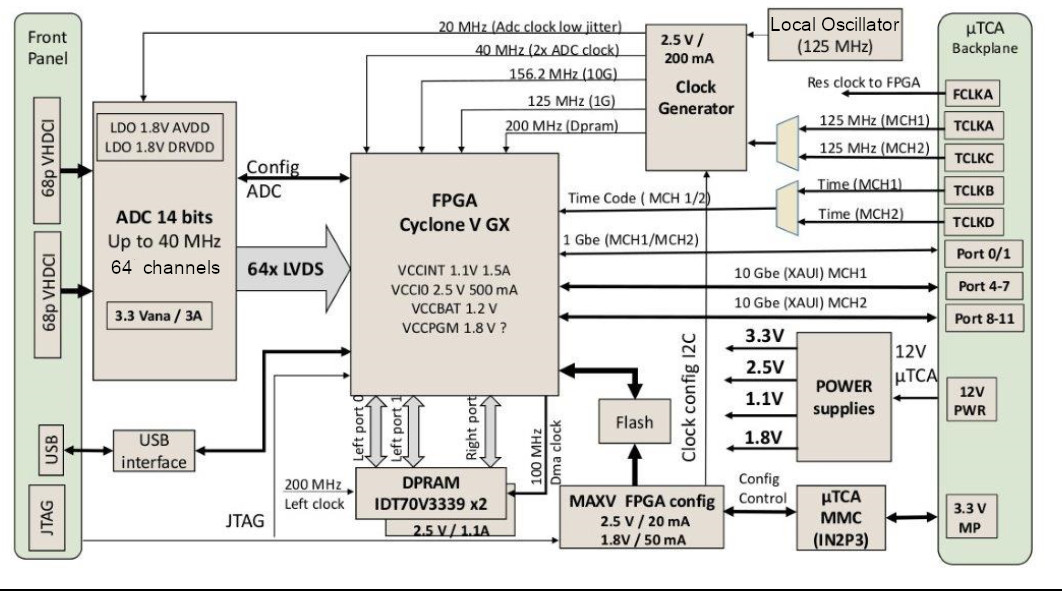}
     \caption{Block diagram illustrating the elements and their functionality of the Advanced Mezzanine digitisation Card (AMC).}
     \label{fig:amc_block}
 \end{figure}

Figure~\ref{fig:amc_block} shows the function block diagram of the AMC digitisation card. The uTCA backplane provides: the power to the card; the access to the XAUI 10\,Gbit/s lanes; the access to the 125\,MHz clock reference and timing data from WR-MCH. The analog data are received via two 68\,pin VHDCI connectors on the AMC front panel. They are fed into analog buffers (ADA4940) at the input of eight octal Analog-to-Digital Converters (ADC). These latter are 14-bit ADCs (AD9257), which digitise the data at a rate of 20\,MHz. 

Digital processing is performed in FPGAs (Altera Cyclone V GX). The data are downsampled to 2.5\,MHz (equivalent to $<1$ mm resolution along the drift coordinate) and only the 12 most significant bits are retained. At a reception of the trigger timestamp packet via WR-MCH, a search for the corresponding ADC sample, trigger-tagged sample, in the internal FIFO buffers implemented in dual-port RAM memories (DPRAM in figure~\ref{fig:amc_block})  is performed. A sequence of \num{10000} ADC samples starting from the trigger-tagged sample are then transmitted. This size of the readout window corresponds to \SI{4}{\milli\second}, a time window that covers the maximum drift time over \SI{6}{\meter} distance in the nominal \SI{500}{\volt\per\cm} drift field. 

\begin{figure}[h!]
     \centering
     \includegraphics[width=0.98\textwidth]{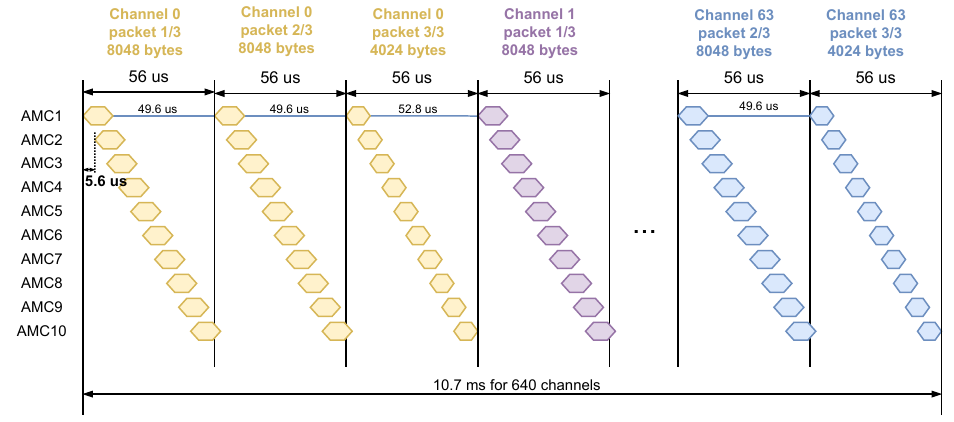}
     \caption{Advanced Mezzanine Card packet pipelined transmission scheduling scheme to operate close to 10 Gbit/s link bandwidth saturation.}
     \label{fig:data_pkt_schdule}
 \end{figure}

The data are transmitted in User Datagram Protocol (UDP) packets and, given the size of the readout window, three jumbo UDP packet frames are required to send the data from each channel. The packet transmission follows a deterministic pipeline scheduling scheme, illustrated in figure~\ref{fig:data_pkt_schdule}, that saturates the MCH 10\,Gbit/s link capacity and transmits event data as quickly as possible to maximize the attainable trigger rate given the available bandwidth. The data packets are sent out every \SI{5.6}{\micro\second} by the ten AMCs in a given crate with each card transmitting a packet every \SI{56}{\micro\second} ({\it i.e.}, equal to $10\times \SI{5.6}{\micro\second}$). The value of the delay between packet transmission is a configurable parameter that is specified at runtime. This flexibility allowed tuning its value to find an optimal point (namely, \SI{5.6}{\micro\second}) where data could exit the MCH switch without any packets being dropped due to the link over-saturation. The transmission of the data for all 640 channels then takes a total of \SI{10.7}{\milli\second} corresponding to a maximum possible trigger rate of \SI{93}{\hertz}. However, in reality the trigger rate was measured to be limited to \SI{70}{\hertz} due to additional latencies stemming from preparation of the packets, treatment of the trigger packets by the DAQ back-end processes, and various handshaking protocols of the AMCs with the DAQ back-end servers. Nonetheless, the scheduled transmission with instantaneous saturation of the  10\,Gbit link bandwidth was stable without any detectable packet loss. 

While operating at a high trigger rate is not mandatory when collecting cosmic ray data, it is essential for the operation with the test particle beam since beam-time is limited. The pipeline transmission scheme was used during most of the detector data taking campaigns with cosmic-ray data. However, an AMC firmware implementing lossless compression using an optimised Huffman encoding scheme has also been developed and deployed near the end of the detector operation. In this firmware version, the data are compressed prior to the transmission by 64 compressor modules, one per each AMC channel, running in parallel to allow data processing in real-time. An order of 5 in data volume reduction was achieved making it possible in the future to exploit the 10\,Gbit/s links at higher trigger rates or even in continuous data streaming mode. 

\subsubsection{Component production and quality control}

Instrumenting the readout of the full \pddp detector with four active CRPs required the production of 120\,FEBs and 480\,ASICs as well as 120 AMCs. This represented a total of \num{15360} analog and digital channels. Such a large-scale production of the electronics components required an implementation of a dedicated Quality Control (QC). Its aim was to systematically verify the functionality of all the analog and digital channels. For this purpose, a dedicated test stand was constructed where both analog and digital cards could be tested. This test stand represented a miniaturized version of the entire DAQ chain of the \pddp detector including the necessary timing/trigger distribution system and a DAQ server running the acquisition from up to two uTCA crates at the same time. A dedicated set of software tools was developed to provide an online analysis of the raw data to quickly assess the functionality of a given electronics component undergoing QC tests. 

\begin{figure}
     \centering
     \includegraphics[width=0.45\textwidth]{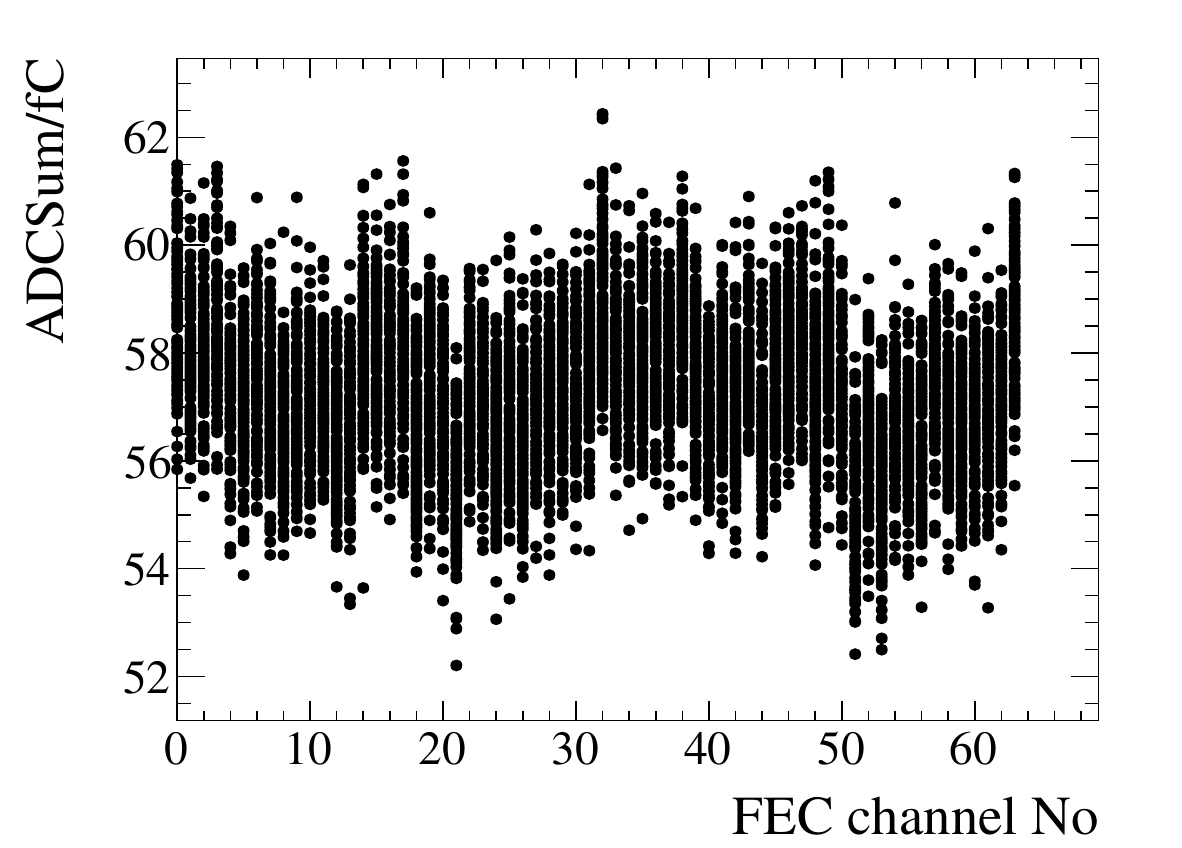}
     \includegraphics[width=0.45\textwidth]{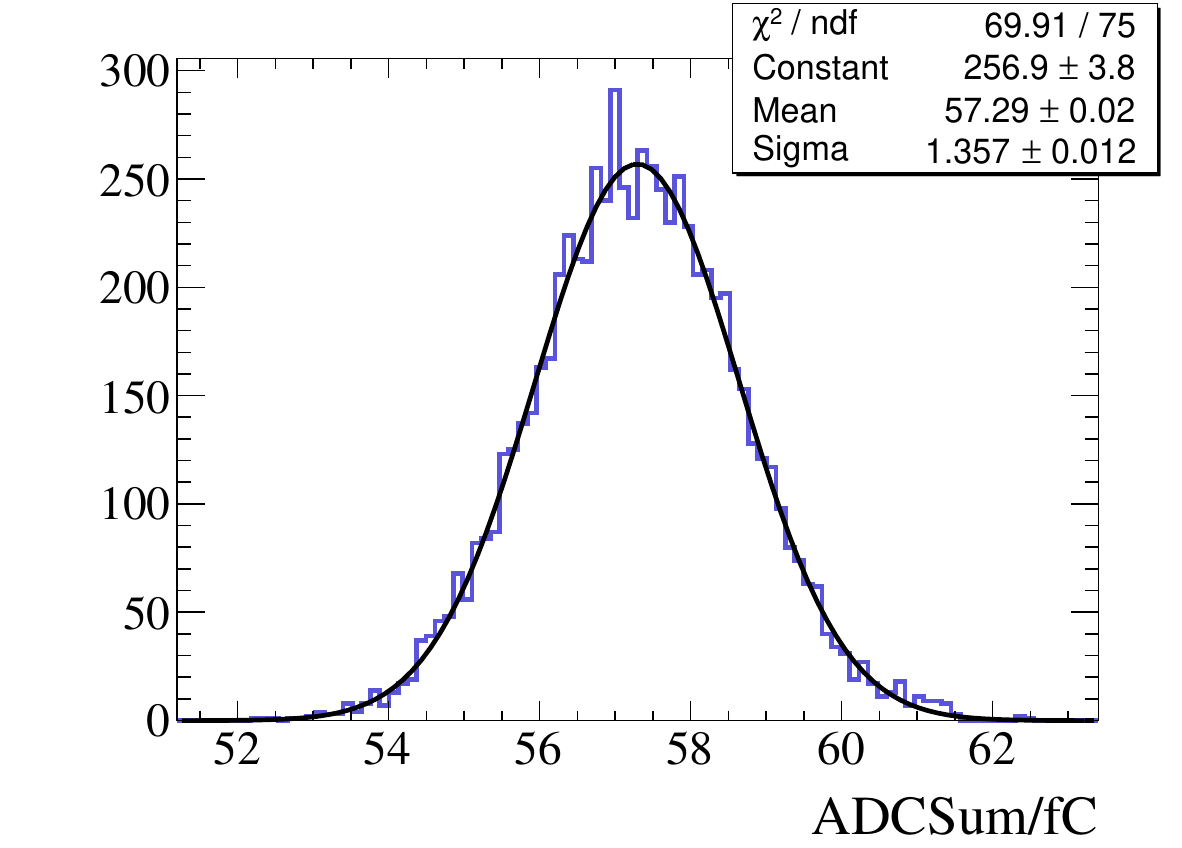}
     \caption{Channel-to-channel gain uniformity measured at room temperature in the ASIC linear response region of the FEB cards produced for \pddp. The plot on the left illustrates the gain (measured as pulse integral, \textit{ADCSum}, per unit of injected charge) as a function of FEB channel number for all tested cards. The plot on the right shows the overall distribution of the obtained gains of all tested analog channel fitted to a Gaussian function.}
     \label{fig:feb_pddp_prod_qc}
 \end{figure}
 
The primary focus of the QC campaign for the electronics was the validation of the FEBs and AMCs. The FEBs were tested with ASICs already mounted as there was no significant evidence of any substantial failure rate for the latter due to, for example, chip-packaging bonding. To test the FEB response as a function of well-calibrated injection charge and verify the overall signal continuity, a dedicated calibration card was developed. It contained a bank of 32 mechanically-tuned injection capacitors, each trimmed to within $<2\%$ from the mean \SI{1.1}{\pico\farad} target value. A precision pulse generator provided an excitation pulse to a given channel on the calibration card selectable by a 32:1 MUX. The latter was controlled via an SPI bus by a Raspberry Pi. 

Figure~\ref{fig:feb_pddp_prod_qc} illustrates the gain response, obtained from measuring the sum of ADC samples of calibration pulses collected for a given FEB channel for several values of injected charge, for the ensemble of the produced FEBs. The systematic variations in the response, visible in the left-hand plot, come from channel-to-channel variations of the calibration card itself. Overall, including this systematic effect, the gain uniformity for all the channels is at a $2\%$ level. 
The QC tests of AMCs consisted in running the acquisition to check:
\begin{itemize}
\item{the transmission of data packets without any losses;}
\item{the integrity of the transmitted data.}
\end{itemize}
The last point was determined by analysing the data to measure the pedestal value of each channel and verify that it was within an acceptable range. 
 
On the mechanical side of the readout system, twelve SFT chimneys were built at CERN to support the readout of the full detector with four active CRPs. Prior to their installation in the detector cryostat, the chimneys were tested for leaks at the cold flange interface as it is the most sensitive part of the assembly given that it is in direct contact with the inner volume of the cryostat. All chimneys were qualified at room temperature using a vacuum system consisting of a vacuum pump and a helium leak detector connected in parallel. A chimney under test was attached to the vacuum system and evacuated until the pressure inside reached a level of $\mathcal{O}(10^{-3})$\,mbar. The leak detector was then activated and helium was sprayed in the vicinity of the cold flange. The chimney tightness was verified down to a leak rate of a couple of \SI{e-9}{\milli\bar \liter \per \second}, which was the level of the background for the measurement. 
 
Three randomly selected chimneys were tested further at cryogenic temperatures. Each chimney from this set was inserted into an open dewar and reconnected to the same vacuum system as used for operation at room temperature. It was evacuated to the \SI{e-3}{\milli\bar} level and LAr was then added to the dewar. The proximity of the cold flange to the liquid surface was adjusted to achieve a temperature at the cold flange of \SI{\sim100}{\kelvin}, similar to the one expected for the operation in the cryostat. A new round of He leak tests was then performed and it was verified that the leak-tightness at the level of \SI{e-9}{\milli\bar \liter \per \second} was still maintained whilst cold for all tested chimneys.
 
\subsection{Charge data acquisition}
\label{ssec:det:daq}

\subsubsection{DAQ back-end system}

\begin{figure}[h!]
\centering
\includegraphics[width=1\textwidth]{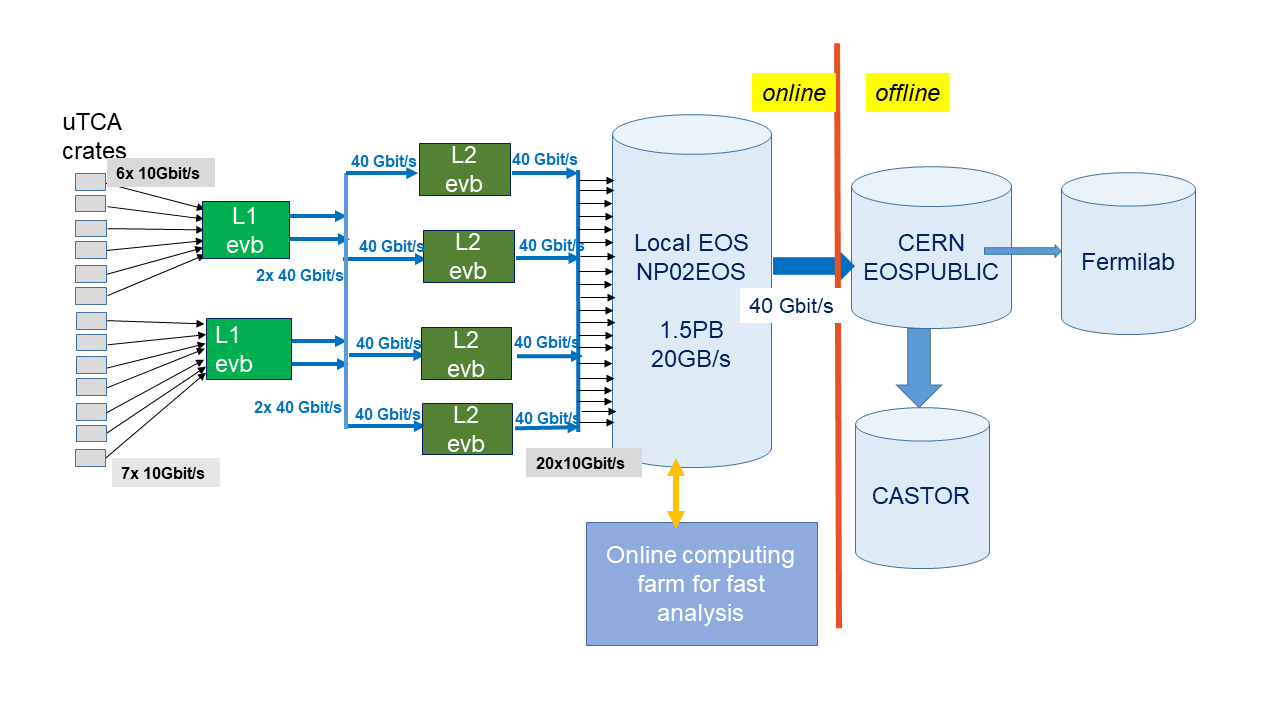}
\caption{\pddp DAQ back-end architecture. The orange vertical line shows the boundary between online and offline.}
\label{fig:be_architecture} 
\end{figure}

Two levels of event builder machines (L1 and L2 event builders), the back-end network infrastructure, the online storage and processing facility are the main elements of the back-end system, as shown in figure~\ref{fig:be_architecture}.

The DAQ back-end system receives the data from the uTCA crates via dedicated Ethernet links, builds from these data the events corresponding to each trigger and assembles several events in files which are stored on the online EOS distributed storage system, integrated in the NP02 back-end. These tasks are supported at a very high data bandwidth of 20\,GB/s, which may support trigger rates up to 100\,Hz, without data compression. 

The data files can be then transferred to the CERN EOS storage and Fermilab by FTS ~\cite{fts} for permanent storage. An online processing farm is also embedded in the \pddp back-end system in order to ensure fast reconstruction of the events for an online assessment of the DAQ data quality and detector operation and performance.

Two L1 event builder machines receive the data flow from the front-end digitisation system (section ~\ref{ssec:det:dstage}) via several dedicated 10\,Gb/s data links, one link from each uTCA crate. Each L1 event builder reads one half of the uTCA crates constituting the digitisation front-end system.
The two L1 machines (DELL R730) have the same technical specifications: 384 GB of RAM; two Intel R710 network interface cards, each supporting four 10 Gb/s data links, and two Mellanox ConnectX-3 network interface cards, each featuring two 40 Gb/s Ethernet QSFP+ ports for connecting the event builder to the back-end infrastructure network; and an Intel Xeon Gold 5122 processor running at 3.6 GHz, with 4 cores and 8 threads.

The task of each L1 event builder is to assemble the data received from the connected uTCA crates corresponding to the same trigger in an event and write this file in its RAM disk in order to make it accessible to the level 2 (L2) event builders. The L2 event builders  put together the two event halves in a single event. Four 40\,Gb/s Ethernet links connect  the two L1 event builders to a router which also interconnects four L2 event builders. In particular the RAM disks of the L1 event builder are mounted on each L2 server as network disks, by using the NFS protocol. Each of the four L2 event builders (DELL R730) is equipped with 192 GB of RAM, two Mellanox ConnectX-3 network interface cards, each featuring two 40 Gb/s Ethernet QSFP+ ports for connection to the back-end infrastructure network, and an Intel Xeon Gold 5122 processor running at 3.6 GHz, with 4 cores and 8 threads.

To ensure the required bandwidth, the four L2 event builders operate in parallel. Each L2 event builder first assembles the raw data in data files containing multiple events in its RAM memory. Once a data file is completed and closed, it is then transferred to the online storage facility (NP02EOS) by a dedicated process (L2EOS) running on each L2 event builder. 



NP02EOS is a high performance EOS based distributed storage system ~\cite{eos}. The system ensures a 20\,GB/s total bandwidth for data storage on disk and it is  composed of 20 storage servers. Each storage server is a  DELL R510 machine with  72\,TB disk space. The 20 storage servers, which are individually connected to a dedicated router with 10\,Gb/s links, provide a total disk space up to 1.44\,PB. The EOS version running on the NP02EOS instance is updated with the one running on CERN EOSPUBLIC. 
The interconnections between the L1 and L2 event builders and the online storage servers are insured by two Brocade routers

The raw data files assembly process by the event builders and their transfer to NP02EOS is performed by a customized software, specifically developed for taking into account the network configuration and the characteristics of the event builders. 

\subsubsection{Timing system}
Timing and synchronisation messages are distributed by a White Rabbit (WR) system ~\cite{WR}, operating on a dedicated synchronous 1\,Gb/s Ethernet network.
A commercial WR switch is fed with 10\,MHz and 1\,PPS (Pulse Per Second) signals provided by a GPS disciplined oscillator. The switch is then used as a Grand Master, providing a common reference clock to all connected nodes. 
The WR system performs periodic self-calibrations to account for propagation delays achieving sub-ns accuracy on the clock synchronisation between distant nodes;  trigger timestamps generated from external sources can also be transmitted through the WR network.

Each uTCA crate is equipped with a timing end node (WR-MCH) which distributes timing and synchronisation messages via dedicated lanes in the crate backplane to each  AMC.
The card contains a WR worker node card, the White Rabbit Lite Embedded Node (WR-LEN), as a mezzanine that runs on a customised firmware enabling it to decode the trigger timestamp data packets (see section~\ref{ssec:det:dstage}).

The trigger timestamps are generated by a White Rabbit Time Stamping Node (WR-TSN) card hosted in the trigger server. The FMC-DIO ~\cite{fmcdio} mezzanine accepts TTL-level signals from external trigger systems.  Once a trigger signal is received, FMC-DIO  generates its timestamp, such that  a dedicated timestamp data packet is sent over the WR network (via the SPEC card ~\cite{spec}) for the connected WR-MCH nodes triggering the detector readout.  

As the light readout data is acquired and stored separately from the anode charge data, the WR SPEC card enables the possibility to synchronize both data acquisitions. This card generates a timestamp on reception of the global trigger signal. This timestamp is incorporated in the light readout data stream for each trigger to enable the matching with the charge data.

\subsubsection{Online data processing system}
\label{ssec:det:prompt}
Once a raw data file is copied to the online storage facility, it is   automatically scheduled for fast online reconstruction on the online processing farm. This farm is composed of 40 Poweredge C6200 servers, corresponding to 450 cores (9270 HES06 computing unit operated under the SLURM workload manager). 

\begin{figure}
\centering
\includegraphics[width=1\textwidth]{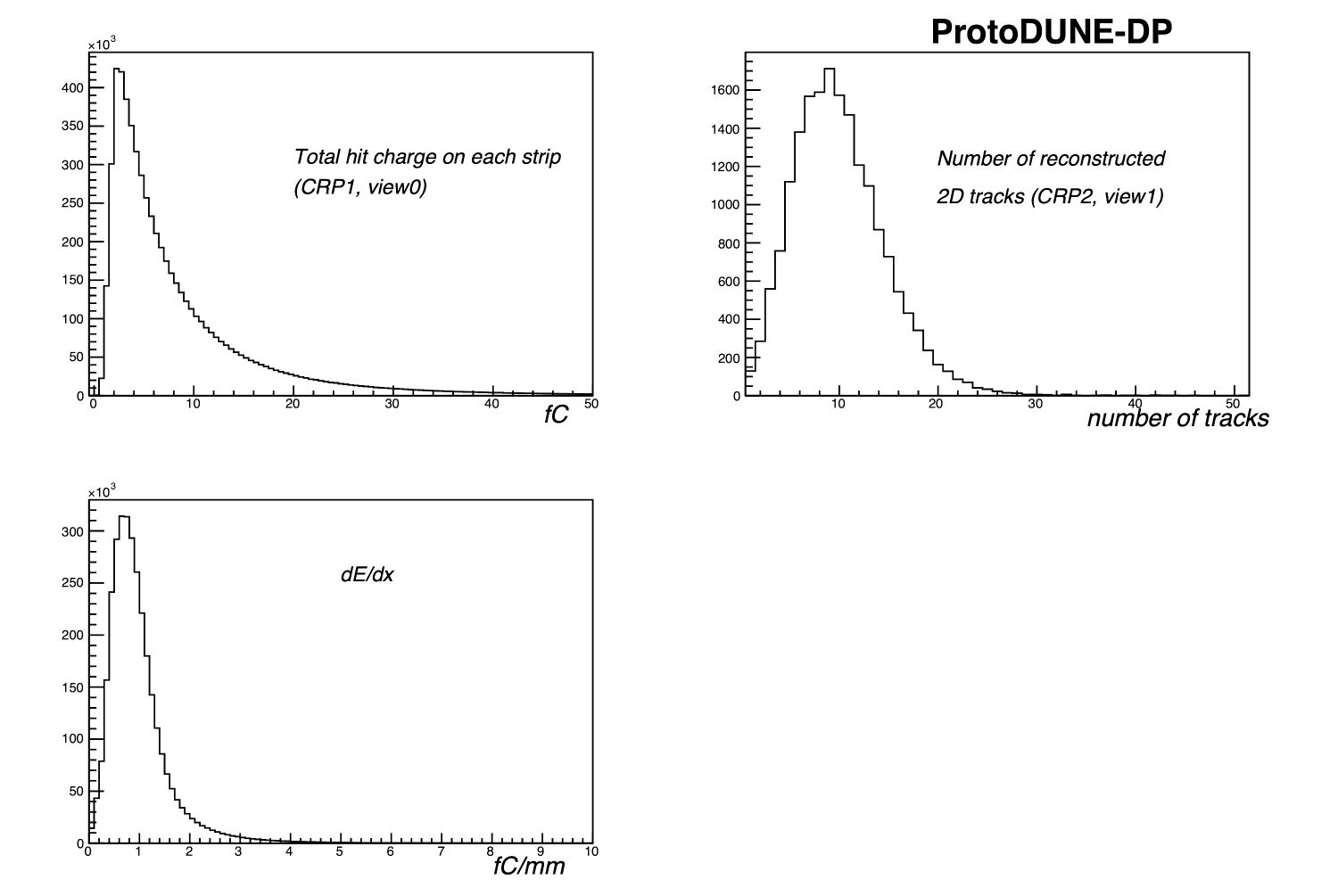}
 \caption{Examples of distributions from data quality monitoring output: total charge on each strip, number of reconstructed 2D tracks, deposited energy per unit pathlength.}
\label{fig:online_dqm} 
\end{figure}

The  time interval between the assembly of a file by an event builder and the availability of the online reconstruction results is of the order of 15 minutes. All events are systematically processed. Hits, 2D tracks and 3D tracks are reconstructed using a fast, simple and robust  reconstruction  software which was already used for the analysis of the 4-tonne demonstrator~\cite{311_technical}  data. The average time to process  one event is around   15 seconds.  The online reconstruction output is used to produce a standard set of distributions for Data Quality Monitoring.  Some examples of DQM distributions are shown in figure~\ref{fig:online_dqm}.

\subsubsection{DAQ software}
 
To operate the DAQ back-end system, several online software applications have been developed to perform the following tasks:
\begin{itemize}
\item Data acquisition and event building: LARGUI, running on the L1 event builders, collects data from the front-end AMC digitisation cards in the uTCA crates and formats data fragments associated with the same trigger into events; L2-merge reads these fragments from the L1 RAM disks and builds complete events on the L2 event builders. 
\item Data storage and transfer: writing of  the raw data files  on the local EOS system and scheduling raw data transfer to CERN and Fermilab
\item Run Control and system synchronisation: supervision of the L1 and L2 servers, control of run start and stop, and synchronisation of the back-end system components
\item Online monitoring: event display using events subsampled from the L2 servers 
\item Online processing management: automatic orchestration of online processing (fast track reconstruction) and submission of dedicated reconstruction jobs to the online farm workers.
\item Online database: logging of all events related to back-end system operation.
\end{itemize}

Intensive testing campaigns performed well in advance of the detector operation ensured smooth handling of the \pddp data by the DAQ software and back-end system.
The delay between the creation of a raw data file and its availability on EOSPUBLIC is 10 minutes. All steps of raw data handling are stored in a dedicated online database. The monitoring of the activity of the DAQ machines, storage and processing farm is performed with two dedicated  Grafana~\cite{grafana}  dashboards 

\subsubsection{Offline data treatment}

All data taken during different campaigns have been copied to Fermilab. The Art framework ~\cite{art}  is the  main framework used by the DUNE collaboration for both ProtoDUNE data reconstruction and simulation, as well as for Far Detector data simulation. Art is used by   several experiments in high energy physics, including  NOvA, MicroBooNE and ICARUS. The Liquid Argon Software (LArSoft) ~\cite{lars},   based on Art, has been developed and maintained by the 
LArSoft Collaboration,  as a shared base of simulation and reconstruction software across LArTPC neutrino experiments.  
All cosmic ray data acquired by \pddp in well defined and stable detector conditions in 2019 and 2020  ($\approx$ 377K events) have been processed with  LArSoft, by performing the reconstruction of hits and 2D tracks. A second processing, including Pandora~\cite{pnd1, pnd2, pnd3, pnd4} reconstruction algorithms, started in spring 2021. The memory footprint of reconstruction jobs is between 1.9 and 2.5\,GB. Data management and job submission was successfully done through the same systems as ProtoDUNE-SP. This allowed NP02 data to be included in the DUNE data catalogue.



\subsection{Cosmic ray taggers}
\label{ssec:det:crt}

Two Cosmic Ray Taggers (CRTs) are suspended on the external walls of the cryostat: one near the roof (Top-CRT) and the other, on the opposite side, near the floor (Bottom-CRT). The CRTs can provide a trigger to the charge and light readout systems for cosmic muons that cross the detector on diagonal trajectories as illustrated in figure~\ref{fig:ProtoDUNEdp_diagram}.

Each CRT is composed of eight scintillator paddles with dimensions 144x14 cm$^2$. The scintillation light in the paddle is detected on each end with a PMT (two per paddle). The relative time difference between signals seen in a PMT on either end is used to estimate the horizontal coordinate of crossing particles, while the location of the hit paddle itself provides the vertical coordinate.  

The 32 PMTs are read by custom electronics integrated in a uTCA system with the trigger logic also implemented in the hardware. An intermediate charge trigger is made from the analogue sum of both PMT signals for each scintillator bar. Then a trigger for crossing particles is generated by a requiring hits in coincidence in each CRT. The TTL trigger signal generated by the CRT readout system is timestamped in a White-Rabbit end-node synchronised to the clock of the charge and light data acquisition systems.  

\subsection{Photon detection system}
\label{ssec:det:pds}

\begin{figure*}[ht]
    \centering
    \includegraphics[width=\textwidth]{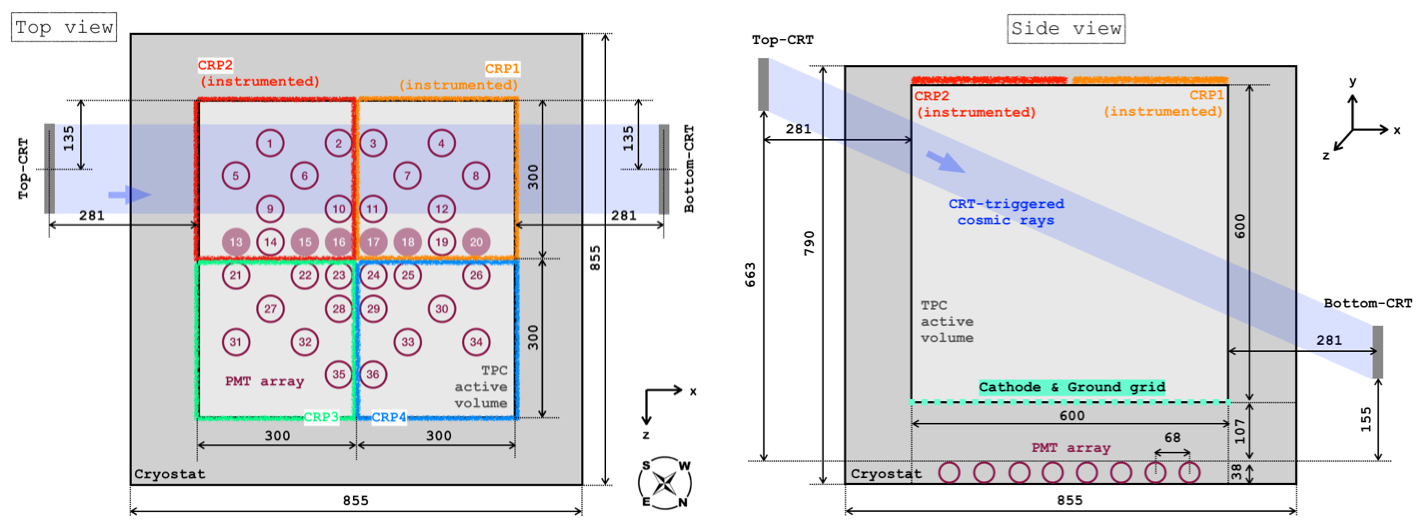}   
    \caption{Views of \pddp. Dimensions and positions of the major elements are indicated (in units of centimeters). The drift direction corresponds to the $y$-axis. The PMTs are represented with circles and, in the top view, the empty circles correspond to PEN PMTs and the filled circles to TPB PMTs.}
    \label{fig:ProtoDUNEdp_diagram}
\end{figure*}

 The Photon Detection System (PDS) of \pddp is formed by 36 8-inch cryogenic PMTs placed below the cathode grid. The goal of the PDS is to provide a trigger for non-beam events, to precisely determine the event time, with possibility to perform calorimetric measurements and particle identification. A diagram with the detailed positioning of the PMTs with respect to the rest of elements inside the detector is shown in figure~\ref{fig:ProtoDUNEdp_diagram}.
 
\subsubsection{Photomultipliers}
\label{sssec:det:pmt}

The 36 Hamamatsu R5912-20Mod PMTs were fully characterised both at room and at cryogenic temperature~\cite{protoDUNEPMTs,Belver:2020qmf} before their installation. 

The PMTs are biased applying a positive HV at the anode while the photocathode is grounded. In this way only one coaxial cable is needed to both feed the HV and read out the signal. A splitter circuit is placed outside the cryostat in order to decouple the HV from the PMT signal. Three CAEN A7030 modules~\cite{CAEN} are used to bias the PMTs with twelve channels each. They are controlled using WinCC, a control and data acquisition interface from Siemens~\cite{WinCC}.

\begin{figure}[ht]
\begin{center}
\includegraphics[width=0.5\textwidth]{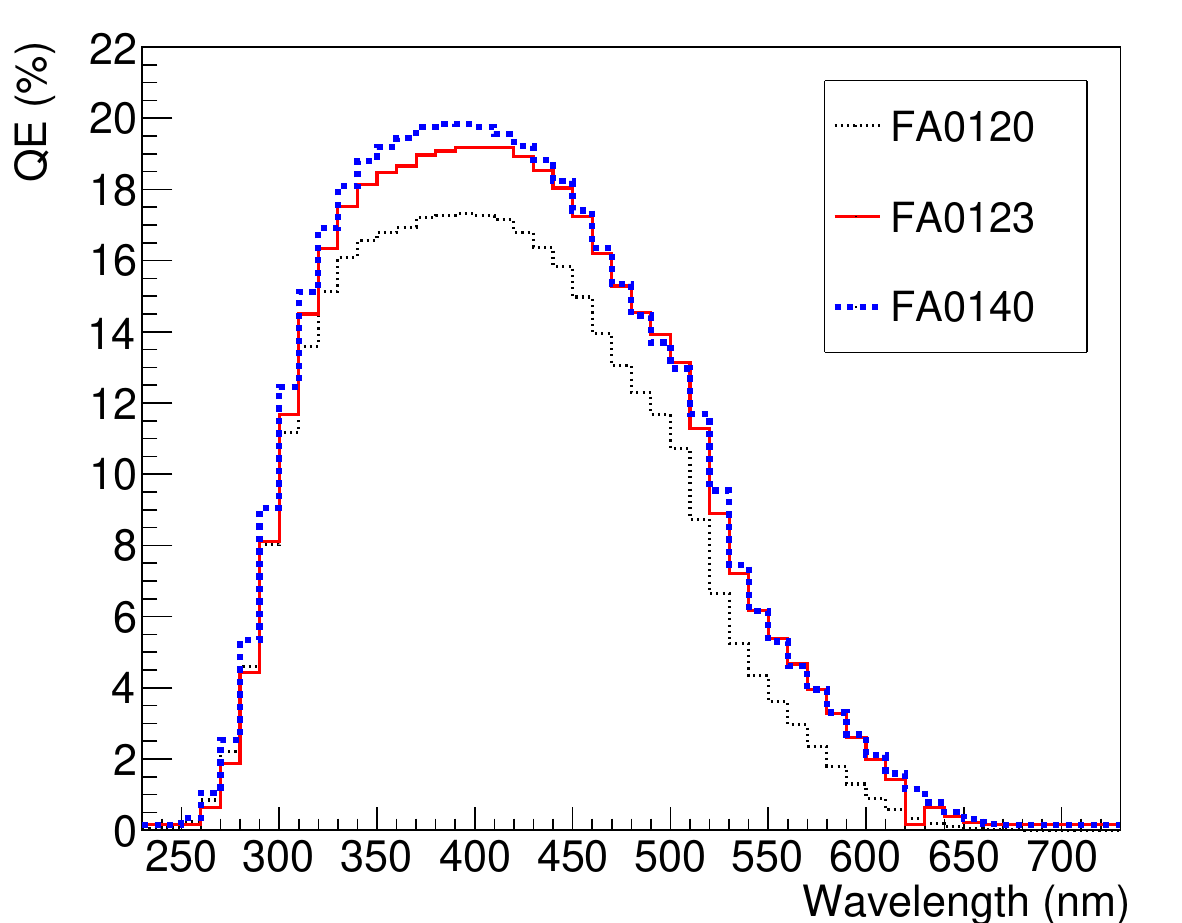}
\caption[Quantum efficiency spectrum for 3 Hamamatsu R5912-20Mod PMTs provided by the manufacturer.]{Quantum efficiency spectrum for 3 Hamamatsu R5912-20Mod PMTs provided by the manufacturer. The maximum value is obtained around 400\,nm.}
\label{fig:PMTQE}
\end{center}
\end{figure}

 The photocathode quantum efficiency ({\it i.e.} the number of electrons emitted by the photocathode per incident photon, QE) versus the wavelength is shown in figure~\ref{fig:PMTQE}, as provided by the manufacturer for three of the PMTs at room temperature. The maximum QE of the PMTs is 20\% at a wavelength of 400\,nm. Additionally, it has been shown that the QE remains stable when going to LAr temperature~\cite{Zhao_2021_PMT_QE_CT}. The main properties of the PMTs are summarised in table~\ref{tab:PMTproperties}.

\begin{table}[ht]
    \centering
    \begin{tabular}{| c | c | }
    \hline
        PMT model & Hamamatsu R5912-20Mod \\
        Number of dynode stages & 14 \\
        Photocathode diameter & 8\,inch \\
        Number of PMTs & 36\\
        Number of of PEN-foil PMTs & 30 \\
        Number of TPB-coated PMTs & 6 \\
        Quantum efficiency at 430\,nm & 0.183$\pm$0.013\\
        Dark current rate at G=10$^{7}$ & 1.7$\pm$0.3 kHz\\
        HV for G=10$^{7}$  & 1324$\pm$100 V \\
    \hline
    \end{tabular}
    \caption{Main properties of the \pddp PMTs.}
    \label{tab:PMTproperties}
\end{table}

\subsubsection{Wavelength shifting}
\label{ssec:det:wls}

\begin{figure}[ht]
    \centering
    \includegraphics[height=0.35\textwidth]{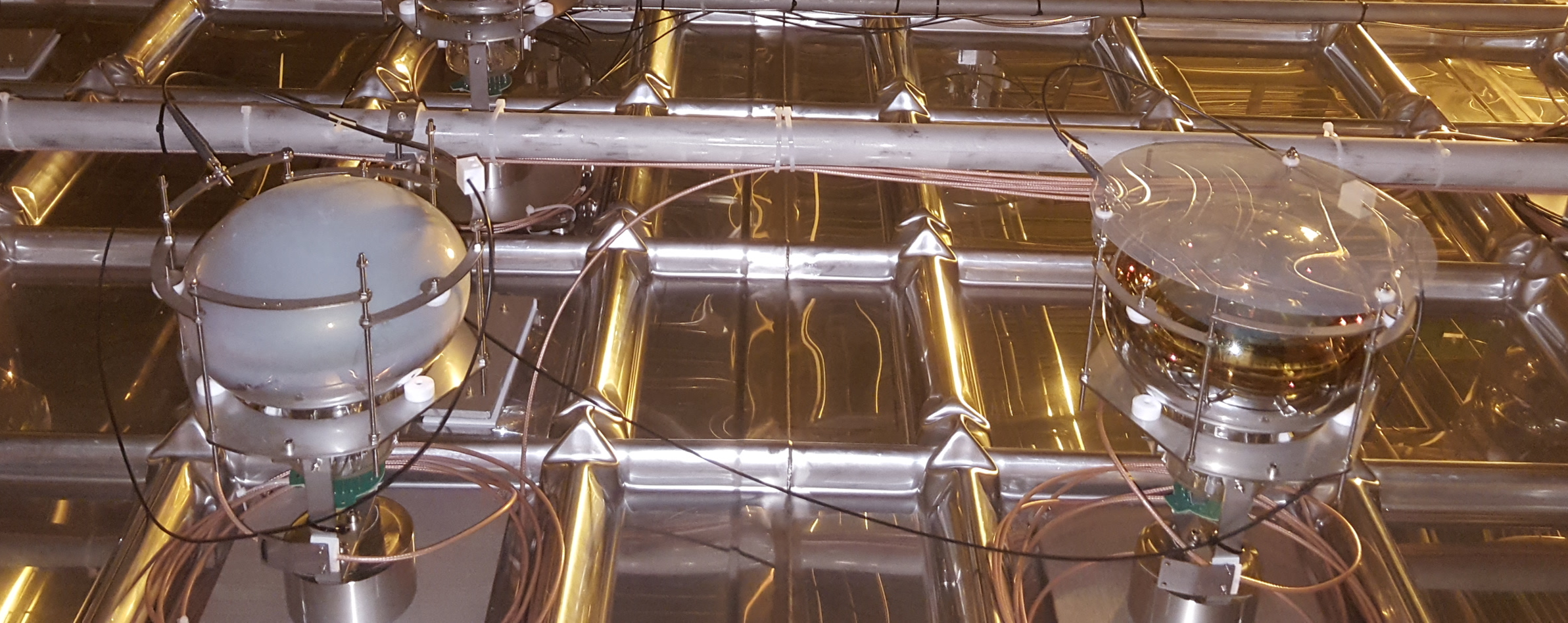}
    \caption[Detail of two PEN and TPB PMTs inside the detector.]{Detail of two PMTs inside the detector. A TPB coated PMT is on the left, a PEN foiled PMT is on the right.}
    \label{fig:PMTpic}
    \end{figure}

Scintillation light in LAr has an emission peak at 127\,nm where most photosensors are not sensitive. To efficiently detect this light, fluorescent materials are introduced to shift the wavelength of the scintillation photons towards the visible range where the PMTs QE is maximal. To do so, \pddp uses polyethylene naphthalate (PEN) foils mounted above 30 PMTs photocathode, and tetraphenyl butadiene (TPB) directly coated on the glass of the remaining 6 PMTs, as shown in figure~\ref{fig:PMTpic}. This design pioneers the use of PEN in a large-scale experiment at the multi-tonne scale, while retaining the ability to compare the performance with TPB.

The PEN sample used in \pddp is transparent and biaxially oriented, manufactured by GoodFellow~\cite{GoodFellow_PEN}, with reference ES361090/6. It has been installed as circle foils of 240\,mm diameter and 0.125\,mm thickness placed tangent to the PMT top glass surface. PEN has a re-emission spectrum around 430\,nm~\cite{PEN-DMary}.

The TPB was deposited over each PMT's polished surface, using a dedicated evaporation system developed for the ICARUS experiment~\cite{Bonesini_2018}. The coating density is 0.2\,mg/cm$^{2}$. Like PEN, TPB re-emits photons in the visible spectrum, with a maximum at around 430\,nm~\cite{TPB-Francini2013}. Additionally, the detection efficiency of TPB-coated PMTs has been measured at the University of Pavia~\cite{Burak:2020scu} by comparing the current given by the PMT under test and by a reference calibrated photodiode in a dedicated setup. Four PMTs were tested, obtaining an average detection efficiency of 0.14$\pm$0.02. The individual values are shown in table~\ref{tab:ProtoDUNE_PMTQE}. This means that 14\% of 127-nm photons arriving at the TPB will produce a photoelectron.

\begin{table}[ht]
    \centering
    \begin{tabular}{|c|c|c|}
        \hline
        PMT    &    QE (430\,nm)  & DE (127\,nm) \\
        \hline
        FA0120 & 0.168 & 0.115\\
        FA0123 & 0.189 & 0.14\\
        FA0140 & 0.192 & 0.145\\
        FA0143 & - & 0.165\\
        \hline
        Average & 0.183$\pm$0.013 & 0.14$\pm$0.02\\
            \hline
    \end{tabular}
    \caption{PMT quantum efficiency (QE) at 430\,nm provided by the manufacturer and TPB-coated PMT effective detection efficiency (DE) for 127\,nm photons measured in the laboratory.}
    \label{tab:ProtoDUNE_PMTQE}
\end{table}

A comparison of the performance of the two wavelength-shifting systems is presented in section~\ref{ssec:ph:wls}  

\subsubsection{Light calibration system}
\label{ssec:det:lcs}

\pddp was equipped with a dedicated light calibration system in order to measure and monitor the gain of the PMTs. The design of the light calibration system was described in detail in a dedicated publication~\cite{Belver:2020qmf}. Having a well-defined gain is important for equalizing the PMT response to provide a uniform trigger, and to perform calorimetric studies.

\begin{figure}[ht]
    \centering
    \includegraphics[height=0.5\textwidth]{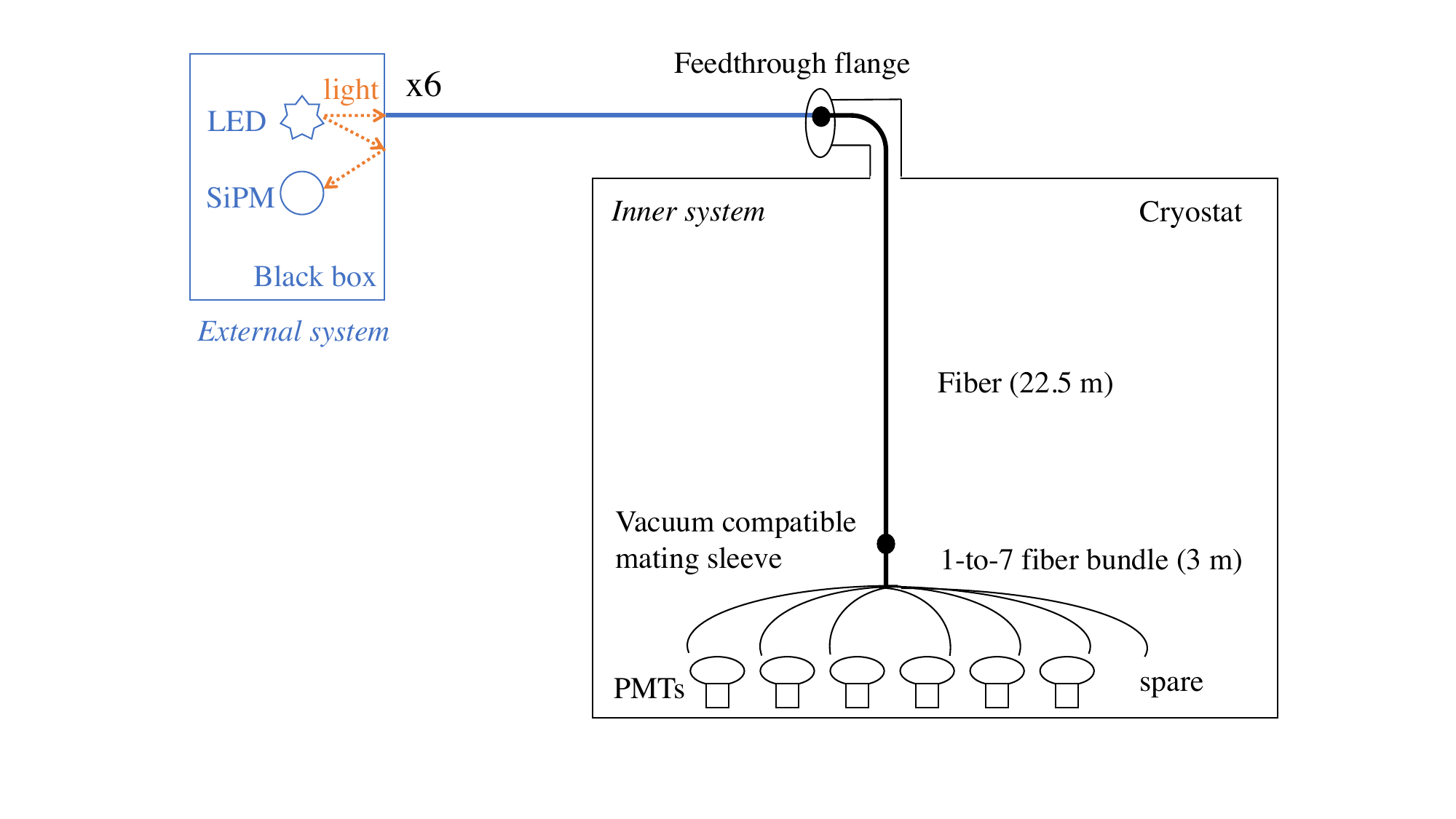}
    \caption[Diagram of the ProtoDUNE-DP light calibration system.]{Diagram of the \pddp light calibration system. The system shown in blue stays at room temperature and the inner system in black is at cryogenic temperature.}
    \label{fig:LCS}
    \end{figure}

The calibration system consists of six blue LEDs of 465\,nm located outside the detector that are driven using a Kapustinsky circuit~\cite{KAPUSTINSKY}. The Kapustinsky circuit provides fast nanosecond pulses with variable amplitude. The LEDs illuminate six fibres and a SiPM is used as a reference sensor. Each fibre is then connected to a 1-to-7 fibre bundle so that one fibre is installed pointing to each PMT. The response of the different components has been fully characterised~\cite{Belver:2019lqm}. The LEDs are pulsed at 1\,kHz with a 30\,ns pulse width, and have a wavelength of 465\,nm where the PMT has nearly maximum quantum efficiency. The Kapustinsky circuit is controlled by a BeagleBone Black board, an open-hardware mini-computer that runs a Linux distribution. A diagram of the light calibration system set-up is shown in figure~\ref{fig:LCS}.

\subsubsection{Light readout system}

The Light Read-Out (LRO) system digitises the PMT signals. It is based on the commercial ADC module v1740 supplied by CAEN. This digitiser has 64 analog input channels with 2\,Vpp dynamic range, and a maximum sampling rate of 62.5\,MS/s. The main specifications are summarised in table~\ref{tab:LRS}. The ADC is connected to a PC via an 80\,MB/s optical link, and driven by a DAQ system based on MIDAS~\cite{midas}, a DAQ framework developed by PSI and TRIUMF. The selected dynamic range and sampling rate allows measuring the time structure of the scintillation light signal which lasts a few nanoseconds at the level of the individual photons.

\begin{table}[ht]
    \centering
    \begin{tabular}{| c | c |}
    \hline
        ADC Model    & CAEN v1740 \\
        Sampling Frequency & 62.5\,MS/s \\
        Dynamic range & 2\,Vpp \\
        Resolution & 12 bits \\
        Channels & 64\\
    \hline
    \end{tabular}
    \caption{Main properties of the \pddp light readout system.}
    \label{tab:LRS}
\end{table}

\subsubsection{Light data acquisition and calibration software}
\label{ssec:pdune_pdsDAQ}

The data taking is performed using a dedicated software~\cite{ProtoDUNEDP_Software} which allows four acquisition modes by driving three different systems: the PMT HV, the light calibration system, and the light readout. The software provides a graphical user interface allowing the user to choose the acquisition  trigger mode, control and define the acquisition settings (front-end and HV), select the light trigger, arrange the calibration mode and settings, and provide the graphical user interface. Individual PMT waveforms are recorded for analysis, so each event contains the information of 36 waveforms.

Four acquisition modes are allowed by the system:
\begin{itemize}
    \item PMT self-trigger: The trigger is defined by looking at the PMT signals. It allows to define a threshold in amplitude during a time window, and ask for coincidences among PMTs.
    \item External trigger: The trigger signal is received by an external system, either the CRT panels (see section~\ref{ssec:det:crt}) or the DAQ global computer.
    \item Calibration mode: An external trigger signal is received from the light calibration system at a tunable frequency synchronized with the calibration light pulse sent to the PMTs. This trigger mode is used to determine the PMT gain.
    \item Random trigger: An external trigger signal is received from the light calibration system at a tunable frequency but keeping the LEDs off.
\end{itemize}

The software communicates with MIDAS to control the light readout, with WinCC to control the PMT voltages, and with the Beaglebone to control the calibration system. The user can control the full system using a graphical user-friendly interface. In order to monitor the data-taking, an event display based on the ROOTANA package~\cite{ROOTANA} was developed.
    
The collected data is first stored in the DAQ computer during the data taking. At the end of the day, a dedicated script saves the data in the backup servers and converts the binary files into ROOT files, that are easier to analyze.

\section{Detector characterisation and performance}
\label{sec:operation}

This section presents the experience from the \pddp commissioning  and operation. Several issues were discovered and are detailed here. The dual-phase CRPs require calm liquid surface without any turbulence. While this was previously the case in the prototype tests (e.g., the 4 tonne demonstrator \cite{311_technical} and the CRP qualification in the coldbox cryostat), the heat input produced bubbling in several location in \pddp cryostat creating surface perturbations that, in turn, impacted the CRP operation. LEMs, having been all stringently qualified (see section~\ref{ssec:det:crd}), started to progressively deteriorate throughout the operation period.

During the commissioning of the \SI{300}{\kilo\volt} high voltage delivery to the cathode (section~\ref{sec:hvcommissioning}) a short developed between the field cage and the HV extender severely impacting the uniformity and strength of the drift field in the detector active volume. An in-situ repair was attempted, but was unsuccessful. A redesigned version was then installed and ultimately successfully tested (section~\ref{sec:fulldrift}). 

The overall timeline of the operations is the following:
\begin{itemize}
    \item First filling of the cryostat completed by August 2019; 
    \item Commissioning of the detector components August - September 2019;
    \item Detector operation from from Fall 2019;
    \item Attempted HV extender repair in June 2020 requiring partial emptying of the detector and then refilling with the Xe-doped LAr from \pdsp;
    \item Complete emptying of the cryostat in September 2020;
    \item Second filling of the cryostat for the test of the redesigned HV delivery system lasting from September 2021 to March 2022.
\end{itemize}
In the rest of the section, the commissioning of several detector components will be presented and the operational aspects, in particular focusing on the challenges encountered, will be given. 

\subsection{Cryostat commissioning}
\label{sec:cryocommissioning}

After the completion of the detector installation, the cryostat leak and pressurisation tests were made. Both liquid and gas re-circulation circuits were purged. The cryostat internal volume was then purged with gas argon injected from the bottom at a rate of about \SI{50}{\gram\per\s}. As argon is denser than air, its resultant laminar flow  towards the top effectively pushed out the air (piston-purge). The gas analysers monitoring the exiting gas for oxygen and water content showed that the bulk of air and moisture were expelled within two days. Once the nitrogen and oxygen levels reached ppm-levels, the cryostat cooling could begin.

In order to limit thermal stresses on the inner membrane as well as the detector elements, the maximum temperature difference between any two points in the cryostat was not to exceed \SI{50}{\kelvin} during the cool-down phase. This was achieved by spraying a cold mist, created from a mixture of liquid and gas Ar, via a network of sprayers located at top of the cryostat over the full detector volume. In additional, small quantities of liquid were injected at the bottom cryostat though the filling line. Once the warmest temperature descended below \SI{130}{\kelvin}, the filling of the cryostat was started.  

The cryostat was filled though the purification filter line from an external storage tank, which was refilled daily on weekdays by truck deliveries to the site. Usually \SI{40}{\tonne} were delivered per day. The contamination levels of LAr were checked for each delivery prior to the transfer to the storage tank. Overall, for more than \SI{900}{\tonne} of LAr delivered, the oxygen content was found to be consistently less than \SI{1}{ppm} and within the acceptance requirements.  

The filling of the \pddp cryostat, with a total of \SI{750}{\tonne} of liquid transferred into the detector volume, took about two months to complete. Once finished, the liquid re-circulation pump was activated in order to start re-circulating argon from purification filter. The cryostat operating pressure was initially regulated at \SI{1010}{\milli\bar}. Since October 2019, the nominal pressure was raised to \SI{1045}{\milli\bar} to match the operational conditions foreseen for the DUNE Far Detectors.

\subsection{High voltage commissioning}
\label{sec:hvcommissioning}

\begin{figure}
\begin{centering} 
\includegraphics[width=0.8\textwidth]{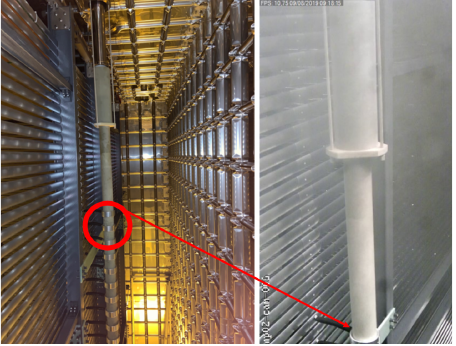}
\caption{\label{fig:dpshort} The region of the short on the HV extender. Left, the extender as seen during construction. Right, a closer view taken by the cryo-camera. The conductor, covered with a black insulator, linking the first degrader ring to the field cage profile is visible in at the bottom of the image}
\end{centering}
\end{figure}

\begin{figure}
\centering
  \includegraphics[width=0.8\textwidth]{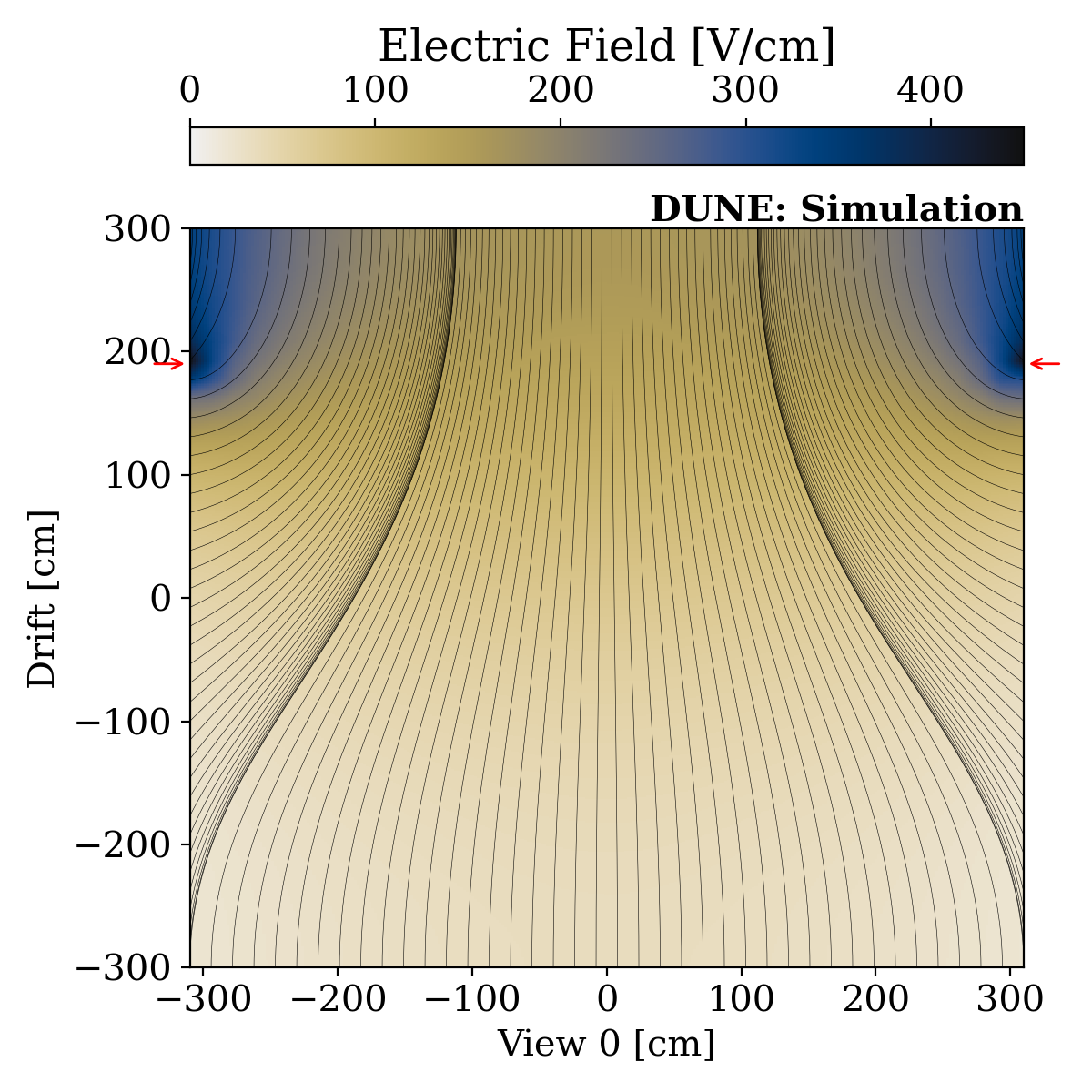}
  \caption{Electric field strength inside the TPC, calculated with COMSOL~\cite{comsol} for \SI{-50}{\kilo\volt} HV setting. The field strength values, the top colour scale in units of \si{\volt\per\cm}, are shown for a vertical plane passing though the centre of the detector active volume with the X-axis (Y-axis) being the second horizontal (vertical cathode-to-CRP) coordinate. 
  The red arrows indicate the vertical position where the short has been simulated. The black lines indicate equi-potential lines.}
  \label{fig:driftefield}
\end{figure}

Once the filling was complete, the commissioning of the TPC HV system began.  The HV delivery to the cathode (and field cage) was ramped down. The HV was stably operated with cathode at \SI{-60}{\kilo\volt} for two days, and then decreased in steps of \SI{50}{\kilo\volt} (with 15-20 minutes pause at each step) down to \SI{-200}{\kilo\volt}. From there, the HV was decreased in steps of \SI{10}{\kilo\volt}, reaching \SI{-250}{\kilo\volt}. Unfortunately, after less than 15 minutes at this voltage, a current trip occurred (set at a level of 500 $\mu$A) on the HV power supply with the cathode voltage dropping immediately to \SI{0}{\volt}.  After this event, it was no longer possible to reach a voltage lower than \SI{-168}{\kilo\volt}: below this value, a high continuous current occurred indicating that the varistors mounted on HVDBs (section~\ref{ssec:det:tpc}) began to be conductive. Detailed analysis of the system monitoring data supported by a dedicated SPICE simulation of the equivalent HV circuit indicated a short developed between the inner conductor of the HV extender and its first degrader ring. The latter is connected to the field cage (see figure \ref{fig:dpshort}) at approximately one quarter of the full drift height.

Even with the HV setting as small as \SI{-10}{\kilo\volt}, the periodic discharges of the system could be seen from the monitoring of the FFS voltage as well as in the data taken with front-end electronics. The frequency of these discharges was increasing with the magnitude of HV going from less than a Hz rate at \SI{-30}{\kilo\volt} to about \SI{4}{\hertz} at \SI{-125}{\kilo\volt}. As a result, the majority of the detector operation was conducted at \SI{-50}{\kilo\volt}: the value that was considered safe enough for a long-term operation and minimally sufficient to allow detecting and reconstructing cosmic ray tracks within the \SI{1.5}{\meter} depth below CRPs. However, short data collection campaigns were also performed with \SI{-110}{\kilo\volt} HV setting. 

Figure \ref{fig:driftefield} shows a COMSOL simulation of the electric field strengths within the TPC active volume, taking into account the electrical short between the field cage and the HV extender. As can be seen the average drift field is much lower than the design target of \SI{0.5}{\kilo\volt\per\cm}, and drops below \SI{0.15}{\kilo\volt\per\cm} for a large fraction (approximately 80\%) of the detector volume. The drift field non-uniformity has severe impact on the reconstruction of cosmic ray tracks not only due to apparent distortions in the track trajectories, but also variations in the recombination as well as diffusion effects and the ionisation electron attachment rates due to longer (slower) drift paths. As discussed later (section~\ref{subsec:tpc_ana}), this effectively limited the usable volume for the reconstruction and analysis to only \SI{50}{\cm} depth below CRPs.

In June 2020, an attempt at fixing the HV short was made by removing the faulty connections from the extender to the field cage. This operation was performed by opening one flange of the cryostat and inserting a cutting tool inside the detector. Unfortunately, the repair was not successful. In the meantime, a new design of the HV extender, discussed in \ref{ssec:det:hvft}, had been also developed. Its test is covered in section~\ref{sec:fulldrift}. 

\subsection{CRP commissioning}
\label{sec:CRPcommissioning}

\begin{figure}
\begin{centering} 
\includegraphics[width=0.8\textwidth]{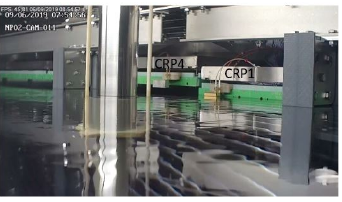}
\caption{View of CRP1 and CRP4 taken with a cryogenic camera with two level meters visible near the corner of each CRP (middle of the image). CRP1 is already in a position with the grid submerged (the LAr level is in the middle of the level meter), while CRP4 is in the process of being lowered position (its level meter is out of the liquid). \label{fig:dpcrplevelmeters} }
\end{centering}
\end{figure}

The CRPs were commissioned after the HV system. For the operation, the CRPs must be moved into position where the extraction grids are completely submerged at or close to their nominal depth of \SI{5}{\mm} below the liquid surface. This alignment has to be performed with the level meters attached to the CRP frames (two of these devices can be seen in the image in figure~\ref{fig:dpcrplevelmeters} taken with the cryogenic camera during CRP commissioning).  Therefore, precise calibration of these level meters is critical prior to the positioning of the CRPs.

The CRP level meters were calibrated in-situ by relying on the precision of the stepper motor encoders of the suspension system. Each CRP was moved vertically in steps of \SI{100}{\micro\meter} allowing to determine the response of the level meters to better than \SI{100}{\micro\meter} precision. After the calibration, the alignment of CRPs relative to the liquid surface was performed. 

The alignment was an iterative process consisting of adjusting a plane defined by the three suspension points with a target of minimizing the spread between the level meter measurements, which is to say the four points on the CRP where the contact with liquid could be measured. The planarity of the CRP was surveyed prior to the installation at room temperature to be better than \SI{1}{\mm} thus defining the alignment target. However, it was not possible to find the configuration for the CRP orientations at cold where the level meter readings would be consistent with such planarity. 

In order to minimize the dead regions in the active volume, no level meters were installed on the CRP sides where they are facing each other. Lacking this additional constraints, a model of a saddle-like CRP deformation was developed where two opposite corners could be lower than the other two by up to \SI{4}{\mm}. Within the model, an orientation were found there grids could be immersed at all points by at least \SI{4}{\mm} (two high corners). However, for the two low corners of the CRPs, the gas layer is reduced to \SI{\sim2}{\mm} instead of the nominal \SI{5}{\mm}. 

The variation in the liquid level above the grid has an effect on the strength of the extraction field responsible for the transfer of the electrons from the liquid to gas. If the extraction field strength could be maintained at or above \SI{2.0}{\kilo\volt\per\cm} with a suitable potential difference between the LEM bottom and the grid over the entire CRP, the impact could be mitigated as the extraction efficiency is largely in saturation at this point (section~\ref{subsec:tpc_gain_measurement}). 

The complete immersion of the CRP grids in the liquid was verified by raising their high voltage. All grids could be powered to \SI{7.5}{\kilo\volt} (\SI{-7.5}{\kilo\volt} applied), while \SI{7.0}{\kilo\volt} is highest grid voltage required for the CRP operation.

\begin{figure}
     \centering
     \includegraphics[width=0.9\textwidth]{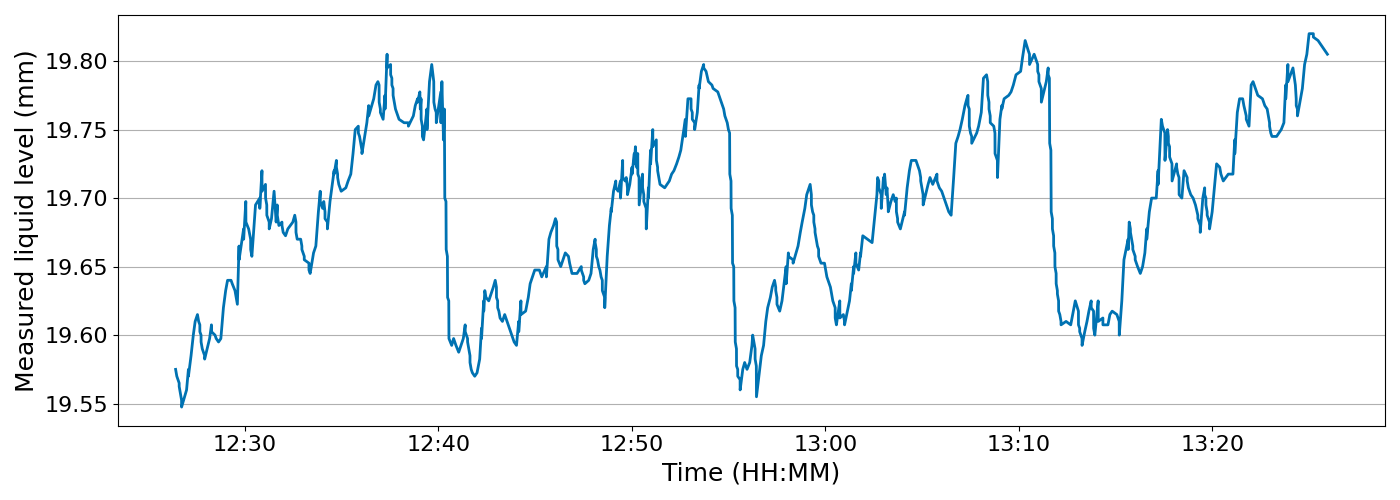}
     \caption{The liquid level monitoring by a level meter on CRP1 as a function of time with an automated level tracking active.  As the liquid level in the cryostat is rising by about \SI{1}{\mm\per\hour}, the CRP positioning system automatically raises the CRP by \SI{250}{\micro\meter}  (the three quick decreases in the level readings) when the liquid rises above a preset target by the corresponding amount.}
     \label{fig:pddp_crplevel}
 \end{figure}

The CRP level meters can also be used by the motor control system for the automatic CRP vertical adjustment relative to the liquid surface (level tracking). The smallest target for the automatic level tracking the CRP position system is capable of is \SI{100}{\micro\meter}. This is essentially determined by the step size of the three motors suspending the CRPs. As an example, figure~\ref{fig:pddp_crplevel} illustrates the automated CRP positioning for one of the CRPs relative to a changing (rising) liquid level. The data shown are the readings from one the CRP (CRP1) level meters as a function of time. In this period the liquid level in the cryostat was rising by approximately $\SI{1}{\mm\per\hour}$. The automated system was set to trigger the CRP position adjustment (the three quick decreases in the level meter reading in the figure) with the level change of \SI{250}{\micro\meter} from the target value.

\subsection{LAr purification process and purity}
\label{sec:purity}

The electron drift time through the \SI{6}{\meter} of LAr takes approximately \SI{4}{\milli\second}, which sets stringent requirements on the LAr purity. Electronegative impurities (such as O$_{2}$) present in the liquid capture drifting electrons resulting in the attenuation of the ionisation charge by the time it reaches the surface where it is collected in the CRPs. The charge attenuation as a function of drift time $t$ follows an exponential dependence given by an electron lifetime $\tau_e$ : 
\[Q(t)=Q_0\exp{(- t/\tau_e)}\]
where $Q_{0}$ is the charge liberated at point of the ionisation. An electron lifetime, for example, of \SI{3}{\milli\second} translates into only 26\% of the charge surviving after \SI{4}{\milli\second} drift. This lifetime corresponds approximately to the concentration of \SI{100}{ppt} of oxygen-equivalent impurities. With the goal of reaching high purities, beyond \SI{100}{ppt}, the liquid in the detector volume must be continuously re-circulated though the purification filter (section~\ref{ssec:det:cryo}) during the operation. With the continuous recirculation, eventually the liquid purity reaches a plateau limit beyond which the purification system would not be able to improve any more. This ultimate purity value depends on how much \ce{O2} permeates inside continuously though various penetrations and piping and flow-rate of the recirculation system that must actively remove these impurities. 

After the filling of the cryostat was completed the pump for the re-circulation of the LAr through the purification circuit was activated and the electron lifetime started to increase from its initial value of a few tens of microseconds. However, following the pump activation the pressure across the input mechanical filter to the purification filter started to rise. To assure the correct operation of the pump, the pressure across filter must not exceed \SI{1}{\bar}. A fraction of LAr had to be thus diverted to bypass the filter leading to some loss in the purification efficiency (the diverted argon was simply re-injected back into the cryostat without passing though the purification filter) in this period. 

\begin{figure}
\begin{centering} 
\includegraphics[width=0.98\textwidth]{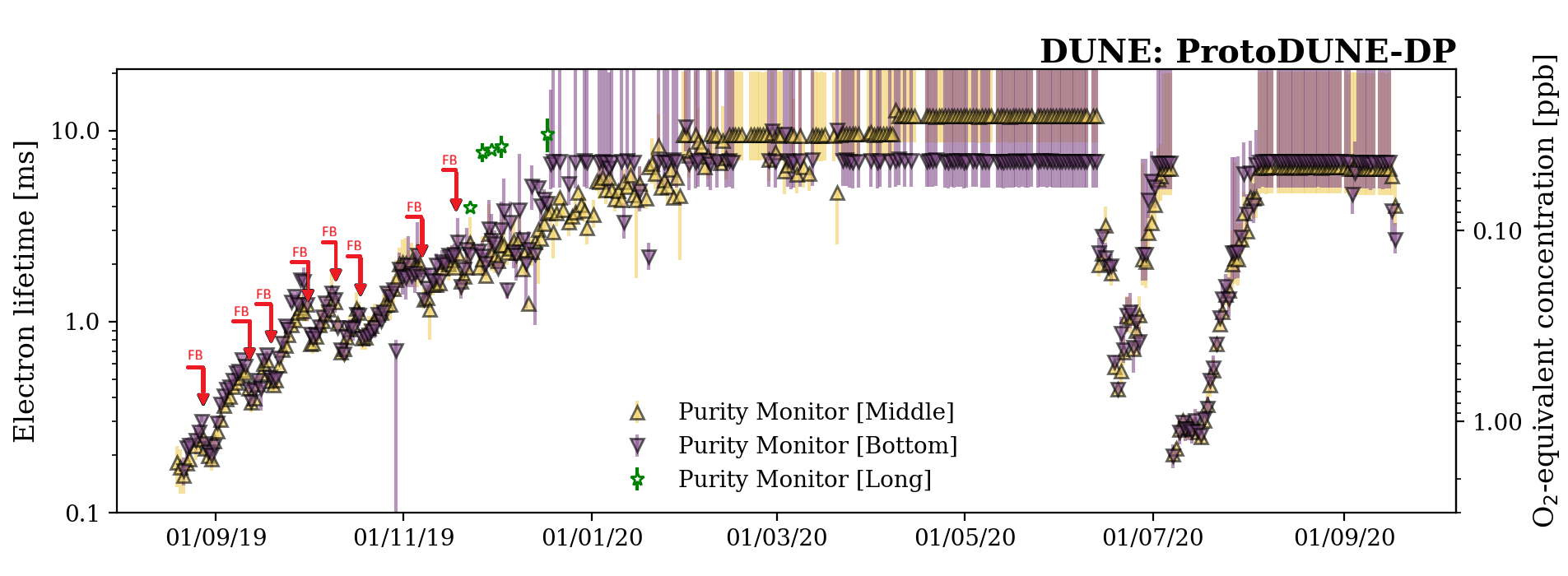}
\caption{Measured electron lifetime  ($\tau_e$) in ProtoDUNE-DP as a function of time from the purity monitors. Red arrows indicate the periods of the filter bypass (FB) for unclogging. The two drops in purity observed during summer 2020 are due to the opening of the cryostat to conduct the HV repair (discussed in \ref{sec:hvcommissioning}) and a technical stop of the circulation system for maintenance. In both cases, an electron lifetime above \SI{5}{\ms} was recovered in less than two weeks. The electron lifetime relates to the concentration of O$_2$-equivalent impurities through: $\tau_e\;\textrm{[ms]}\approx0.3/\rho\;\textrm{[ppb]}$~\cite{BUCKLEY1989364}. \label{fig:electronlifetime} }
\end{centering}
\end{figure}

The input mechanical filter was eventually warmed up and inspected with an endoscope camera. Dust-like material was found and upon its removal the pressure returned to nominal. However, the pressure began to rise again requiring another filter unclogging. The filter unclogging procedure was then established as as a regular maintenance operation when the pressure across the filter was starting to reach the \SI{1}{\bar} limit. This operation would take about 18 hours to complete. During this time the liquid pump would continue to operate, while the filter was bypassed and partially warmed. It was then purged with a reverse flow of argon gas from outlet to inlet. After the purge the bypass was closed, the filter cooled, and the liquid purification resumed. This procedure was performed six times until the end of November 2019. Since then, the pressure increases across the filter were no longer observed until the end of the run. Chemical and microscopic analyses was performed on the debris extracted from the filter, but no clear source of the contamination could be established.

The daily evolution of the electron lifetime measured by the two short purity monitors is shown in figure~\ref{fig:electronlifetime} (figure~\ref{fig:prm_waveform} shows an example of the cathode/anode signals from one of these detectors) starting after the completion of the cryostat filling until the end of the run in 2020. The periods corresponding to the filter unclogging operations in the Fall of 2019 are indicated by red arrows. These coincide with observed drops in LAr purity due to the purification filter by-pass described above. The large drops in purity in the summer 2020 are due to the intervention (opening of the cryostat) for the attempted HV repair followed by a technical stop of the re-circulation system for maintenance. 

The rate of the observable increase in liquid purity after the filling and activation of the pump for the LAr re-circulation is much slower than what was seen during the operation of the NP04 cryostat (\cite{abud:hal-03519598},\cite{Bremer_2022}) where the electron lifetime could improve approximately by one order of magnitude every ten days (two complete volume changes) until ultimately reaching extremely high, \SI{>30}{\milli\second}, values. The downtime in the purification imposed by the periodic filter unclogging could only partially account for this discrepancy. On the other hand, the rate of the recovery of the electron lifetime observed after the re-circulation system maintenance (last dip in figure~\ref{fig:electronlifetime}) is approximately a factor of 7 per two volume changes (10 days). While this shows a somewhat lower efficiency in comparison to the NP04 cryostat, it still illustrates a high capability of the purification system.  

The long purity monitor took periodic data since November 2019, and measured electron lifetime values above \SI{7}{\milli\second} increasing to \SI{10}{\milli\second} by January 2020 (these data are also included in figure \ref{fig:electronlifetime} for reference). On average, the purity reported by the long monitor was a factor $\sim$3 greater than the values obtained from the two short ones. While the former was indeed designed to better measure the electron lifetime in high purity argon, the cause of such a large discrepancy between the two monitor types has not been understood and this instrumental issue remains unresolved. Furthermore, the electron lifetime measurements obtained from the analysis of the cosmic ray data, albeit at the depth of no more than \SI{50}{\cm} (section~\ref{subsec:tpc_purity}), were in an agreement with the short purity monitor results. 

This discrepancy makes the quantification of the ultimate liquid purity achieved in \pddp less certain. The short purity monitors are expected to loose sensitivity with the electron lifetime beyond a several milliseconds resulting in the plateauing of the values they report. However, both types of monitors should also give consistent results in their overlap region around \SI{2}{\milli\second} electron lifetimes. As a consequence, the electron lifetime of \SI{>5}{\milli\second} could be reported here as a likely lower limit.

\subsection{LAr surface stability}
\label{sec:larsurface}

The Dual-Phase concept requires on a still liquid surface, such that the electric field above the extraction grids (in both liquid and gas) can be stably maintained. Surface turbulence inducing variations in the thickness of the liquid layer above the extraction grid leads to changes of the LEM-grid capacitance that in turn can trigger LEM discharges. Sufficiently large ripples in the surface can also result in an instantaneous exposure of the parts of the extraction grid in gas triggering discharges that can have a destructive impact on the LEMs as well the front-end analogue electronics connected to the anode plane above.

The liquid surface quality in the \pddp detector could be surveyed with the cryogenic cameras viewing the surface outside of the TPC active volume (covered by CRPs). As soon as filling was completed, bubbling could be observed from two identifiable sources. The first one was the clips which join together the aluminium field cage profiles. The bars of the field cage have an open profile with an overhang, and at the point of the clip a small hole is present. Bubbles forming within the profile are trapped by the overhang and escape via this hole. These bubbles break the surface between the outer edges of the CRPs. The second source was the HV feedthrough, from which large bubbles were visible breaking the surface once every few seconds.

\begin{figure}
\includegraphics[width=7cm]{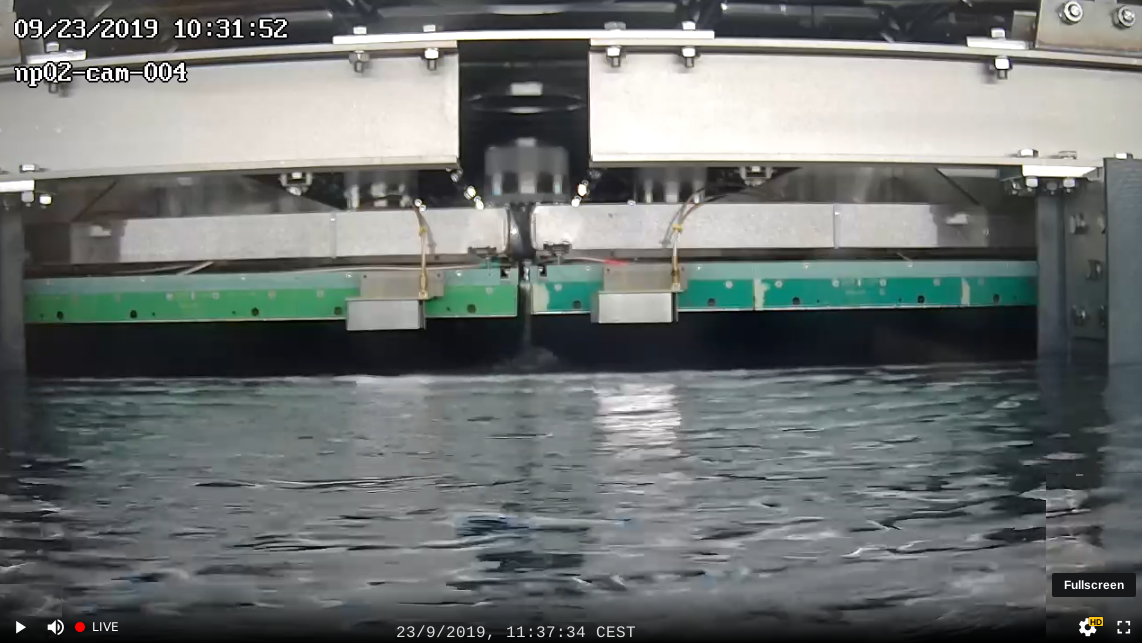}
\includegraphics[width=7cm]{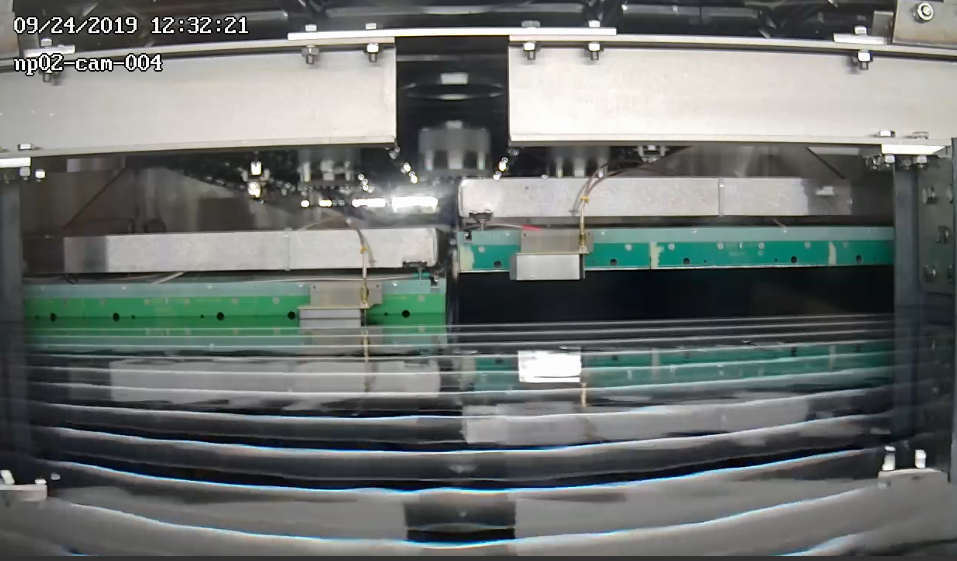}
\caption{Cryo-camera pictures of the LAr surface exterior of the TPC active volume. The left image illustrates the surface turbulence resulting primarily in this view from bubbles emerging from the field cage profile clip below.  The right image shows the calmness of the surface in the same location following a pressure bump.} \label{fig:dpfilled}
\end{figure}

The left image of figure~\ref{fig:dpfilled} illustrates the typical liquid surface conditions with the turbulence caused by the ripples of the rising bubbles at various locations. The CRPs could not be stably powered with this high level of surface turbulence without taking significant risk of quickly damaging detector components due to grid discharging in gas. Thus a strategy to quench or reduce the bubble formation in order to the improve the surface stability had to be developed. 

Increasing the pressure raises the saturation temperature of the liquid. The external heat input then goes towards raising the bulk LAr temperature until a new thermal equilibrium (boiling restarts) can be re-established. This approach was pursued and led to an implementation of pressure cycles where the pressure in the cryostat would be raised by \SI{35}{\milli\bar}, from \SI{1010}{\milli\bar} to \SI{1045}{\milli\bar}, and then lowered back to the nominal value after the bubbling restarted. The period at high pressure (until the bubbles would reappear) was found empirically to last between 8 to 12 hours. The CRP commissioning and the first active data taking campaigns in August - September 2019 (table~\ref{tab:charge_data_summary}) were performed in these high-pressure time windows. The image on the right of figure~\ref{fig:dpfilled} shows the typical liquid surface quality when CRP could be operated. 

The high pressure cycles were executed on a daily basis as lowering the pressure had the reverse effect of increasing the bubbling and the surface turbulence (the CRPs were powered off at this point and raised in order to avoid wetting LEMs) requiring approximately another 12 hours to re-establishing the equilibrium at the lower pressure (saturation temperature) before the cycle could be successfully repeated. In addition to changing the bulk liquid argon temperature, the liquid level would also be affected; during high pressure cycle it would slowly rise at a rate of about \SI{1}{\mm\per\hour} and then decrease afterwards. Given that and the required high periodicity of the pressure cycles (with low, $<50\%$, duty cycle for the CRP running) more stable operating conditions for the cryogenic system had to be found and a different approach was attempted. The pressure was increased only for a short, a couple of hours, period and then lowered back to \SI{1010}{\milli\bar}. Following this pressure bump, the bubbling did not re-appear for several days at the nominal (lower) pressure. This procedure could then be successfully repeated as needed as a function of the re-appearance of the bubbling with the high pressure bump duration period lasting only 2 hours. 

In mid-October 2019, the nominal operating pressure was increased to \SI{1045}{\milli\bar} in line with the expected operating environment for DUNE Far Detector modules. This resulted in significantly longer periods, lasting up to two weeks, of stable surface conditions for the detector operation between pressure bumps (to \SI{1070}{\milli\bar} at this point). By the end of 2019, a more preventive approach was taken and the regular pressure bumps were executed once a week. All the detector technical tests and data collection runs took place in the interim periods between pressure bumps. 

While the turbulence due to the surface bubbling was visibly suppressed with the pressure bump (high pressure cycles) procedure, a small residual surface ripple, not discernable in the still image of figure~\ref{fig:dpfilled}, was observed. The maximum total height of these ripple waves was estimated from an analysis of the camera video feed to be less than \SI{500}{\micro\meter}. The ripple had a periodic structure with the frequency of about \SI{5}{\hertz}. 

In general, the exact surface conditions within the CRP active area could not assessed during operation as this part of the surface was invisible to the installed cameras. Before emptying of the detector in September 2020, a dedicated investigation was performed to address this. A camera with an integrated LED bar was introduced into the cryostat via a spare feedthrough outside of the detector volume, but close a corner of CRP1. The camera could be rotated on its support from outside allowing to conduct a visual inspection both above and below the liquid surface. The observations were made before and after the pressure bump procedure in order to confirm the bubble disappearance. This inspection revealed the existence of persistent (even after the pressure bump) bubbling from the clips in corner of the first (top) and second field cage profiles, bubbling from several specific points on the cryostat walls, and isolated bubbles emerging at the surface below CRPs without any identifiable source. 

\subsection{CRP high voltage stability}
\label{sec:LEMstability}

The CRP and LEM stability is closely connected to the turbulence of the liquid surface. In particular formation of the gas bubbles at the liquid-gas interface region covered by or near to the grids and the LEMs impacts the HV stability of the entire CRP system and makes it difficult to operate. In these conditions, there is a possibility to get local liquid movement exposing the extraction grid to gas. The grid would then spark and could damage not only LEMs but also FEBs reading the anodes. For the latter, if the damages were limited to a couple non-functionating channels per card, these cards (16 in total) could be left in place as the total number of non-functioning channels they represented for the detector readout was <1\%. However, at the beginning the operation, when the impact of the surface turbulence was not completely appreciated, eight FEBs were simultaneously damaged and could not be powered any more necessitating their replacement. Three additional cards were damaged in a similar way at a later date thus bringing the total number of cards for the mandatory replacement to 11. 
\begin{figure}
\begin{centering}
\includegraphics[width=0.9\textwidth]{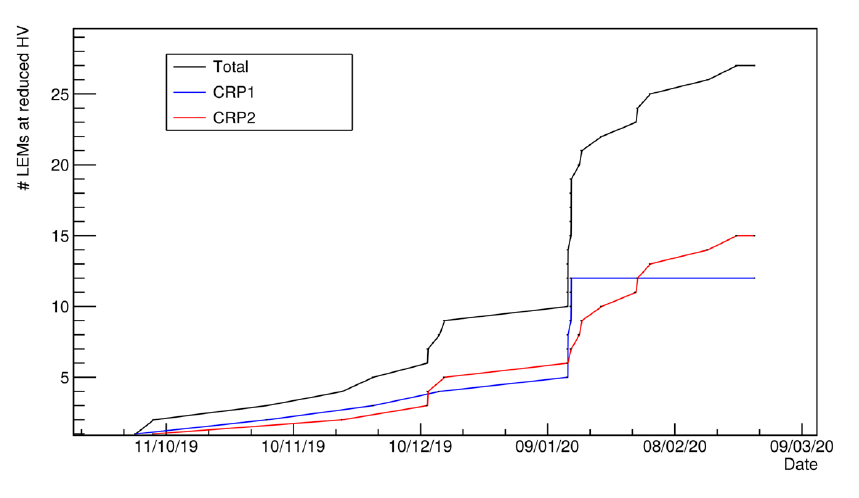}
\caption{Evolution of the number of LEMs that could not hold $\dvlem=\SI{2.9}{\kilo\volt}$ voltage across the two electrodes for CRP1 (blue), CRP2 (red), and the sum total for both CRPs (black). 
\label{fig:LEMreducedHV} }
\end{centering}
\end{figure}

In order to study and assess the stability of the CRPs, between mid-August 2019 and mid-March 2020, CRPs were operated for a total of more than 2700 hours, corresponding to about 50$\%$ of the time.  The LEM top electrodes were nominally powered to \SI{-0.5}{\kilo\volt} giving an induction field to the anode of \SI{2.5}{\kilo\volt/\cm}. The voltage setting  for the bottom electrodes of the LEMs ranged between \SI{-3.4}{\kilo\volt} and \SI{-3.9}{\kilo\volt} corresponding to \dvlem of \SI{2.9}{\kilo\volt} and \SI{3.4}{\kilo\volt}, respectively. The full description of different CRP tests could be found in \cite{Autiero:2714256}. The principal result, in view of a large-scale deployment of the dual-phase CRPs in a DUNE Far Detector module, is the observation of an increasing number of LEMs developing instabilities (discharging continuously or frequently) requiring them to be powered at lower voltages.  

Figure~\ref{fig:LEMreducedHV} shows the evolution of the number of LEMs as a function of time that could not be brought up to $\dvlem=\SI{2.9}{\kilo\volt}$, the nominal voltage difference where LEMs can provide some (a factor of 2-3) charge amplification. Over the course of 8 months, the tally reached 27 LEMs (38\%) of which 16 (22\%) could only be operated at or below $\dvlem = \SI{2}{\kilo\volt}$. At this level, the resultant average electric field, $\SI{20}{\kilo\volt\per\cm}$, in the LEM holes is not sufficient for the charge amplification and these devices simply introduce an additional stage for the transfer of the extracted charge to the anode. No clear correlation between the deterioration of given LEM and the cumulative number of sparks a given the unit withstood has been established. On the other hand, visual inspection of CRPs after the cryostat opening showed the presence of the carbonisation spots in the corners of several problematic LEMs.

The gradual LEM deterioration or ageing serves to highlight a challenge of extrapolating a successful operation in the testbench-size setups to large detectors where multiple readout elements have to be integrated to provide the necessary detector coverage. All LEMs installed in \pddp were qualified at higher voltages (\dvlem = \SI{3.5}{\kilo\volt}) in a systematic way as part of their quality control procedure. In general, the deterioration of the dielectric strength of the micro-pattern devices, in absence of obvious pollution, is due to a formation of local areas preferred by the current flow. Based on the detailed scanning of the damaged LEMs in the coldbox tests (section~\ref{ssec:coldbox}) and a visual inspection of the CRPs after the cryostat opening, these areas are likely to be found in the corner or border regions where the holes could have rim micro-defects as a consequence of the global etching process used in their manufacturing. On the other hand, in \pddp the deterioration process, independently of these defects, could have been further aided by the discharges occurring in the LEM-grid region caused by the liquid surface instabilities. 

\subsection{Muon trigger with cosmic ray taggers}
\label{sec:fulldrift}
\begin{figure}[ht!]
    \centering
    \includegraphics[width=0.7\textwidth]{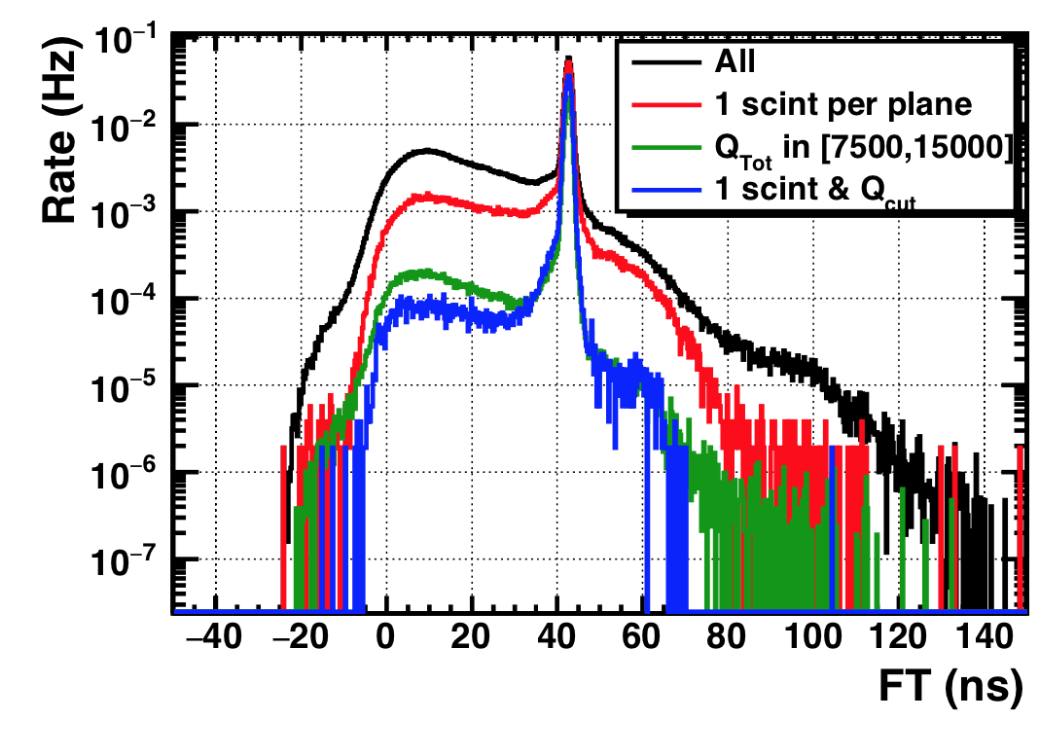}
    \caption{Distribution of the time differences (FT) between events in the two CRTs. The coincidence event rate before any selection (black) is \SI{0.3}{\hertz}. It is reduced to the rate of \SI{0.11}{\hertz} (blue) after requiring a single bar hit in each CRT (red) and placing a selection cut on visible charge (green).}
    \label{fig:CRTcuts}
\end{figure}

The CRT trigger is defined based on coincident hits in between both CRTs suspended on the opposite sides of the cryostat warm structure. Figure~\ref{fig:CRTcuts} shows the event rate versus a calculated time of flight between between the CRTs. The prominent peak in the distribution corresponds to the expected time of flight of \SI{\sim42}{\nano\second} for the relativistic particles. The total event rate of \SI{0.3}{\hertz} (distribution in black) reduces after a series of selection cuts to find clean coincidences corresponding to crossing muons to a trigger rate of \SI{0.11}{\hertz} (blue distribution). 

The trajectory of the crossing muon though the detector can be reconstructed from the location of the hit scintillator bar (vertical coordinate) and the time difference between the two PMT signals on each end of the bar (horizontal coordinate) in each CRT. This information is timestamped and matched offline with the charge and light data.  

\subsection{Tests of the redesigned high voltage delivery system}
\label{sec:fulldrift}

Demonstration of the HV delivery system capable of sustained delivery of $\sim$\SI{300}{\kilo\volt} was one of the key elements in the technical program of \pddp. It also represents a critical development for the evolution of the DUNE Far Detector module design towards the vertical drift TPC detector. To this end, a new HV delivery system (section~\ref{ssec:det:hvft}) featuring in particular a substantially modified HV extender design was built and installed in the detector in the summer of 2021. In order to prevent the bubbling from the field cage observed previously, the top three profiles were also dismounted during this period. The cryostat was refilled and the test of this HV system began in September 2021.  The first few weeks of operation confirmed the absence of bubbling from the field cage and HVFT. Once the LAr purity achieved electron lifetimes over \SI{1}{\milli\second}, the HV was ramped to \SI{-300}{\kilo\volt}. 

During operation, leakage current spikes occurred at rate of about \num{0.5} per hour. The HV power supply maximum current was set to \SI{30}{\micro\ampere} resulting in the output voltage drop during these current spike events with a subsequent recovery at ramp-up speed of \SI{3}{\kilo\volt/\second}. The current spikes were found to be accompanied by the sparks recorded with the cryo-cameras propagating along the top face of the UHMWPE support disk. An electrostatic simulation performed of this region indicated that a possible cause is the free charge accumulation on the polyethylene near the spherical receptacle (see figure~\ref{fig:hvnewpics}) from the LAr ionisation by cosmic rays. This charge can then migrate radially creating the observed streamer events. This issue is to be mitigated in the future by reshaping the receptacle near the disk and also possibly adding grooves in the support disk surface to minimize charge migration. However, the current spikes were never associated with sparks alongside the extender indicating that the diameter of the conductor as well as its distance from the field cage and the cryostat walls were sufficiently large. 

The test was stopped in early November 2021 due to a failure on the HVFT warm-side. Examination identified that faults developed in both the HVFT insulation and the HV cable bringing the power from the power supply. Examination of the radial cut of the feedthrough insulation in the vicinity of the fault with the optical microscope indicated a presence of a (partially burned) fibrous polymeric material embedded on the polyethylene body suggesting possible contamination during the manufacturing of raw material. Examination of the cross-section cuts of the cable in the damaged region suggested a mechanical wear of the dielectric from the cable (over)bending in this region. Both of these problems could be addressed in the future with a more stringent quality control after feedthrough production and stricter cable handling guidelines. The cable and HVFT were replaced with spares and the HV system operation was re-established.
 
A  continuous operation of the HV delivery system at \SI{-300}{\kilo\volt} in the high purity LAr, \SI{>3}{\milli\second} electron lifetime, lasted about two months (Jan. 2022 to Mar. 2022). Apart from the occasional streamer events (about 1 every 2 hours), described above, the system was remarkably stable throughout this entire test period. Taking into account the downtime due to the HV recovery after these events, the integrated system up-time over the entire period was $>99.9\%$.

In this period the CRPs were also lowered into the operation position, with the extraction grid immersed, in order to collect cosmic ray data with the full TPC \SI{6}{\meter} drift.  However, CRP1 and CRP2 could not be successfully biased to operating conditions. This could be due to an interplay of a variety of underlying causes: damages incurred on LEMs in the course of the first \pddp run; debris trapped in the grids after emptying and refilling the cryostat; grid loosening due to thermal cycling. Cosmic ray data were therefore collected by reading CRP4 only. The latter, however, was operated in the dual-phase mode with the extraction grid immersed in liquid and the anodes collecting the extracted charge in the gas phase. Over 1 million of events were collected with a high rate random trigger as well as the CRT trigger. However, as discussed in section~\ref{subsec:tpc_6m}, the analysis of these data is not reported here.

\section{TPC performance measurements}
\label{sec:tpcresp}

About 1.2 million cosmic ray events were recorded in \pddp using a random trigger. The data-taking periods consisted of a few hours of operation over four periods: September, October and November 2019; and January 2020. Moreover, a few runs taken during the commissioning period in early September 2019 could also be used for physics analyses. Table~\ref{tab:charge_data_summary} summarizes the detector conditions of the different data-taking periods.\\

The limited statistics available for each data-taking period, together with the drift-field non-uniformities, the time-dependent degradation of the LEMs, and the other effects described in this section, significantly limited the quality and stability of the data. The available data therefore did not allow for a meaningful quantitative assessment of the Dual-Phase system performance, nor for a reliable evaluation of the associated systematic uncertainties. Nevertheless, by applying corrections derived either from simulations or from theoretical models, the data were sufficient to perform qualitative studies of the anode and amplification systems presented in this section. These studies demonstrate that, despite the various limitations affecting the detector, the charge extraction, amplification and collection system of the detector exhibited the expected behaviour, at least to a qualitative extent. \\

The collected events during stable operation have been reconstructed and analysed using two methods. LArSoft~\cite{LarSoft_Snider:2017wjd, PierreThesis}, the DUNE framework for simulation and reconstruction, has been adapted from previous work on the 4-tonne demonstrator \cite{311_performance} and ProtoDUNE-SP~\cite{Abi_2020}. An alternative reconstruction code LARDON~\cite{lardon,PabloThesis}, entirely based on python, has also been used. Both reconstruction chains are described in section~\ref{subsec:tpc_reco}. The selection of muon-like tracks, and the strategy employed to mitigate the drift field non-uniformities are explained in section~\ref{subsec:tpc_ana}. \\
From this sample, the evolution of the electron lifetime, which is inversely proportional to the contaminant concentration in the LAr, 
is extracted and discussed in section~\ref{subsec:tpc_purity}. The effective gain of the Dual-Phase LArTPC system is explained in section~\ref{subsec:tpc_gain}, where the effects of the LEM thicknesses and the extraction field are discussed. The effective gain evolves with time which could be partly due to the charging-up effect, described and observed in section~\ref{subsec:tpc_chargingup}. The dependence of the effective gain on the LEM voltage is shown in section~\ref{subsec:tpc_gain_measurement}. Finally, the data taken with the full drift is briefly discussed in section~\ref{subsec:tpc_6m}.

\begin{table}[ht]
    \centering
    \begin{tabular}{l|ccccc}
    Run Period & Dates &E$_{extr,l}$ & \dvlem & E$_{ind}$ & Pressure\\
    \hline
    Commissioning &06$\sim$16/09/2019 & 1.25$\sim$2\,kV/cm & 2.9\,kV & 2.5\,kV/cm & 1045\,mbar\\
    September & 18/09/2019 & 2\,kV/cm & 2.9$\sim$3.1\,kV & 2.5\,kV/cm & 1045\,mbar\\
    October &2$\sim$3/10/2019 & 1.25$\sim$2\,kV/cm & 2.9$\sim$3.2\,kV & 2.5\,kV/cm & 1010\,mbar\\
    November&21$\sim$22/11/2019 & 2\,kV/cm & 2.9$\sim$3.2\,kV & 2.5\,kV/cm & 1045\,mbar\\
    January&13$\sim$14/01/2020 & 1.9\,kV/cm & 2.5$\sim$3.0\,kV & 2.5\,kV/cm & 1045\,mbar\\

    \hline
    \end{tabular}
    \caption{Summary of the cosmic data run periods recorded by \pddp. In a given run period, the range explored for extraction field in LAr and the voltage difference across the LEMs are highlighted.   }
    \label{tab:charge_data_summary}
\end{table}

\subsection{Event reconstruction}
\label{subsec:tpc_reco}

\subsubsection{Data preparation}
\label{subsec:tpc_reco_preparation}

\begin{figure}[h]
    \centering
    \includegraphics[width=0.8\textwidth]{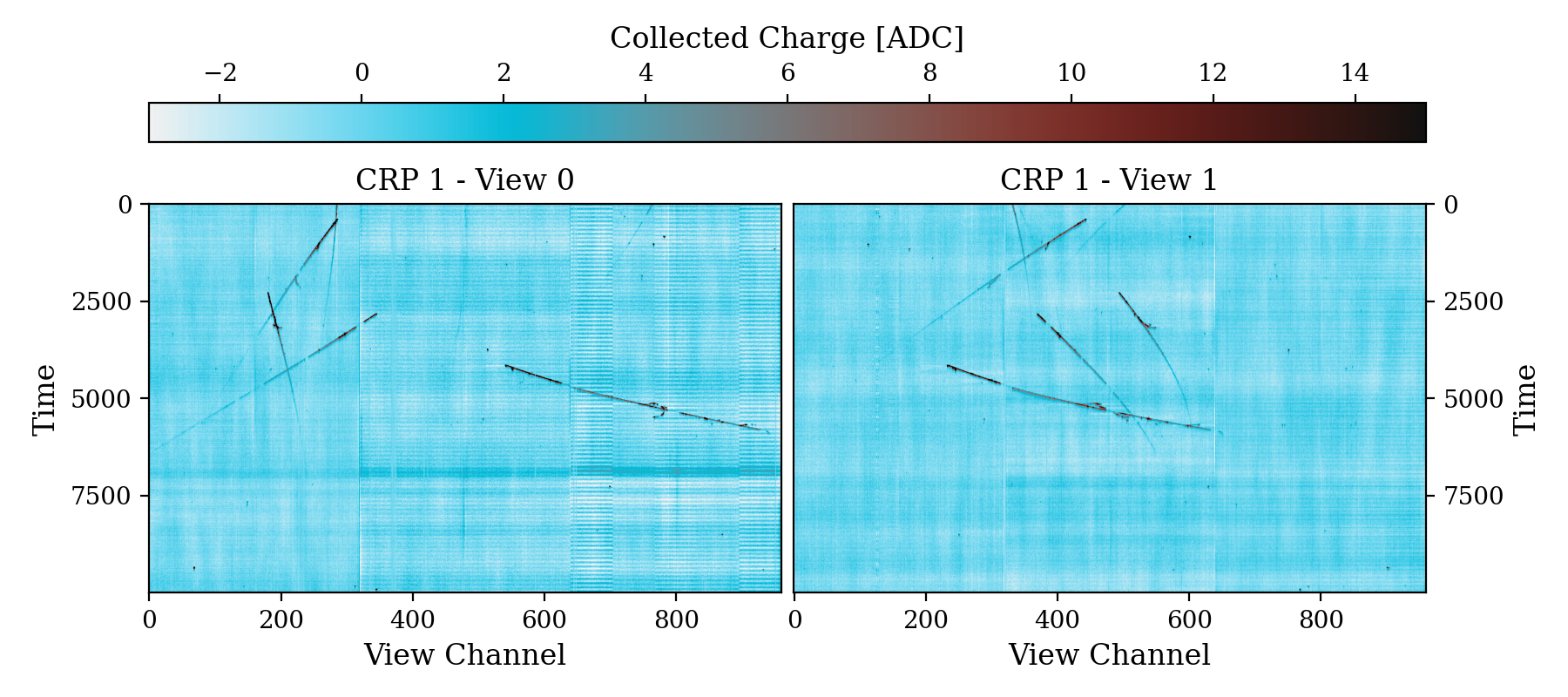}\\
     \includegraphics[width=0.8\textwidth]{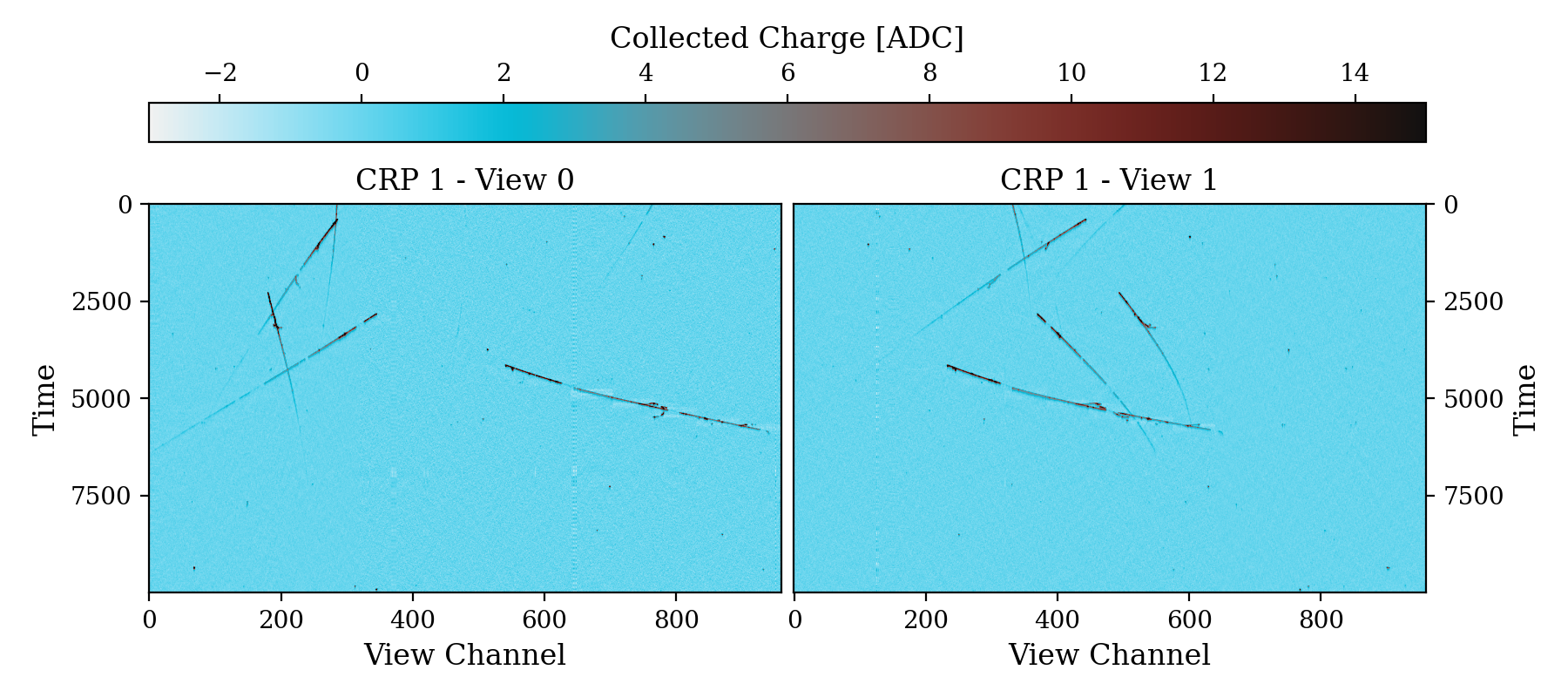}\\
    \caption{Cosmic tracks recorded in \pddp. For clarity, only data recorded in the first CRP is shown. The upper plot shows the raw event, where only the mean pedestal is subtracted for each channel. The lower plot shows the same event after noise filtering. This event was taken with the LEMs at a nominal amplification field set at 31~kV/cm in September 2019.}
    \label{fig:event_display_raw_filt}
\end{figure}

\begin{figure}[h]
    \centering
    \includegraphics[width=1\textwidth]{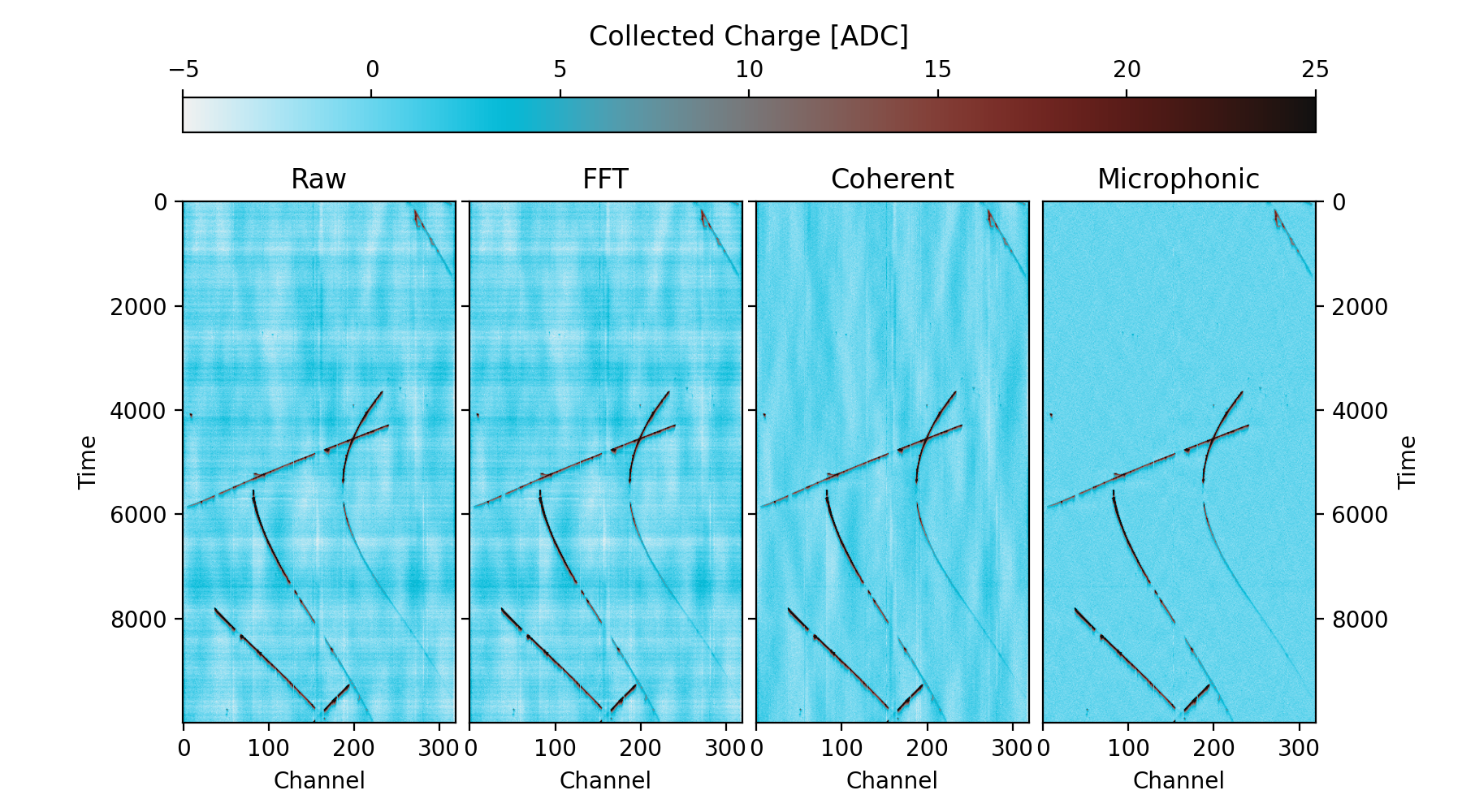}
    \caption{Effect of the three noise filters (low pass FFT filter, coherent noise, and microphonic filter) applied to the raw \pddp data. For clarity, only a subset of the event is shown. This run was taken in October 2019 with an amplification field set to 32~kV/cm.}
    \label{fig:event_display_noise_filters}
\end{figure}

\begin{figure}[h]
    \centering
    \includegraphics[width=1\textwidth]{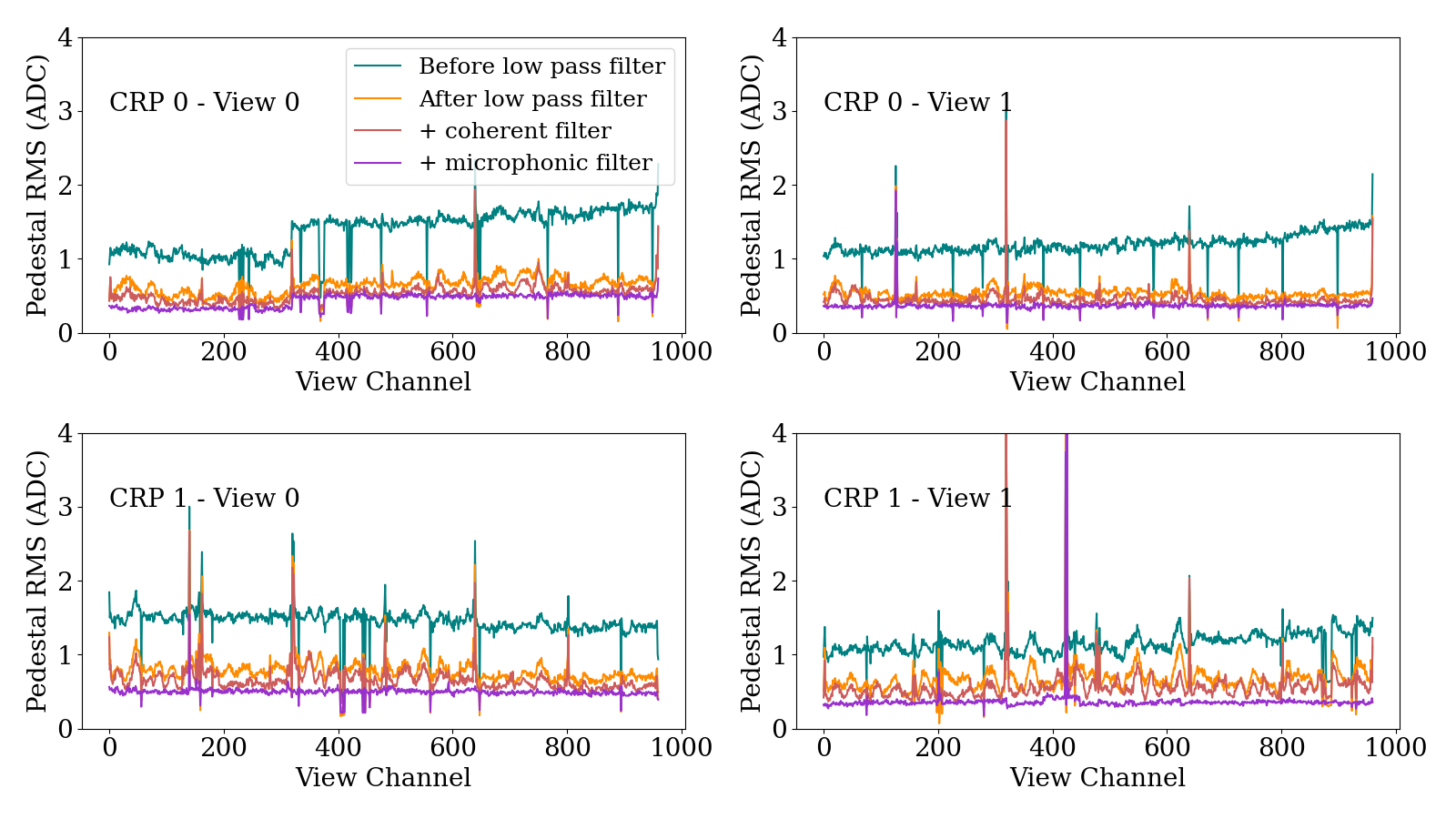}
    \caption{Evolution of the pedestal RMS for each channel (broken down by view and CRP) with the LARDON noise filtering algorithms. This run was taken in January 2020.}
    \label{fig:pedrms_evo_filter}
\end{figure}

An event in \pddp consists of 4480 waveforms\footnote{There are 960 channels per view for the fully equipped CRP1 and CRP2, and 320 channels per view for the anode-only CRP4} of 10000 samples each. Each raw waveform has the mean pedestal subtracted. Figure~\ref{fig:event_display_raw_filt} shows a 2D display of a recorded cosmic event: the upper plot shows the raw event. In the two software packages used, LArSoft and LARDON, the first step in the data processing chain is the filtering of the noise. Three consecutive filters are applied on data. The first one is a low-pass filter to remove the high-frequency noise inherent to the electronics. The second filter removes the so-called coherent noise, a perturbation seen by multiple strips at the same time. This noise comes from slow control connections not well grounded at the flanges, which act like antennas and inject environmental noise inside the cryostat that is picked up by the CRP strips. 
In a group of channels connected to the same electronic card, the mean noise per time sample is computed and subtracted. Finally, small baseline distortions caused by acoustic vibration modes of CRPs (microphonic noise) are removed for each channel by subtracting the mean noise computed in a sliding time window. Figure~\ref{fig:event_display_raw_filt} shows an event containing a number of cosmic-ray muon tracks before (top) and after (bottom) the noise filtering procedure, while figure~\ref{fig:event_display_noise_filters} shows the effect of each filter on a subset of waveforms. The three filters reduce the pedestal RMS of the waveforms by a factor of ~3, as shown in figure~\ref{fig:pedrms_evo_filter}. \\

\begin{figure}[h]
\centering
\includegraphics[width=9cm]{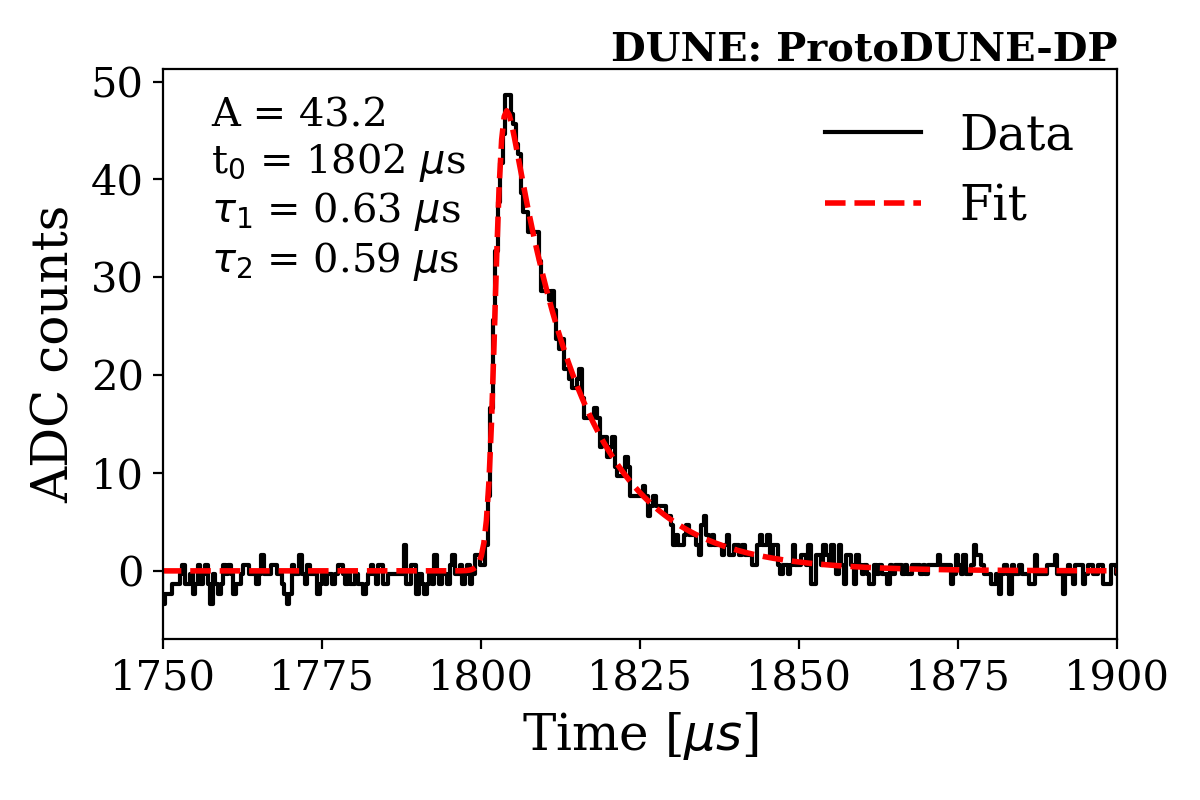}
\caption{Example of a hit fitted with equation \ref{eq:pulsefit} by LArSoft.} \label{fig:pulsefit}
\end{figure}

Once the noise is filtered, the final step of the data preparation is the search for hits in each waveform. The algorithms in the two software packages look for a number of consecutive samples over a threshold, where the latter is linked to the amount of remaining noise in the waveform. A hit is then defined by the total charge, peak time and duration. In LARDON, these quantities are computed from the hit shape. In LArSoft, they are extracted from a fit using an empirical parametrisation of the hit shape as a function of time $t$ modelled as: 
\begin{equation}
    f(t)=A\frac{e^{\frac{t-t_0}{\tau_1}}}{1+e^\frac{t-t_0}{\tau_2}},
    \label{eq:pulsefit}
\end{equation}
where $A$ is the hit amplitude, $t_0$ is the peak time, and $\tau_{1,2}$ are the parameters describing the shape. An example of a hit fitted by LArSoft with this function is shown in figure~\ref{fig:pulsefit}.

For each view, a 2D image can be formed by representing hits
as points with coordinates given by the strip ID and peak time. The collection of hits is then used to perform the 3D reconstruction of the event.

\subsubsection{Track reconstruction with LArSoft and Pandora}
\label{subsec:tpc_reco_pandora}
In LArSoft, the track reconstruction is performed using Pandora, a powerful multi-algorithm framework for automated event reconstruction \cite{pnd1} also used by other LArTPC-based experiments \cite{pnd4,pdsp_pandora_paper}. 

\paragraph{Reconstruction overview}
Pattern recognition in Pandora starts with 2D clustering of the reconstructed hits in each of the detector's readout planes (views).




2D clusters in different views are then compared to find matching combinations that correspond to the same particle.
Based on the matched clusters, a 3D trajectory is constructed for each particle.  
Finally, the reconstructed particles are linked together into hierarchies: in the case of cosmic rays, delta rays are attached to the parent muon.

While the existing reconstruction, later referred to as “standard Pandora”, was initially applied to \pddp without requiring modifications, a number of features were specifically developed for this detector with the aim of bringing the reconstruction performance close to that of a three-readout plane detector.

\paragraph{Hit width cluster merging}
Particles that travel along the readout channels of the detector will produce hits with large widths, spanning long regions of time, which were not clustered efficiently by Pandora.
A new algorithm was developed to inject knowledge of the hit width into the procedure. 
Clusters with wide hits are identified in the event, and 
the hits in these clusters are broken into smaller, constituent hits which are fed into a linear fit to estimate the cluster direction. 
The cluster edges are defined by the extremal constituent hit positions.
Cluster merges are made if successive extremal hits are close to one another in the 2D plane, the cluster directions align and the final cluster direction after the merge agrees with that before.

The hit width cluster merging algorithm found one of its first uses at \pddp where significant improvements to the reconstruction were seen, as discussed later. 

\paragraph{2D $\to$ 3D matching with calorimetry at \pddp}

In a three-view detector, Pandora's standard procedure is to match clusters across views, making use of 2D cluster coordinates and knowledge of the readout orientation: all valid cluster combinations in any pair of views are used to predict where a cluster should be found in the third view. For three-view detectors, the third view provides redundancy, thus making this method robust against ambiguous cases, and allowing the correction of pattern recognition mistakes in one or two views.
With only two views available, as in \pddp, utilising purely spatial information only permits matching of clusters based on their end points in the drift coordinate. However, this method cannot resolve the ambiguities that arise when at least two clusters have the same end points in a given view. In order to overcome these limitations, a novel approach to 2D~$\to$~3D matching, which utilises calorimetric information, was developed.

Pairs of clusters that overlap in time are identified and fractional profiles of the charge measurements for the two clusters are constructed within the overlap region.
 The similarity of the charge profiles is then used to determine which clusters belong to the same particle. A pair of clusters from different views under consideration for matching is referred to as a \textit{matching candidate}.
 
\begin{figure}[h]
\begin{subfigure}[b]{0.5\textwidth}
\includegraphics[width=0.9\textwidth, height=4cm]{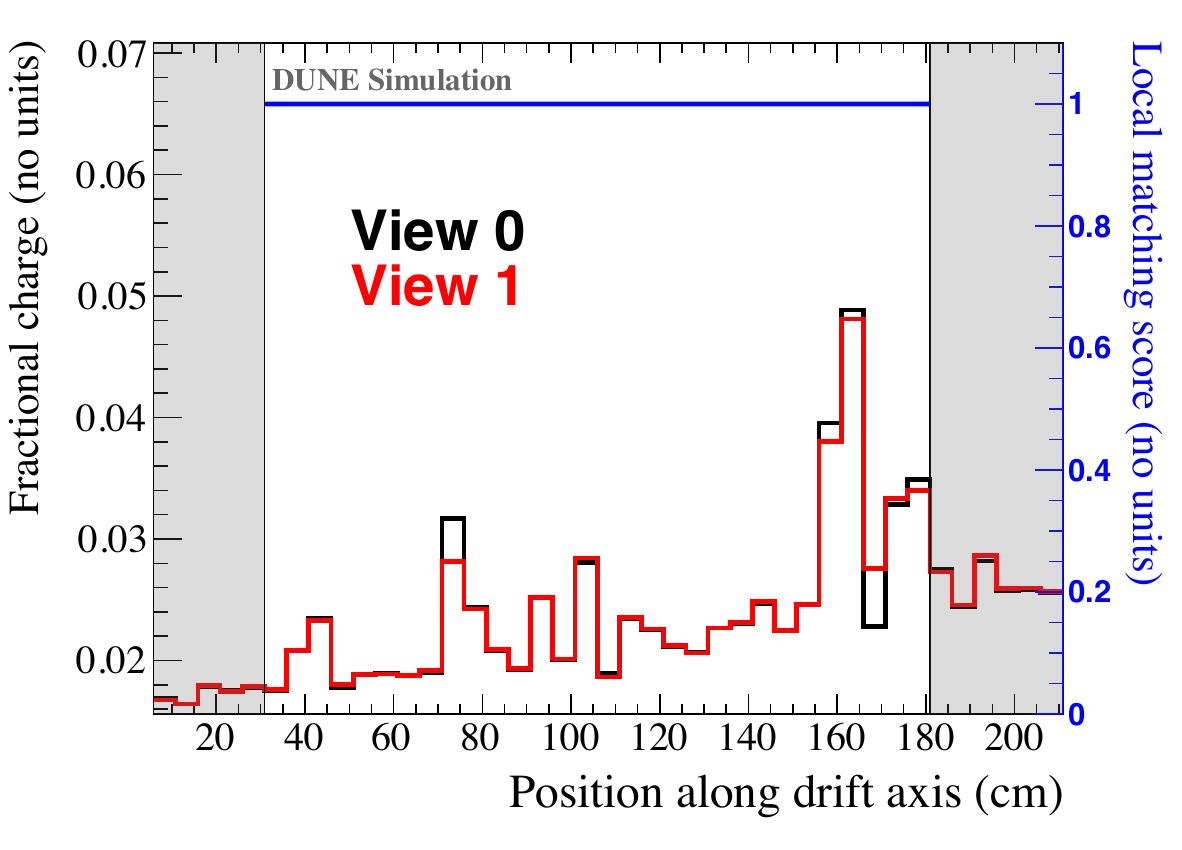}
\caption{A correct match.}
\label{fig:subim1}
\end{subfigure}
\begin{subfigure}[b]{0.5\textwidth}
\includegraphics[width=0.9\textwidth, height=4cm]{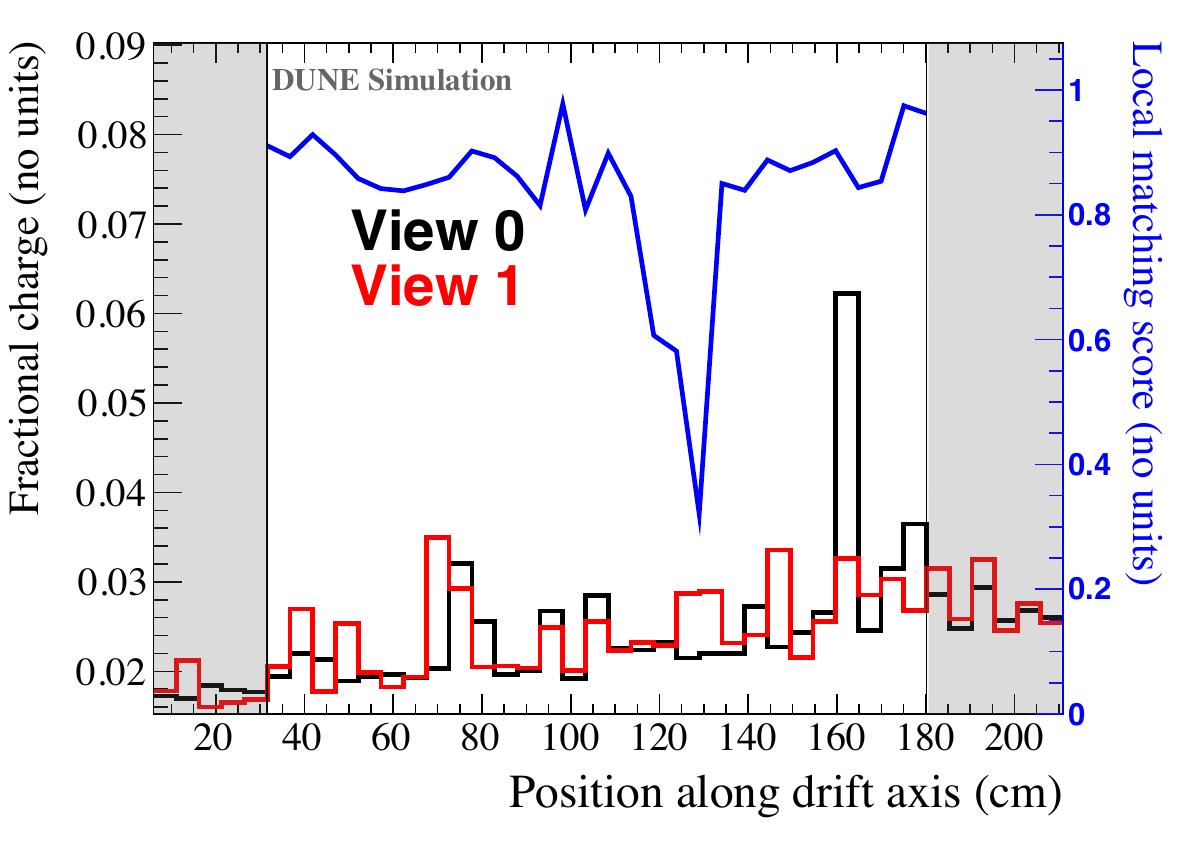}
\caption{An incorrect match.}
\label{fig:subim2}
\end{subfigure}
\begin{subfigure}[b]{0.5\textwidth}
\includegraphics[width=0.9\textwidth, height=4cm]{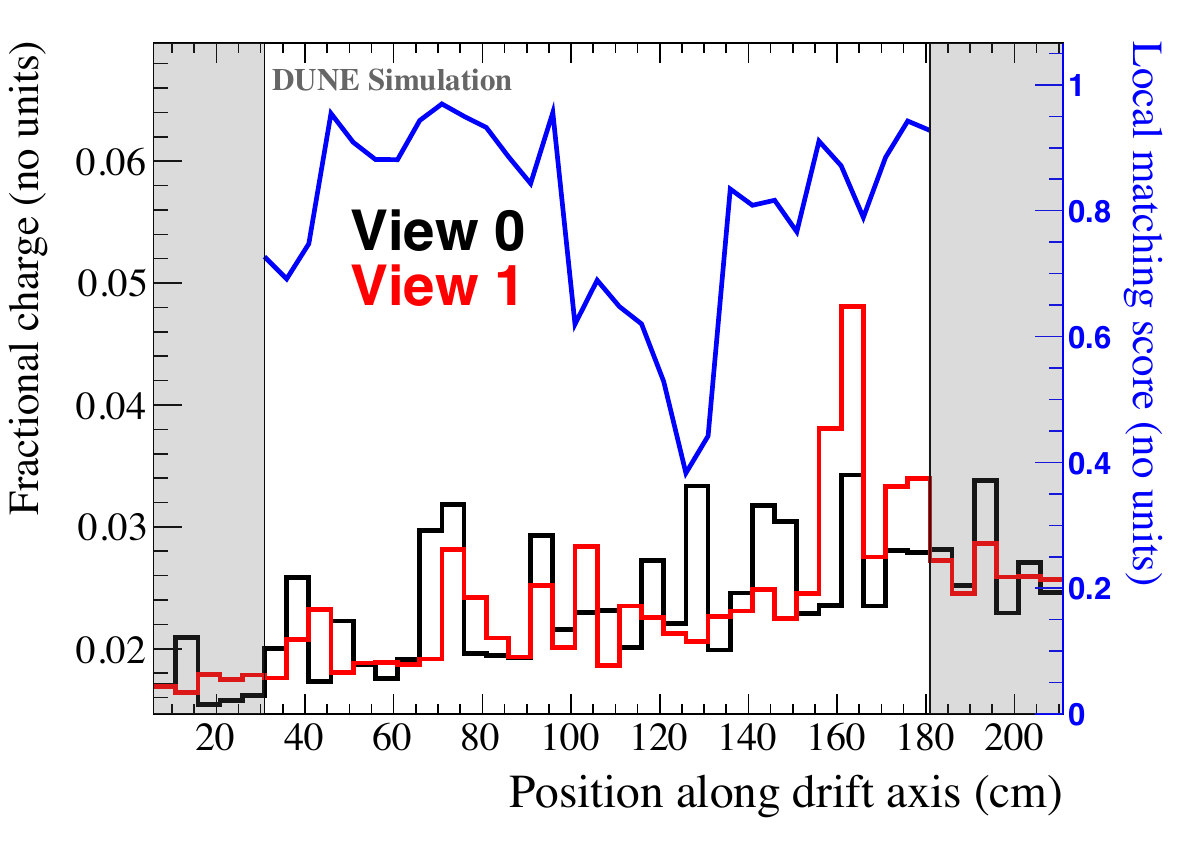}
\caption{An incorrect match.}
\label{fig:subim3}
\end{subfigure}
\begin{subfigure}[b]{0.5\textwidth}
\includegraphics[width=0.9\textwidth, height=4cm]{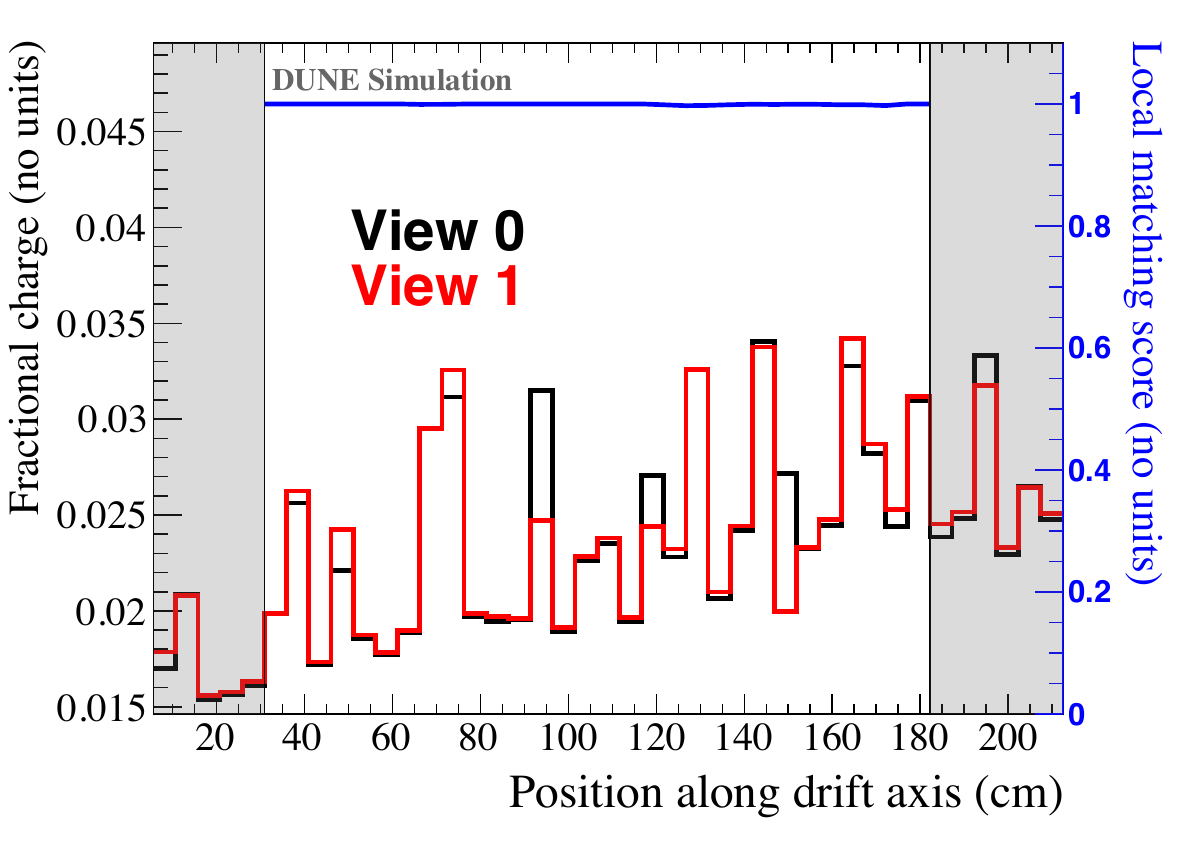}
\caption{A correct match.}
\label{fig:subim4}
\end{subfigure}
\caption{Calorimetry-based similarity scores for the 2D clusters in the two \pddp views for a di-muon particle-gun event produced with LArSoft. Only two of the four combinations correspond to a pair of clusters made by the same particle. The fractional charge profiles are shown in black (view 0) and red (view 1). The leading edge of the bins in the fractional charge profiles indicates each charge sampling position. The local matching score is shown in blue. A local matching score that is consistently very close to 1 across the whole position range indicates a correct match. The shaded region highlights where too few points were available to calculate a local matching score.}
\label{fig:caloprofiles}
\end{figure}

Example profiles for a di-muon particle gun Monte Carlo event, produced with the \pddp LArSoft simulation, where the two muons originate from the same vertex and fully overlap in the drift region, are shown in figure \ref{fig:caloprofiles}: each muon produces a cluster in one of the readout views, thus four possible combinations of two clusters, or matching candidates, are to be considered (the black and red histograms in each of the sub-figures represent the fractional charge profiles for view 0 and view 1 clusters respectively, for one of the four pairings under consideration). 

In order to assess the similarity of profile pairs, a time window is slid across them bin-by-bin, and a local matching score is repeatedly calculated at each step (blue line in sub-figures). The score is defined as $L = 1-p$, where $p$ is the $p$-value for measuring a correlation coefficient $r$, of the bins contained in the window, assuming the true $r$ is 0. A local matching score with values consistently around 1 indicates a good match. A locally matched fraction is then defined as the fraction of $L$ values above a specified threshold.

\paragraph{Performance studies on cosmic ray simulation}

\begin{figure}

\includegraphics[height=5cm]{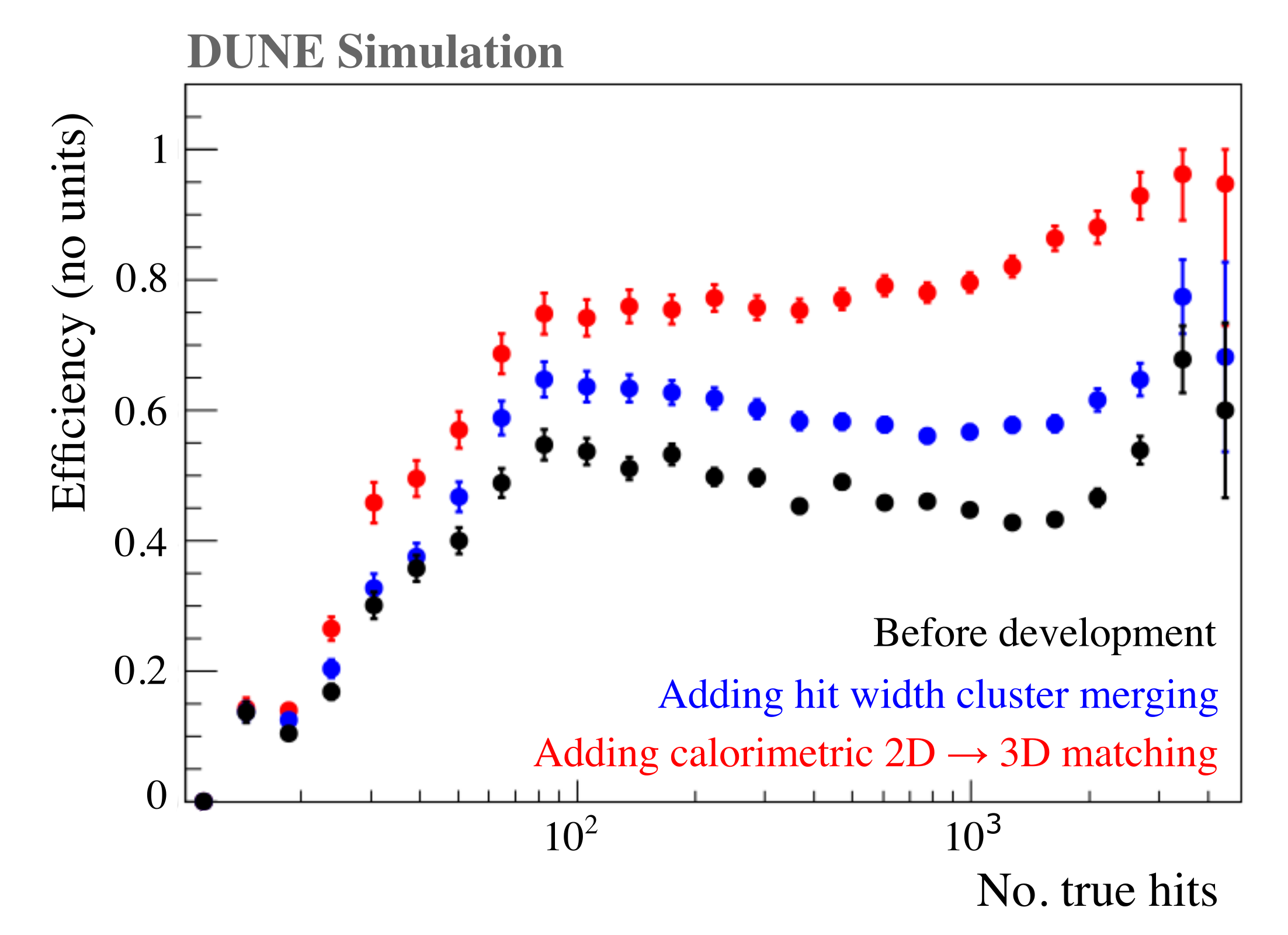}
\includegraphics[height=5cm]{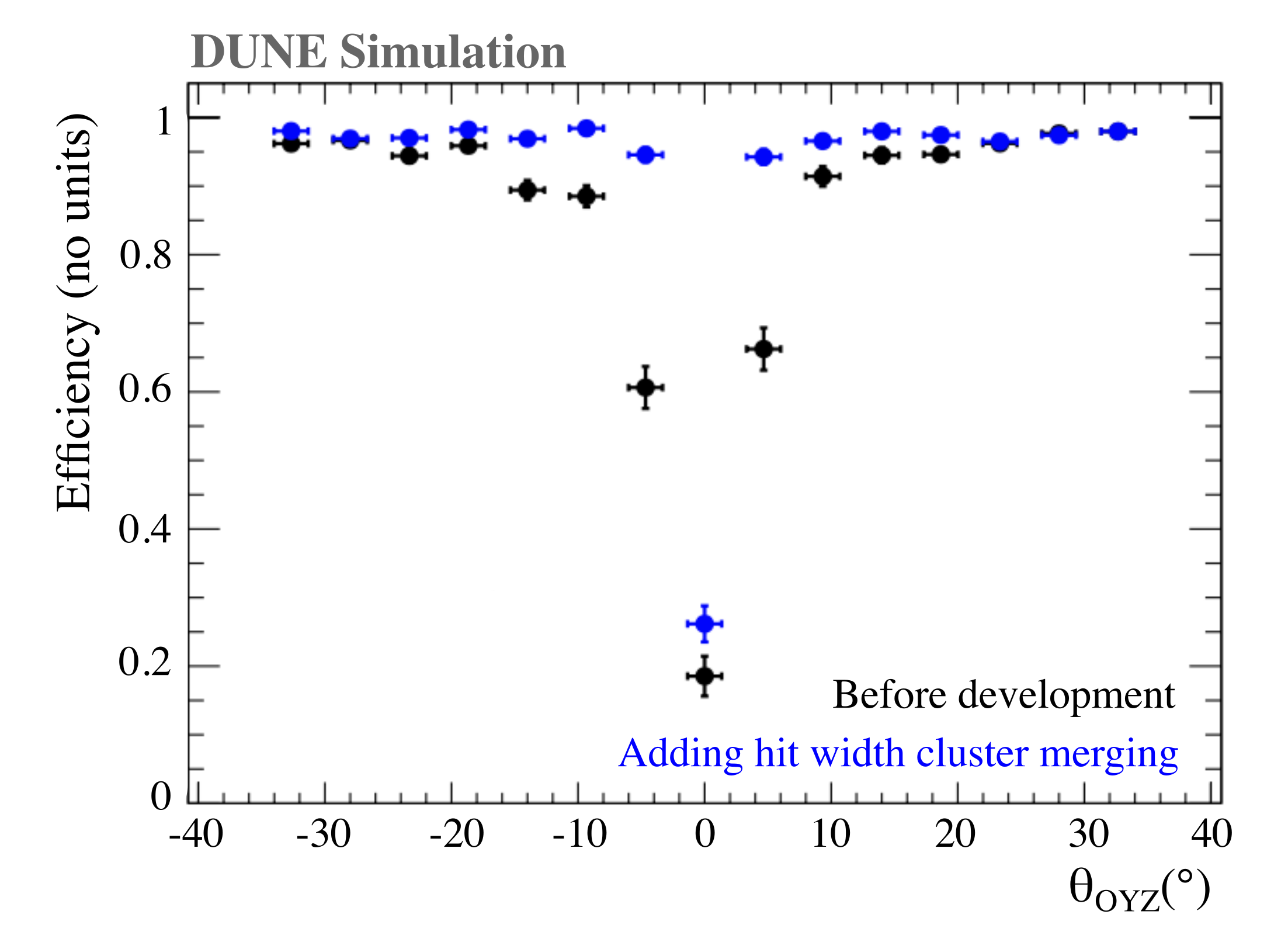}
\caption{(Left) Reconstruction efficiency for cosmic muons as a function of the total number of true hits, with the standard Pandora reconstruction (black), with the addition of the new hit width-based cluster merging algorithm (blue), and with the addition of the new calorimetric 2D~$\to$ ~3D matching (red). (Right) Reconstruction efficiency, as a function of the angle in the horizontal plane OYZ, for a simulated sample of single 2\,GeV/c muons, with the standard Pandora reconstruction (black) and with the addition of the new hit width-based cluster merging algorithm (blue).}
\label{fig:hitWidthAlgo_effiency}
\end{figure}

The performance of the new hit width cluster merging algorithm and the new 2D~$\to$~3D calorimetric matching developed for \pddp were tested on simulated events that contained cosmic rays in the full readout window, with a few tens of tracks per event~\cite{EtienneThesis}. 10000 cosmic muons were simulated in a \pddp detector operation with a nominal drift field (500~V/cm) without electronic noise nor space-charge. A 30 ms electron lifetime was set up according to the values obtained in ProtoDUNE-SP. 
Three different reconstruction configurations were tested:
\begin{enumerate}
    \item Standard Pandora
    \item Standard Pandora, adding the hit width cluster merging algorithm
    \item Standard Pandora, adding both the hit width cluster merging algorithm and the novel 2D~$\to$~3D calorimetric matching.
\end{enumerate}
Figure~\ref{fig:hitWidthAlgo_effiency} shows the reconstruction efficiency as a function of the number of hits associated with the simulated cosmic muons. For a given particle type, efficiency is defined as the fraction of reconstructible simulated particles that are successfully matched to at least one reconstructed particle.  A particle is deemed reconstructible if it consists of a minimum of 15 hits in total, with at least 5 hits per plane.  Matching between simulated and reconstructed particles is solely determined by assessing deposited energy contributions to the constituent hits. A match is considered successful if it has a purity greater than 10\% and a completeness greater than 50\%. Purity measures the fraction of hits in the reconstructed particle that originate from the matched simulated particle, while completeness measures the fraction of hits from the simulated particle that are included in the reconstructed particle. Thus, purity quantifies the degree to which a reconstructed particle is free from contamination by hits originating from other true particles, whereas completeness quantifies how much of the true particle has been reconstructed. These performance figures are obtained using nominal simulation and therefore do not fully include detector effects present in real data. 
The reconstruction efficiency as a function of the angle in the horizontal plane was also studied separately, using a simulated 2\,GeV/c single-muon sample, as shown in figure~\ref{fig:hitWidthAlgo_effiency}, 
demonstrating the improvements achieved by the new hit width cluster merging algorithm in recovering performance for tracks parallel to the strip direction (at $\theta_{OYZ}$=0), where the standard Pandora typically finds a single reconstructed hit. Tracks parallel to the strip direction remain a difficult topology to recover with current matching algorithms as hit patterns reduce to one long pulse train along a single strip, whose shape is non-trivial to interpret and provides limited description of a particle's trajectory.

Overall, the reconstruction efficiency rises from 46\% for the Standard configuration to 57\% when adding the hit width cluster merging algorithm, and to 76\% when also including the new 2D~$\to$~3D calorimetric matching.

As shown in~\cite{Brailsford:2021htz}, a two-view detector under-performs with respect to a three-view detector in terms of track reconstruction efficiency. However, further development of the reconstruction, in particular on the calorimetric 2D~$\to$~3D matching, may reduce the performance difference, targeting failure modes where one of the views has poor quality. Conversely, the overall performance for neutrino flavour tagging using whole-event information via a CNN is largely unaffected by the number of readout planes.

\subsubsection{Track reconstruction with LARDON}
\label{subsec:tpc_reco_lardon}

From the collection of identified hits described in~\ref{subsec:tpc_reco_preparation}, the LARDON software first builds 2D tracks in each view, and then associates the reconstructed 2D objects in 3D.
In each view, the hits are ordered in a R-tree~\cite{Rtree} which is a tree data structure used for indexing multi-dimensional information. In LARDON, the R-tree is indexed with the hit position along the view and hit peak time converted into a distance using an assumed electron drift velocity. From a hit not yet attached to a 2D track, a collection of spatially close-by hits is extracted from the R-tree. A 2D track is seeded and its parameters are updated with the addition of nearby hits using a Kalman-like algorithm~\cite{BILLOIR1984352}. Once this step is done for all the hits, a stitching algorithm connects tracks potentially broken due to un-instrumented areas of the detector based on track directions compatibility and distances. A minimum spanning tree graph is then built for each 2D track to identify the delta-rays from the vertices of the graph.
The 3D tracks are built from the association of two 2D tracks, one in each view. The 2D tracks must overlap in time, and in the overlapping region the amount of charge deposited in each view must be equal within 25\%. Each track has a best-match track in the other view, and when the matches are reciprocal, a 3D track is built: the missing coordinate in one view is computed from the interpolated path of the other track. LARDON's reconstruction has been tested on simulations and also compared to the LArSoft results where similar performance was established in terms of number of tracks reconstructed, track orientation and measured calorimetry.

\subsection{Muon selection and field correction}
\label{subsec:tpc_ana}

In order to assess the performance of \pddp with cosmic ray tracks, the analysis relies on the selection of muon tracks as they have well defined long trajectories and deposit a known amount of energy in LAr.  

A set of criteria are applied to select a sample of muon-like tracks for both the LArSoft~\cite{PierreThesis} and LARDON~\cite{PabloThesis} reconstruction chains.
These selections are described in the following section. 

\subsubsection{Muon selection}
\label{subsec:tpc_ana_muon}
To remove noise tracks, tracks are required to have a minimum 3D length of 20\,cm and contain at least 15 associated hits in each view. Only tracks entering the detector through the anode (CRP) after the start of the data taking are selected such that their arrival time could be computed. 
Unfortunately, due to the short between the field cage and the HV extender, the drift field was not uniform in the detector. In particular, as seen in figure~\ref{fig:driftefield}, the field had a large horizontal component near the field cage. Further requiring that the starting position of each track is at least 30\,cm away from the field cage ensures that the tracks did cross the CRP plane. 
Furthermore, as the LEMs have an un-instrumented border (see figure~\ref{fig:lem_carbon_cb}) and some LEMs were operated at low voltage, it is also required that the track begins at least 5\,cm away from its closest LEM border. All analyses presented in this section uses track sample from this muon selection.\\

\begin{figure}
    \centering
    \includegraphics[width=0.6\textwidth]{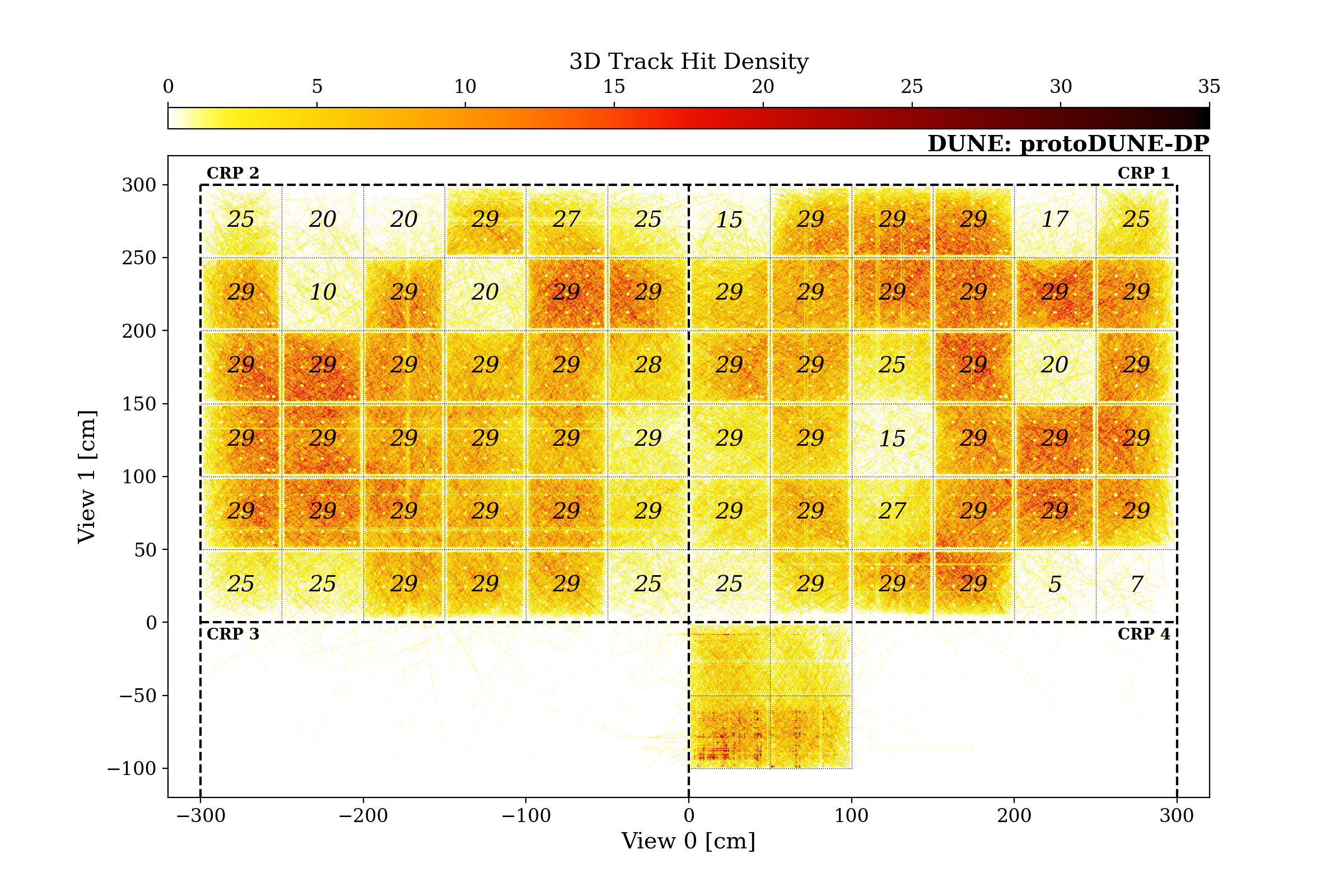} 
     \includegraphics[width=0.38\textwidth]{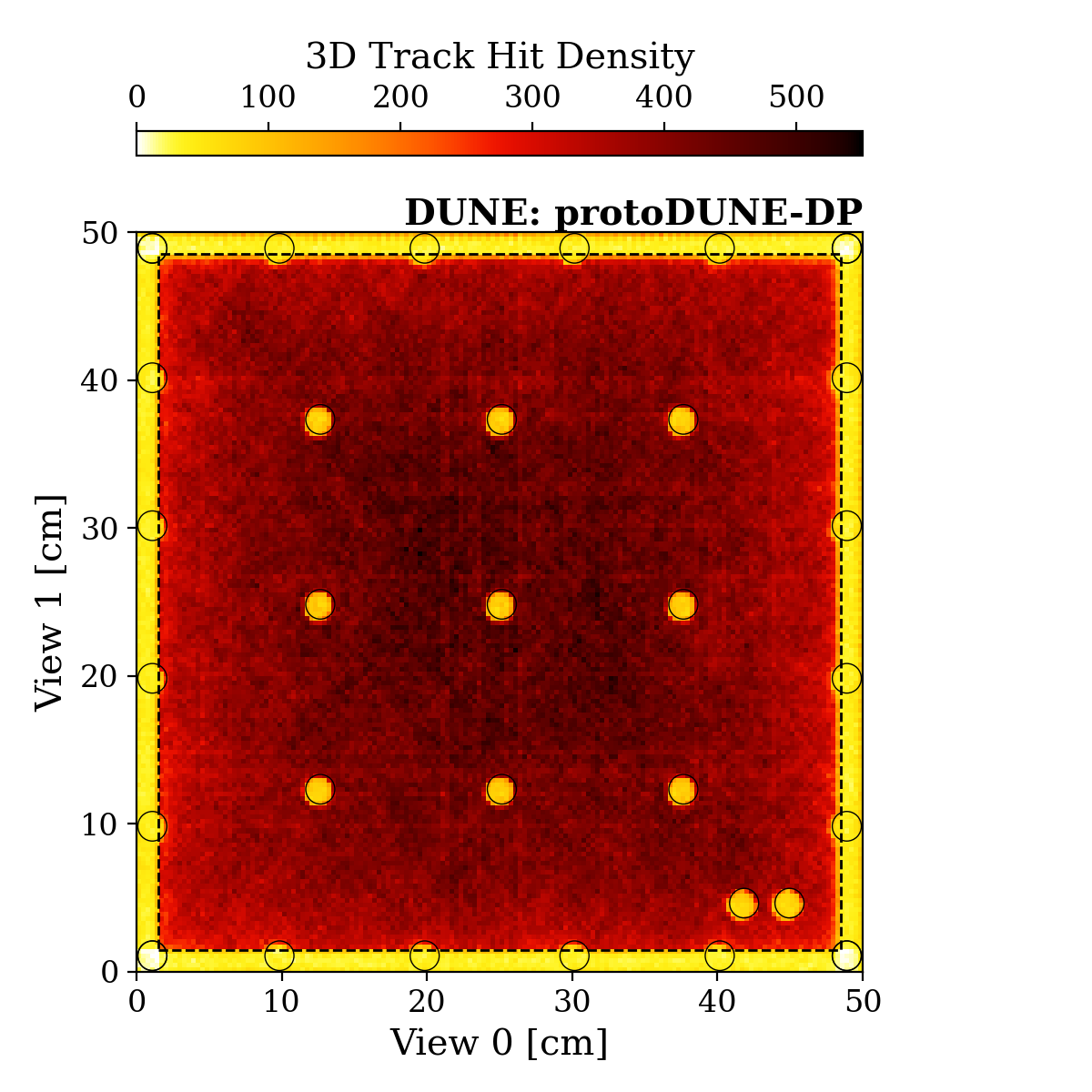}  
    \caption{Projection of reconstructed 3D tracks along the two views. The color scale represents the density of hits at a given position. (Left) The 4 CRPs are highlighted in solid dashed lines, the LEMs/Anodes with thin dashed lines. The amplification field applied to each LEM (in kV/cm) is also indicated. (Right) All the tracks are summed into one effective LEM. The theoretical location of un-instrumented area (protective border, screws and HV connectors) of the LEMs are indicated. This run was recorded in January 2020, and the plots are made using the LARDON reconstruction.}
    \label{fig:ana_3Dhit_density}
\end{figure}

In figure~\ref{fig:ana_3Dhit_density} (left), all 3D tracks reconstructed and selected with the cuts described previously are superimposed and projected along the two collection views. The color scale represents the density of hits attached to a reconstructed 3D track at a given point. In this run, not all LEMs were set at the same voltage -- those are indicated in the figure. It is clear from the figure that the amplification field has a direct impact on the number of tracks reconstructed. In figure~\ref{fig:ana_3Dhit_density} (right), a similar plot is shown, but all the hits are gathered into a single effective LEM. From the density, one can clearly identify the un-instrumented protective border, the location of the screws and the HV connectors of the LEMs. 

\subsubsection{Recombination and field distortion corrections}
\label{subsec:tpc_ana_field}

\begin{figure}
    \centering
    \begin{tabular}{cc}
    \includegraphics[width=0.4\textwidth]{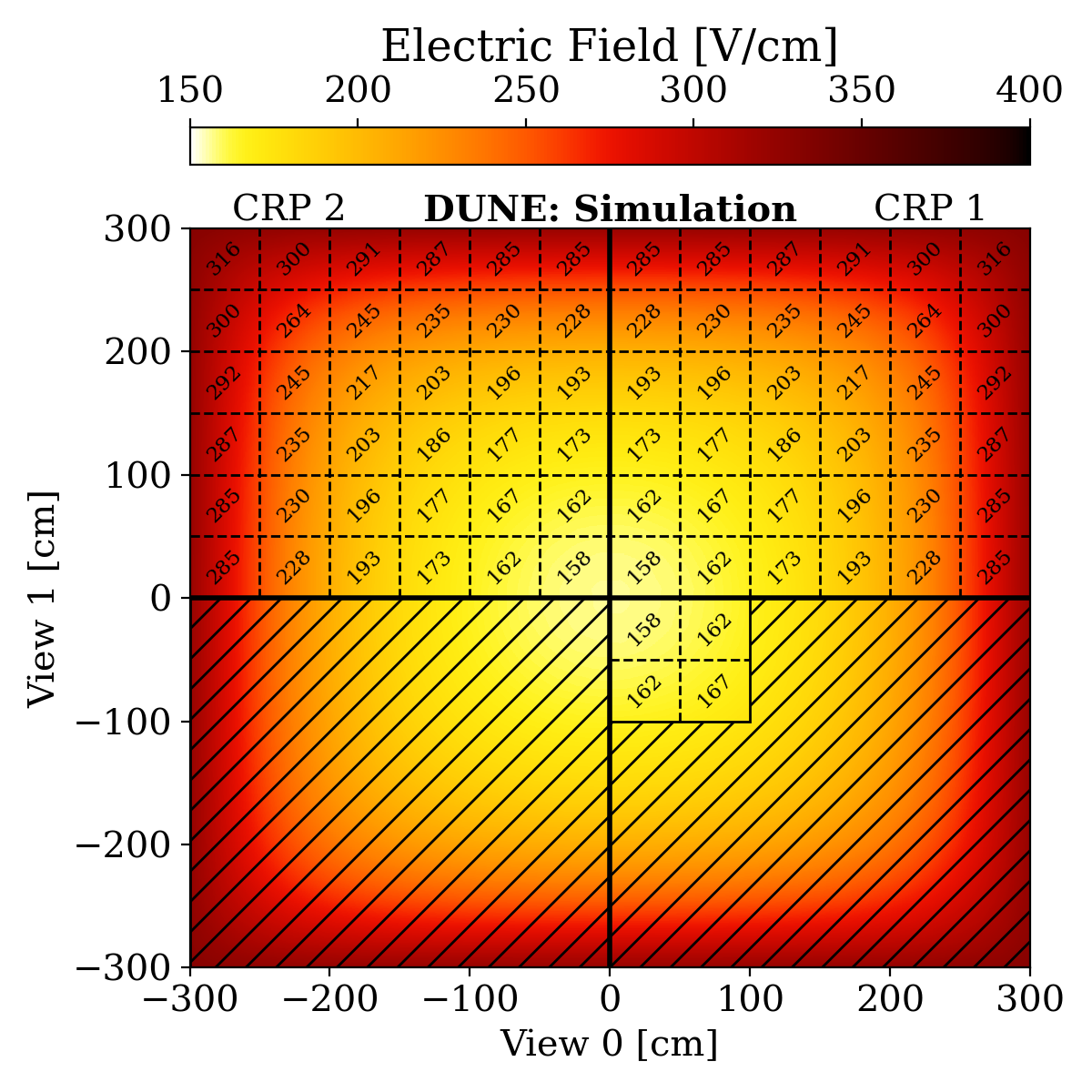}& 
    \includegraphics[width=0.5\textwidth]{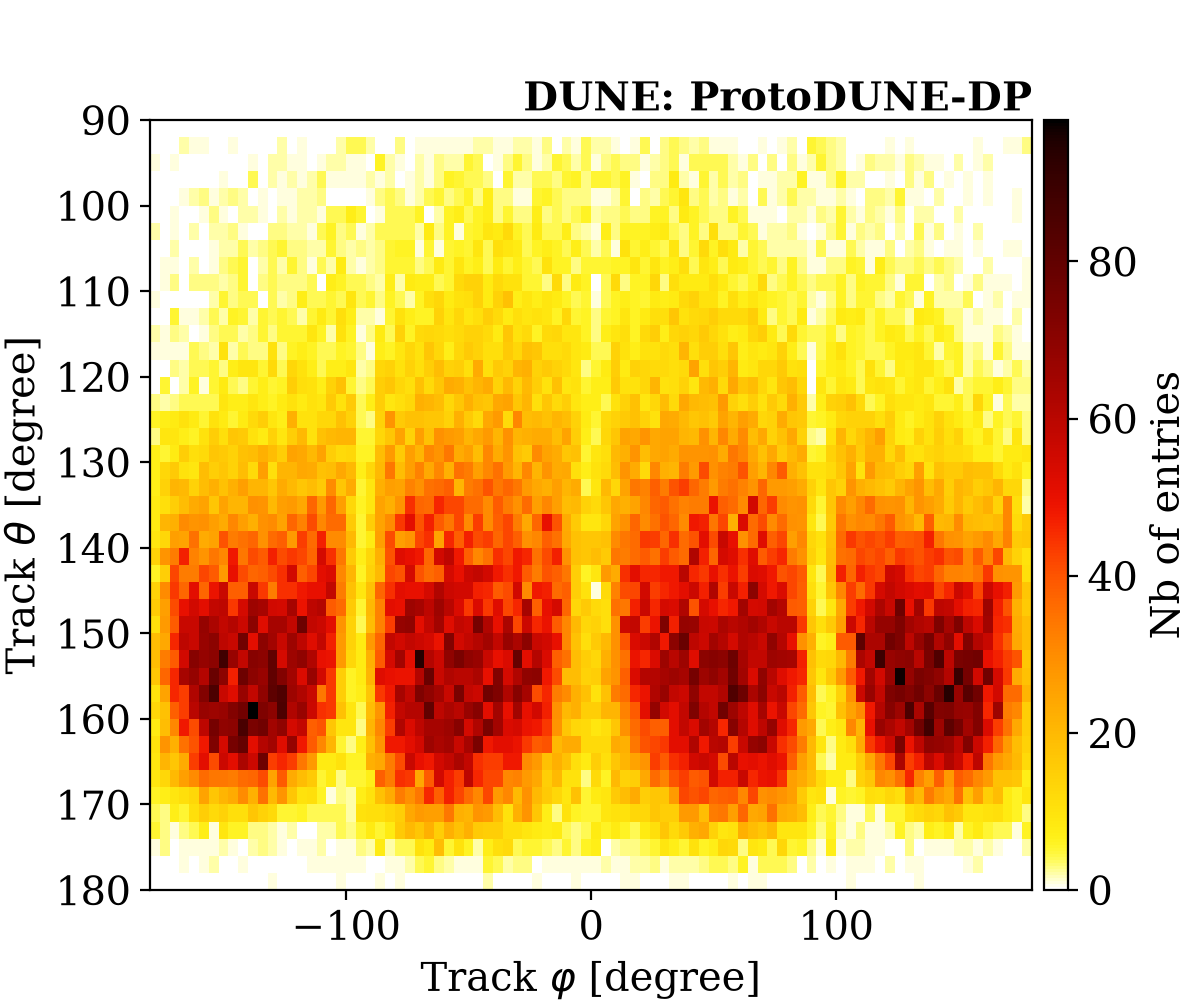}
    \end{tabular}
    \caption{(Left) Mean value of the simulated drift field over 50\,cm of drift from the anode (see text) using COMSOL~\cite{comsol}, shown from above the detector. The mean value over each LEM/Anode is written. (Right) LARDON-reconstructed, selected and field-corrected muon-like  angular phase-space. The azimuthal angle $\varphi$ is computed with respect to the View 0 axis; the polar angle $\theta$ to the drift axis. The loss of nearly horizontal tracks for $|\varphi| \gtrsim 145^\circ$ is attributed to the effect of the nearby Jura mountains. This run was recorded in September 2019 at a $\dvlem =$ 2.9\;kV.}
    \label{fig:ana_drift_field_phase_space}
\end{figure}

The charge collected per unit length, $dQ/ds$ measured in fC/cm, is directly related to the energy deposition of the particle:
\begin{equation}
\label{eq:dqds_formula}
    \frac{dQ}{ds} = G_{\mathrm{eff}}\times\frac{dE}{ds}\times\frac{q}{W_i} \times r(E_{\mathrm{drift}}),
\end{equation}
where $G_{\mathrm{eff}}$ is the effective gain of the detector, $dE/ds$ = 2.1\,MeV/cm on average for muons at the minimum ionizing potential (MIP), $W_i$ = 23.6\,eV/pair is the LAr ionisation energy, $q$ the electron charge and $r(E_{\mathrm{drift}})$ is the recombination factor.
In both software packages, LArSoft and LARDON, the reconstruction is performed with an assumed constant value of the electric field.
As discussed in section~\ref{sec:operation}, the drift field is not uniform in \pddp: the recombination factor and the electron drift velocity are position dependent. This non-uniformity is clearly visible in the events, where most of the tracks appear to be bent, as seen in figure~\ref{fig:event_display_raw_filt}. It is therefore mandatory to correct the collected charge, $dQ$, and the unit length, $ds$, of all the hits associated to the selected tracks to have a consistent measurement of the deposited charge. 
Using COMSOL~\cite{comsol}, a simulation of the field inside the TPC has been carried out, reproducing the data taking conditions with a nominal voltage of $V_{\mathrm{cath}} = 50$\,kV delivered to the cathode and the shorted Field Cage ring. From this simulation, a maximum depth has been computed over the discretised collection plane with the condition that the recombination factor (affecting $dQ$) and the drift velocity (affecting $ds$) do not vary by more than 2\% with respect to their value at the collection plane. The conservative maximum depth value used in the analyses is 50\,cm for the whole detector, which is smaller than the depth computed from the criteria defined previously. Figure~\ref{fig:ana_drift_field_phase_space} (left) shows the mean value of the electric field over 50\,cm of drift from the collection plane, and the mean value per LEM/Anode.

\begin{figure}
    \centering
\includegraphics[width=0.9\textwidth]{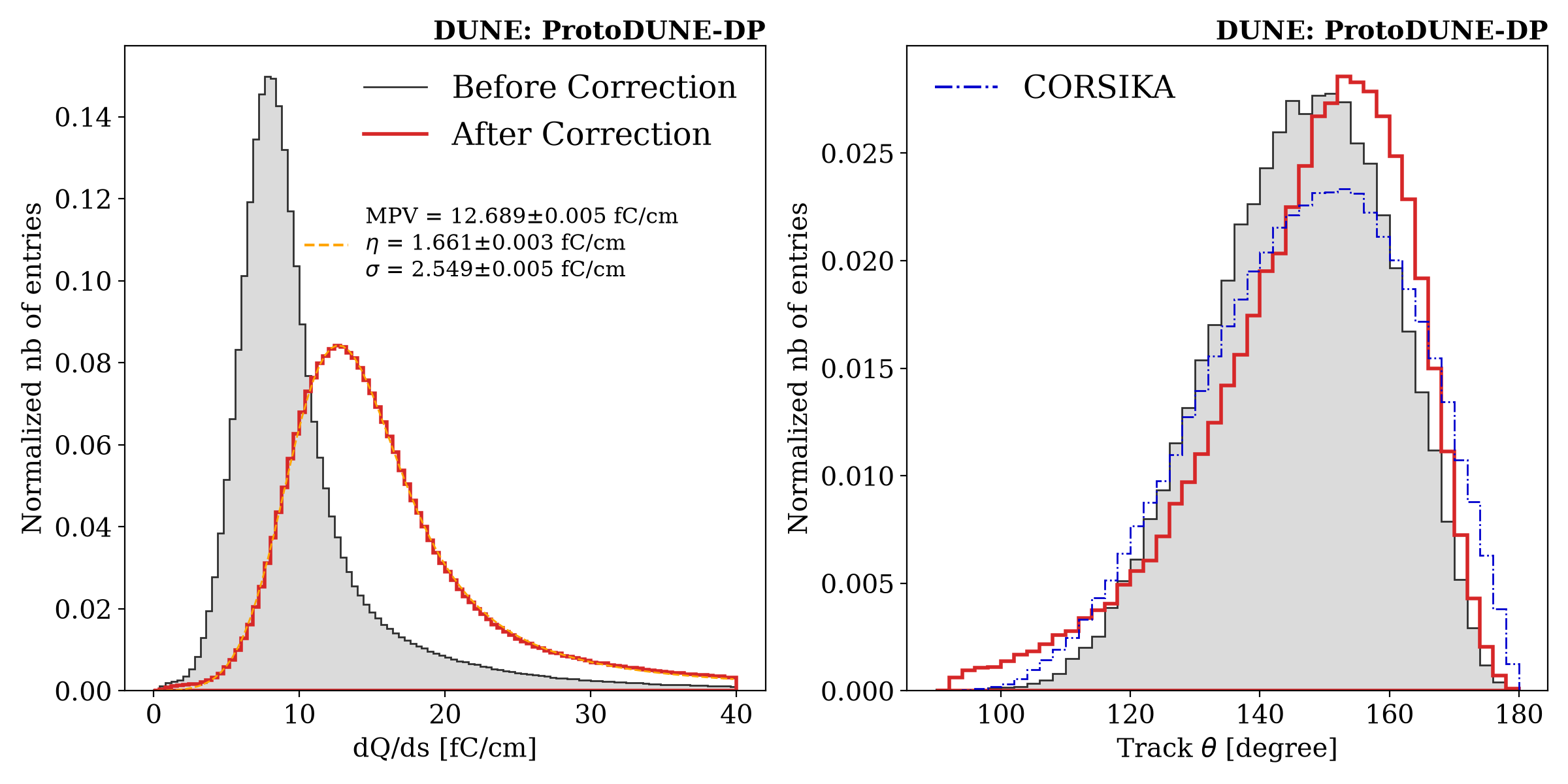}
\caption{Effect of the recombination and field distortion corrections on the collected charge (left) and polar angle (right) muon-like tracks using LARDON. The corrected distribution of $dQ/ds$ is fitted with a Landau (MPV, $\eta$) convolved with a gaussian (MPV, $\sigma$) function. The expected polar angle distribution cosmic flux at sea-level from CORSIKA is also indicated. This run was recorded in September 2019 at a $\dvlem =$ 2.9\;kV.} 
\label{fig:ana_field_corr}
\end{figure}

The depth of the hits associated to the reconstructed 3D tracks is corrected with the drift velocity estimated from the field simulation~\cite{AMORUSO200468}. The $ds$ of each hit is then recomputed with the updated track 3D position. The amount of charge collected $dQ$ is also corrected using the local recombination factor~\cite{AMORUSO2004275}. Figure~\ref{fig:ana_field_corr} shows the effect of the field corrections on the collected charge and on the polar distributions of selected tracks. 




The transverse displacement of the electron cloud due to the drift field non uniformity is not taken into account in the correction process. 
Therefore, azimuthal angle $\varphi$ of the track is not affected by the field correction. Figure~\ref{fig:ana_drift_field_phase_space} (right) shows the angular phase-space of the reconstructed, selected and corrected tracks. The absorption of nearly horizontal tracks by the nearby Jura mountains is clearly visible in the distribution. 


\subsubsection{Charge sharing between views}
\label{subsec:tpc_reco_quality}
\begin{figure}[h]
\centering
\includegraphics[width=0.9\textwidth]{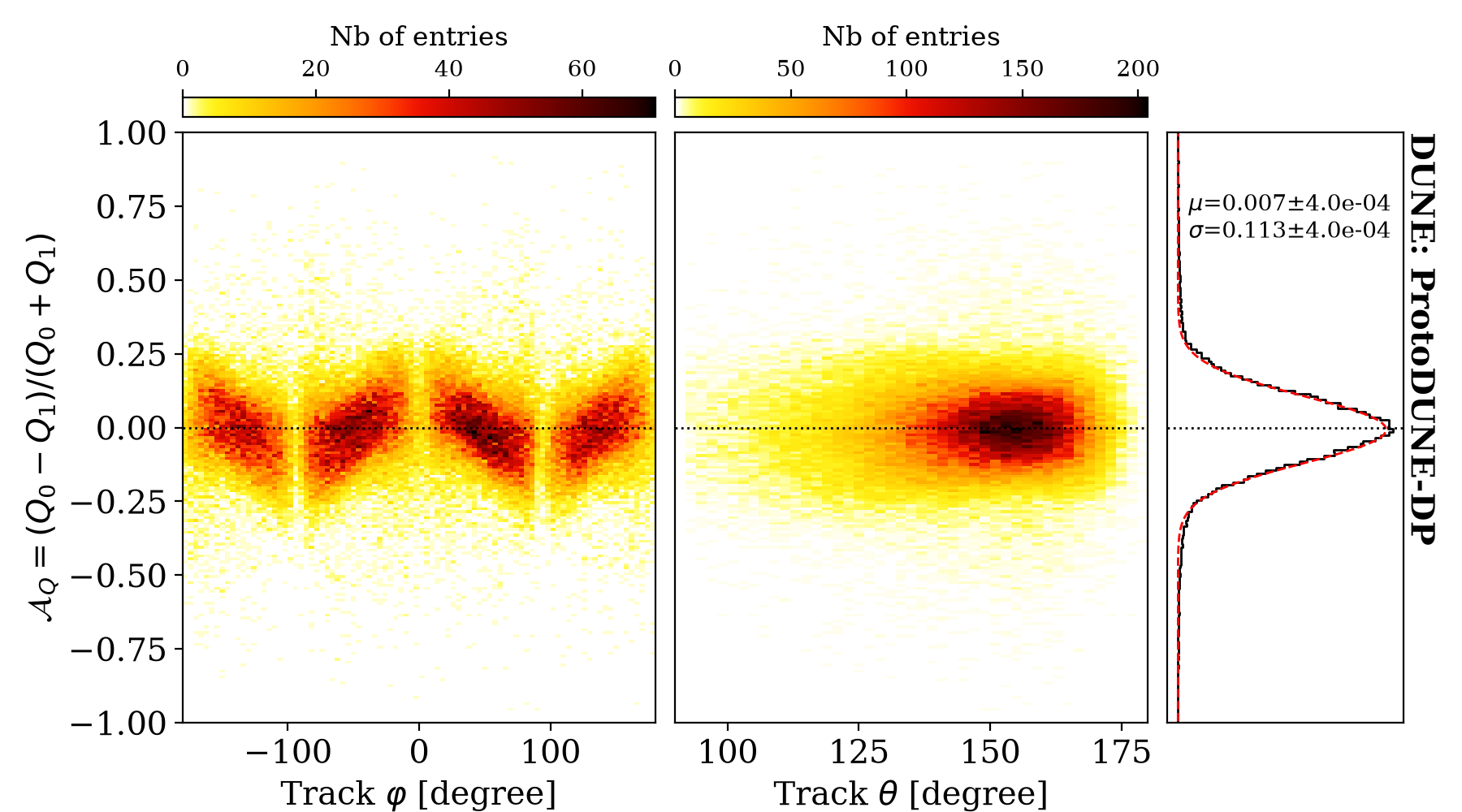}\\
\includegraphics[width=0.9\textwidth]{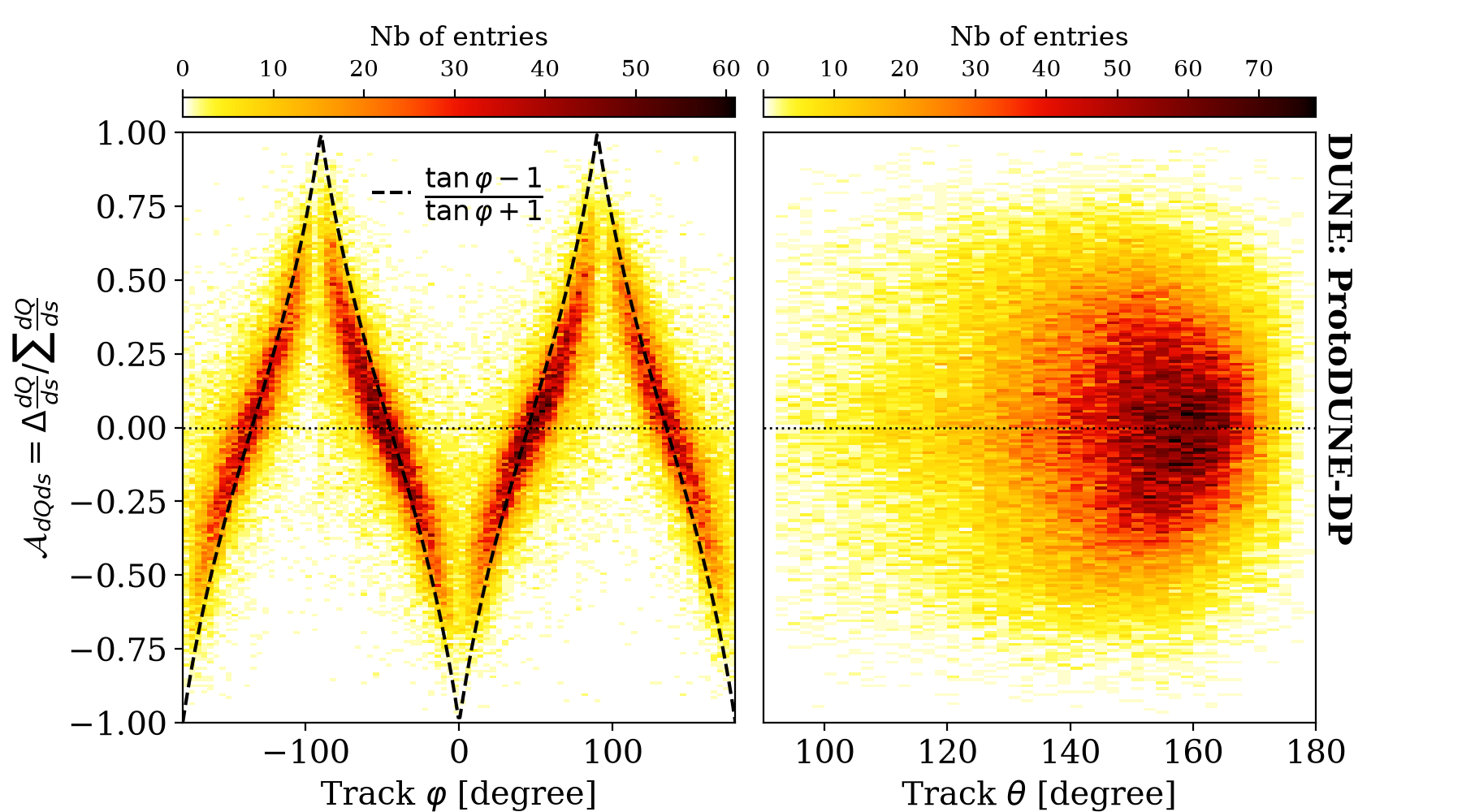}
\caption{Charge sharing asymmetries between the two collection views: $\mathcal{A}_{q}$ (top) and $\mathcal{A}_{dQds}$ (bottom). The asymmetries are presented as a function of track $\varphi$ (left) and $\theta$ (right). This run was recorded in September 2019 at a $\dvlem =$ 2.9\;kV and reconstructed with LARDON. }
 \label{fig:charge_asym}
\end{figure}

The anode in \pddp is designed for equal charge sharing between the two collection views~\cite{Cantini_2014}. This key requirement facilitates the reconstruction of charge signals, and thus measurements of physical quantities such as the effective gain of the detector. To study this equal sharing one can define a charge asymmetry coefficient $\mathcal{A}_q$ as
\begin{equation}
   {\mathcal{A}_q} = \frac{Q_0 -Q_1}{Q_0 + Q_1},
\end{equation}
where $Q_i$ represents the sum of the charge deposited per tracks in view $i=0, 1$.
The distributions of $\mathcal{A}_q$ are shown in figure~\ref{fig:charge_asym} (top) as a function of the azimuthal and polar angles of the tracks. The global $\mathcal{A}_q$ follows a gaussian distribution centred at $\mu=0.007$ as expected from the anode design, with a spread of $\sigma=0.113$. The $\mathcal{A}_q$ deviates from its central value for tracks with directions nearly parallel to the strips of the two views : $\varphi=0^\circ,\pm 90^\circ, \pm180^\circ$. For these tracks, the reconstruction is difficult as the amount of collected charge is shared on just a few strips. The hit finding algorithm has trouble to reconstruct single and long charge deposits and disentangle them from noise within the waveform.

Another asymmetry coefficient that can be looked at to assess the good charge sharing as well as the quality of the reconstruction is $\mathcal{A}_{dQds}$ defined as~\cite{311_performance}:
\begin{equation}
   {\mathcal{A}_{dQds}} = \frac{dQ/ds|^{tot}_0 -dQ/ds|^{tot}_1}{dQ/ds|^{tot}_0 + dQ/ds|^{tot}_1},
\end{equation}
where $dQ/ds|^{tot}_i$ is the sum of the reconstructed charge deposited per unit length per track in view $i=0, 1$. The sum can be approximated as $dQ/ds|^{tot}_i\approx Q_i/\ell_i$ where $\ell_i$ is the projected track length on the strips of the views:
\begin{eqnarray*}
    \ell_0 &= L\sin\theta\cos\varphi\\
    \ell_1 &= L\sin\theta\sin\varphi\\
\end{eqnarray*}
and $L$ is the track total length. If the anode equally splits the charge among the two views, then $\mathcal{A}_{dQds}$ has only a dependence on the azimuthal angle of the track: $$\mathcal{A}_{dQds}= \frac{\tan\varphi-1}{\tan\varphi+1}.$$  The distributions of $\mathcal{A}_{dQds}$ as a function of tracks $\varphi$ and $\theta$ are shown in figure~\ref{fig:charge_asym} (bottom). The $\varphi$ dependence is clearly seen and it is in good agreement with the expectation. Small deviations are nevertheless observed, and is likely due to the drift field non-uniformity. As expected, there is no correlation between $\mathcal{A}_{dQds}$ and $\theta$. For tracks with $\varphi=\pm45^\circ, \pm135^\circ$, the projected lengths are the same on the two views and $\mathcal{A}_{dQds} = 0$ which is observed in the figure.
The study of the two asymmetries indicates the good performance of the anode and the 3D track reconstruction.

\subsection{Measurement of the liquid argon purity}
\label{subsec:tpc_purity}

As primary ionisation electrons move along electric drift field lines towards the anode, they have a non-zero probability to be attached to electronegative contaminants such as oxygen, nitrogen, water or carbon dioxide.
The total amount of charge collected thus decreases with the time $t$ the electrons have drifted before being collected at the anode. This loss of charge follows an exponential 
law:
\begin{equation}
\frac{dQ}{ds}(t) = \frac{dQ}{ds}(t_0)\times \exp(-\frac{t-t_0}{\tau_e}),
\end{equation} 
where $dQ/ds(t_0)$ is the initial ionisation charge created at the time $t_0$ of interaction in LAr and $\tau_e$ is called the electron lifetime, related to the equivalent oxygen concentration through the parametrisation~\cite{BUCKLEY1989364}:
\begin{equation*}
    \tau_e\;[\mathrm{ms}] = \frac{0.3}{\rho_{O_2}\;[\mathrm{ppb}]} .
\end{equation*}

\begin{figure}[htb]
\centering
\begin{tabular}{cc}
  \includegraphics[width=0.45\textwidth]{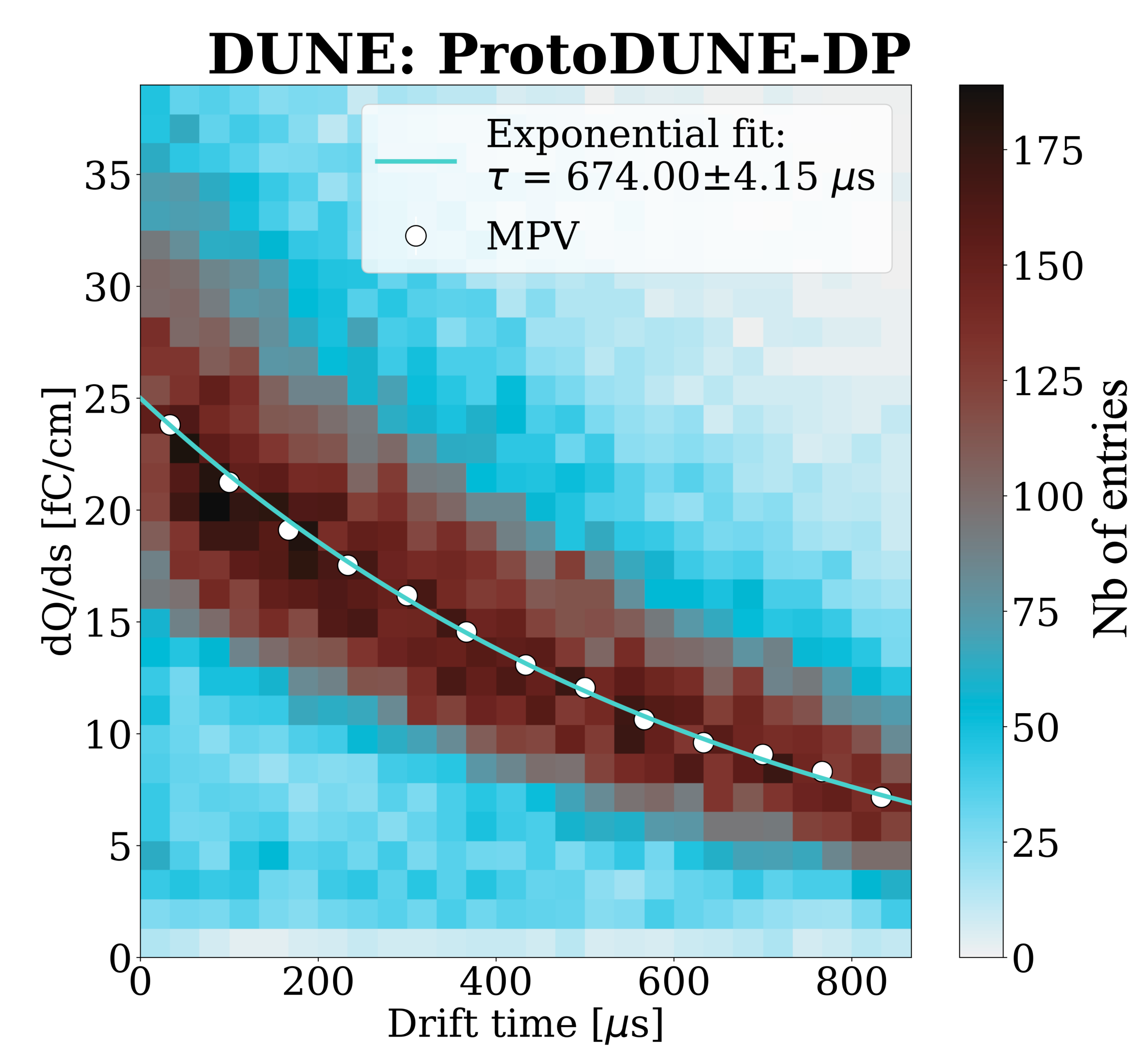} &
      \includegraphics[width=0.45\textwidth]{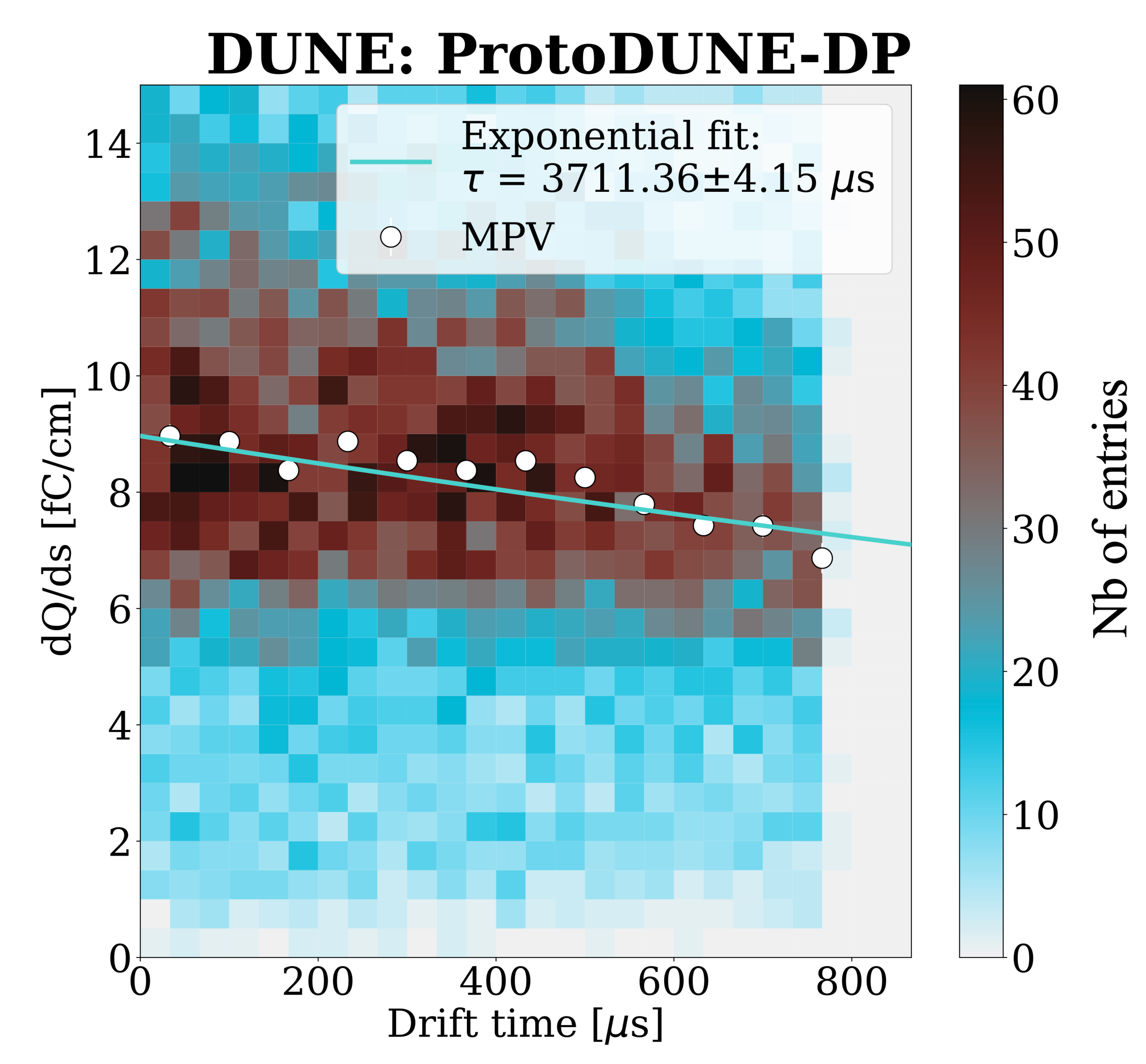}
  \end{tabular}
\caption{$dQ/ds$ as a function of drift time in a given LEM. (Left) run taken in October 2019 with a low LAr purity. (Right) run taken in January 2020 where the purity is much higher. The exponential fit to the MPV to extract the lifetime is superimposed, the fitted error are shown. Both runs were taken with $\dvlem =$ 3.1\;kV reconstructed with LARDON.}
 \label{fig:dqdxvst}
\end{figure}

\begin{figure}
    \centering
    \begin{tabular}{cc}
    \includegraphics[width=0.45\textwidth]{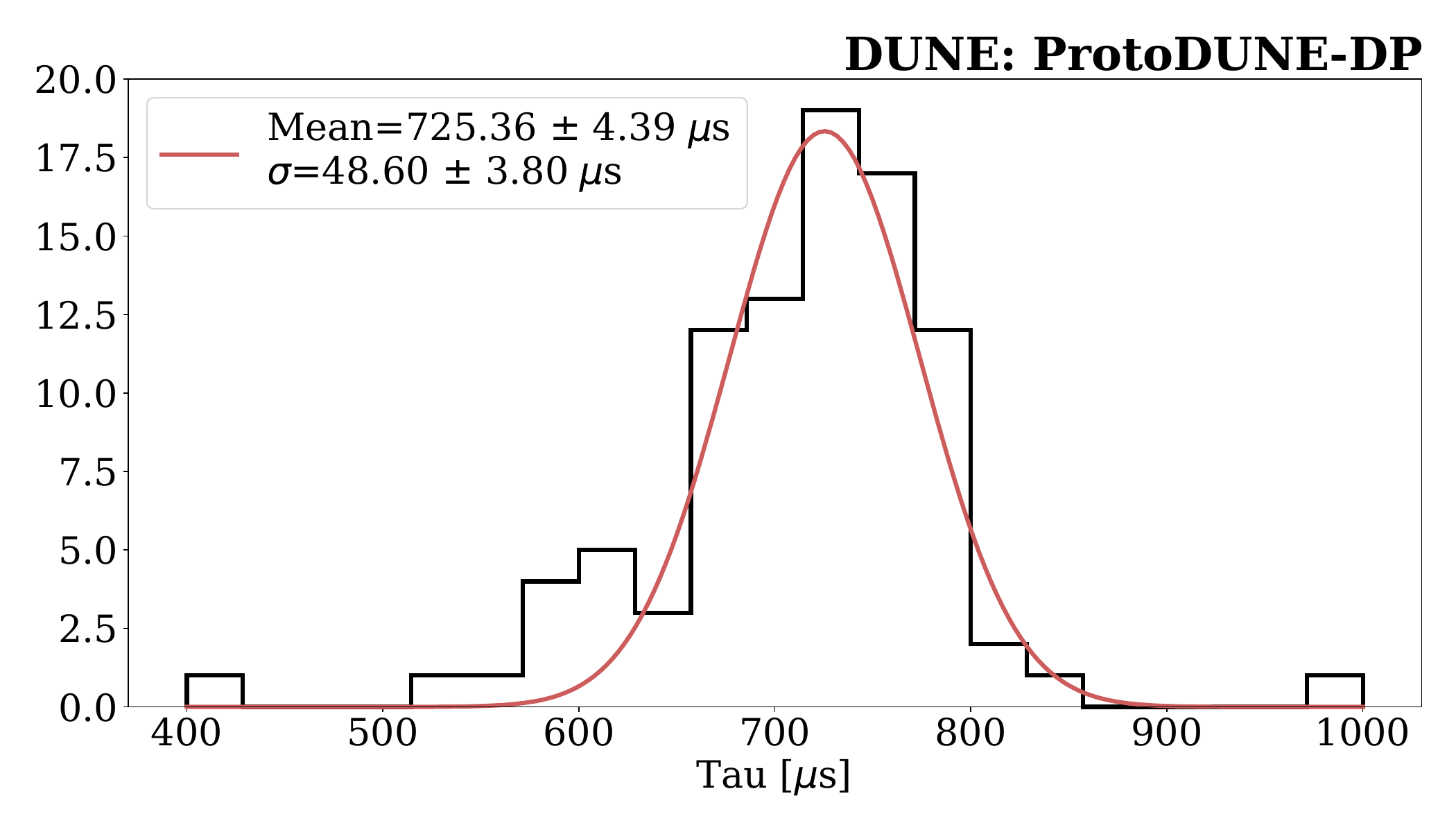} &
    \includegraphics[width=0.45\textwidth]{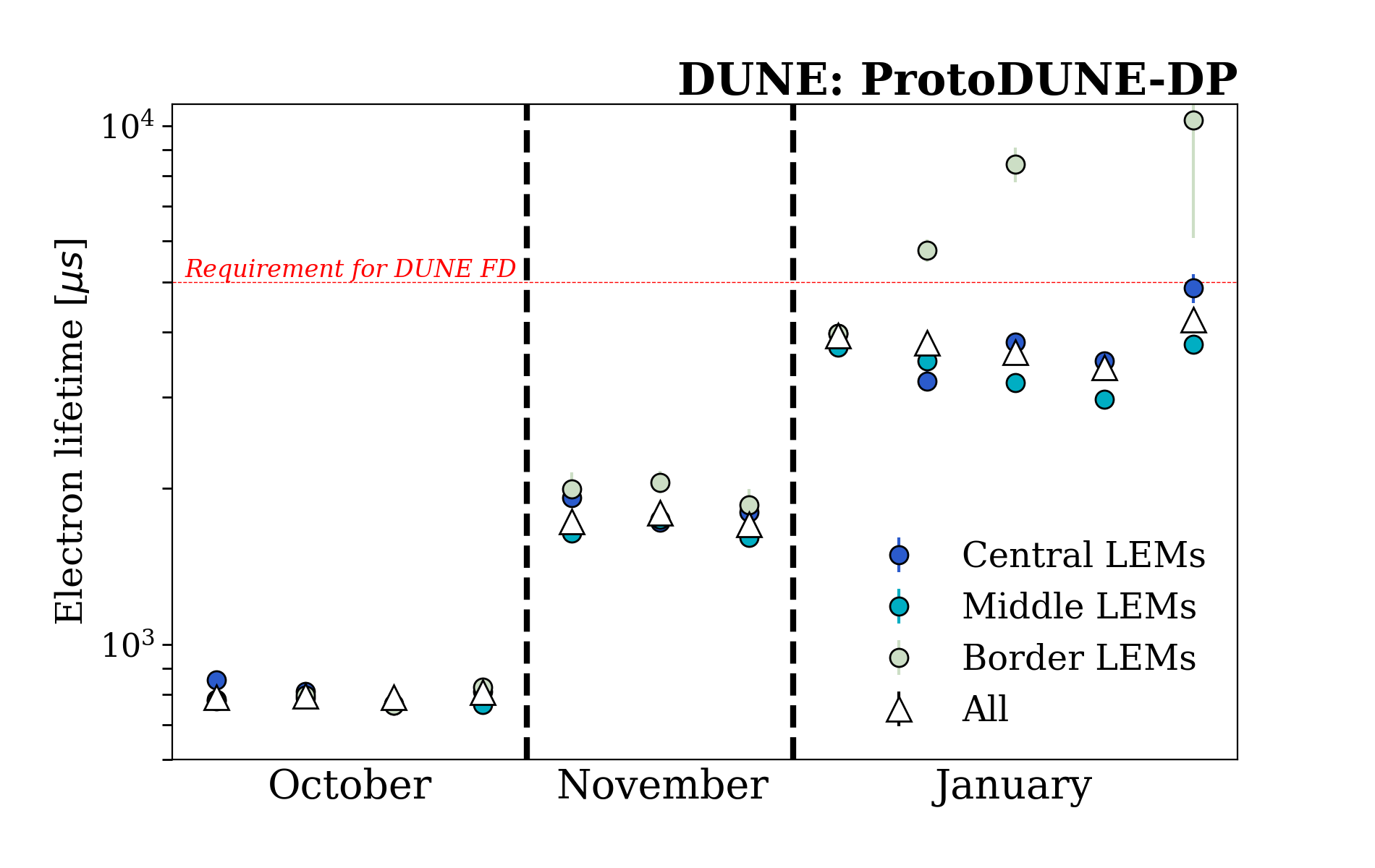}
   
    \end{tabular}
     \caption{(Left) Distribution of the fitted lifetime values in all the LEMs in a given run for the analysis with LARDON reconstruction. (Right) Comparison of the purity measurements using the LArSoft reconstruction in different zones of the TPC at three data-taking periods.}
    \label{fig:tau_gauss_zones}
\end{figure}

From the tracks selected by the cuts described in section~\ref{subsec:tpc_ana_muon}, the electron lifetime can be retrieved from the evolution of the collected charge at the anode per unit drift length, $dQ/ds$, as a function of the drift time. Some typical distributions of $dQ/ds$ and its dependence on the electron drift time are shown in figure~\ref{fig:dqdxvst}. The most probable value (MPV) of the distributions are extracted from a fit with a Landau function convolved with a Gaussian. The $dQ/ds$ MPV as a function of drift time is then fitted with an exponential function to extract $\tau_e$. In the LARDON analysis, the fit is performed for each LEM and view separately (the LEMs on the border of the detector are excluded) and the average $\tau_e$ value from a gaussian fit of the distribution is considered, as shown in figure~\ref{fig:tau_gauss_zones} (left). To ensure that field inhomogeneities induce no bias on the electron lifetime measurement, the uniformity of the electron lifetime estimation with respect to the (X;Y) position on the TPC anode is verified. The TPC is divided into three regions corresponding to an increasing mean drift field (selecting center, middle and border LEMs). The electron lifetime analysis has been reproduced in the three different zones and the results are shown in figure~\ref{fig:tau_gauss_zones} (right). The results are consistent within 3$\sigma$, except in the borders, where smaller uncorrected effects could affect the obtained results, such as the dependency of the attachment rate of the electrons on the drift field, or imperfect estimation of the drift field inhomogeneities. This justifies why those LEMs are excluded from the analysis.

\begin{figure}[h]
    \centering
    \includegraphics[width=\textwidth]{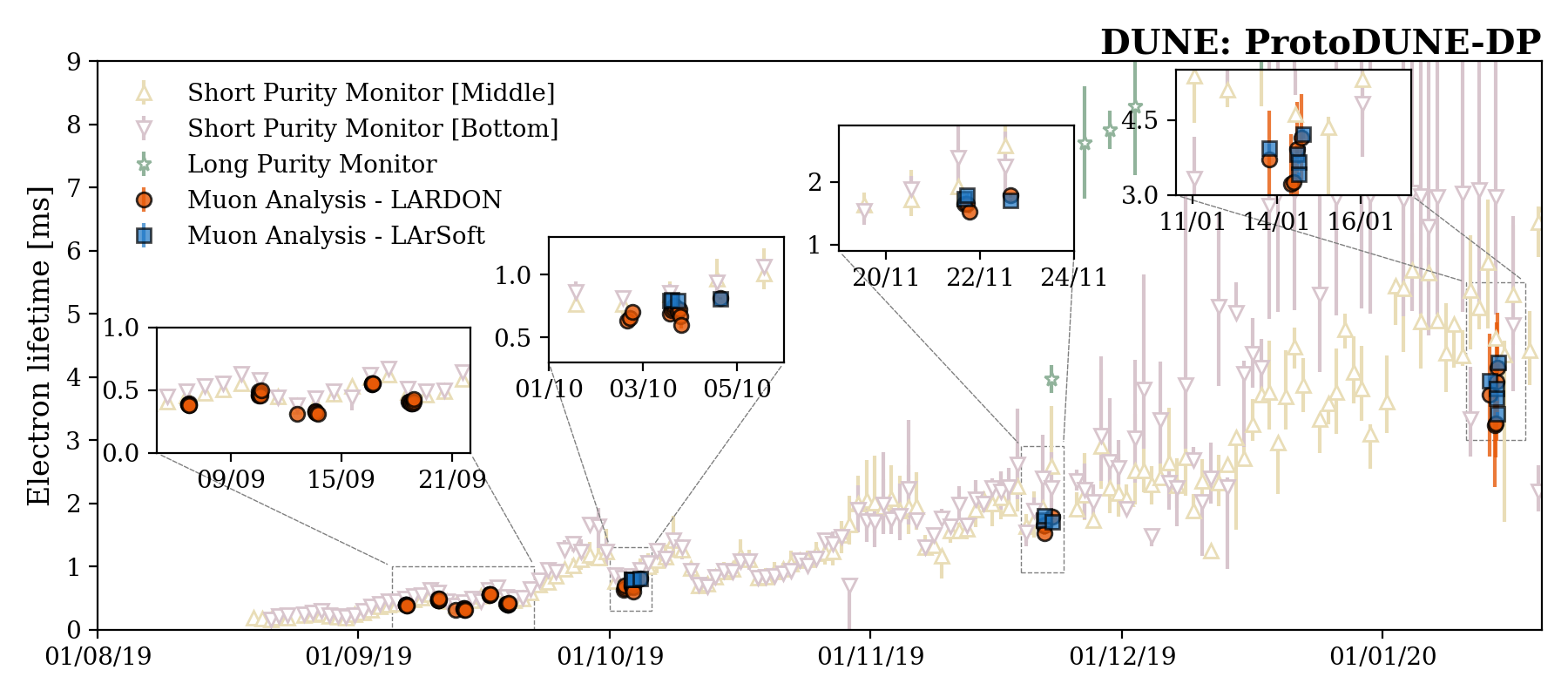}
    \caption{Measurement of the electron lifetime over the first 6 months of detector operation. In open triangles, the daily measurements of the purity monitors. The results obtained from the two analyses are shown: in circles using the LARDON reconstruction and analysis chain, in squares using the standard LArSoft software.}
    \label{fig:ana_purity}
\end{figure}

The fit results on data for the four cosmic runs taken in September (including some data taken during the commissioning period), October, November, and January (see table~\ref{tab:charge_data_summary}) can be seen in figure~\ref{fig:ana_purity}. The two analyses provide consistent results. As the LAr purity improves between September and January, the electron lifetime
$\tau_e$ is also increasing by a factor $\sim$8. 
The evaluation of the liquid argon purity with the cosmic ray data was limited by the short depth of the effective active volume and, more generally, by the non-uniformity of the drift electric field in the detector. In particular, when the electron lifetime becomes higher than a few~ms, the $dQ/ds$ evolution with the drift time becomes marginal over 50~cm of drift, as illustrated in figure~\ref{fig:dqdxvst}-right. Consequently, the values extracted from the January run should be interpreted as indicative trends rather than quantitative measurements.
In the figure, the measurements from the purity monitors are overlaid. The results obtained from cosmic-ray tracks are in good agreement with the short purity monitor measurements in terms of the observed trends, while the agreement in absolute values is reasonable, given the limitations of the analysis. As already discussed in section~\ref{sec:purity}, the disagreement with the long purity monitor measurement could not be resolved.

The achieved purity matches the order of magnitude required for a long drift dual phase module: at an electron lifetime of 5~ms, the amount of charge collected after 6~m (resp. 12~m) of drift is reduced by 53\% (resp. 78\%). On the other hand, the achieved electron lifetime in ProtoDUNE-DP is substantially lower than the $\sim\SI{100}{\milli\second}$ 
reported in~\cite{Abi_2020} for the ProtoDUNE-SP detector using the same cryostat technology. 
The origin of this difference is not understood. Some hypotheses are considered: detector material outgasing in the ullage space or contamination by chemical compounds of unknown origin that can not be purified. Moreover, only short purity monitors were continuously operated whose sensitivity was saturating at 10\,ms.\\

\subsection{Effective gain of the system}
\label{subsec:tpc_gain}

The effective gain $G_{\mathrm{eff}}$ of the detector can be expressed as:
\begin{equation}
     G_{\mathrm{eff}} = \mathcal{T}\times G_{\mathrm{LEM}},
     \label{eq:geff}
\end{equation}
where $\mathcal{T}$ is the electron transparency of the system and $G_{\mathrm{LEM}}$ is the LEM amplification factor.

\begin{figure}[h]
\centering
\includegraphics[width=0.95\textwidth]{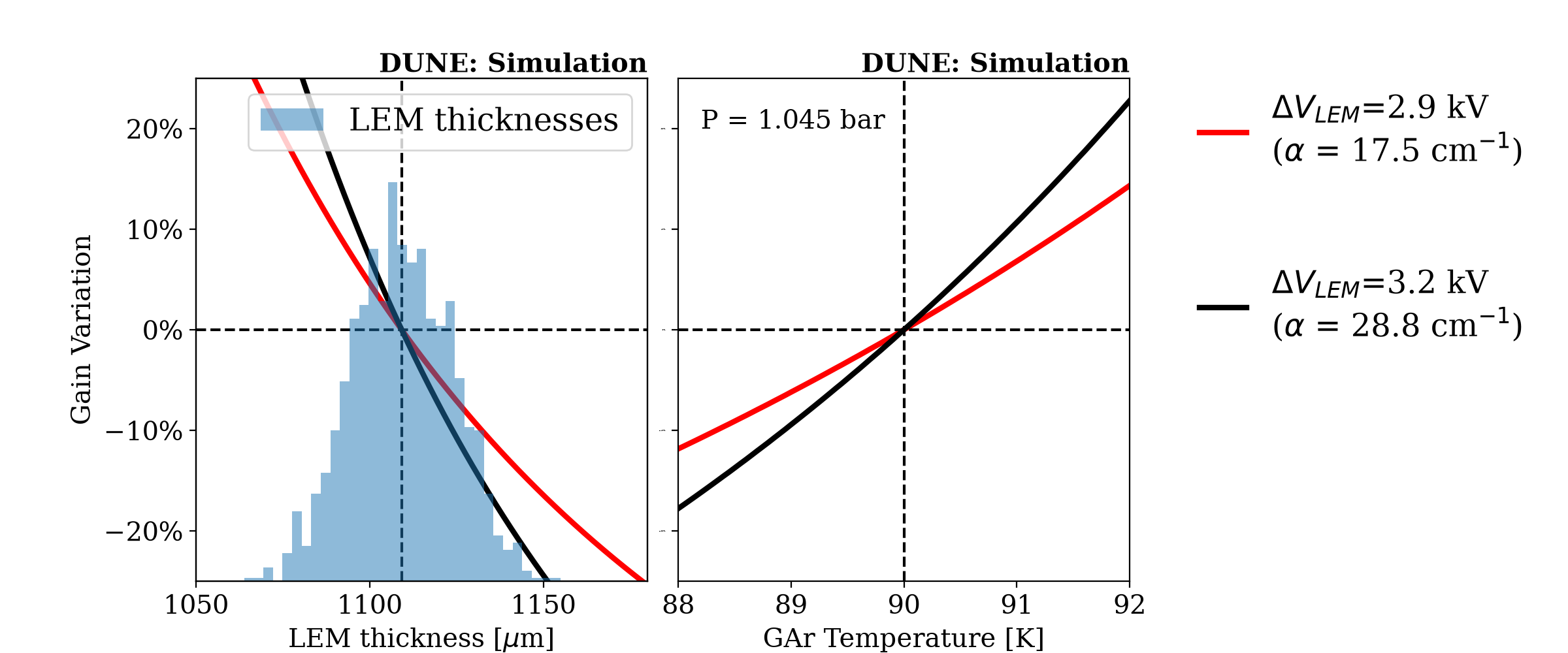}
\caption{Impact of the LEM thickness (left) and gas temperature (right) on the gain at $\dvlem=$ 2.9 and 3.2\,kV according to Equation~\ref{eq:glem}. The values of $\alpha$ are extracted from MAGBOLTZ~\cite{magboltz}. The vertical dashed lines show the respective nominal values. The distribution of the measured thicknesses of the LEMs installed in \pddp are overlaid.}
 \label{fig:gain_variation}
\end{figure}

The gain of the LEM can be parametrized by:
\begin{equation}
    \label{eq:glem}
    G_{\mathrm{LEM}} = \exp(\alpha\times d),
\end{equation}
where $d$ is the amplification length and $\alpha$ the first Townsend ionisation coefficient, which is analytically described as~\cite{AOYAMA1985125}: 
\begin{equation}
\label{eq:alpha}
    \alpha = A_\rho\exp(-B_\rho/E_{LEM}),
\end{equation}
where $A_\rho$ and $B_\rho$ are density-scaled constants that depend on the gas and $E_{LEM}=\dvlem/d$ is the amplification field inside the LEM holes. Figure~\ref{fig:gain_variation} shows a sensitivity study of the LEM thickness and the argon gas temperature impact on the gain of the LEM at two different voltages according to equation~\ref{eq:glem}. The mean value of the thickness of the installed LEMs is 1.109\,mm with a spread at $1\sigma$ of 16\,$\mu$m, which would corresponds to a gain variation of $\pm$7\% at $\dvlem = 2.9$\,kV. As explained in section~\ref{sec:larsurface}, the nominal pressure of the cryostat during data-taking is 1045\,mbar, except during the October run where it was set to 1010\,mbar. The temperature of the gas at the level of the LEM holes cannot be monitored directly, and is extrapolated from the temperature gradient measured above the CRP plane. The nominal value of the argon gas is 90\,K, and is assumed to be constant over the collection area. As seen in figure~\ref{fig:gain_variation} (right), a variation of 0.5\,K would result in a gain variation of $\pm3$\% at $\dvlem = 2.9$\,kV.

\begin{figure}[h]
\centering
\includegraphics[width=0.95\textwidth]{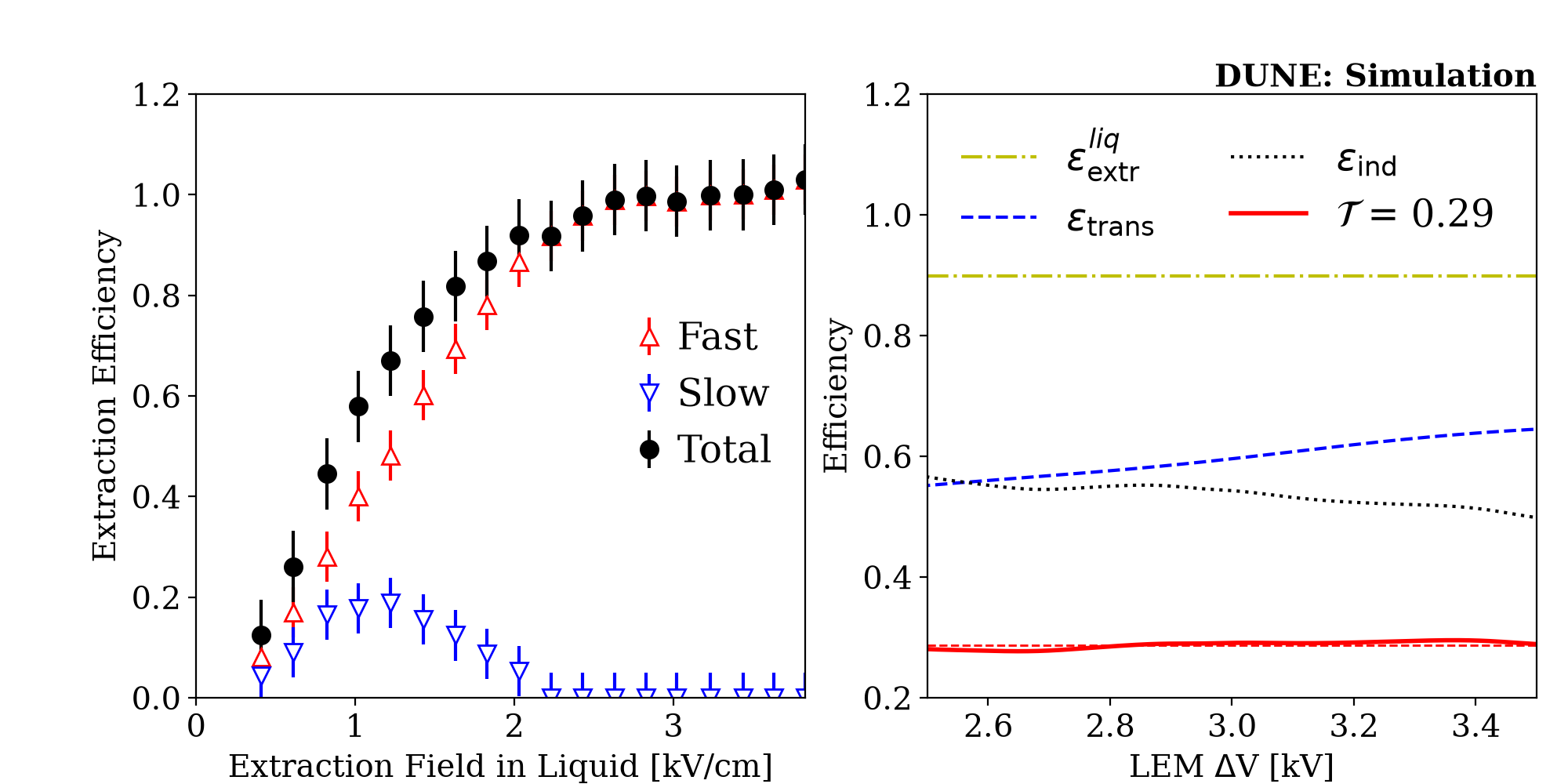}
\caption{(Left) Electron extraction efficiency from liquid to gas as a function of the extraction field in liquid. The slow and the fast extraction components are shown. The data are taken from~\cite{Gushchin:1982}. (Right) Electron transparency $\mathcal{T}$ of the data taken in \pddp as a function of the voltage applied on the LEM from simulations~\cite{cotte:tel-02382815}. The transparency is found constant at a value $\mathcal{T}=0.29$ over the whole range of fields.}
 \label{fig:extraction_transparency}
\end{figure}

The transparency $\mathcal{T}\leq1$ is defined as the product of the  efficiencies in the extraction and induction regions:
\begin{equation}
    \mathcal{T} = \varepsilon_{extr}^{liq} \times \varepsilon_{trans}^{LEM} \times \varepsilon_{ind}.
\end{equation}
The electron extraction efficiency from the liquid to the gaseous phase, $\varepsilon_{extr}^{liq}$ has been measured in similar conditions by Gushchin {\it et al}~\cite{Gushchin:1982}. The electron collection efficiency on the bottom ($\varepsilon_{trans}^{LEM}$) and on the top ($\varepsilon_{ind}$) of the LEM are estimated from simulation~\cite{cotte:tel-02382815, 311_performance}.
As shown in figure~\ref{fig:extraction_transparency} (left), the efficiency to extract electrons from liquid to gas depends on the strength of the local extraction field in liquid argon. Above 2\,kV/cm, the efficiency is greater than 90\%, and the extraction occurs in a timespan of less than 100\,ns, referred to as the ``fast'' component in the figure. When the field is lower, fewer electrons are transmitted to the gaseous phase, and a fraction, up to 20\%,  of the electrons are transmitted via thermionic emission with a characteristic time around 10\,ms~\cite{Gushchin:1982}, referred to as the ``slow'' component in the figure. We consider being only sensitive to the fast extraction component in the data. \\
Otherwise mentioned, the data was taken with an extraction field at 2\,kV/cm in the liquid phase.
During the data-taking of \pddp, the LEMs were operated at voltages ranging from $\dvlem=2.5$ up to 3.2\,kV and extraction grid voltages $V_{grid}$ from -5.4 to -6\,kV. All data were taken with an induction field of 2.5\,kV/cm.
From transmission simulations using MAGBOLTZ and Garfield++~\cite{cotte:tel-02382815}, the expected electron transparency of \pddp is estimated to be constant at $\mathcal{T}=0.29$, as shown in figure~\ref{fig:extraction_transparency} (right).

In the analyses presented in following sections, the effective gain is computed as the sum of the MPV of the charge collected per unit length in both views over the expected most probable amount of charge deposited by a muon at MIP:
\begin{equation}
    G_{\mathrm{eff}} = \frac{dQ/ds|^{\mathrm{MPV}}_{0}+dQ/ds|^{\mathrm{MPV}}_{1}}{dQ/ds|^{\mathrm{MPV}}_{\mathrm{MIP}}}.
    \label{eq:eff_gain_meas}
\end{equation}

The two views provide independent measurements of the charge deposited along the same track. Their sum therefore provides an estimate of the total reconstructed charge per unit length, accounting for equal contributions measured by both views, as discussed in section~\ref{subsec:tpc_reco_quality}. The MPVs are extracted from independent fits to the charge distributions in each view, where each $dQ/ds$ measurements are corrected for the drift field non-uniformity (through the recombination and drift velocity) and the purity.\\
This approach differs from previous dual-phase analyses, in which the sum of mean of the $dQ/ds$ distribution in each view was used instead~\cite{Cantini_2015, 311_performance}. We use the MPV because it is less sensitive than the mean to the high-charge tail of the distribution and to reconstruction effects that can produce outliers. An example of a $dQ/ds$ distribution and its corresponding fit with a Landau distribution convoluted with a Gaussian, to account for spread of the distribution due to detector effects, is shown in figure~\ref{fig:ana_drift_field_phase_space} (left).
The expected MPV of a MIP track in \pddp is estimated using the Landau-Vavilov formula: $dQ/ds|^{\mathrm{MPV}}_{\mathrm{MIP}} = 12.3$\,fC/cm before recombination~\cite{PabloThesis} where the angular phase-space of the reconstructed and selected tracks, shown in figure~\ref{fig:ana_drift_field_phase_space} (right), was taken into account in the computation, as the Landau-Vavilov formula depends on the amount of LAr crossed.

\subsubsection{Effect of the LEM thickness}
\label{subsec:tpc_gain_lem}

\begin{figure}[h]
\centering
\includegraphics[width=0.75\textwidth]{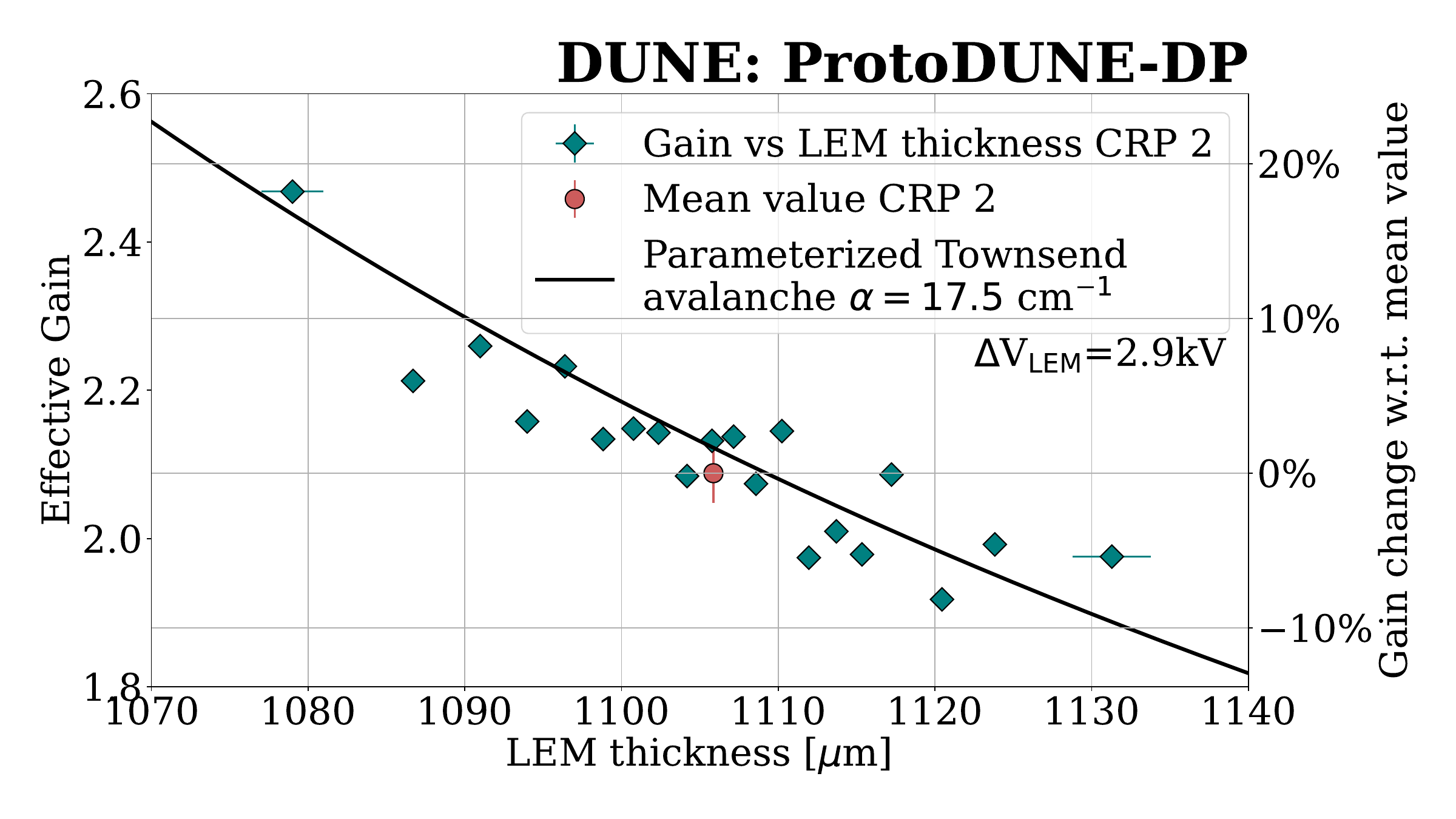}
\caption{Effective gain measured as a function of the LEM thicknesses of CRP2 at a LEM operation voltage of 2.9\,kV and reconstructed with LARDON. The red circle point indicates the value extracted over all LEMs of CRP2 with a mean thickness of 1106\,$\mu$m. The axis on the right shows the expected relative gain variation from that measurement. }
 \label{fig:gain_lem_thickness}
\end{figure}

During the qualification process of the LEMs prior to their installation in the detector, a detailed measurement of their thickness was performed in bins of $10\times 10$\,cm$^2$ (section~\ref{ssec:det:crd}). By using a run with stable and continuous operation for almost two hours taken in September 2019, sufficient statistics in each corresponding thickness bin are obtained. The effective gain as a function of the LEM thickness is presented in figure~\ref{fig:gain_lem_thickness}, with an operating voltage of the LEM at $\dvlem=2.9$\,kV. For visibility, the data points with similar thicknesses are averaged together. Compared to the results obtained over all LEMs, the gain varies in agreement with the predictions shown in figure~\ref{fig:gain_variation} (left).

\subsubsection{Effect of the extraction field}
\label{subsec:tpc_gain_extraction}
\begin{figure}[h]
\centering
\includegraphics[width=0.75\textwidth]{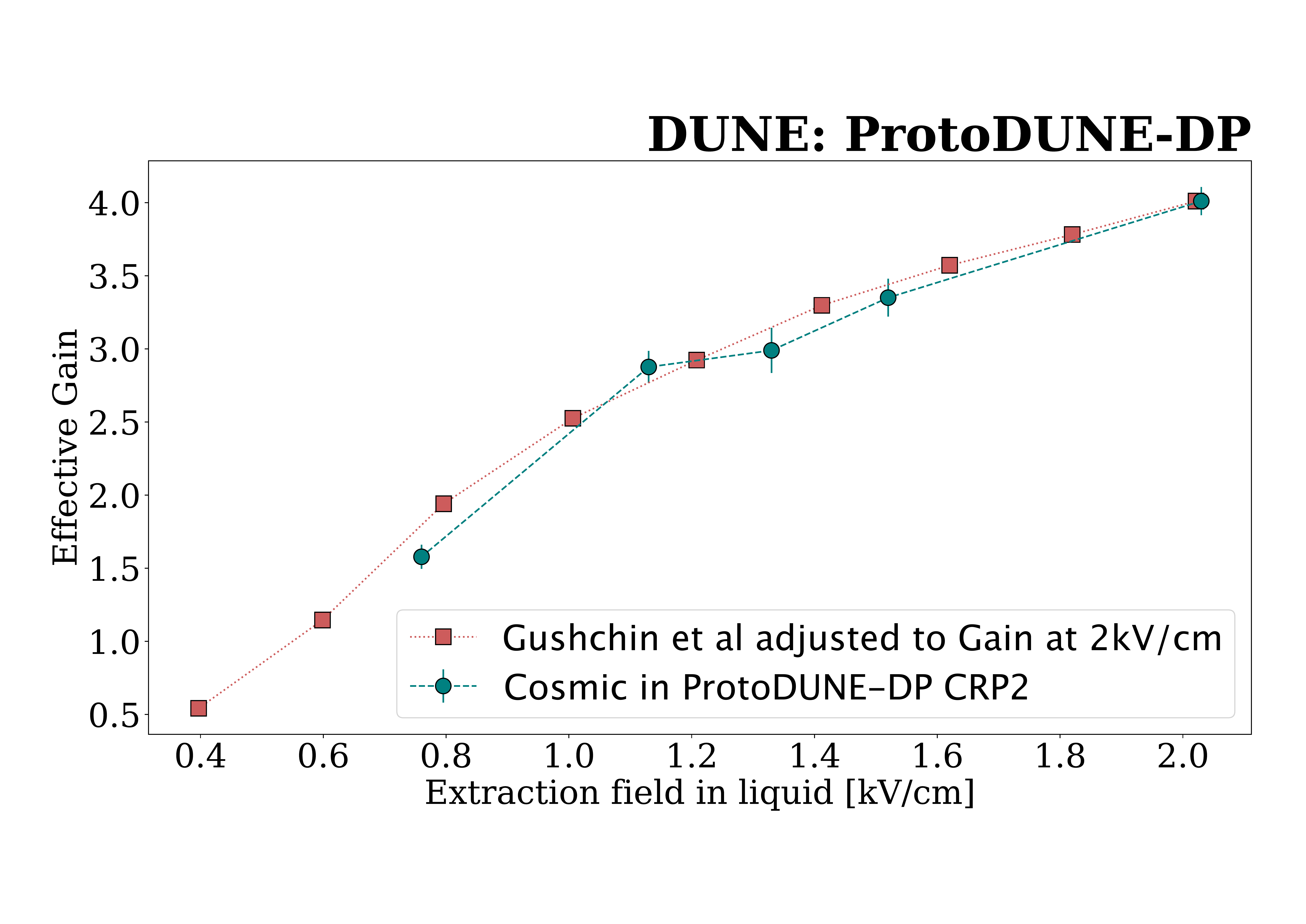}
\caption{Effective gain measured as a function of the extraction field in liquid during the October run, at a LEM operation voltage of 3.1\,kV and reconstructed with LARDON. The efficiencies measured by Gushchin {\it et al}~\cite{Gushchin:1982} are overlaid and adjusted to the measured point at $E_{extr}^{liq}=2$\,kV/cm.}
 \label{fig:gain_lem_extraction}
\end{figure}

During the data-taking period of October 2019, an extraction field scan was performed. The LEM operation voltage was set to $\dvlem=3.1$\,kV and the induction field at 2.5\,kV/cm. Five dedicated runs were taken with extraction fields in liquid ranging from 0.8 to 2~kV/cm. Following the procedure described in with~\ref{eq:eff_gain_meas} for each run, the evolution of the effective gain as a function of the extraction field in liquid is shown in figure~\ref{fig:gain_lem_extraction}. 
The measured relative field dependence agrees with the shape of the published fast extraction efficiency curve by Gushchin {\it et al}~\cite{Gushchin:1982} after normalisation to the measured value at $E_{extr}^{liq}=2$\,kV/cm.
~\cite{Gushchin:1982}.

\subsection{Charging-up effect}
\label{subsec:tpc_chargingup}

\begin{figure}[h]
\centering
\includegraphics[width=0.95\textwidth]{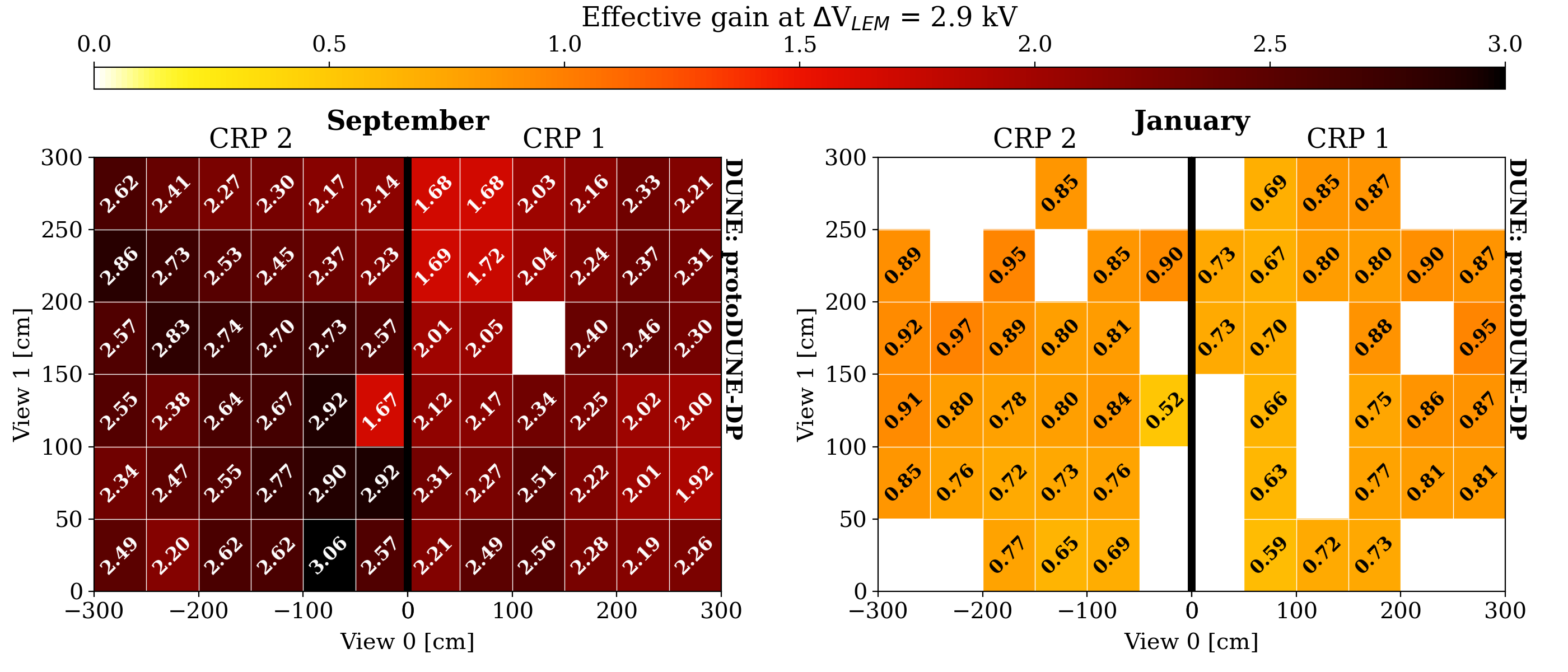}
\caption{Effective gains of all LEMs for two cosmic runs taken in September (left) and in January (right) in the same HV conditions reconstructed with LARDON. Only the LEMs that have sustained a potential difference $\Delta{V}_{LEM}$ within +/- 100V around the initial nominal voltage of 2.9~kV during the whole run duration are shown here.}
 \label{fig:gain_map_charging_up}
\end{figure}

Figure~\ref{fig:gain_map_charging_up} shows a map of the effective gains in each LEM in two runs taken at different run period with the same operating conditions. A significant reduction of the effective gain, by a factor $\sim2.5$, over time is visible: this decrease in $G_\mathrm{eff}$ with time is likely due to the known LEM charging-up effect. As the LEMs are operated, charges (e$^-$ and Ar$^+$) accumulate on the insulator inside and around the LEM holes. These charges will build-up and modify the electric field strength and lines inside the LEM holes. 

In~\cite{Cantini_2015} (referred to as the 3L detector) the charging-up effect was studied on LEMs with similar design as the one installed in \pddp. From continuous operation of the 3L detector for up to 3 days, it was shown that the LEM gain exponentially decreases  with a characteristic time $\tau$ until it reaches a plateau. The charging-up characteristic time $\tau$ depends on \dvlem: the stronger the field inside the LEM, the more charges are created and accumulated, hence the faster the charging-up occurs. Their measurements indicate that once the LEM gain is stabilized, the gain is reduced by a factor of $\sim3$. The effect was also observed on the LEMs designed for \pddp \cite{Eurin:2025jqt}.

\begin{figure}[h]
\centering
\includegraphics[width=0.85\textwidth]{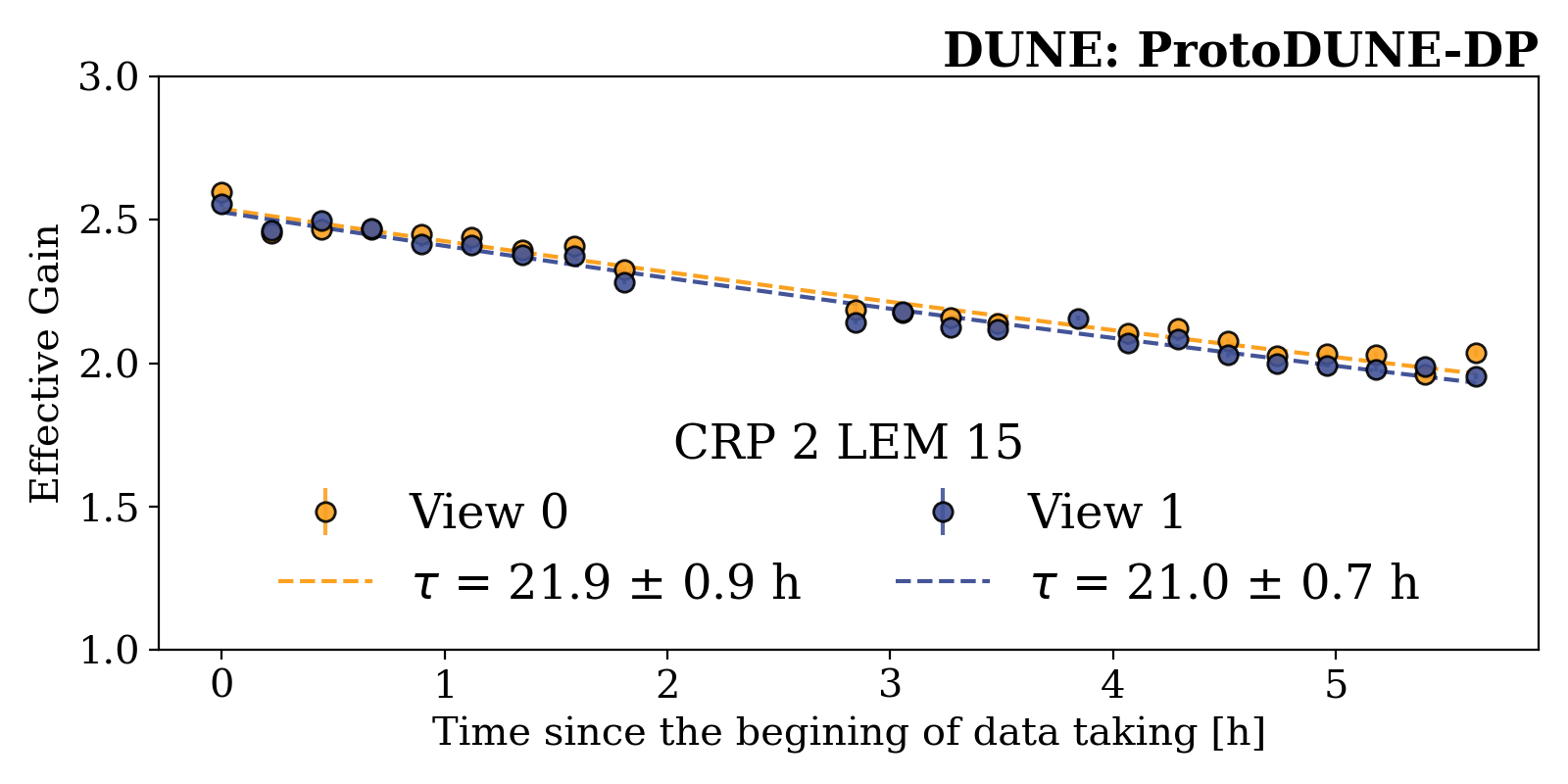}
\caption{Evolution of the effective gain of one LEM with time since the data-taking run started. The fit of the exponential decrease is also shown for the two views. The 1\,h long absence of data after 2\,h of operation is due to a photon detection system DAQ trigger test. In this trigger condition not enough tracks were recorded to perform the effective gain measurement. This run was recorded in September 2019 at a $\dvlem = 2.9$\,kV and reconstructed with LARDON.}
 \label{fig:charging_up_onelem}
\end{figure}

\begin{figure}[h]
\centering
\includegraphics[width=0.95\textwidth]{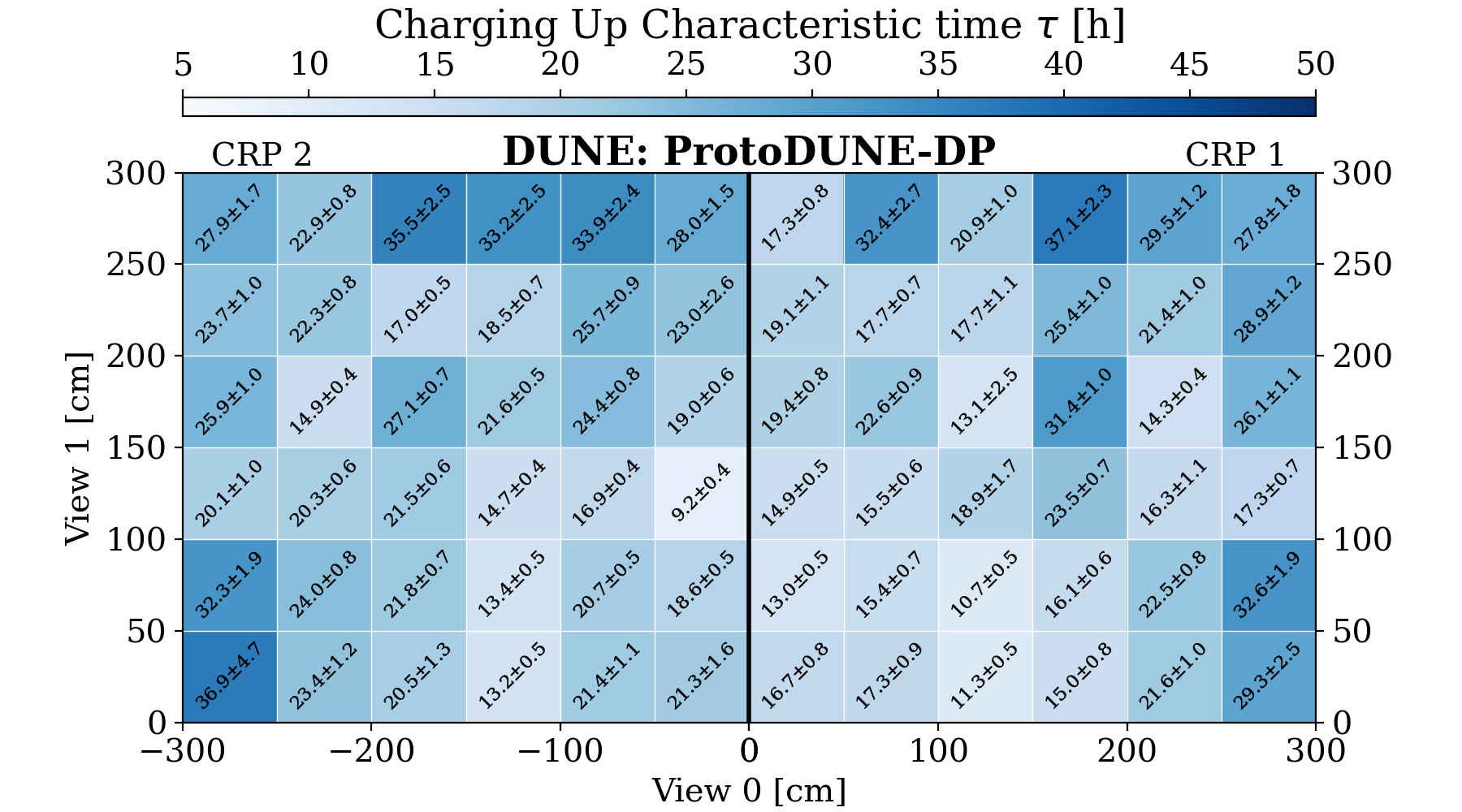}
\caption{Extracted charging-up characteristic time for each LEM of the two CRPs.}
 \label{fig:chargingup_alllem}
\end{figure}

In the beginning of \pddp operations in September 2019, the nominal operation voltage of the LEMs was set to $\dvlem=2.9$\,kV. 
In particular, on September 18$^{th}$, the detector took continuous and stable operation for almost 6\,h. The data was divided in periods of $\sim$20\,min, and the effective gain during each period for each LEM and each view was measured. In figure~\ref{fig:charging_up_onelem}, the effective gain evolution with time is shown for one LEM. The effective gain decrease with time is fitted with a simple exponential function:
\begin{equation}
    G_{\mathrm{eff}}(t_{run}) = A\exp(-\frac{t_{run}}{\tau}),
    \label{eq:charging_up_expo}
\end{equation}
where ($t_{\mathrm{run}}$) denotes the time elapsed since the beginning of the run. Given the short time span covered by the available data, a more sophisticated function, such as the one used in~\cite{Cantini_2015}, could not be reliably constrained.
Figure~\ref{fig:chargingup_alllem} shows the extracted charging-up characteristic time of each LEMs with the same method. Excluding the LEMs near the field cage (where the drift field varies strongly), the average value over the two CRPs is $\tau = 18.9$~h, with a standard deviation of 4.5~h across the considered LEMs. The average is consistent with the characteristic times reported in~\cite{Cantini_2015}.
The extracted characteristic times exhibit a substantial spread across the LEMs. Several factors that influence the LEM gain may also affect its charging-up behaviour, including the LEM thickness, the gas density, and the drift field strength through its effect on the recombination factor. Since these factors may act simultaneously, it is difficult to disentangle their individual contributions to the observed variations.

In the early stage of \pddp commissioning, specific detector operation tests required to set temporarily some LEMs at lower voltages or even to disconnect them. For example, in figure~\ref{fig:gain_map_charging_up}, the run used in September was taken after a 4h-long period where all CRP2-LEMs were switched off. We can see that in general the effective gains of CRP1-LEMs are lower than the ones of CRP2, even though all the LEMs were set at similar voltages in this run.


This is the only measurement of the charging-up characteristic time that could be done with \pddp data. Indeed, during the commissioning and in between two cosmic data-taking periods, the detector underwent various tests to understand and improve the LEMs behavior. This implies that each LEM has its own operation history, and therefore is at a different charging-up state. We could not operate the detector for a long-enough stable condition such that all the LEMs would have reached a charging-up equilibrium.
Consequently, no attempt was made to correct the LEM gain for the charging up effect. 
The observed gain variations between LEMs in a given run may thus reflect a combination of differences in LEM thickness, possible gas temperature gradients, and different charging-up states resulting from their respective operating histories. Furthermore, the observed decrease in effective gain over time cannot necessarily be attributed solely to charging-up: other effects, such as LEM ageing, may also contribute. The available \pddp data do not allow these possible contributions to be disentangled or independently quantified.

\subsection{Effective gain with LEM voltage}
\label{subsec:tpc_gain_measurement}

\begin{figure}[h]
\centering
\includegraphics[width=0.9\textwidth]{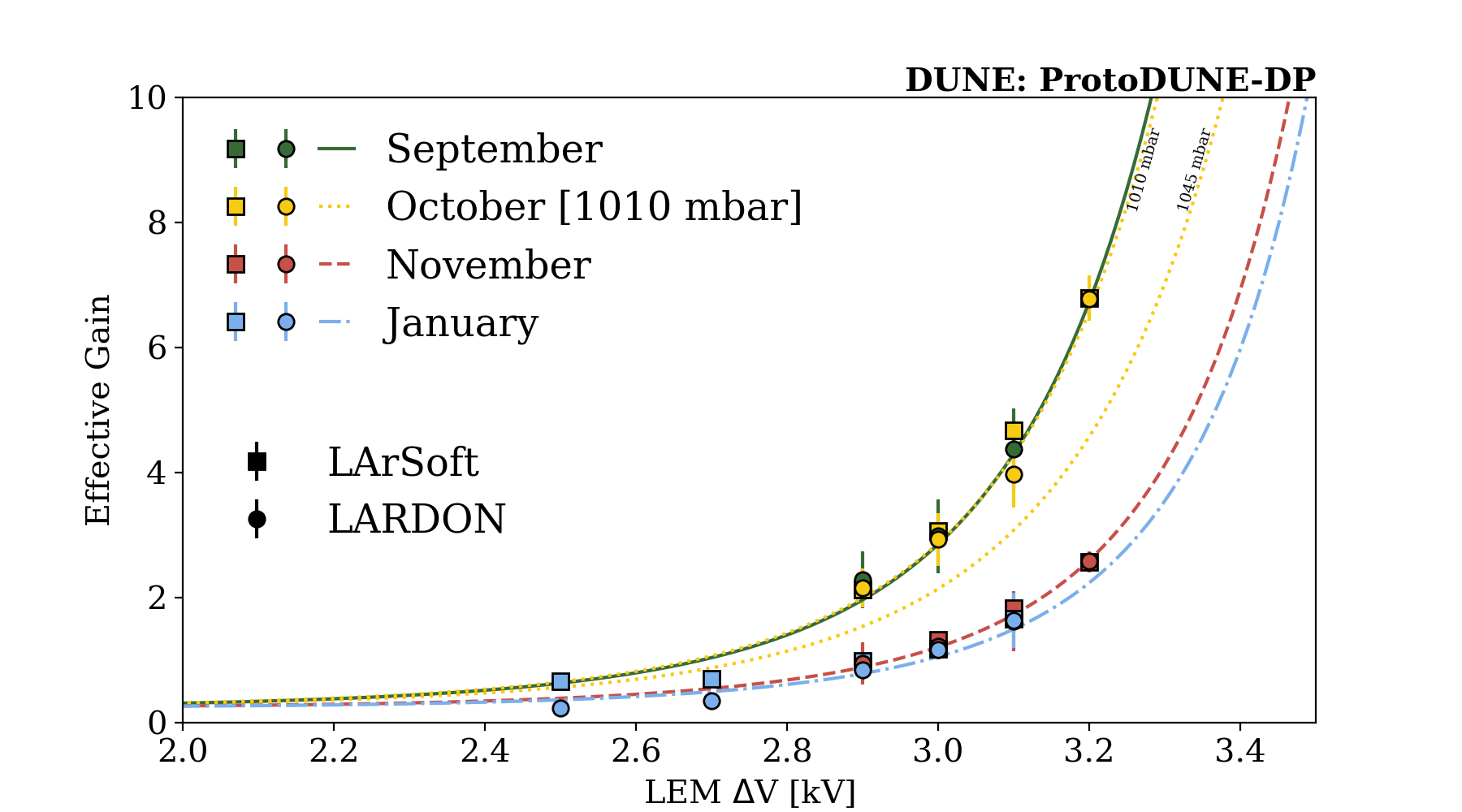}
\caption{Effective gain as a function of the LEM potential difference for different periods measured by the two analyses methods. The results from the LARDON analysis chain are shown in circles, and in squares using LArSoft. The lines show the results of the global fit, where the transparency is a common parameter for the effective gain fit of the 4 periods. The fit of the October data, taken at a pressure of 1010\,mbar, is rescaled at the nominal pressure for comparison.}
 \label{fig:gain_data_fit}
\end{figure}

In each of the data-taking periods, a scan of the LEM operating voltage has been performed. All scans were performed with similar extraction and induction fields. As explained previously, many factors influence the effective gain measured in each LEM: their thickness, the gas temperature and their charging-up state. To extract the CRP effective gain in a given run, the following strategy was used:
\begin{itemize}
    \item Only the LEMs of CRP2 are considered. As explained in the previous section, a CRP1-only long operation in the beginning of \pddp commissioning made their LEMs potentially more charged up than the ones of CRP2. In this period, not all LEMs of CRP1 were continuously operated at the nominal voltage, so the charging-up state of CRP1 LEMs may not be uniform.
    \item The LEMs near the field cage are also excluded from the analysis as the drift field varies too quickly near the boundary of the detector, and the field-correction applied to the measured $dQ/ds$ in those LEMs is considered to be not reliable.
    \item From the remaining 25 LEMs, only those at the nominal operation voltage are considered. The average value of all measured effective gains is taken.
\end{itemize}
The evolution of the effective gain as a function of the LEM potential difference is presented in figure~\ref{fig:gain_data_fit}. The results from the LARDON and LArSoft analyses are in agreement within 5\% for most data points.  At very low \dvlem for the January run, the two analyses differ: in those runs, the effective gain is very small such that the hits are less distinguishable from the noise. The noise filtering process becomes very important in the computation of the collected charge. Although the two reconstruction chains have the same noise removal strategies, there are small differences which are visible for these points. 
The highest effective gain achieved was $G_\mathrm{eff} = 6.8 \pm 0.03$ in the September run at $\dvlem = 3.2$\,kV ; at a given \dvlem the effective gain decreases in later data-taking periods. This behaviour can be attributed to the charging-up effects, although the aging of the LEMs with time could also impact their performance.
It is important to recall that for the data taken in October the cryostat was set to a pressure of 1010\,mbar: the argon gas density was lower which resulted in higher effective gain. From November to January runs, the effective gain change is fairly small, which can indicate that the charging-up may be almost complete; unfortunately, no further runs could be taken after that period. The highest effective gain reached in the January run is $G_\mathrm{eff}= 1.5 \pm 0.09$ at a $\dvlem = 3.1$\,kV.

To fit the evolution of $G_{\mathrm{eff}}$ with \dvlem for each data-taking period, the equation of the effective gain (Equation~\ref{eq:geff}) needs to be modified, following the procedure described in~\cite{Cantini_2015,Badertscher_2011}:
\begin{equation}
    \label{eqn:geff_xkappa}
    G_\mathrm{eff} = \mathcal{T}\times\exp \left[ A_\rho x_\mathrm{eff} \exp\left(-\frac{B_\rho}{\kappa E_0}\right)\right],
\end{equation}
where $x_\mathrm{eff}$ represents an effective phenomenological amplification path length inside the LEM of nominal thickness $d_\mathrm{LEM} = 0.1$\,cm and $\kappa E_0$ is the electric field inside the LEM, with $E_0 = \Delta V_\mathrm{LEM}/d_\mathrm{LEM}$.\\
The constants $A_\rho$ and $B_\rho$ are estimated using \textsc{magboltz}~\cite{magboltz} simulation, where the Townsend coefficients $\alpha$ are extracted in the range of 15 to 50~kV/cm and fitted to the electric field. The extracted values are: 
\begin{eqnarray*}
    A_\rho &= 3568.7 \;\mathrm{cm}^-1\\
    B_\rho &= 154.2 \;\mathrm{kV/cm}\\
\end{eqnarray*}
using the argon gas density $\rho$ at a pressure of 1.045\,bar and a temperature of 90\,K. \\ 

Figure~\ref{fig:extraction_transparency} shows that the transparency $\mathcal{T}$ should be independent from the voltages applied on the LEMs. 
Accordingly, the four data-taking periods are fitted with Equation~\ref{eqn:geff_xkappa} with a common value of $\mathcal{T}$. The results of the global fit are given in Table~\ref{tab:Geff_fit_pddp}. The extracted transparency of the detector $\mathcal{T}$ = $0.252\pm0.011$ is slightly lower than predicted by the simulations. 
While $x_{\mathrm{eff}}$ increases with time, reaching values larger than the physical thickness of the LEMs, $\kappa$ decreases. These trends may be consistent with charging-up effects modifying the electric field inside the LEM holes. However, they may also reflect the influence of electric-field non-uniformities within the holes and other detector effects that could not be identified or independently quantified with the available data. 
The parameterization provides an empirical description of the effective-gain dependence on the LEM voltage for the different data-taking periods. However, the data are affected by substantial detector-related variations and uncertainties that are difficult to quantify reliably. The fit results should therefore be interpreted primarily in terms of the observed trends, rather than as precise measurements of the model parameters. In particular, the parameter uncertainties returned by the fit do not account for all potential sources of systematic uncertainty.

\begin{table}[htbp]
\centering
\caption{\label{tab:Geff_fit_pddp} Fitted values and errors of $\mathcal{T}$, $x_{\mathrm{eff}}$ and $\kappa$ for each data-taking period. }
\smallskip
\begin{tabular}{|c|c|c|c|}
\hline
Period & x$_{\mathrm{eff}}$ (cm) & $\kappa$ & $\mathcal{T}$\\
\hline
September & 0.086 $\pm$ 0.007 & 1.062 $\pm$ 0.009 & \multirow{4}{*}{0.252 $\pm$ 0.011}\\
October & 0.077 $\pm$ 0.005 & 1.059 $\pm$ 0.004 & \\
November& 0.252 $\pm$ 0.032 & 0.810 $\pm$ 0.004 & \\
January & 0.337 $\pm$ 0.068 & 0.764 $\pm$ 0.004 & \\
\hline
\end{tabular}
\end{table}

\begin{figure}[h]
\centering
\includegraphics[width=0.9\textwidth]{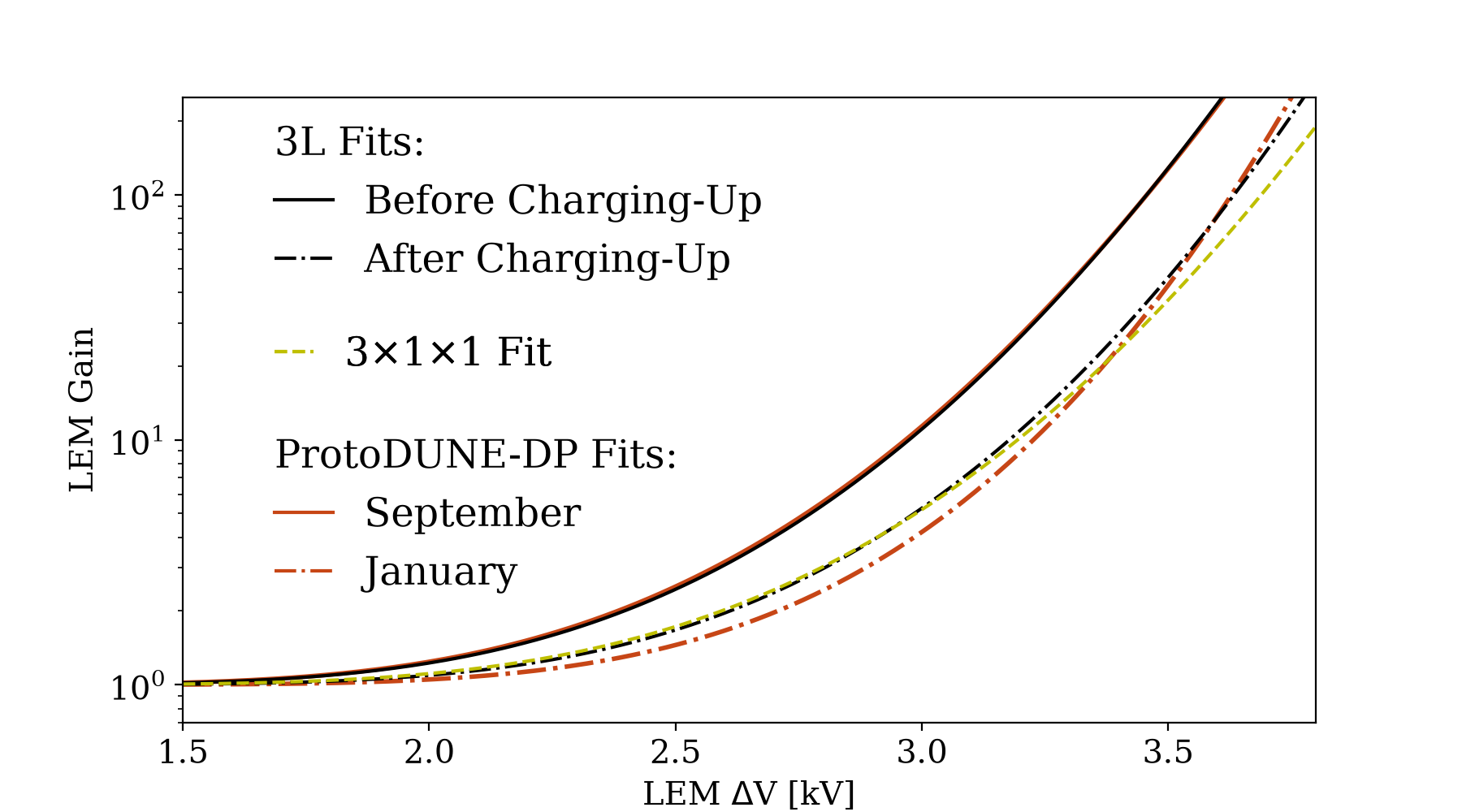}
\caption{Fits of the LEM gain ({\it i.e.} $\mathcal{T}=1$) as a function of the LEM potential difference as measured in \pddp, $3\times1\times1$~\cite{311_performance} and 3L~\cite{Cantini_2015} detectors. For the latter, the measurements before and after the charging-up are shown, while the charging-up state is unknown for the other two experiments. The fits of the previous detector are re-scaled to an argon gas pressure of 1045\,mbar and temperature of 90\,K.}
 \label{fig:gain_comp}
\end{figure}

Figure~\ref{fig:gain_comp} compares the fitted LEM gain ({\it i.e.} $\mathcal{T}=1$) evolution with the LEM potential difference for different dual-phase experiments. The results from the previous measurements are rescaled to the \pddp argon gas temperature and pressure. The metric of the LEM gain, instead of the effective gain, was chosen for the comparison as many parameters can influence the extracted value of CRP transparency: the data-taking conditions, the reconstruction and the analysis procedure. This makes the extrapolation of $\mathcal{T}$ difficult among experiments. \\
In the 3-litre LArTPC dual-phase setup~\cite{Cantini_2015}, a long-term operation of different LEM layout was performed. The effective gain as a function of the LEM potential difference before and after the completion of the charging-up of LEMs with the same design as the one installed in \pddp was done. The data are fitted with the same global procedure as described previously. 
In the $3\times 1\times 1$\,m$^3$ 4-tonne demonstrator~\cite{311_performance}, only one scan of the LEM operating voltage could be performed because of the technical limitation of the experiment. As a consequence, the charging-up state of the extracted gain evolution could not be assessed. 

The three experiments show qualitatively consistent trends in the gain evolution, both during the early stages of charging-up and as the gain approaches its plateau. The comparison with the $3\times 1\times 1$,m$^3$ 4-tonne demonstrator suggests, retrospectively, that its data may have been collected when charging-up was nearly complete. However, this interpretation remains tentative, as the charging-up state of the demonstrator at the time of data taking is not independently known.


\subsection{Data with full drift}
\label{subsec:tpc_6m}

\begin{figure}[h]
\centering
\includegraphics[width=0.4\textwidth]{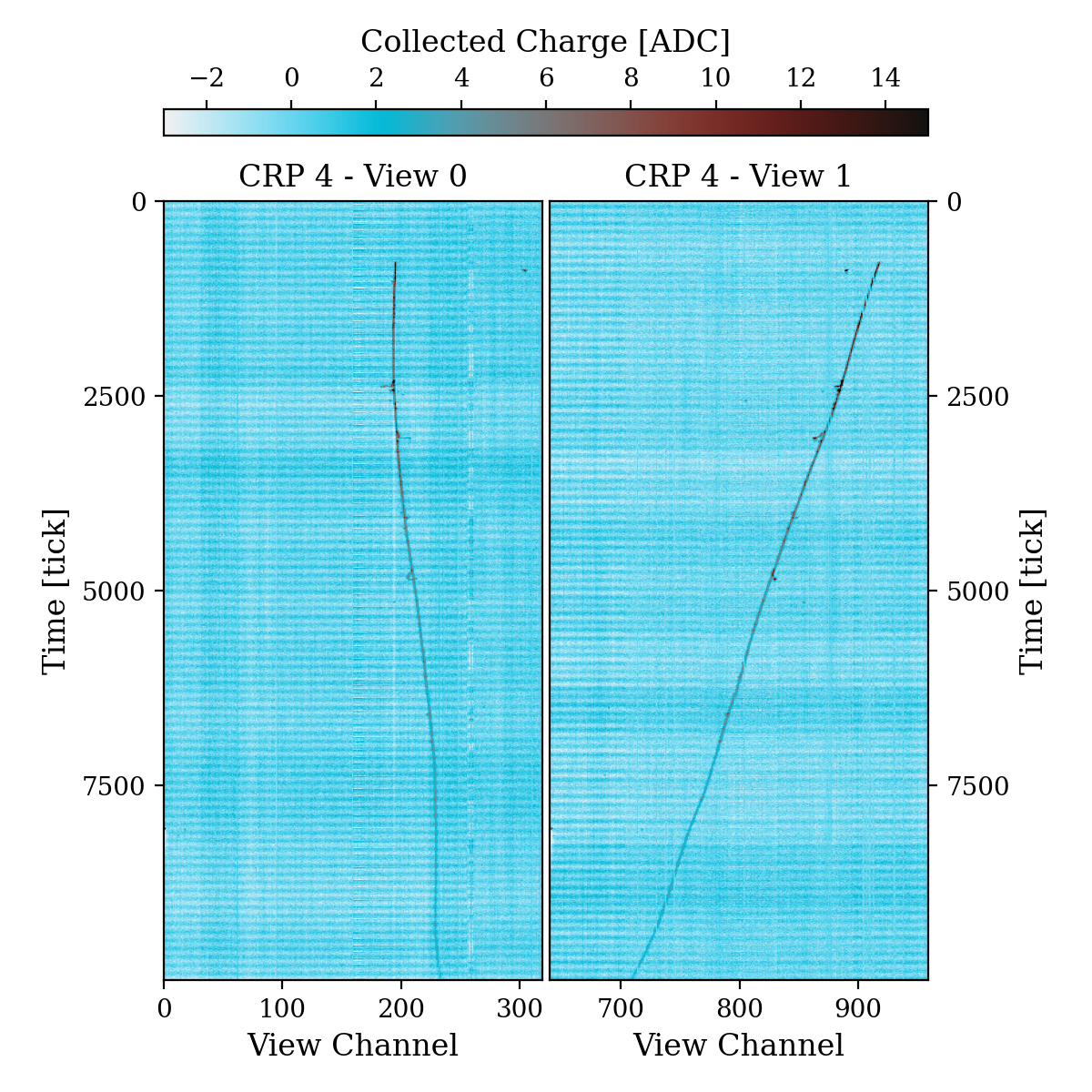}
\includegraphics[width=0.4\textwidth]{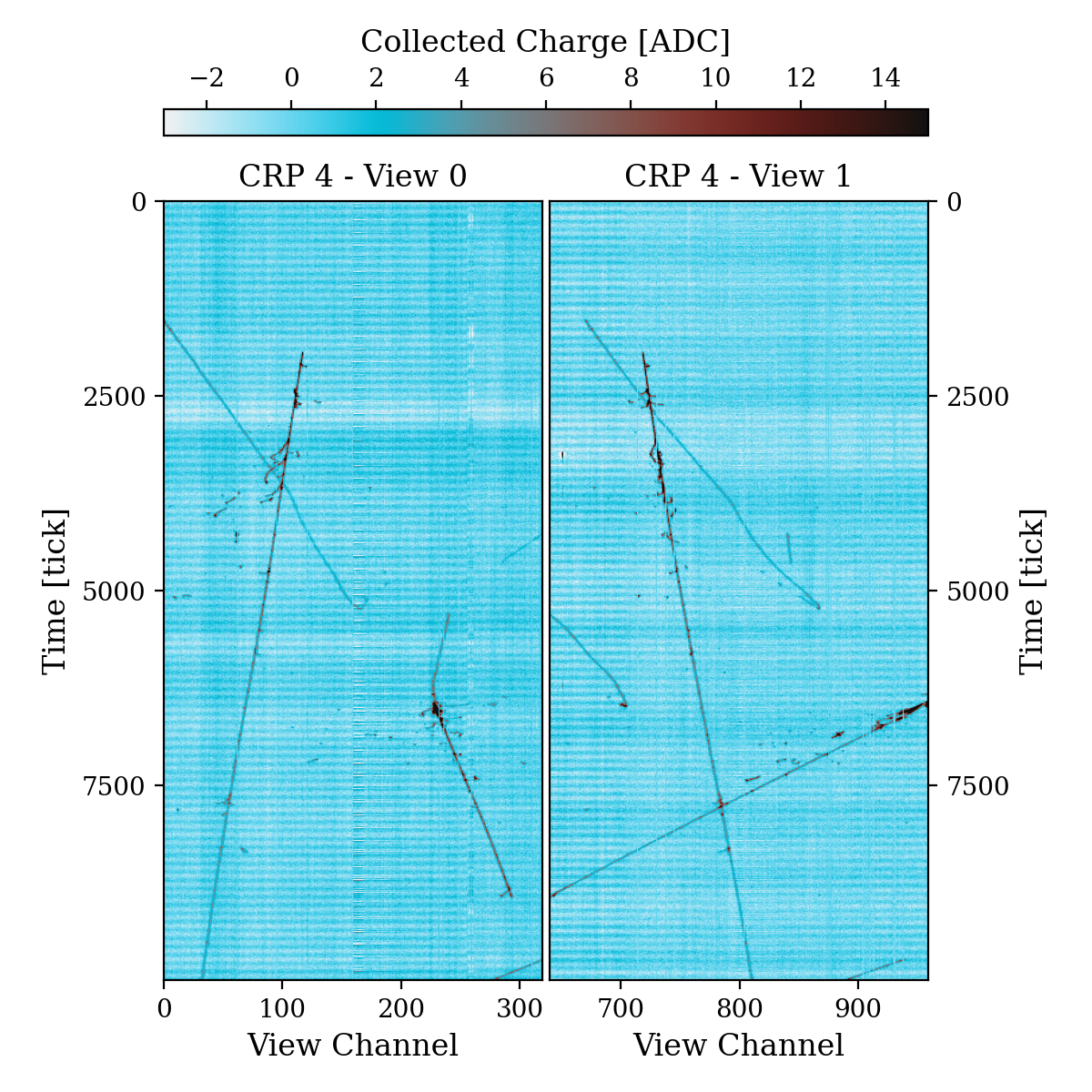}

\caption{Long cosmic track events recorded in \pddp with the full 6\,m drift seen in the two views of CRP4 (only the instrumented channels are shown for clarity). The event on the left contains a single long muon track, while the event on the right has also a stopping muon and two other cosmic tracks. For both events, only the mean pedestal of each channel is subtracted. One time tick corresponds to 0.4~$\mu$s.} 
 \label{fig:ED_6m}
\end{figure}

As discussed in~\ref{sec:fulldrift}, a new extender and HVFT were installed in \pddp in 2021, which allows to reach the nominal drift field of 500~V/cm over 6~m. In January 2022, once the level of LAr impurities was below 100~ppt, corresponding to an electron lifetime of 3~ms, a cosmic run started with this new configuration. However, the CRPs equipped with the amplification system could not be operated during this period. Only CRP4, instrumented with four anodes over an area of 1~m$^2$ without LEM amplification, could collect the ionization charge.
Figure~\ref{fig:ED_6m} shows cosmic-ray tracks for which ionization charge was transported and read out after drifting over 6~m. This demonstrates successful charge transport and readout over the full drift distance in the limited configuration described above. It does not demonstrate full-area operation of the nominal \pddp charge-readout system with LEM amplification. To the best of our knowledge, this is the longest drift distance demonstrated for charge transport and readout in a liquid-noble-gas TPC, exceeding the 1~m achieved in the 4-tonne demonstrator~\cite{311_technical} and 3.5~m achieved in \pdsp~\cite{Abi_2020}. 

The specific detector configuration in which the data was taken, with only 1\,m$^2$ collection surface over 6\,m drift, makes the analyses very challenging for different reasons:
\begin{itemize}
    \item Most recorded tracks cannot be $t_0$-corrected with the charge reconstruction only, {\it i.e.} they enter the TPC in un-instrumented areas. While technically the $t_0$ of all tracks can be retrieved using the PDS system, this analysis has not been much explored for technical reasons -- light and charge data are taken with different DAQ systems, and due to an expected low efficiency of the charge-light matched tracks: in this configuration, the PDS system is detecting many more tracks than what the charge system can record. 
    \item The sample of anode-cathode crossing tracks, which is self $t_0$-corrected, is on one hand statistically limited and on the other hand very challenging to reconstruct as those tracks appear to be nearly vertical in this detector configuration.
\end{itemize}
As a consequence, the collected data with the 6~m-drift configuration were not analyzed in this work.

\section{Photon detection system performance}
\label{sec:photon}

The Photon Detection System (PDS) collected cosmic-ray data for 18 months in stable conditions with all 36 PMTs in operation. The prompt scintillation light (usually referred to as S1 signal) in LAr has two components: S1 fast with a lifetime of $\sim$6\,ns, and S1 slow of $\sim$1.6\,$\mu$s. In addition, the electro-luminescence secondary scintillation light, called S2, is produced in the gas phase of the detector when electrons, extracted from the liquid, are accelerated in the electric fields between the liquid phase and the anode. 
The PDS took data during the full detector operation from June 2019 until November 2020, and dedicated measurements with the cathode at -300\,kV were taken in autumn 2021.

 The ProtoDUNE-DP PDS performance was evaluated, the scintillation light production and propagation processes were analysed, and a comparison of simulation to data was performed, improving the understanding of liquid argon properties~\cite{ProtoDUNE_DP_light}. The following features of the PDS performance show the successful operation of the PDS: 

\begin{itemize}
    \item Signal-to-noise ratio: a signal-to-noise ratio greater than 11 at a gain of $10^{7}$ was obtained, which is important to measure low energy signals at the level of the Single PhotoElectron (SPE). 
    \item The LAr scintillation time profile is described as a sum of three exponential decays. The average value of $\tau_{slow}$ is $1.45\pm0.20$\,$\mu$s, remaining stable during the stable detector operation, which indicates  LAr purity at the ppb level. A third decay component, $\tau_{int}$, is observed and related to delayed emission from the wavelength shifting material~\cite{Segreto, Whittington}.
    \item The complete PDS (36 PMTs) detects about 0.7\,PE/MeV and 2.36\,PE/MeV for CRT-trigger and random-trigger muon tracks, respectively. The first case corresponds to an average track-PMT distance of 5\,m 
    with an average deposited energy of 1.9\,GeV, 
    whereas the second case corresponds to muons crossing the detector at all distances from the PMTs (from close to distant tracks, up to 7\,m away from the PMTs) with an average deposited energy of 0.8\,GeV. 
    \item The electroluminescence light, S2, produced in the gas phase about 7\,m away from the PMTs is observed in all 36 PMTs implying a high efficiency for the PDS. 
    \item Xe doping: ProtoDUNE-DP data demonstrated an improvement in the light detection efficiency and uniformity in large LArTPCs thanks to Xe doping. A low doping level of 5.8\,ppm of Xe doubles the collected light at large distances (3$-$5\,m from the PMTs) even with the presence of 2.4\,ppm of N$_{2}$. However, it must be considered that the reduction observed in the fast signal amplitude could affect the performance of a light-based trigger. Detailed Xe doping studies using ProtoDUNE-DP data can be found in~\cite{SotoOton:2812306} and ProtoDUNE-SP data in~\cite{dunecollaboration2024doping}.
\end{itemize}

In this section, additional and updated studies are presented, as well as results with the cathode at -300\,kV. 
These studies correspond to data taken without electroluminescence signal. 
The relative timing alignment among PMTs is evaluated in section~\ref{ssec:ph:time}. A dedicated estimate of the PEN wavelength-shifting efficiency is carried out in section~\ref{ssec:ph:wls}. Finally, the light yield suppression and the SPE rate as a function of the drift field are studied with the full drift operation of the detector in sections~\ref{ssec:ph:ly} and~\ref{ssec:ph:spe}. 

\subsection{Time alignment among PMT signals}
\label{ssec:ph:time}

The relative timing alignment among PMTs is tested by studying the time alignment of all PMT signals with respect to a reference PMT for the same physics event, $\delta t = t_{0,ref}-t_{0,i}$. A relative timing alignment among PMTs below 100\,ns, together with other requirements, will allow the PDS to group signals from different PMTs produced by the same physics event. As a result, the PDS will be able to provide a trigger for a supernova burst happening within the galaxy~\cite{GallegoPhD2021}, and an event time for proton decay searches~\cite{SotoOton:2812306}.


ProtoDUNE-DP PDS is designed to avoid any source of time misalignment. The average delay introduced by every PMT cable is $496.8\pm0.4$\,ns, with a maximum difference between PMTs of 2\,ns~\cite{Hamamatsu}. Additionally, the PMTs introduce a jitter given by the Transit Time Spread (TTS) of 3\,ns. These two factors are negligible compared to the digitizer sampling frequency of 16\,ns. The topology of the particle track can introduce time differences due to light propagation, as the light arrives earlier at PMTs placed close to the photon production site where the energy is deposited. By selecting a PMT placed at the centre of the matrix as reference, the maximum difference in light propagation distance between PMTs is 3\,m, corresponding to a time difference of $\sim$22\,ns, which is less than 2 samples.

\begin{figure}[ht]
    \centering
    \includegraphics[width=0.7\textwidth]{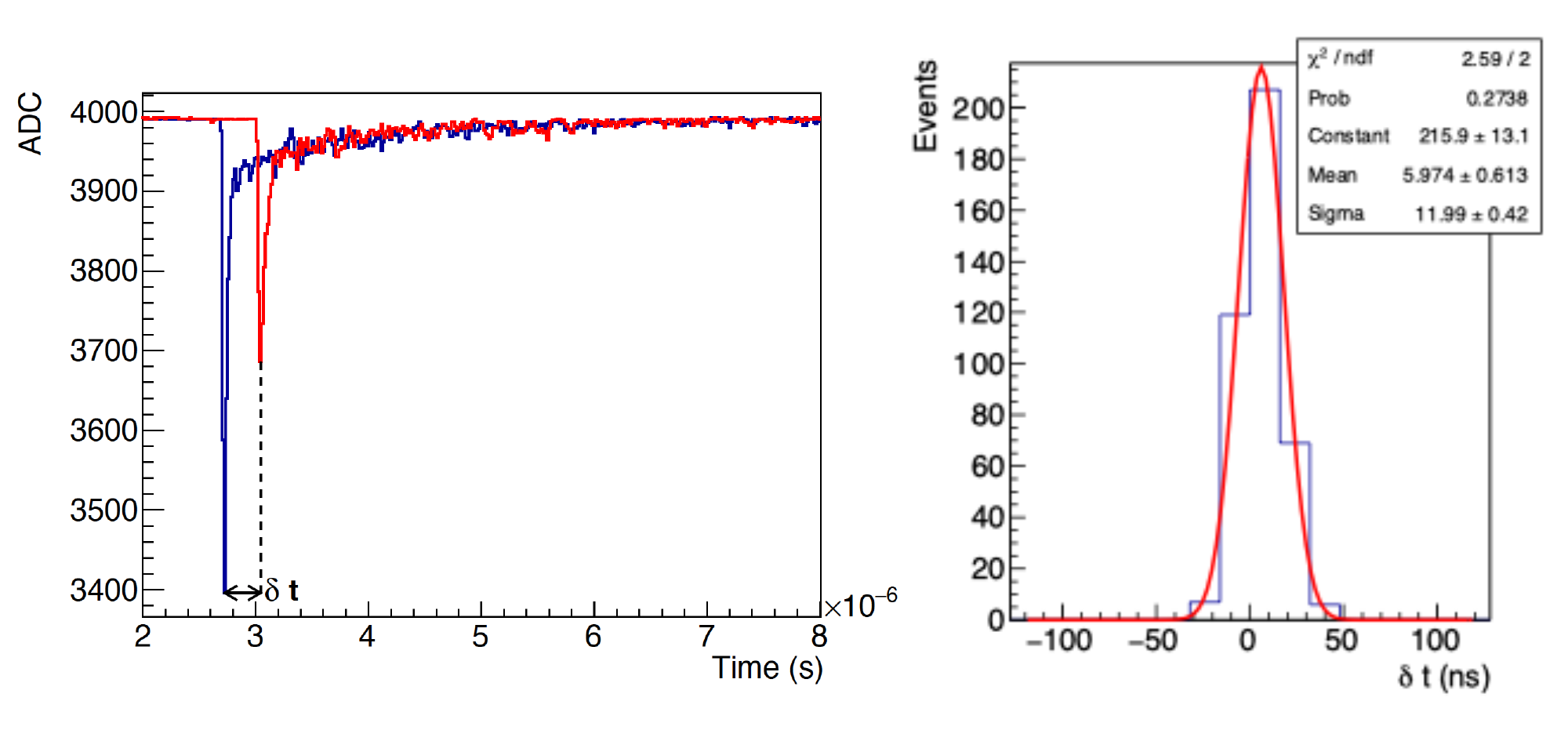}
    \caption{(Left) Example of waveforms from two PMTs for the same event. By comparing the timing of both S1 scintillation light signals we can measure the relative timing alignment between these two PMTs. S1 signal on the second PMT has been manually shifted to illustrate the example. (Right) $\delta$t distribution for one PMT.}
    \label{fig:Timing_DeltaTDistribution}
\end{figure}

The data are triggered on a PMT placed at the centre of the detection plane with a threshold of $\sim$25 PEs. For completeness, the analysis is repeated using data triggered on the CRTs, selecting the same PMT as reference. An algorithm identifies S1 signals within each waveform so that a collection of pulses is obtained in every waveform for each PMT, where each pulse is aimed to represent a S1 signal. 
The time of the pulse with maximum amplitude ($t_{0,i}$) in the waveform is compared to the time of the pulse with maximum amplitude in the triggering channel $t_{0,ref}$ event by event. The delay between these two times is $\delta$t, as shown in figure~\ref{fig:Timing_DeltaTDistribution}.


\begin{figure}[ht]
    \centering
    \includegraphics[width=0.7\textwidth]{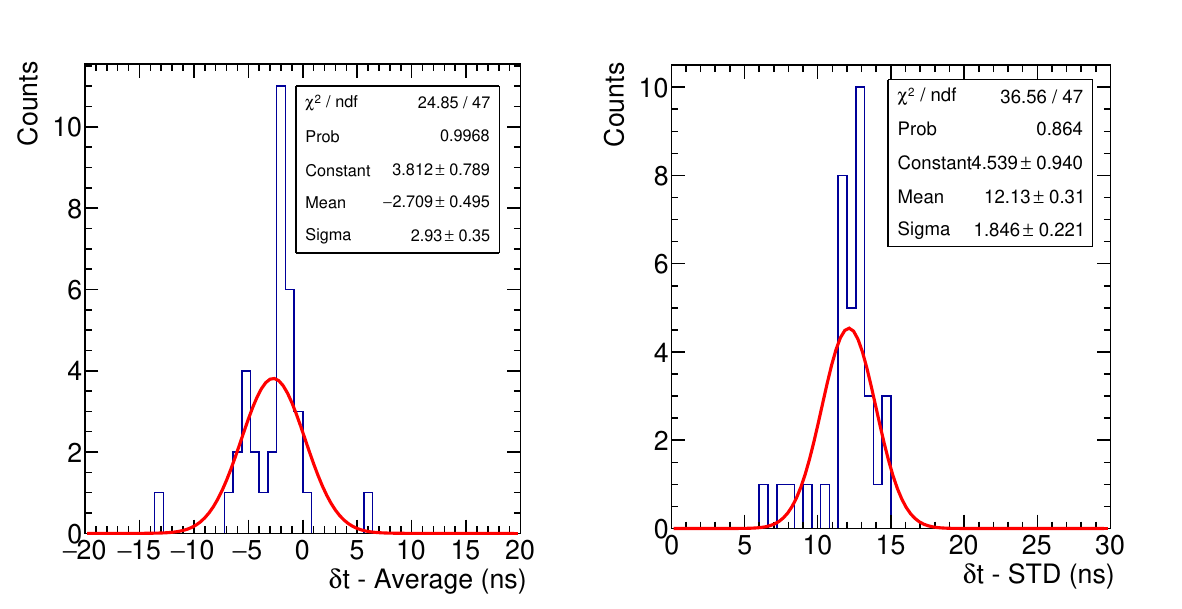}
    \caption{Average (left) and standard deviation (STD, right) of the $\delta$t distribution for all of the PMTs.}
    \label{fig:timing}
\end{figure}

Figure~\ref{fig:timing} shows the distribution of the $\delta$t mean and $\sigma$ for all PMTs with respect to the reference PMT, fitted to a Gaussian function. The mean value of $\delta$t is $-3\pm3$\,ns and $\sigma$ is $12\pm2$\,ns. Both values are below the sampling frequency of the data acquisition system (16\,ns). The results of the data taken triggering on CRT panels are consistent with those stated above, below the 16\,ns sampling frequency of the data acquisition system, showing a mean value of $\delta t = -3\pm4$\,ns and $\sigma = 15\pm3$\,ns.

The relative timing alignment has been shown to be below the sampling frequency of the acquisition system in ProtoDUNE-DP.

\subsection{Measurement of the PEN wavelength-shifting efficiency}
\label{ssec:ph:wls}

 To efficiently detect the LAr scintillation light, wavelength-shifting (WLS) materials are typically placed in front of the photosensors. Tetraphenyl butadiene (TPB) is an organic compound broadly used in particle physics as a WLS~\cite{311_technical,ICARUS_TPB,TPB-Benson2018}. However, TPB is usually deposited on surfaces with a dedicated evaporation system, which is hard to scale to large surfaces as required by large-scale detectors; TPB coatings are also very delicate, as they can be detached~\cite{Burak:2020scu} or affected by VUV light. For these reasons, other easier to handle materials are being explored, such as polyethylene naphthalate (PEN), and alternative techniques such as the use of Xe doped LAr. 

Six PMTs are coated with TPB in \pddp and the other 30 PMTs have a PEN foil on top. PEN, a fluorescent  thermoplastic similar to PET~\cite{PEN-DMary}, is a promising alternative with a photoluminescence spectrum similar to the one of TPB peaking at $\sim$430\,nm, used in ProtoDUNE-DP for the first time in a large scale experiment.


The TPB wavelength shifting efficiency at 127\,nm remains controversial, with reported values going from 50\% up to 200\%~\cite{TPB-Benson2018,Lally_TPB,TPB_Graybill,Burak:2020scu}. We calculate the TPB wavelength shifting efficiency in subsection~\ref{ssec:ph:wls:cal} based on a simple model and laboratory measurements of the ProtoDUNE-DP PMTs.

The PEN wavelength shifting efficiency is not well established yet, as values in the range from 0.4\%-47\% are reported~\cite{Kuzniak,Abraham_PEN_2021,Boulay_PEN_2021} pointing to a dependence on the sample. Also, the measurements are relative to the TPB, whose efficiency is not clear yet.


\subsubsection{TPB efficiency calculation with lab data}
\label{ssec:ph:wls:cal}



To compute the wavelength shifting efficiency of the material itself, a model separating the different contributions to the light signal is proposed in equation~\ref{eq:pen_model1}. In equation~\ref{eq:pen_model1}, NPE is the number of detected photoelectrons, $\gamma$ is the number of VUV photons arriving to the WLS, $\epsilon$ is the wavelength shifting efficiency at 87\,K, defined as the number of visible photons re-emitted per incident VUV photon, $\Delta$ represents the photon transport losses from the WLS (either the coating or the foil) to the photocathode, and it is defined as the number of photons arriving to the PMT photocathode per visible photon emitted by the WLS, and QE is the PMT quantum efficiency.

\begin{equation}
  \text{NPE} = \gamma \ \epsilon \ \Delta \ \text{QE}
  \label{eq:pen_model1}
\end{equation}

An effective PMT photon-detection efficiency (DE) is defined in equation~\ref{eq:pen_de}, as the PEs detected by the PMT per incident VUV photon. 

\begin{equation}
  \text{DE} = \frac{\text{NPE}}{\gamma} =
  \epsilon \ \Delta \ \text{QE}
  \label{eq:pen_de}
\end{equation}

DE was measured for TPB-coated PMTs at room temperature at the University of Pavia for four of the PMTs by comparing the current given by the PMT under test with a reference calibrated photodiode in a dedicated setup. They measured an average value of DE$_{\mathrm{TPB,Lab}} = 0.14 \pm 0.02$~\cite{Burak:2020scu}. This measurement is compatible with the value obtained by ICARUS of $0.12 \pm 0.01$ using the same setup~\cite{Bonesini_2018}.

The QE was measured at room temperature by the manufacturer (Hamamatsu) for three of the PMTs. They measured a value of QE$ = 0.183 \pm 0.013$ at 430\,nm, which is assumed to be constant at cryogenic temperatures~\cite{Bueno_2008,Zhao_2021_PMT_QE_CT} and no variation among PMTs is considered.

As the TPB is directly coated over the PMT glass, it is estimated that 50\% of the photons will reach the photocathode, since the re-emission is isotropic assuming a simple model of $\Delta_{\mathrm{coat}} = 0.5$. As the PEN foil is placed tangentially instead of covering the PMT glass like the TPB, these transport losses are computed using a Monte Carlo simulation~\cite{SotoOton:2812306}. It is found that 24.7\% of the photons re-emitted by the PEN foil would arrive at the PMT photocathode ($\Delta_{\mathrm{foil}} = 0.247$).

Considering Eq.~\ref{eq:pen_de} and the values of DE$_{\mathrm{TPB,Lab}}$, QE and $\Delta_{\mathrm{coat}}$ described above, a value for the TPB wavelength shifting efficiency at room temperature of $\epsilon_{\mathrm{TPB,Lab}} = 1.5\pm0.3$ is obtained. The TPB ionisation could be the reason for having an efficiency greater than one~\cite{Segreto}. In this study, it is assumed that $\epsilon_{\mathrm{TPB,Lab}}$ does not change when going from room to cryogenic temperature.

\subsubsection{Measurement of the PEN efficiency with PMT-trigger muon data}
\label{ssec:ph:wls:pmt}


 The absolute PEN wavelength shifting efficiency ($\epsilon_{\mathrm{PEN}}$) is obtained from the relative measurement $\epsilon_{\mathrm{PEN}}/\epsilon_{\mathrm{TPB}}$, equation~\ref{eq:pen_epsilon_pen}, computed with ProtoDUNE-DP data corrected for the different geometry of the WLSs, and the TPB wavelength shifting efficiency calculated with laboratory data ($\epsilon_{\mathrm{TPB,Lab}}$) in the previous section.

\begin{equation}
  \epsilon_{\mathrm{PEN}} = \frac{\epsilon_{\mathrm{PEN}}}{\epsilon_{\mathrm{TPB}}} \epsilon_{\mathrm{TPB,Lab}}
  \label{eq:pen_epsilon_pen}
\end{equation}

The relative WLS efficiency of both materials (PEN and TPB) at 87\,K is obtained by dividing equation~\ref{eq:pen_model1} for PEN and TPB, as shown in equation~\ref{eq:pen_model2}. NPE is measured directly in the data, while $\gamma$ and $\Delta$ account for the different geometries of the WLSs and are determined using simulations

\begin{equation}
  \frac{\epsilon_{\mathrm{PEN}}}{\epsilon_{\mathrm{TPB}}} =
    \frac{\textit{NPE}_{\mathrm{PEN}} \ \gamma_{\mathrm{coat}} \ \Delta_{\mathrm{coat}}}
       {\textit{NPE}_{\mathrm{TPB}} \ \gamma_{\mathrm{foil}} \ \Delta_{\mathrm{foil}}}
  \label{eq:pen_model2}
\end{equation}


In PMT-trigger data, the track event geometry is not known as data were taken without drift field, and the comparison between PEN-foil and TPB-coated PMTs is done using PEN-TPB PMT pairs placed symmetrically with respect to the trigger PMT. 
Two TPB-coated PMTs placed at the centre of the ProtoDUNE-DP detector are selected as triggering PMTs, and the response of five PMT pairs is compared at different gains (10$^7$, 2$\cdot$10$^7$, 5$\cdot$10$^7$ and 10$^8$). 
The average light collected on the PMTs for the selected events is $\sim$200\,PEs on TPB-coated PMTs, and $\sim$50\,PEs on PEN-foil PMTs. The $NPE_\mathrm{PEN}/NPE_\mathrm{TPB}$ ratio is stable for each pair at different gains. An average $NPE_\mathrm{PEN}/NPE_\mathrm{TPB}$ ratio of $0.25\pm0.03$ is obtained, where the uncertainty is the STD among PMT-pairs, which agrees with the expected error due to the QE variation between PMTs. 
On average, ProtoDUNE-DP TPB-coated PMTs detect four times more photons than PEN-foil PMTs. 

The ratio of photons arriving to the TPB coating over the PEN foil is, on average, $\gamma_\mathrm{coat}/\gamma_\mathrm{foil} = 0.69\pm0.16$, as calculated using the cosmic-muon simulation~\cite{ProtoDUNE_DP_light} where the error is estimated as the STD of the simulation. This factor depends on the selected sample, and the simulation of a different event topology would vary its value. 
The fact that the PEN foil receives 30\% more photons than the TPB coating can be explained as its two faces are exposed to LAr while the TPB only has one.

From equation~\ref{eq:pen_epsilon_pen}, and considering the TPB wavelength shifting efficiency at room temperature calculated in section~\ref{ssec:ph:wls:cal} ($\epsilon_{\mathrm{TPB,Lab}}=1.5\pm0.3$), a relative $\epsilon_{\mathrm{PEN}}/\epsilon_{\mathrm{TPB}} = 0.35\pm0.09$ and an absolute PEN wavelength shifting efficiency of $\epsilon_{\mathrm{PEN}} = 0.50\pm0.17$ are derived. 

\subsubsection{Measurement of the TPB and PEN wavelength shifting efficiency with CRT-trigger muon data and simulation}
\label{ssec:ph:wls:crt}

In a different approach, the PEN and TPB WLS efficiencies ($\epsilon_{\mathrm{PEN}}$ and $\epsilon_{\mathrm{TPB}}$) are obtained directly by comparing muon data from ProtoDUNE-DP with simulations according to equation~\ref{eq:pen_de}. The NPE are obtained from CRT-trigger muon data, and the number of photons arriving to each PMT is computed in the simulation using the position of the muon tracks provided by the CRT trigger system. 

Five datasets of $\sim5000$ events each ($\sim$15\,h data taking) with the PMTs operating at a gain of at least $2\times10^{7}$ are selected. A data selection is applied to avoid noisy events and the pile-up of signals. Additionally, a sample of events with the same geometry as the CRT trigger events are simulated in LArSoft~\cite{SotoOton:2812306}, in order to obtain the number of VUV photons arriving to the WLS of each PMT. An estimate of the number of photons is obtained from the mean of a Gaussian fitted to the peak of each distribution. This value is a better observable to compare both distributions than the mean, since the latter is biased by high energy vertical showers that trigger the CRTs and are not included in the simulation~\cite{GallegoPhD2021}. The value is very stable when comparing different data runs, with a variation below 5\%.


The ratio between the NPEs and the number of VUV photons arriving to each PMT provides the effective detection efficiency (DE) for each PMT, as described in equation~\ref{eq:pen_de}. An average of DE$_{\mathrm{PEN}}=0.018\pm0.002$ and DE$_{\mathrm{TPB}}=0.11\pm0.02$ is obtained, where the uncertainty is the standard deviation for all the PMTs. 
PEN-PMTs detect $\sim$50\,PEs in the data from $\sim$3000 photons arriving to the foil in the simulation whereas TPB-PMTs detect $\sim$200\,PEs in the data from $\sim$2000 photons arriving to the coating in the simulation.

Finally, by correcting the photon-transport losses from the WLS to the photocathode ($\Delta$) and QE, the absolute wavelength shifting efficiency is obtained ($\epsilon_{\mathrm{PEN}}$ and $\epsilon_{\mathrm{TPB}}$), as described in equation~\ref{eq:pen_model2}. The average of the distributions is $\epsilon_{\mathrm{PEN}}=0.39\pm0.05$ and $\epsilon_{\mathrm{TPB}}=1.2\pm0.2$, where the uncertainty is the standard deviation of the distributions, suffering from QE variations PMT by PMT. The stability of $\epsilon_{\mathrm{TPB}}$ and $\epsilon_{\mathrm{PEN}}$ is studied for different runs and the variations are compatible with the errors. 

The values obtained for these absolute efficiencies ($\epsilon_{\mathrm{PEN}}$ and $\epsilon_{\mathrm{TPB}}$) are in agreement with the values found in the literature. However, they depend highly on the simulation parameters. For example, a hypothetical over/under estimation of the reflected light, would reduce/increase this wavelength shifting efficiency values accordingly. Therefore, a better estimator would be the relative value to reduce any systematic bias introduced by the simulation. The resulting relative value is $\epsilon_{\mathrm{PEN}}/\epsilon_{\mathrm{TPB}} = 0.32\pm0.07$, which is in agreement with the value computed in the previous section and with the values reported in the literature for a similar sample~\cite{Abraham_PEN_2021}.

\subsubsection{Discussion of the results}
\label{ssec:ph:wls:res}

In this work, the performance of the PEN foils as WLS is measured for the first time in a large LArTPC detector, ProtoDUNE-DP. The performance of PMTs covered with PEN foils on the top is compared with PMTs with TPB directly coated over the glass. The results, shown in Table \ref{tab:pen_results} as the weighted average, indicate that PEN-foil PMTs detect only 25\% of the 127\,nm photons detected by TPB-coated PMTs at 87\,K.

    \begin{table}[h]
        \centering
        \begin{tabular}{|c| c |c |c|}
        \hline
           Analysis & Value &  Result & Comment\\
    \hline
        Lab measurement & $\epsilon_{\mathrm{TPB}}$  & $1.5\pm0.3$ & \\
    \hline
        \multirow{2}{*}{PMT-trigger data} & $\epsilon_{\mathrm{PEN}}/\epsilon_{\mathrm{TPB}}$ &  $0.35\pm0.09$ & \\
        & $\epsilon_{\mathrm{PEN}}$ & $0.50\pm 0.17$  & Assuming lab meas. \\

    \hline
        \multirow{3}{*}{CRT-trigger data} 
            & $\epsilon_{\mathrm{PEN}}/\epsilon_{\mathrm{TPB}}$ & $0.32\pm0.07$ & \\
            & $\epsilon_{\mathrm{PEN}}$ & $0.39\pm0.05$  & From simulation\\
            & $\epsilon_{\mathrm{TPB}}$ & $1.2\pm0.2$  & From simulation\\
    \hline
       Average & $\epsilon_{\mathrm{PEN}}/\epsilon_{\mathrm{TPB}}$ & $0.33\pm0.05$&\\
        \hline
        \end{tabular}
        \caption{Summary of WLS efficiencies obtained in ProtoDUNE Dual-Phase. The measurement of $\epsilon_{\mathrm{PEN}}$ using PMT-trigger data assumes the measurement of $\epsilon_{\mathrm{TPB}}$ in the laboratory. The measurement of $\epsilon_{\mathrm{PEN}}$ and $\epsilon_{\mathrm{TPB}}$ using CRT-trigger data depends directly on the simulation. }
        \label{tab:pen_results}
    \end{table}

Taking into account geometrical differences due to the position of the foil and the coating, a relative PEN/TPB wavelength shifting efficiency of $\epsilon_{\mathrm{PEN}}/\epsilon_{\mathrm{TPB}}=0.33 \pm 0.05$ is obtained, meaning that TPB would emit three times more photons than PEN for the same amount of incident VUV photons. This value is in agreement with some measurements found in the literature~\cite{Abraham_PEN_2021}. 


Using CRT triggered data and a simulation to compute the number of photons arriving to the wavelength shifters, the absolute wavelength shifting efficiency of $\epsilon_{\mathrm{PEN}} = 0.39 \pm 0.05$ and $\epsilon_{\mathrm{TPB}} = 1.2 \pm 0.2$ are obtained.

A system based on TPB as WLS is recommended to maximize the detection efficiency. Nevertheless, PEN can be taken into account as an alternative when the detection efficiency is not critical compared to the benefits of an easy installation.

\subsection{Light yield suppression with the drift field}
\label{ssec:ph:ly}

The presence of an electric field suppresses the recombination of electrons and ions. As a result, the light production is reduced with the increasing drift field. Comparing the detected light with a drift field of 0.5\,kV/cm (provided by a cathode voltage of -300\,kV) and without drift field, the amount of photons produced by the electron-ion recombination can be estimated. For this, CRT-trigger events taken at full drift field during autumn 2021 (see Sec.~\ref{sec:operation}) have been analyzed. Gain calibration is performed by measuring the SPE on each sample, to suppress any possible gain variation. The selection of the PMTs is the same if both samples, and since it is a ratio, any PMT dependence is also suppressed. The data were taken the same day, during a dedicated scan, to avoid variation in the detector conditions.

\begin{figure}[ht]
    \centering
    \includegraphics[width=0.55\textwidth]{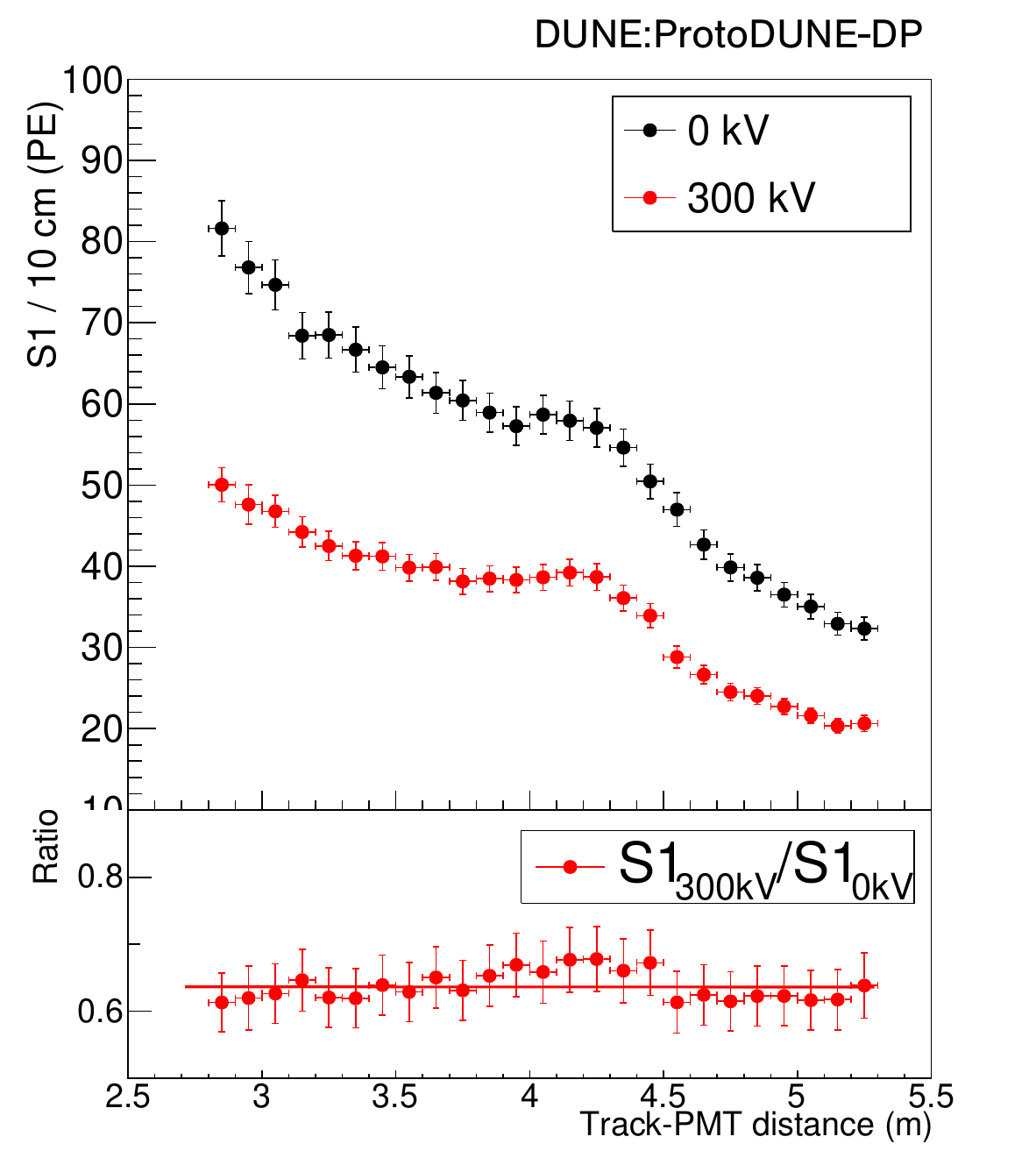}
    \caption[S1 charge as a function of the track-PMT distance for CRT trigger muons at the nominal cathode voltage (-300\,kV) and at 0\,kV.]{Top: S1 charge of PEN PMTs as a function of the track-PMT distance for CRT trigger muons at the nominal cathode voltage (-300\,kV in red) and at 0\,kV (in black). Bottom: Ratio of the -300\,kV over 0\,kV curves. }
    \label{fig:per_crt_driftfield}
\end{figure}

Figure~\ref{fig:per_crt_driftfield} shows the dependence of the detected light ({\it i.e.} S1) for CRT-trigger events with the track-PMT distance with and without drift field for all PEN PMTs. The track-PMT distance is computed as the closest distance from the PMT position to the CRT-trigger track position which is provided by the CRT panels. The ratio of the S1 at the nominal cathode voltage (-300\,kV) and at 0\,kV as a function of the track-PMT distance is fitted to a constant to obtain the average ratio of  $0.62\pm0.01$, which means that at least 38\% of the scintillation light detected in the absence of a drift field comes from electron-ion recombination.  The error is the statistical uncertainty from the fit. Recombination could be slightly higher since recombination is not completely suppressed at the nominal drift. field, and the PMTs would detect some light produced outside the active volume where there is no drift field. 
The correlation of both curves is lost above 5.3\,m, which corresponds with PMTs placed far away from the track, and the signal is dominated by background signals. To avoid this distortion, the fit is performed up to 5.3\,m. The flat region around 4\,m is due to the uneven layout of PMTs on the floor.

\begin{figure}[ht]
    \centering
    \includegraphics[width=0.65\textwidth]{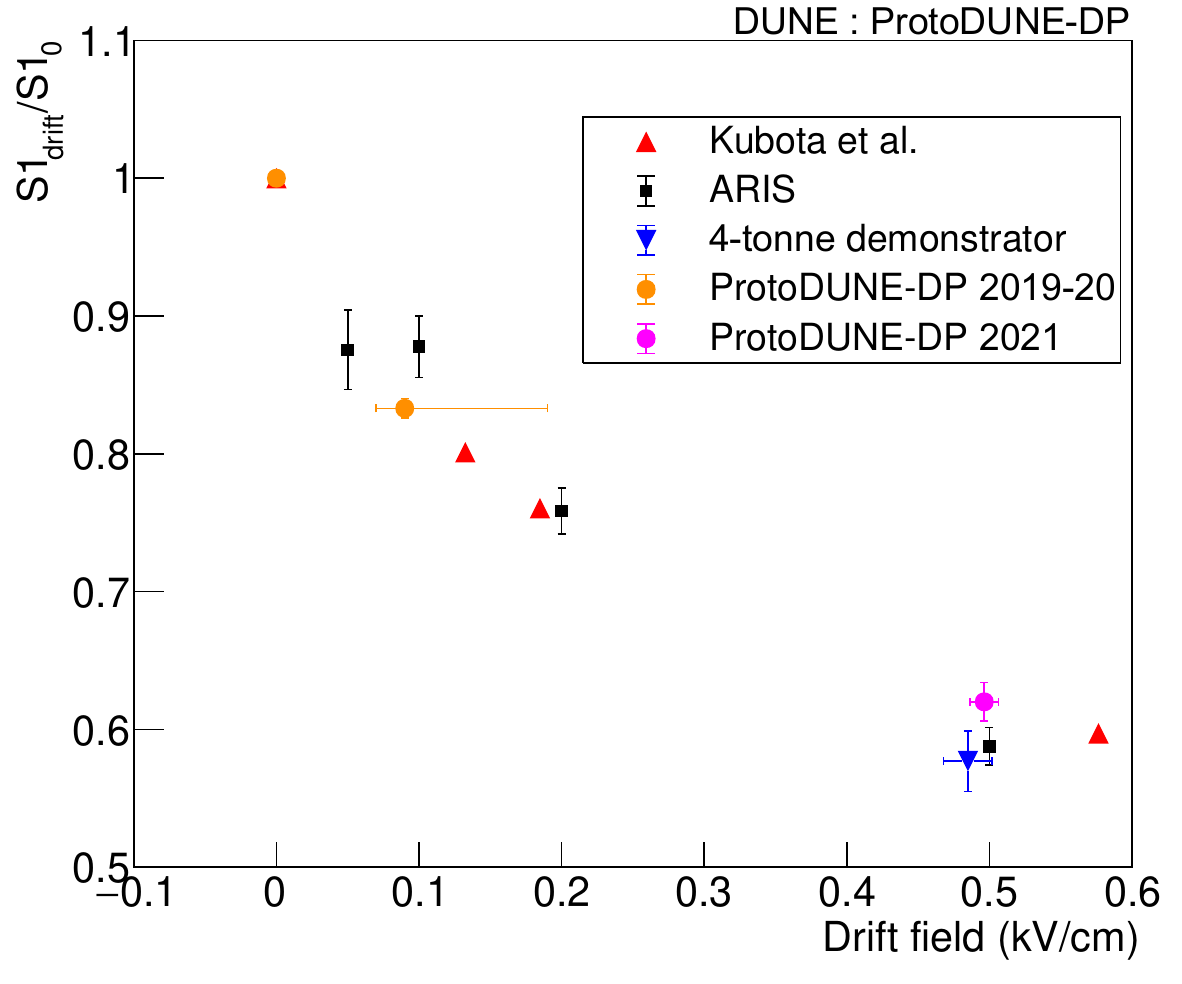}
    \caption[Comparison of the measured value for S1$_{300\text{kV}}$/S1$_{0\text{kV}}$ with other measurements in the literature]{Comparison of the measured value for S1$_{300\text{kV}}$/S1$_{0\text{kV}}$ with the previous ProtoDUNE-DP result~\cite{ProtoDUNE_DP_light} in orange and other measurements in the literature:~\cite{Kubota} in red,~\cite{Aris} in black,~\cite{311_light} in blue.}
    \label{fig:per_recomb}
\end{figure}

The reduction of the S1 signal as a function of the drift field is shown in figure~\ref{fig:per_recomb} for this result, the value obtained during the operation of ProtoDUNE-DP in 2019-2020~\cite{ProtoDUNE_DP_light}, and with other values in the literature obtained for electrons~\cite{Kubota,Aris} and muons~\cite{311_light}. This measurement is in good agreement with the values reported in the literature.

\subsection{Dependence of the Single Photoelectron rate on the drift field}
\label{ssec:ph:spe}

The effect of the presence of the drift field in the Single Photoelectron (SPE) rate is characterised by considering data taken during 2021 when the cathode was operated at its nominal voltage of -300\,kV. Data taken using a random trigger was used to avoid any threshold bias. PMT gain is corrected on each PMT by identifying SPEs on the same sample, to suppress any gain variations. A hit finding algorithm is applied that identifies and isolates any pulse in the waveform of an amplitude larger than half PE. A clustering algorithm is applied to group the signals close in time. Pile up during clustering is avoided by considering the amplitude of the signals (if a new hit has an amplitude 3 times larger than the last cluster hit, a new cluster is created). All SPEs in a cluster with amplitude below 2PEs are considered for this rate measurement.

Table~\ref{tab:protodune_spe_rate} summarizes the SPE rate results for PEN and TPB PMTs with the cathode operated at 0\,kV and at -300\,kV. While the model of Ar$_{2}^{+}$-ion recombination proposed in~\cite{Luo:2020itx} predicts a smaller rate with the drift field, the SPE rate increases a factor of 2.7 on PEN PMTs and 1.9 on TPB PMTs when activating the drift field.

\begin{table}[!ht]
\begin{center}
\begin{tabular}{|c | c |c |}
\hline
      & SPE rate at 0\,kV (kHz) & SPE rate at -300\,kV (kHz) \\
        \hline
    PEN    &  $106\pm15(\delta_{\mathrm{PMTs}})\pm20(\delta_{\mathrm{Runs}})$&  $284\pm55(\delta_{\mathrm{PMTs}})\pm71(\delta_{\mathrm{Runs}})$   \\
    TPB    &  $306\pm46(\delta_{\mathrm{PMTs}})\pm50(\delta_{\mathrm{Runs}})$&  $581\pm60(\delta_{\mathrm{PMTs}})\pm95(\delta_{\mathrm{Runs}})$ \\
            \hline
\end{tabular}
\caption[SPE rate for PEN and TPB PMTs with the cathode at 0k\,V and 300\,kV.]{SPE rate for PEN and TPB PMTs with the cathode at 0k\,V and -300\,kV. $\delta_{PMTs}$ is the STD among PMTs (average of the different runs), and $\delta_{Runs}$ is the STD among runs (average of PMTs).  }
\label{tab:protodune_spe_rate}
\end{center}
\end{table}

A compatible SPE rate in absence of drift field  was measured in~\cite{ProtoDUNE_DP_light} during the operation of ProtoDUNE-DP in 2019-2020. There are several sources that can contribute to the SPE rate. First, the radioactivity of natural argon. Argon isotopes such as $^{39}$Ar decay depositing a small amount of energy of the order of keV and producing small scintillation light signals that contribute to the SPE background. Second, the cosmic particles crossing far away from the PMTs or depositing a small amount of energy can also contribute to the SPE rate detected by the PMTs. The PMT dark current also contributes to the SPE rate, with an average value of 1.7\,kHz per PMT at cryogenic temperature~\cite{protoDUNEPMTs}. 

\begin{figure}[ht]
    \centering
    \includegraphics[width=0.99\textwidth]{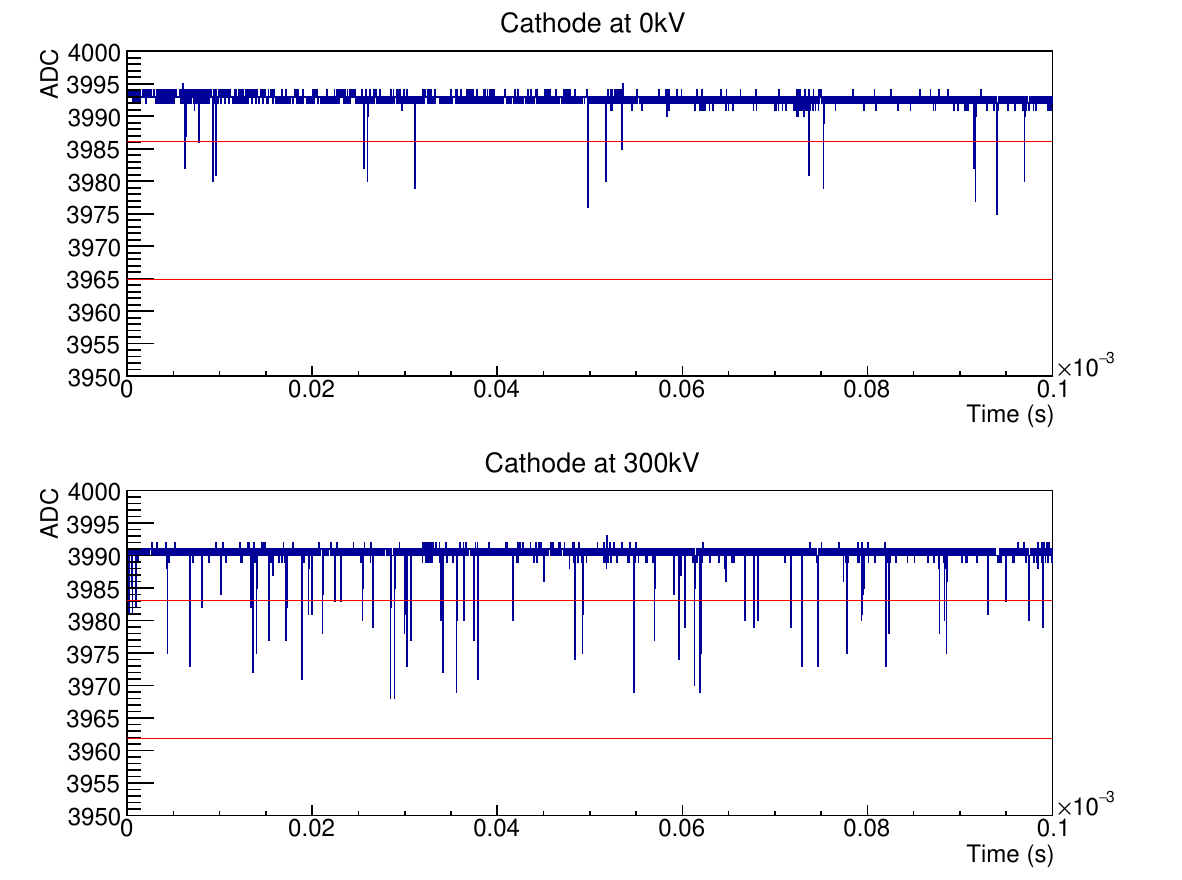}
    \caption[PMT waveforms with and without drift field.]{Example of two waveforms taken with random trigger in a TPB PMT. Top panel shows a waveform taken with the cathode off, bottom panel shows a waveform taken in the same PMT with the cathode at its nominal voltage of -300\,kV. Red lines show the range for SPE identification.}
    \label{fig:pdune_waveform_cathodeonoff}
\end{figure}

The increase in the SPE rate when activating the cathode voltage is clearly visible in figure~\ref{fig:pdune_waveform_cathodeonoff}. ProtoDUNE Single-Phase reported a drop of the SPE rate from 270\,kHz to 180\,kHz when activating the drift field~\cite{Dante_Lidine2019}. MicroBooNE also reported a drop in the SPE rate from $\sim$340 to $\sim$240\,kHz when activating the drift field~\cite{Dante_muboone}. The different result in ProtoDUNE-DP is not understood, but could be explained by the special operation of the detector as data were taken during a HV test, without extraction field. In this situation, the ionisation electrons could recombine at the top of the active volume as they are not collected, producing more photons that are detected as an increase of the SPE rate. 

\section{Summary and conclusions}
\label{sec:conclusion}

This paper presents an overview of the operational experience and performance of ProtoDUNE-DP, the largest Dual-Phase liquid argon TPC ever operated, conceived as a large-scale prototype for a DUNE Far Detector module.

The principal phase of the detector operation lasted approximately one year from the summer 2019 to the summer 2020. Much of this operation time was dedicated to understanding the stability of the Charge Readout Planes (CRPs) and quantifying the sparking rates of the Large Electron Multipliers (LEM) in different HV configurations. 
It was observed that the number of LEMs unable to maintain the nominal operating HV 
due to a continuous rate of discharges increased with time. 

The data acquisition system operated smoothly since the beginning of operation in 2019 until a new phase of data taking in 2022.  Trigger rates ranged  between 10 and 50\,Hz and no data compression was applied.  A total of 1.9 million events were collected, resulting in a total data volume of about 260\,TB. The online processing facility was systematically running track reconstruction in real time on the acquired data, ensuring data quality monitoring. Raw data were transferred from CERN to FNAL and were incorporated into the DUNE data catalogue. Reconstruction was performed on a subsample of these raw data using the official DUNE software framework, LArSoft, and with LARDON. 

The performance of the Dual-Phase charge readout was studied with cosmic muon tracks. 
The performance of the anode and the reconstruction could be assessed by looking at the charge sharing asymmetries between the two orthogonal views. The electron lifetime measured in different data-taking periods was in good agreement with the results of the short purity monitors, but not the long one that reported consistently much higher LAr purity. Many factors influencing the effective gain were studied with the \pddp data: the extraction field strength, the LEM thickness, the evolution with time through the charging-up effect and the voltages applied on the LEM faces. The results obtained are in agreement with the literature. The transparency of the CRP was estimated from data taken at different \dvlem to be $\mathcal{T}=0.252$, a value slightly lower than the one, $\mathcal{T}=0.29$, predicted by simulations. The highest effective gain measured was $G_{\mathrm{eff}} = 6.8$ for $\dvlem = \SI{3.2}{\kilo\volt}$. The LEM gain was found to progressively decrease with time. While this was expected for such devices because of the charging of the dielectric material in the holes, it was not possible to follow precisely and systematically the time-evolution of the phenomenon due to the complicated operating conditions. In the last data-taking period, the highest effective gain measured was $G_{\mathrm{eff}} = 1.5$ for $\dvlem = \SI{3.1}{\kilo\volt}$. 



After the failure of the initial HV extender, a new one with an improved design was operated at \SI{-300}{\kilo\volt} and allowed to collect a sample of cosmic ray data with the full \SI{6}{\meter} depth of the drift volume, the longest drift distance for the charge in a liquid argon detector. This demonstrated charge transport and readout over the full 6 m drift using the detector configuration available at the time.

The photon detection system collected cosmic-ray data for 18 months in stable conditions, with all 36 PMTs in operation. The calibrations performed during the detector operation allowed to monitor the PMT gain, which showed a stable behaviour. The relative timing alignment among PMTs is below \SI{16}{\nano\second}, the sampling frequency of the acquisition system, far surpassing the requirement and validating the PDS design for the DUNE Dual-Phase  Far Detector module. A dedicated estimate of the PEN wavelength-shifting efficiency is carried out, yielding a relative PEN/TPB wavelength shifting efficiency of $\epsilon_{\mathrm{PEN}}/\epsilon_{\mathrm{TPB}}=0.33 \pm 0.05$. The reduction of the S1 signal with a drift field of \SI{0.5}{\kilo\volt/\cm} (provided by a cathode voltage of \SI{-300}{\kilo\volt}) is in agreement with the values reported in the literature. Additionally, the effect of the drift field in the SPE rate was characterised.

The long-term operation stability of the CRPs in ProtoDUNE-DP was hindered by the turbulence of the liquid surface. With that and the observed rapid LEM deterioration, the scaling of this Dual-Phase readout technology to the dimensions needed for the DUNE Far Detector modules is currently challenging.  
Nonetheless a number of important milestones for the development of the liquid argon TPCs as large neutrino detectors were attained.
In particular, the operational experience from ProtoDUNE-DP informed the subsequent development of the Vertical Drift (VD) technology~\cite{VD_TDR}.

This new LArTPC design was developed by expanding the idea of reading the charge with the PCB-based electrodes to a concept of perforated anodes that can be operated directly in the liquid. This allows for a considerable simplification of the Dual-Phase CRP by removing the extraction grid and LEMs, which are very sensitive to environmental aspects in the cryostat (even if in principle the insulation and cleanliness could be improved), and  eliminates the impact from any instability of the liquid surface. The elements responsible for several of the principal operational limitations discussed in this paper are thus removed.
The signal of the drifting charge is recorded when passing through a stack of perforated anode PCBs thereby inducing signals on the electrode strips printed on top and bottom faces of each board. A desired number of views can be thus achieved by stacking multiple anodes in a CRP structure. 

This new detector concept 
allows placing a single cathode plane at higher voltages than in the horizontal drift single phase LArTPC technology and
maximizing the active volume achievable in the DUNE cryostats.
The VD LArTPC technology was actively developed over the past few years and has been adopted in DUNE for the first far detector module (VD module). 
The maximum drift distance for each cell is similar to that of \pddp, therefore the successful validation of the \SI{-300}{\kilo\volt} delivery system and operation of the detector at this HV in ProtoDUNE-DP fulfilled important milestones for the future of the DUNE VD detector module. ProtoDUNE-VD has now been constructed in the same cryostat and has been collecting data since 2025.

\bibliographystyle{JHEP}
\bibliography{sections/bibliography}
\end{document}